%% file: thesismaster.tex
\documentclass[12pt,letterpaper]{report}
\pdfoutput=1
\usepackage{graphicx}

\newcommand{\thesistitle}{On the Dynamics of Glassy Systems}
\newcommand{\thesisauthor}{Le Yan}
\newcommand{\thesisadvisor}{Matthieu Wyart}
\newcommand{\graddate}{September 2015}

\input{definitions}

\begin{document}

%% Produces a test "layout" page, for "debugging" purposes only.
%% Comment out for final version.
%\layout % requires package layout (see above, on this same file)

    %%%%%%Preliminary pages
    %%%%%%%%%%%%%%%%%%%%%%%%%%%%%%%%%%%%%%%%%%%%%%%%%

    %% Sets page numbering to "roman style" i, ii, iii, iv, etc:
    \pagenumbering{roman}

    %%%%%% Title page %%%%%%%%%%%
    \include{titlepage/titlepage}
    \newpage
    %%%%%% Copyright %%%%%%%%%%%
    \input{Copyright}
    \doublespacing
    \newpage
    %%%%%%%%%%%%% Blank page %%%%%%%%%%%%%%%%%%
    \thispagestyle{empty}
    \vspace*{0in}
    \newpage
    %%%%%% Copyright %%%%%%%%%%%
    \addcontentsline{toc}{chapter}{Dedication}
    \include{dedication/dedication}
    \newpage
    %%%%%% Acknowledgements %%%%%%%%%%%
    \addcontentsline{toc}{chapter}{Acknowledgements}
    \include{acknowledgements/acknowledgements}
    \newpage
    \addcontentsline{toc}{chapter}{Preface}%
    \include{preface/preface}

    \newpage
    %%%%%% Abstract %%%%%%%%%%%
    \addcontentsline{toc}{chapter}{Abstract}%
    \include{abstract/abstract}

    \tableofcontents
    \newpage
    \listoffigures
    %\newpage
    %\listoftables
    \newpage
    \addcontentsline{toc}{chapter}{List of Appendices}%
    \listofappendices
    \newpage
 
    %%%%%%The main pages %%%%%%%%%%%
    %%%%%%%%%%%%%%%%%%%%%%%%%%%%%%%%%%%%%%%%%%%%%%%%%%%%%%%%
    % Each chapter should be a separate file
    \pagenumbering{arabic}
    \include{introduction/introduction}

    \include{thermodyn/thermodyn}
    \include{rigidity/rigidity}

    \include{adaptive/adaptive}
    \include{erosion/erosion}
    \include{marginal/marginal}

    \include{future/future}

    \include{conclusion/conclusion}
    \begin{appendices}
    \include{appendices/appd_thermodyn}

    \include{appendices/appd_rigidity}
    \include{appendices/appd_adaptive}
    \include{appendices/appd_marginal}
    \end{appendices}

    %%%% Input bibliography file %%%%%%%%%%%%%%%
    \addcontentsline{toc}{chapter}{Bibliography}
    \include{bibliography/bibliography}
   
\end{document}

%% file: definitions.tex
\usepackage[nottoc]{tocbibind}
\usepackage[toc,page]{appendix}
\usepackage[backend=bibtex,natbib=true,style=numeric-comp,sorting=none]{biblatex}
\usepackage{geometry}               % See geometry.pdf to learn the
\usepackage{rotating}

\usepackage{caption}
\usepackage{color}
\usepackage{subcaption}  % Emma added it
\usepackage{amsmath,textcomp}
\usepackage{amssymb}
\usepackage{amsthm} %Jan added
\usepackage{multirow} %Jan added

\usepackage{bm}

\usepackage{latexsym}
\usepackage{amsfonts}
\usepackage{hhline}
\usepackage[]{units}

\usepackage{url}
\usepackage{indentfirst}
\usepackage{suffix}

\usepackage{parskip}
\usepackage{setspace}
\usepackage{hyperref}
\newcommand{\be}{\begin{equation}}
\newcommand{\ee}{\end{equation}}
\newcommand{\bea}{\begin{eqnarray}}
\newcommand{\eea}{\end{eqnarray}}
\newcommand{\bml}{\begin{multline}}
\newcommand{\eml}{\end{multline}}

\newcommand{\real}[1]{\Re\left\{ #1 \right\}}

\newcommand{\tr}{\mathop{\mathrm{tr}}}

\newcommand{\mh}{\mathcal{H}}
\newcommand{\mf}{\mathcal{F}}
\newcommand{\md}{\mathcal{D}}
\newcommand{\mg}{\mathcal{G}}
\newcommand{\mk}{\mathcal{K}}
\newcommand{\ms}{\mathcal{S}}
\newcommand{\mt}{\mathcal{T}}
\newcommand{\mi}{\mathcal{I}}

\newcommand{\mpp}{\mathcal{P}}
\newcommand{\mm}{\mathcal{M}}

\newcommand{\mz}{\mathcal{Z}}

\newcommand{\rd}{{\rm d}}

\newcommand{\dH}{\Delta\mathcal{H}}
\newcommand{\mjs}{\langle Jss\rangle_{m'}}
\newcommand{\ml}{\langle\lambda\rangle_{m'}}

\newcommand{\lp}{\left(}
\newcommand{\rp}{\right)}
\newcommand{\lm}{\lambda_{m'}}
\newcommand{\rw}{\rm w}

\newcommand{\vecr}{\left(\vec{r}\right)}

\newcommand\vectorpotential{\vec{A}\left(\vec{r},t\right)}
\WithSuffix\newcommand\vectorpotential*{\vec{A}^\ast\left(\vec{r},t\right)}
\newcommand{\electricfield}{\vec{E}\left(\vec{r},t\right)}
\WithSuffix\newcommand\electricfield*{\vec{E}^\ast\left(\vec{r},t\right)}
\newcommand{\magneticfield}{\vec{H}(\vec{r},t)}
\WithSuffix\newcommand\magneticfield*{\vec{H}^\ast\left(\vec{r},t\right)}

\newcommand{\pol}{\varepsilon}
\newcommand{\polj}{\pol_j}
\newcommand\polarization{\hat{\pol}\vecr}
\WithSuffix\newcommand\polarization*{\hat{\pol}^\ast\vecr}
\newcommand{\polarizationj}{\polj\vecr}
\WithSuffix\newcommand\polarizationj*{\polj^\ast\vecr}

\newcommand{\mytilde}{\raise.17ex\hbox{$\scriptstyle\mathtt{\sim}$}}

\theoremstyle{definition}

\newcommand\listappendixname{List of Appendices}
\newcommand\appcaption[1]{%
  \addcontentsline{app}{section}{#1}}
\makeatletter
\newcommand\listofappendices{%
  \chapter*{\listappendixname}\@starttoc{app}}
\makeatother

\newcommand{\nocontentsline}[3]{}
\newcommand{\tocless}[2]{\bgroup\let\addcontentsline=\nocontentsline#1{#2}\egroup}

\DeclareRobustCommand{\gobblesix}[9]{}
\newcommand*{\SkipTocEntry}{\addtocontents{toc}{\gobblesix}}

%% file: titlepage/titlepage.tex
\thispagestyle{empty}
\begin{center}
        
        {\large\textbf{\thesistitle}}
        
        \vspace{0.7in}
        by
        
        \vspace{0.7in}
        
        \text{\thesisauthor}
        
        \vspace{0.7in}
        
        \begin{doublespace}
           A dissertation submitted in partial fulfillment\\
           of the requirements for the degree of\\
           Doctor of Philosophy\\
           Department of Physics\\
           New York University\\
           \graddate
        \end{doublespace}
\end{center}
\vspace{.7in}
\noindent\makebox[\textwidth]{\hfill\makebox[2.5in]{\hrulefill}}\\
\makebox[\textwidth]{\hfill\makebox[2.5in]{\hfill\thesisadvisor\hfill}}

%% file: Copyright.tex
%Copyright

\thispagestyle{empty}
\hbox{\ }

\vfill
\renewcommand{\baselinestretch}{1}
\small\normalsize

\vspace{-.65in}

\begin{center}
%\large
\normalsize{\copyright \hbox{ }
%Copyright by\\
Le Yan %Type your name as it appears in University records
\\
All Rights Reserved, 2015}
\end{center}

\vfill

%% file: dedication/dedication.tex
\chapter*{Dedication}
To my mother.%, who gives me life.

%To my father, who brings me thoughts.

%To all my friends, who make what I am.

%% file: acknowledgements/acknowledgements.tex
\chapter*{Acknowledgements}
% I am thankful to ... especially ... and last but not least ...
% Matthieu Wyart, Markus Muller, Alexander Grosberg, David Pine, Paul Chaikin, Daniel Stein, Grier, Jasna, Marc, Alexandra, PIs at CSMR
First of all, I would like to thank my advisor, Matthieu Wyart, for his mentoring and guidance in science and in career. His deep insight into the physical world always enlightens me and leads to the right way to solve the scientific problems. %and enthusiasm has inspired many young physicists, including myself, and lead to many fantastic projects and new discoveries. 
I would like to thank Markus M\"uller, for his thoughtful direction in our collaboration on the spin glass project. I would like to thank my dissertation committee of Alexander Grosberg, David Pine, Paul Chaikin and Daniel Stein for their helpful suggestions on both science and career. I would also like to acknowledge all other principal investigators in the Center for Soft Matter Research at NYU, David Grier, Jasna Brujic, Marc Gershow and Alexandra Zidovska, for their discussions and comments that helped me make better science.

% Gustavo During, Edan Lerner, Eric DeGiuli, Jie Lin, Lang, Jan, Cato, Payam, Henrique, Yin, Chen, Wenhai, Guolong, people at CSMR
It was a great pleasure to work in the Center for Soft Matter Research, and I want to thank all current and former members for their support over the past four years. In particular, I want to thank, Gustavo D\"uring, Edan Lerner, Eric DeGiuli and Jie Lin in Wyart's group, Marco Baity-Jesi, Antoine Barizien and Alexis Front visited to the group, collaborations with whom are extremely pleasant and enthusiastic; Lang Feng, who introduced me to the Center; Payam Rowghanian, Jan Smrek, and Cato Sandford, who created a casual atmosphere in the office room we shared and taught me a lot in language and culture; and Yin Zhang, Chen Wang, Wenhai Zheng, and Guolong Zhu, who continually ``bothered'' me with their research problems. %I greatly appreciate the discussion with all the members in Wyart's group: John Royer, Corinna Maass, Kun-Ta Wu, Qin Xu, Jeremie Palacci, Gray Hunter, Colm Kelleher and especially Bezia Laderman who assisted me with my

% Yanchao, Hong, Shuang, Yuqian, Lan, Tao, Hongliang, Daniel, Patrick, Stefano, Victor, friends
Last but not least, I am thankful to all my friends who helped me and encouraged me in my life at New York University, without whom I could not finish my Ph. D. study. Especially, I want to thank, Yanchao Xin, Hong Zhang, Shuang Li, and Yuqian Liu, who were my roommates and closest friends in New York; Tao Jiang, Lan Gong, and Hongliang Liu, who guided me as senior Ph. D. students; and Patrick Cooper, Daniel Foreman-Mackey, Victor Gorbenko, Shahab Kohani, Henrique Moyses, and Stefano Storace, who are my same-year fellows survived in the end.

%% file: preface/preface.tex
\section*{Preface}
% A preface generally covers the story of how the book came into being, or how the idea for the book was developed; this is often followed by thanks and acknowledgments to people who were helpful to the author during the time of writing.
% mystery of time on life, on career, we'll die but leave wealth to descendants, that is knowledge. on science, vital scientific questions.
Life is transient. %The harsh reality always arouses people's blues. 
%``Equating life and death is ridiculous.'', 
%``The future generations will look upon us just like we look upon our past.'', 
%a quote from Chinese calligrapher Wang Xizhi's most famous work -- {\it Langtingji Xu}, or ``Preface to the Poems Collected from the Orchid Pavilion'', laments this sympathy. To leave the evidence of their livings was the motivation for their collection. 
What should we do in our limited time to make some differences? %The question came up to me when I still dreamed. 
Some figures, like Euclid, are ``alive'' in people's minds even after thousands of years, %still live in other's minds after they past, 
because their masterpieces keep inspiring and benefiting people. Knowledge of nature and ourselves is a wealth of all mankind, a contribution to which would surely make my life meaningful. That is my initial motivation for doing science. 
% static not evolving or dynamic. To change the world.

%contents of the thesis,
%This work is based on the following articles that were or will be published elsewhere..
Followed a typical education path, I became a Ph. D. student at New York University, standing at the gate to my research career dream and looking for questions that I could make use of my lifetime to work on. Fortunately, I met my second mentor (after my father as the first), Matthieu Wyart. 
He introduced me to the question of transients in nature -- the dynamics of glassy systems. 
 
``The flying arrow is motionless'', the famous paradox of ancient Greek philosopher Zeno, has a similar modern version in the physics of glasses: ``the flowing liquid is solid''. Zeno's paradox is deeply related to calculus and the concept of limit, nonetheless, the physics lying behind the glass connects to a long dynamical time scale. %Above and below the time scale, the system behaves very differently. 
Below the time scale, a glass appears to be static and solid, while above, it flows as a sticky liquid, like honey. 
It is also a physical phenomenon closely related to life. Imagine you are stuck in a traffic jam: if you are in a hurry, staring at the second hand on your watch, you may probably curse the jammed traffic; but if you are relaxed, enjoying the music and the views, you may happily drive to your destination after momentary waitings. These dynamical systems characterized by long time scales are glassy systems. 

In glassy materials, the dynamical time scale can be tuned by temperature and driving forces like gravity. %Experiments suggest a dynamical phase transition show that there exist a critical temperature and a critical driving force in glassy systems, at which the time scale diverges in the thermodynamic limit. 
%Many of them show a transition from a static state to a dynamical state near a critical temperature or a critical force. 
For example, the wax flows more and more easily if one heats it up; the sandpile slides faster and faster when the bed of the dump truck is tilted more and more steeply. 
However, how the time scale depends on temperature and driving force may vary from system to system, and there are not yet universal rules like Newton's Laws to describe different dynamics. We are curious to know what determines the dynamical properties of the glassy systems from microscopic level; Is there any dynamical universality among systems that possess certain microscopic features? 
%The question is both of fundamental importance, as glassy systems are common in from physics to biology, and of practical value for materials science and engineering. 
%We have studied the dynamics and the correlated thermodynamics of glasses under cooling: what aspect of microscopic structures determines the temperature dependence of dynamics, and how microscopic structures may evolve under cooling correspondingly. We have also studied the dynamical phase transition of an athermal glassy system under a driving force -- erosion of a granular bed: what microscopic mechanism makes the transition in a special universality class. 
%What makes a glass different from a crystal and other ordered system is its randomness: the microscopic structures are not predictable after knowing a few atoms 
One prominent concept of dynamical systems is self-organization: the trend that the system spontaneously evolves to states where the dynamical time scale diverges. So the dynamical transitions are usually relevant in glassy systems. The dynamics in this situation become rich: phenomena of all different time scales accompanied with different sizes appear. For example, in the Earth crust as a glassy system, earthquakes varying in intensity from unnoticeable to devastating can happen. %in the ecosystem on Earth, life forms vary from bacterias, micrometers in size and hours in lifespan, to whales, hundreds of meters in size and dozens of years in lifespan. However, this critical dynamics are barely understood even in the simplest systems. 
But how our nature spontaneously evolves to these critical states is still lack of a mathematical description. %Are the components of the system independent or closely correlated in this special state? %We have studied one category of the simplest critical dynamics: what are the non-trivial correlations of individuals and how they emerge spontaneously in the system. 

I have studied these general questions in specific systems and models during my Ph. D. and conclude them in this dissertation. The dissertation may never be eternal, %nor this preface like Wang Xizhi's. 
but to me, it is a first small but important step towards my scientific career, which, I hope, will leave some valuable thoughts that continually inspire and benefit others.

%% file: abstract/abstract.tex
\section*{Abstract}
\label{sec_abstract}

Glassy systems are disordered systems characterized by extremely slow dynamics. Examples are supercooled liquids, whose dynamics slow down under cooling. The specific pattern of slowing-down depends on the material considered. %We poorly understand this dependence, in particular, which aspects of the microscopic structures control the dynamics and other macroscopic properties is unclear. 
This dependence is poorly understood, in particular, it remains generally unclear which aspects of the microscopic structures control the dynamics and other macroscopic properties.
Attacking this question is one of the two main aspects of this dissertation. 
We have introduced a new class of models of supercooled liquids, which captures the central aspects of the correspondence between structure and elasticity on the one hand, the correlation of structure and thermodynamic and dynamic properties on the other. These models can also be resolved analytically, leading to theoretical insights into the question.  Our results also shed new light on the temperature-dependence of the topology of covalent networks, in particular, on the rigidity transition that occurs when the valence is increased.  Observations suggested the presence of a nearly critical range in the proximity of the rigidity threshold. 
Our work rules out the predominant explanation for this phenomenon by a ``rigidity window'' where the rigidity is barely satisfied. % and the system is near criticality. 

Other questions appear in glassy systems at zero temperature, when the thermal activation time is infinitely long. In that situation, a glassy system can flow if an external driving force is imposed above some threshold. Near the threshold, the dynamics are critical. To describe the critical dynamics, one must understand how the system self-organizes into specific configurations. 

The first example we will consider is the erosion of a river bed. Grains or pebbles are pushed by a fluid and roll on a disordered landscape made of the static particles. Experiments support the existence of a threshold forcing, below which no erosion flux is observed. Near the threshold, the transient state takes very long and the flux converges very slowly toward its stationary value. In the field, this long transient state is called ``armoring'' and corresponds to the filling up of holes on the frozen landscape by moving particles. The dynamics near the threshold are relevant for geophysical applications -- gravel river beds tend to spontaneously sit at the threshold where erosion stops, but are poorly understood. 
In this dissertation, we present a novel microscopic model to describe  the erosion near threshold. This model makes new quantitative predictions for the erosion flux {\it vs} the applied forcing and predicts that the spatial reparation of the flux is highly non-trivial: 
it is power-law distributed in space with long range correlation in the flux direction, but no correlations in the perpendicular directions. We introduce a mean-field model to capture analytically some of these properties. 

To study further the self-organization of driven glassy systems, we investigate, as our last example, the athermal dynamics of mean-field spin glasses. 
Like many of other glasses, such as electron glasses, random close packings, etc., the spin glass self-organizes into the configurations that are stable, but barely so.  Such marginal stability appears with the presence of  a pseudogap in soft excitations -- a density of states vanishing as a power-law distribution at zero energy. 
How such pseudogaps appear dynamically as the systems are prepared and driven was not understood theoretically.  We elucidate this question, by introducing a stochastic process mimicking the dynamics, and show that  the emergence of a pseudogap is deeply related to very strong anti-correlations emerging among soft excitations.

%% file: introduction/introduction.tex
\chapter{Introduction}
\label{intro}
\begin{figure}[h!]
\centering
\includegraphics[width=.8\columnwidth]{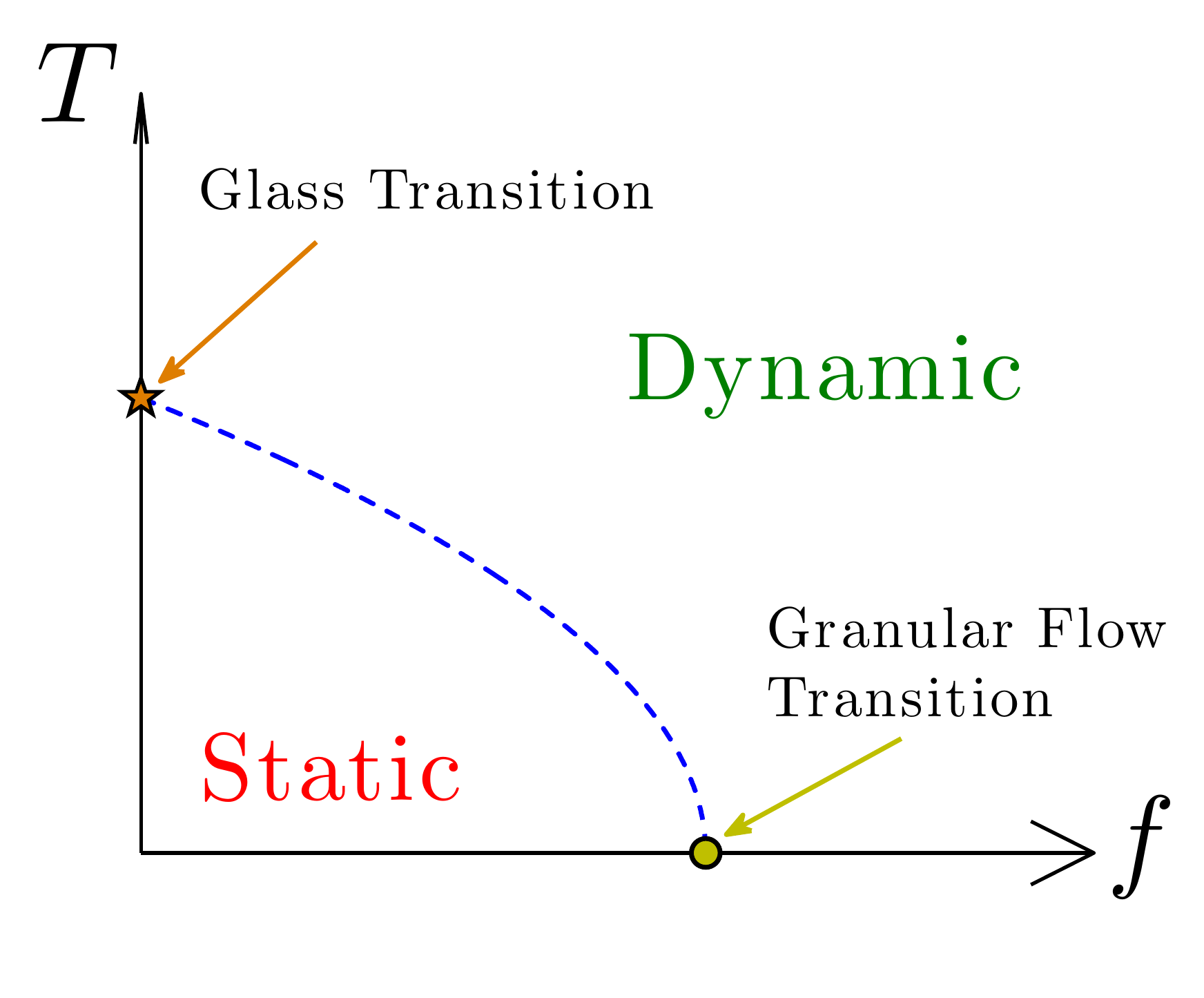}
\caption{\small{A phase diagram of static and dynamic phases of glassy systems. $T$ is the temperature, $f$ is the driving force.
}}\label{intro_phase}
\end{figure}

Statistical physics, one of the cornerstones of modern physics and chemistry, sets the theoretical framework to describe systems in thermal equilibrium. However, more often, systems are not equilibrated. This is the case for open systems which receive energy fluxes from their environments, such as the living or social systems. Other examples are glassy systems, whose thermal activation time is so long that they cannot equilibrate on practical time scales. Examples are the structural glasses that make our windows, whose dynamics are dominated by a glass transition between an equilibrated liquid state and an out-of-equilibrium glass state with an extremely long relaxation time. Other examples are granular materials, where the temperature effect is too small to be relevant. % is essentially absent
These systems are yield stress materials, which can flow if a sufficient forcing is applied. The transition between the solid and liquid phases is a non-equilibrium dynamical transition. The two transitions are represented in the phase diagram in Fig.~\ref{intro_phase}. Understanding them is a long-standing topic of soft condensed matter physics. 

%The glassy systems are composed of highly interactive elements.  
The thermal relaxation of a liquid can be enormously prolonged under cooling~\cite{Ediger96,Debenedetti01}. %The glassy systems are dynamic if they are of sufficiently high temperature or under strong enough external drive. %For example, glass materials flow as liquids when the temperature is above glass transition temperature; the sand grains drift if the slope of the sand pile is greater then some critical value~\cite{}. %  
Classical examples are structural glasses, which can be elegantly shaped by blowing them. 
Blowing is possible because these materials are very viscous and flow on the scale of seconds near the glass transition, compared with picoseconds in liquid phase at high temperature. 
In light of the long history of glass manufacture~\cite{Douglas72}, it is surprising how little is understood of the microscopic cause for their slow dynamics.% the microscopic cause for their slow dynamics is still not well understood. 
In particular, the specific pattern of slowing down near the glass transition depends on the material considered~\cite{Bohmer92,Wang06}. There is no theoretical framework to predict that dependence. 

At zero temperature, the thermal activation time of a glassy system is absent. The system can however transition from a static state to a dynamic flowing state %characterized by a finite time scale, 
once a sufficient external forcing is applied. For instance, rocky river beds and sand dunes can flow when pebbles and sand grains are eroded by the water flow and the wind respectively~\cite{Bagnold66}. 
Many experiments support that erosion occurs only when the stress applied by the fluid %on the particles 
exceeds a threshold. Understanding this threshold is important for geophysical applications, as river beds tend to spontaneously evolve toward this threshold where erosion stops~\cite{Parker07,Bak13}. No compelling theoretical framework has been proposed to describe the dynamics near the threshold. One interesting effect is the ``armoring'' phenomenon: near the threshold, there are long transients where the average number of mobile particles flowing above the frozen bed slowly decays. This slow decay corresponds to the filling up of holes by mobile particles: the system self-organizes into configurations where a minimal number of particles remain mobile. 
%One of the central questions about glassy systems is to understand the slow dynamics and the long time scale at the microscopic level. % in the situations of different dynamic sources, temperature or ordered forces. %Specific to the glassy systems driven by the entropic force 
%Another fundamental question deals with the presence of a diverging time scale -- how to interpret the dynamics in these glassy systems. We explore these two general questions about the dynamics of glassy systems in this dissertation.

Another kind of self-organization occurs in glassy systems at zero temperature, when interactions are effectively long-range. Examples include electron glasses, spin glasses or packings of hard particles  where elastic interactions dominate. Such systems spontaneously self-organize into configurations that are stable, but barely so. As a consequence, rich dynamics occur when a perturbation is applied. Typical examples are Barkhausen noises in magnetic spin systems~\cite{Perkovi95,Zapperi98,Doussal10} and avalanches in sand and snow packings. Currently, we lack a dynamical description of how the marginality is reached in these systems. %We make progress on this question in a model system, fully connected spin glasses, known as the Sherrington-Kirkpatrick model. 
 
 \section{Dynamics under cooling -- glass transition}
 \label{intro_1}
 %no universality in glass, MCT, activation, thermodynamic transition (RFOT), properties materials dependent, related to rigidity of structure, intermediate phase, reproduce the experimental results
 A glass is a solid which forms when a supercooled liquid -- a liquid at a temperature below its melting point -- falls out of equilibrium under cooling. In the practical sense, glass transition is essentially different from other equilibrium phase transitions, where the two phases are both in thermal equilibrium and the transition is determined by symmetry breaking~\cite{Chaikin00,Cardy96}. For example, the translation invariant symmetry breaks in crystallization~\cite{Chaikin00}. By contrast, no obvious symmetry is found broken in the glass transition problem. So there is no comparable theory based on symmetry explaining the glass transition. 
 %According to the renormalization group theory~\cite{Wilson83,Cardy96}, second-order transitions determined by the same symmetry thus fall in the same universality class, which is characterized by the same critical exponents of power-law diverging quantities including characteristic length scales and order parameters. 
 %\bea
 %\xi\sim|\Sigma-\Sigma_c|^{-\nu};\\
 %q\sim(\Sigma_c-\Sigma)^{\beta};
 %\label{intro_length}
 %\eea
 %where $\xi$ is the length scale, and $q$ is the order parameter; $\nu$ and $\beta$ are corresponding critical exponents; $\Sigma$ and $\Sigma_c$ are the control parameter and the critical value, which is the temperature in dynamical glass transition.  
 %The dynamical processes in equilibrium phase transitions are also featured by the power law diverging time scales, 
 %\be
 %\tau\sim\xi^{z}\sim|\Sigma-\Sigma_c|^{-\nu z}
 %\label{intro_tau}
 %\ee
 %where $\tau$ is the characteristic time scale, and $z$ is the corresponding critical exponent~\cite{Hohenberg77}. 
 
 %A mode-coupling theory of liquids does predict universal power-law diverging length and time scales and critical exponents at a dynamical glass transition $T_c$~\cite{Leutheusser84,Bouchaud96,Biroli04, Biroli06}. However, the dynamics in real glass transitions contradicts with the prediction from traditional renormalization group theory on phase transition and the mode-coupling theory. 
 The most significant features of glasses are their extremely slow dynamics, which appear to be associated with collective behaviors characterized by a increasing correlation length of the dynamics~\cite{Hohenberg77,Leutheusser84,Bouchaud96,Biroli04, Biroli06}. 
 This increase is however moderate, since the size of collectively rearranging regions in a liquid increases only by four to five times, in contrast to a $10^{16}$-fold growth in the relaxation time~\cite{Ediger96,Debenedetti01,Dalle-Ferrier07,Glotzer00,Bouchaud04,Berthier05,Kurchan09}. Experimentalists often fit this fast rise of the relaxation time by the Vogel-Fulcher-Tammann law~\cite{Vogel21,Fulcher25,Tammann26},
 \be
 \tau(T)=\tau_0\exp\lp\frac{A}{T-T_0}\rp,
 \label{intro_vft}
 \ee 
 where $\tau_0$, $A$ and $T_0$ are fitting parameters for different materials. 
 The exponential form, Eq.(\ref{intro_vft}), indicates that the dynamics in glasses slow down much faster than a typical thermal activation process, where the relaxation time is captured by Arrhenius law: 
 \be
 \tau(T)\propto\exp\lp\frac{\Delta F}{T}\rp,
 \label{intro_tf}
 \ee
 where $\Delta F$ is the free energy barrier of relaxation. The two formulas are consistent only if the free energy barrier $\Delta F$ depends on temperature and becomes singular at $T_0$. In most fragile liquids, this $\Delta F$ can increase by 6 to 7 fold under cooling. % and becomes singular at a finite $T_0$, which is proposed to be the Kauzmann temperature $T_K$, where the entropy crisis happens~\cite{Kauzmann48,Adam65}. The entropy crisis occurs when the entropy in the supercooled liquid becomes less than the entropy of its crystal counterpart, that is to say, the configuration entropy vanishes. A thermodynamic transition is shown in random energy model~\cite{Derrida81} at $T_K$. 

\begin{figure}[h!]
\centering
\includegraphics[width=.8\columnwidth]{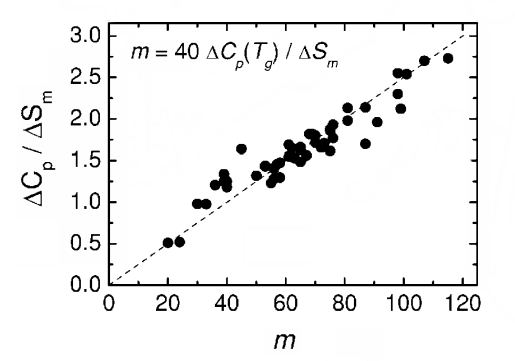}
\caption{\small{Scatter plot of the jump of specific heat $\Delta C_p/\Delta S_m$ and the fragility $m$ of different glassy materials. The dashed line is given by $m=40\Delta C_p/\Delta S_m$, where $\Delta S_m$ is the entropy gain in melting of the same material. The plot is reproduced from Reference~\cite{Wang06}.
}}\label{intro_cpm}
\end{figure}

 %Thanks to the fast diverging time scale, all the properties of glasses are observed at a finite glass transition temperature $T_g$. 
 %Despite the buried universality class of glass transition, some material dependent properties are revealed to be correlated. 
 %The thermal activation law, Eq.(\ref{intro_tf}), connects the dynamics with the thermodynamics in glasses, as revealed in experiments. 
 Experiments reveal a connection between the dynamics and thermodynamics in supercooled liquids.
 For instance, the jump of specific heat $\Delta c_p$ and the fragility $m$ are linearly correlated for different kinds of glass-forming materials, ranging from network to polymer glasses, as shown in Fig.~\ref{intro_cpm}~\cite{Martinez01,Wang06}. The fragility, characterizing the temperature dependence of the relaxation time in different materials, is defined as,
 \be
 m\equiv\left.\frac{\partial\ln\tau(T)/\tau_0}{\partial(T/T_g)}\right|_{T=T_g}=\ln\frac{\tau(T_g)}{\tau_0}-\left.\frac{\partial\Delta F(T)}{\partial T}\right|_{T_g}.
 \label{intro_m}
 \ee
 The liquids following the Arrhenius law, Eq.(\ref{intro_tf}) with a temperature independent $\Delta F$, are termed as ``strong'' with small $m_0=\ln\tau(T_g)/\tau_0$; while those very non-Arrhenius liquids that $\Delta F$ increases significantly under cooling are termed as ``fragile'' with large $m$ values. 
 The jump of specific heat characterizes the number of degrees of freedom contributing to the configuration entropy -- the degeneracy of metastable states in the liquid phase. These degrees of freedom are frozen at the glass transition. Specifically, the jump of specific heat is defined as the capacity difference between the liquid phase and the glass phase, 
 \be
 \Delta c_p\equiv\frac{1}{N}\left.\frac{\partial H(T)}{\partial T}\right|^{T_g^+}_{T_g^-},
 \label{intro_cp}
 \ee
 where $H(T)/N$ is the enthalpy density of the system. Both $m$ and $\Delta c_p$ capture the temperature dependence of energy measures of the glasses, must thus be correlated.  
 Most glass theories%, including the random first order transition theory
 ~\cite{Kirkpatrick89,Bouchaud04,Lubchenko07,Biroli12} have concentrated on reproducing the linear correlation $\Delta c_p\propto m$, however, only a few~\cite{Hall03,Bevzenko09} have tried to provide an explanation on how they are determined microscopically. 
 
\begin{figure}[h!]
\centering
\includegraphics[width=.8\columnwidth]{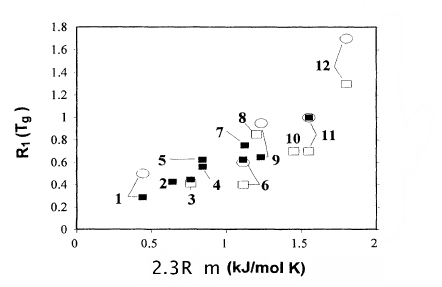}
\caption{\small{Scatter plot of boson peak intensity ratio $R_1$ and fragility $m$ of different glasses. The plot is reproduced from Reference~\cite{Ngai97}.
}}\label{intro_rm}
\end{figure}

 Experiments hint that the elasticity of the structures may be the key factor and both the energy barrier $\Delta F$ and the enthalpy $H$ are purely manifestations of this. The elasticity of a structure is featured by its vibration spectra. Glasses are distinguished by  the boson peak, a large number of low-frequency modes additional to phonons, in their spectrum. It is found~\cite{Ngai97,Novikov05} that the fragility of a glass-forming material is inversely proportional to its intensity of the boson peak, defined as, 
 \be
 In\equiv R_1^{-1}\equiv{\rm max}(D(\omega)/\omega^2)/{\rm min}(D(\omega)/\omega^2),
 \label{intro_boson}
 \ee
 where $D(\omega)$ is the number density of vibrational modes of frequency $\omega$, shown in Fig.~\ref{intro_rm}. 
 
\begin{figure}[h!]
\includegraphics[width=.45\columnwidth]{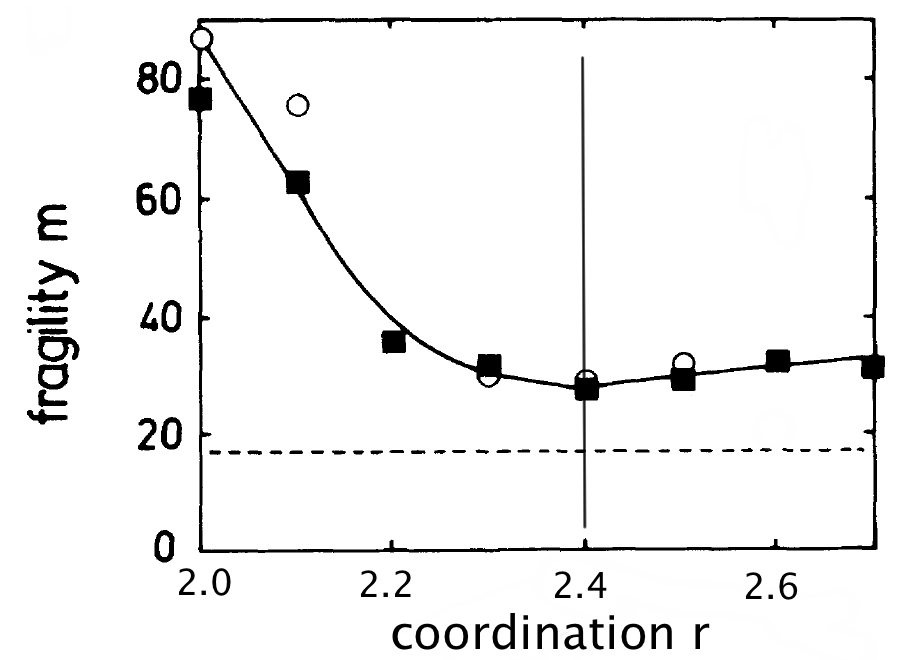}
\includegraphics[width=.55\columnwidth]{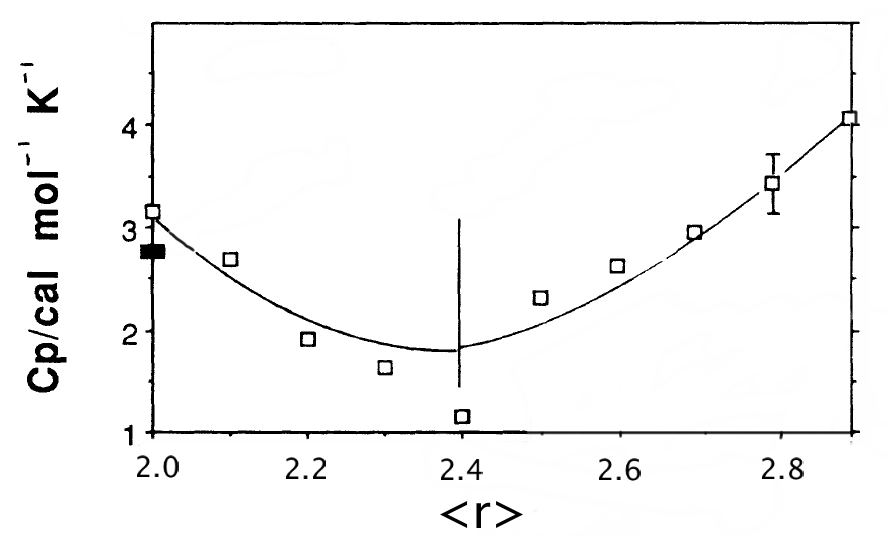}
\caption{\small{The fragility $m$ (Left) and the jump of specific heat $\Delta c_p$ (Right) of chalcogenides with coordination number $r$. The plots are reproduced from References~\cite{Tatsumisago90,Bohmer92}.
}}\label{intro_mcp}
\end{figure}

 Another line of evidence originates from chalcogenides, a kind of network glasses, where particles interact prominently through specialized covalent bonds, whose number can be experimentally tuned by changing the component ratio~\cite{Ota78,Mahadevan83,Sreeram91}. The mechanical stability theory developed by Maxwell~\cite{Maxwell64} predicts that the rigidity of a structure can change by simply tuning the number of constraints in the structure. Applying this to network glasses and counting both the stretching and bending constraints of covalent bonds, Phillips and later Thorpe~\cite{Phillips79,Phillips85} pointed out that the rigidity of the covalent network sets on at a critical coordination number $r_c=2.4$, where the coordination number $r$ is the average number of covalent bonds per atom. Experiments~\cite{Tatsumisago90,Bohmer92} indicate a special correlation between the glass properties and this rigidity transition: both the fragility and the jump of specific heat vary non-monotonically when tuning the coordination number and are minimal at the proximity of the rigidity threshold, as shown in Fig.~\ref{intro_mcp}. Moreover, some recent experiments~\cite{Selvanathan99,Georgiev00,Wang00,Chakravarty05,Wang05,Chen08,Bhosle12,Bhosle12a} suggest that there exists even a range of coordination number around $r_c$, where the network glass is strong and the stress distribution is homogeneous. % and the structure is lack of any median-range order~\cite{} identified by a vanishing non-reversible heat in 
 This range near $r_c$ is termed as Intermediate Phase~\cite{Selvanathan99}. 
 
 However, no theory has been developed to successfully rationalize these observations connecting the elasticity of the microscopic structures and the dynamic and thermodynamic properties of the glasses. We develop such a theory and give a quantitative prediction on the thermodynamics of the glasses based on their structures.
 
 %\textcolor{red}{question?}
 
 \section{Rigidity of a structure}
 \label{intro_2}
 %Jamming and formalism, Maxwell, rigidity percolation (percolation), rigidity window
 The two sets of hints on the elasticity of the microscopic structures, the boson peak intensity and the rigidity transition of the interaction network, are in fact two sides of a coin. Recent observations~\cite{Trachenko00, OHern03, Chen08, Chen10, Ghosh10} and theories~\cite{Wyart05a, Brito09, DeGiuli14b, DeGiuli15, Franz15b} on various amorphous materials including jammed packings and random elastic networks indicate that the structures near a rigidity threshold display large boson peaks. At the jamming point~\cite{Liu98,Liu10,Liu10a}, the hard particle packings become incompressible, which corresponds to the onset of the rigidity of the contact networks where the number of forced contacts equals to the number of degrees of freedom~\cite{Maxwell64}. The densities of states in both jammed packings and critically rigid random networks are filled with low-frequency anomalous modes contributing to strong boson peaks. Therefore, it is essential to investigate the rigidity of a structure in order to study later the correlation of the elasticity and the dynamics of supercooled liquids. 
 %These low-frequency anomalous modes arise from the marginally rigid nature of the interacting network~\cite{Wyart05a}. By breaking a finite number of bonds, one can generate the same number of floppy modes, which cost no energy~\cite{Wyart05a, DeGiuli}. 
 
\begin{figure}[h!]
\centering
\includegraphics[width=.9\columnwidth]{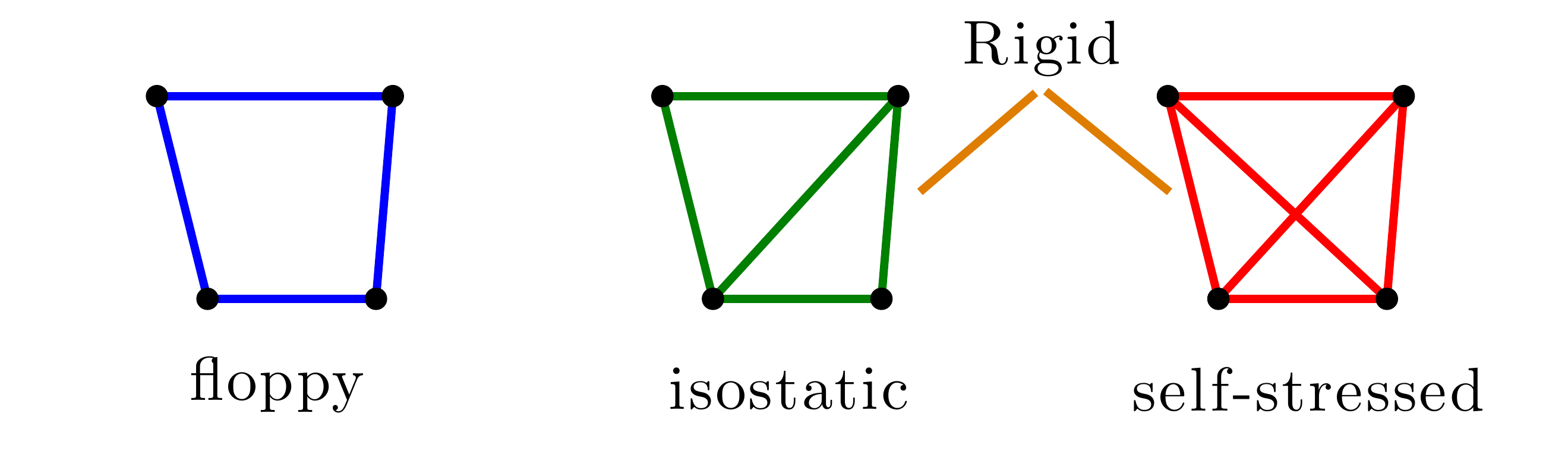}
\caption{\small{An illustration of the rigidity transition of a network by adding pairwise constraints from floppy (Left) to isostatic (Middle) and to self-stressed (Right).
}}\label{intro_isostatic}
\end{figure}

 The rigidity of a structure can arise in a purely topological scenario: a network of stiff bonds becomes rigid as the number of bonds increases (independent of the specific geometry), illustrated with a four-joint network in Fig.~\ref{intro_isostatic}. Maxwell~\cite{Maxwell64} first proposed a criterion on the critical number of constraints: when the number of degrees of freedom overwhelms the number of constraints, the structure is floppy with some deformation modes that cost no elastic energy; on the contrary, when the number of constraints exceeds the number of degrees of freedom, the structure is ``more than'' rigid -- it contains some redundant constraints, removing of which does not affect the rigidity. To characterize these under-constrained and over-constrained features, we define the number of floppy modes (or ``zero modes''), $F$, and the number of redundant constraints (also termed ``self-stress states''), $N_R$. The Maxwell counting indicates~\cite{Maxwell64}, 
 \bea
 F=dN-N_B\qquad{\rm if}\quad dN>N_B;\\
 N_R=N_B-dN\qquad{\rm if}\quad dN< N_B,
 \label{intro_maxwell}
 \eea
 where $N$ is the number of particles in the network, $d$ is the spatial dimension, and $N_B\equiv zN/2$, is the number of bonds connecting the particles, with $z$ as the coordination number. In mean-field, the rigidity switches on simultaneously throughout the system at the critical coordination number $z_c=2N_B/N=2d$, where there is no floppy modes nor redundant bonds. %(\textcolor{red}{picture to illustrate F and Nr})
 In general, a random network can violate the special Maxwell counting, Eqs.(\ref{intro_maxwell}), but must satisfy~\cite{Laman70,Jacobs95},
 \be
 F-N_R=dN-N_B.
 \label{intro_gmax}
 \ee 
 This topological counting is rooted in the linear elasticity of structures~\cite{Calladine78,Lutsko89,Karmakar10b}. We introduce a linear elasticity formalism to give a robust mathematical definition of these numbers and derive the general Maxwell counting, Eq.(\ref{intro_gmax}).  
 
\begin{figure}[h!]
\centering
\includegraphics[width=.65\columnwidth]{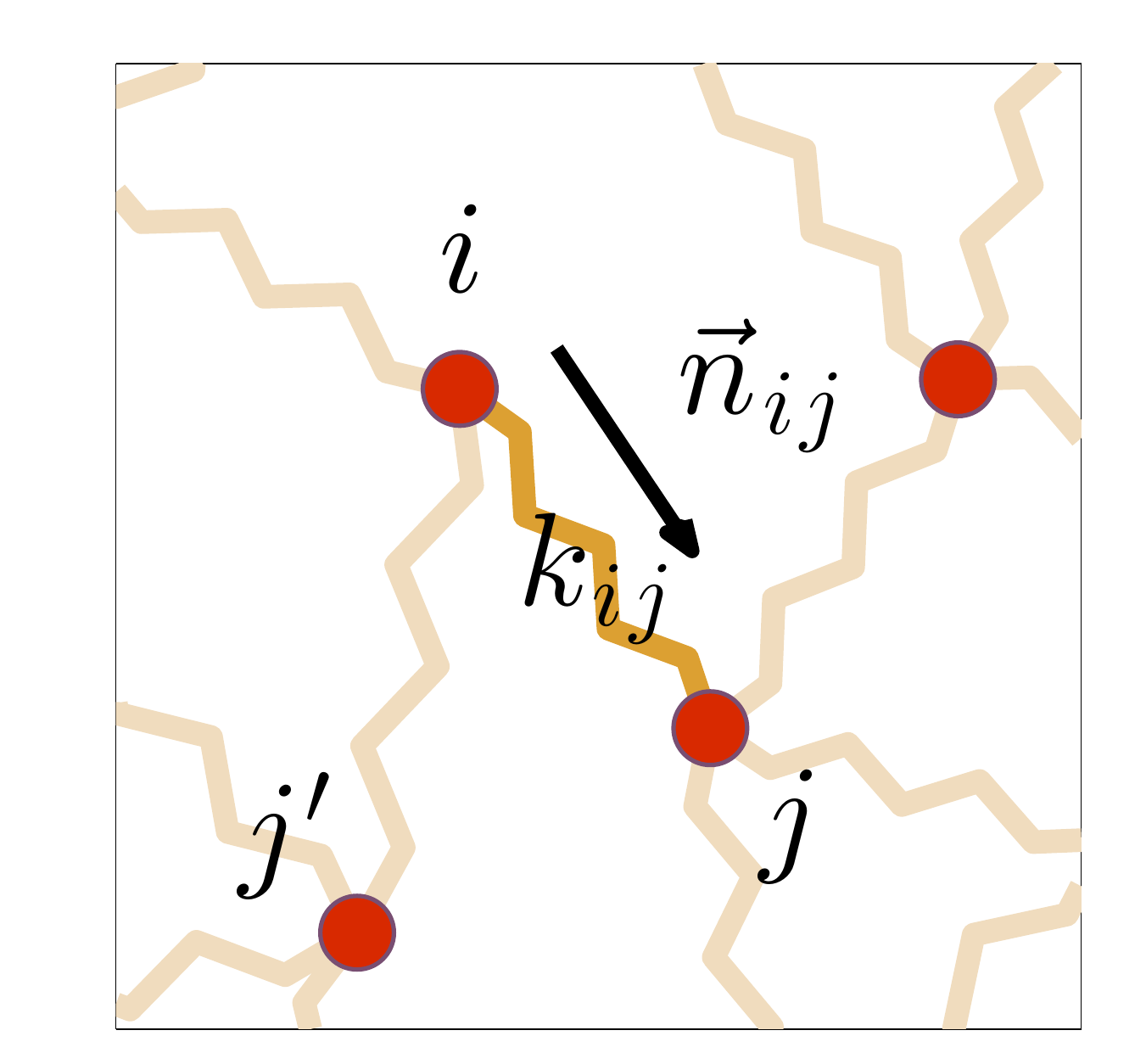}
\caption{\small{Zoom-in illustration of a spring network with a spring of stiffness $k_{\langle i,j\rangle}$ connecting particle $i$ and $j$. 
}}\label{intro_network}
\end{figure}

 Consider a generic  spring network~\cite{Jacobs95} with no nonlinear structures like two springs adjacent in a straight line, shown in Fig.~\ref{intro_network}. The positions of the particles define a vector $|\vec{R}\rangle$ in $d\times N$-dimension configuration space, where we use bra-ket notation for vectors. If particle $i$ connects to particle $j$, the distance between the two is given by $r_{\langle i,j\rangle}\equiv||\vec{R}_i-\vec{R}_j||$, and $N_B$ springs define a vector $|r\rangle$ in $N_B$-dimension contact space. If we perturb the system by a small displacement field $|\delta\vec{R}\rangle$, the distance between particles $i$ and $j$ changes by,
 \be
 \delta r_{\langle i,j\rangle}=(\delta\vec{R}_j-\delta\vec{R}_i)\cdot\vec{n}_{ij}+o(\delta\vec{R}^2),
 \label{intro_displace}
 \ee
 where $\vec{n}_{ij}$ is the unit vector pointing from $i$ to $j$. %The displacement difference projected onto the direction of the two particles
 In the linear response region, we neglect all higher order effects, and define a displacement independent matrix, $\ms$, to connect the displacement field $|\delta\vec{R}\rangle$ and the distance change field $|\delta r\rangle$,
 \be
 \ms_{\langle i,j\rangle,k}=\vec{n}_{ij}(\delta_{ik}-\delta_{jk}),
 \label{intro_smat}
 \ee
 where $\delta_{ik}$ is the Kronecker delta function. Equation (\ref{intro_displace}) can then be abbreviated as $\delta r_{\langle i,j\rangle}=\sum_{k}\ms_{\langle i,j\rangle,k}\cdot\delta\vec{R}_k$, or simpler as $|\delta r\rangle=\ms|\delta\vec{R}\rangle$. The structure matrix, $\ms$, also termed as the compatibility matrix, is $N_B$ by $dN$. 
 
 The relation between the contact tensions and the vector forces on particles is dual to the relation between the distance changes of contacts and the displacements of particles. If there is a tension field $\{f_{\langle i,j\rangle}\}$ on the springs, positive for being stretched and negative for being compressed, the force $\vec{F}_i$ on particle $i$ is 
 \be
 \vec{F}_i=\sum_{j}f_{\langle i,j\rangle}\vec{n}_{ij}=\sum_{\langle j,k\rangle}\vec{n}_{jk}(\delta_{ij}-\delta_{ik})f_{\langle j,k\rangle}.
 \label{intro_force}
 \ee
 This formula can also be abbreviated as $|\vec{F}\rangle=\mt|f\rangle$, where $\mt$ is a force-independent $dN$ by $N_B$ matrix. $\mt$ and $\ms$ are transposes of each other, $\mt^t=\ms$. 
 
 A floppy mode is a vector in configuration space, a displacement field along which does not change the distances of contacts in the linear order. In this formalism, it corresponds to a nontrivial solution of equations, $\ms|\delta\vec{R}\rangle=0$. The number of floppy modes, $F$, is then equal to the dimension of the kernel of matrix $\ms$~\cite{Lay11}, 
 \be
 F\equiv\dim({\rm ker}\ms).
 \label{intro_f}
 \ee
 Similarly, a self-stress state corresponds to a vector in contact space, a tension field along which does not change the mechanical stability, that is to say, $\mt|f\rangle=0$. Accordingly, the number of redundant constraints satisfies
 \be
 N_R\equiv\dim({\rm ker}\mt).
 \label{intro_nr}
 \ee
 As $\mt^t=\ms$, ${\rm rank}(\ms)=dN-F={\rm rank}(\mt)=N_B-N_R$ immediately leads to Equation (\ref{intro_gmax}). % is just a trivial result of these definitions Eqs.(\ref{intro_f},\ref{intro_nr}). 
 
 Another benefit from this formalism is that the linear elastic energy corresponding to any small %strain $|\delta\vec{R}\rangle=|\vec{u}\rangle$ or mismatch of spring rest lengths 
 distortion field can be calculated to the linear order without doing relaxation for the mechanical equilibrium. For a given mismatch of springs $|\delta r\rangle=|\epsilon\rangle$, the mechanical equilibrium of the network is achieved with a non-affine response $|\delta\vec{R}_{\rm n.a.}\rangle$,
 \bea
 |\vec{F}\rangle=\ms^t|f\rangle=\ms^t\mk|\epsilon\rangle-\ms^t\mk\ms|\delta\vec{R}_{\rm n.a.}\rangle=0,\\
 |\delta\vec{R}_{\rm n.a.}\rangle=\mm^{-1}\ms^t\mk|\epsilon\rangle,
 \label{intro_na}
 \eea 
 where $\mk$ is a diagonal matrix with $\mk_{\gamma\gamma}=k_{\gamma}$ the spring constant of spring $\gamma$, and $\mm\equiv\ms^t\mk\ms$ is the dynamic matrix. %, which is inverted in its non-singular subspace. 
 The elastic energy equals formally to,
 \be
 \mh(|\epsilon\rangle)=\frac{1}{2}\langle\epsilon-\ms\delta\vec{R}_{\rm n.a}|\mk|\epsilon-\ms\delta\vec{R}_{\rm n.a}\rangle=\frac{1}{2}\langle\epsilon|\mk-\mk\ms\mm^{-1}\ms^t\mk|\epsilon\rangle.
 \label{intro_elastic}
 \ee 
 
 As the topological counting, Eq.(\ref{intro_gmax}), captures the essentials of linear elasticity, several linear rigidity transition scenarios relying only on the topology of the interaction networks have been intensively studied. (See Fig.~\ref{3_f1} for illustration.) %it implies that the Maxwell's rigidity criterion that the structure becomes marginally rigid when the number of constraints $N_B$ equals to the degrees of freedom $dN$ may fail to determine the rigidity onset in a complex interacting network structure of a glass. 
 First, the rigidity percolation scenario~\cite{Feng84,Feng85a,Garboczi85,Arbabi93,Moukarzel95,Jacobs95} searches for a rigid backbone, a rigid cluster spanning over the system, in the networks generated by randomly diluting bonds on lattices. In this scenario, the critical transition is second order at a threshold below $z_c$~\cite{Jacobs95}, characterized by a continuous order parameter $P_{\infty}$ -- the fraction of bonds in the rigid backbone continuous in $z$. This indicates that the rigid backbone is a fractal object composed of a vanishing number density of bonds. In addition, the local stresses appear with redundant constraints in rigid islands at a connection density $z$ far below the rigidity threshold, due to the spatial fluctuations of connections. 
 
 Based on the assumption that the local stresses cost elastic energy and the networks self-organize to release the energy at low temperature, Thorpe and his followers~\cite{Thorpe00,Thorpe01,Chubynsky06} proposed a peculiar rigidity transition scenario, known as ``rigidity window''. The self-organized networks are generated by redistributing some of the connections from randomly diluted networks to avoid redundant constraints. The resulting networks contain rigid backbones with finite probabilities~\cite{Chubynsky06} from a coordination number $z_{\rm iso}<z_c$, while the stress only appears when $z>z_c$. The coordination number range from $z_{\rm iso}$ to $z_c$ of rigid non-stressed networks is termed as ``rigidity window''. %The fluctuations in the spatial distribution of the connections make the ``window'' possible. 
 Moreover, this peculiar window has been proposed as a candidate for the intermediate phase observed in chalcogenides~\cite{Thorpe01a,Boolchand05}.
 
 The last scenario, found in the contact networks of jammed packings of soft spheres, obeys the special Maxwell counting: both the rigidity and the stress sets on at the same coordination number. 
 The spatial fluctuation of the connections is so insignificant that a mean-field theory~\cite{Maxwell64,Wyart05a,During13,DeGiuli14b} captures the transition. Both the probability of being rigid and the order parameter $P_{\infty}$ are first-order step functions, that is to say, almost every contact is in the rigid cluster at $z_c$ when the rigidity percolates. Moreover, the mean-field theory predicts that the networks in this scenario possess a shear modulus linear in the coordination number, $G\propto|z-z_c|$, and a flat density of states of anomalous modes, $D(\omega)\sim \omega^0$, above a characteristic frequency $\omega^*\sim|z-z_c|$. %, whose density $D(\omega)\sim\omega^{d-1}$ for $d$ dimensional random networks. %in various jammed packings from granular media to colloidal suspensions and in 
 In 3D, the boson peak intensity, $In\sim\omega^{*-2}$, is thus peaked at the jamming point when the network is marginally rigid. 
 
 Not only does the rigidity transition scenario control the dynamics of the glasses, but the rigidity problem as a non-equilibrium phase transition phenomenon is of fundamental importance in itself. Therefore, it is significant to understand how the network topology evolves under cooling and which of the three rigidity transition scenarios applies to the real glasses. 
 
 \section{Non-equilibrium dynamical phase transition}
 \label{intro_3}
 %Directed percolation, depinning transition, vortex latex, yielding transition (unjamming under shear)
 At zero temperature, a static glassy system can transition to an absorbing dynamical state under a certain dynamical driving. %For example, a sandpile, a random packing of sand grains, starts to slide when its slope is so large that the gravity drives the grains to flow, as shown in Fig.~\ref{intro_surface}. 
%\begin{figure}[h!]
%\centering
%\includegraphics[width=.8\columnwidth]{introduction/avalpink}
%\caption{\small{Surface flow of a sand grains driven by gravity. The plot is reproduced from Reference~\cite{Jaeger1}.
%}}\label{intro_surface}
%\end{figure}
 %The erosion problem is poorly understood
 For instance, a flow of pebbles or grains occurs when a viscous fluid shears a substrate of sedimented particles, which are repulsive in short-range. %constrained by the gravity. %The side view of the granular flow is the same as Fig.~\ref{intro_surface} after a rotation. 
 This phenomenon is commonly known as erosion. %, which happens when a fluid exerts a sufficient shear stress on the sedimented layer. % and slope of a sandpile (Fig.~\ref{intro_surface}). 
 %Hydrodynamics is involved in the erosion problem at strong driving limit with resuspension~\cite{Bagnold66,Leighton86,Leighton87}, where particles are no longer constrained to the surface. 
 Water flow and wind shape the Earth's landscape through the erosion effect, which has thus long been the central topic of geophysics~\cite{Bagnold66}. %However, there are dozens of theories in the field fitting different experimental data but inconsistent with each other. Especially near the threshold of the erosion, where 
 Many theories about the erosion have focused on a continuous description of the particle flux versus certain fluid speed~\cite{Leighton86,Leighton87,Ouriemi09,Aussillous13}. However, this description, which applies when the resuspension of sedimented particles happens, fails near the erosion threshold. %what are the critical exponents and what determines the universality class of the transition are fiercely debated far from agreement~\cite{Charru04,Lobkovsky08,Ouriemi09,Aussillous13,Houssais15}. 
 Predicting the flux of particles is difficult in the latter case, even though this situation is relevant in gravel rivers, where the river beds self-organize until the fluid stress approaches the threshold value and the erosion stops~\cite{Parker07,Bak13}.
 
\begin{figure}[h!]
\centering
\includegraphics[width=.8\columnwidth]{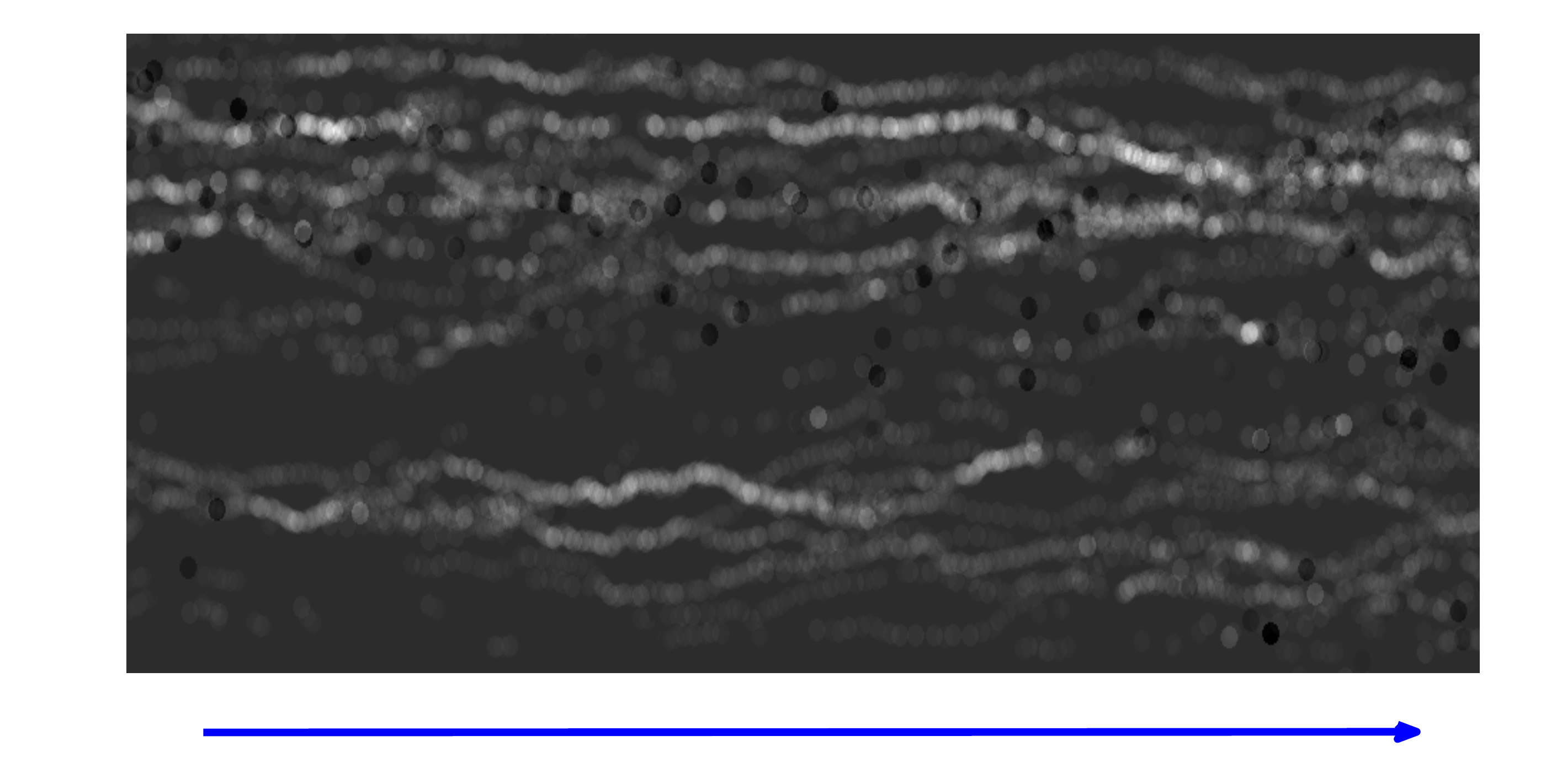}
\caption{\small{Particle trajectories (top view) of a granular bed erosion driven by water flow. The water is injected from the left to right, as indicated by the blue arrow. The plot is reproduced from Reference~\cite{Zou15}.
}}\label{intro_pattern}
\end{figure}
 
 % The universality is unclear
 A granular material flows at a certain stress anisotropy, $\theta^*=\Sigma/p\sim1$. In the low Reynolds number region, under a laminar driving flow, the stress anisotropy on the substrate is quantified by a dimensionless Shields number~\cite{Shields36}, $\theta\equiv\Sigma/(\rho_p-\rho)gd$, 
 where $\Sigma$ is the shear stress from the fluid, and $(\rho_p-\rho)gd$ quantifies the pressure due to the gravity on one layer of sedimented particles of mass density $\rho_p$ and typical size $d$ immersed in the fluid of density $\rho$. % quantifies the strength of the stress exerted by the fluid relative to the gravity. %: $\Sigma$ is the shear stress from the fluid; $\rho_p$ and $\rho$ are mass density of the particle and the fluid; $d$ is the typical size of the particles. 
 From hydrodynamics, the relative pressure on particles at depth $H$ below the surface is, $p=(\rho_p-\rho)gH$. The moving particles in the flowing boundary should meet the stress anisotropy requirement, $\Sigma/(\rho_p-\rho)gH\gtrsim\theta^*$, thus, the depth $H$ of the flowing boundary~\cite{Bagnold66} is proportional to the Shields number, $H\lesssim d\theta/\theta^*$. 
 Therefore, there is a threshold $\theta_c\sim\theta^*\sim1$, near which, $\theta-\theta_c\ll\theta_c$, the moving particles are localized in a layer of a few particle-size thick near the boundary and crawl on the rough surface made of other static particles~\cite{Bagnold66}. 
 
 As shown by trajectories in the Fig.~\ref{intro_pattern}, in this region, the particles do not simply follow the laminar flow: they roll around in the perpendicular directions, due to an interplay with the random surface of static particles and the interactions of active particles. 
 Sometimes, the mobile particles may even be trapped and become inactive. 
 Only those particles moving along the laminar flow contribute to the net flux of sedimented particles. 
 A steady flux, $J$, switches on when the fluid flow drives stronger than $\theta_c$~\cite{Shields36,White70}. ({\it i}) Under a constant shear stress near the threshold, %predicting the flux J of particles as a function of  is difficult, both for turbulent and laminar flows [4, 8]. %stay on the river bed without resuspension~\cite{Bagnold66,Leighton86,Leighton87,Ouriemi09,Aussillous13} and move when net forces exert on them. 
 %It is a glassy system in the sense that the static particles form a rough random landscape that the mobile particles have to evolve long before the current converges to a steady state. 
 %In these glassy systems, the existence of a critical driving force has been controversial without a microscopic dynamical model for a long time. Until recent argument of self-organized criticality~\cite{Bak87,Bak13} reveals that the natural river beds and sandpiles sit close to the critical point. The erosion of the landscape usually widens and flattens the river channels, which in turn decreases the water speed and thus the shear stress until the erosion becomes extremely slow.
 the system shows critical dynamics in experiments~\cite{Charru04,Lobkovsky08,Ouriemi09,Aussillous13,Houssais15},  
 \be
 J\sim(\theta-\theta_c)^{\beta};
 \label{intro_order}
 \ee
 where %$J$ is the dynamical order parameter; $\Sigma$ and $\Sigma_c$ are the control parameter and the critical value, which quantifies the driving force in a dynamical transition; 
 $\beta$ is the critical exponent characterizing the transition. Some works~\cite{Charru04,Ouriemi09,Aussillous13,Houssais15} show $\beta=1$, while other values also fit well in some experiments~\cite{Lobkovsky08}. 
 ({\it ii}) The typical speed of the particles is, however, not critical near the threshold~\cite{Charru04,Lajeunesse10,Duran14}. It is rather the number of active particles that vanishes at the threshold. 
 ({\it iii}) Before entering the steady state, the system undergoes a transient process, known as ``armoring'' or ``leveling'', where some of the active particles get trapped and shape the landscape. At the vicinity of the threshold, this transient process is characterized by a typical time scale $\tau$ that diverges, 
 %The dynamical processes in equilibrium phase transitions are also featured by the power law diverging time scales, 
 \be
 \tau\sim|\theta-\theta_c|^{-z},
 \label{intro_tau}
 \ee
 where $z$ is another critical exponent. %In addition, the eroding fluxes are usually correlated in space with diverging length scales,
 %\bea
 %\xi_{\parallel}\sim|\theta-\theta_c|^{-\nu_{\parallel}}\\
 %\xi_{\perp}\sim|\theta-\theta_c|^{-\nu_{\perp}}
 %\label{intro_length}
 %\eea
 %where $\xi_{\parallel}$ is the correlation length scale in the direction parallel to the direction of the driving flow, and $\xi_{\perp}$ is the scale in the perpendicular direction. $\nu_{\parallel}$ and $\nu_{\perp}$ are corresponding critical exponents. 
 Surprisingly, though the divergence of the transient time scales and the ``armoring'' processes in the transients are reported~\cite{Houssais15}, quantitative studies of the divergence~\cite{Clark15} and the spatial organization of the flux have barely been done. 
 
 Two distinct theoretical views have been proposed for the erosion near the threshold. Bagnold and followers~\cite{Bagnold66,Chiodi14} introduced the concept of ``a moving flow boundary'', where active particles forming a layer carry a fraction of the shear stress from fluid such that the lower layers remain static under the critical shear which is balanced by friction. The depth of the flow boundary, or the amount of moving particles, is thus proportional to $\theta-\theta_c$. In this view, the dynamic response recovers the critical flux ({\it i}) and the vanishing number of moving particles ({\it ii}). However, this hydrodynamic treatment that the active particles in the boundary move in 
an average manner captures no transients ({\it iii}) nor spatial organization of flux and applies only when active particle interactions are irrelevant. %without the geometric details of the granular bed nor particle interactions applies only when moving particles are far apart. 
 
 By contrast, the other view raised in erosion/deposition models~\cite{Charru04} emphasizes the slow ``armoring'' of the particle bed. Models assume that initially active particles, moving on a frozen static background, may be trapped by ``holes'', which are energy depressions in the landscape. Consequently, the number of active particles contributing to the steady flux is less than the initial number of them. The shear stress $\theta$ tunes the number of these energetic ``holes'', and $\theta_c$ corresponds to the critical stress where the number of holes matches the number of initially active particles. 
 Although this view captures ({\it i,ii,iii}) qualitatively well, that some of the active particles have to fill all holes and the rest contributes to the steady flux is an implicit assumption, which is highly non-trivial. In fact, the disorder of the static bed will lead the mobile particles to follow favored paths which eventually lead to a few channels, thus exploring only a small fraction of the space, as found in river channels and the aggregations of adhesive particles~\cite{Scheidegger67,Takayasu88,Dhar06,Rinaldo14}. Some plastic-depinning models of vortex dynamics in dirty Type-II superconductors~\cite{Watson96,Watson97} have argued that the fact that the active particles explore a vanishing fraction of the surface at the threshold implies a vanishing number of holes to be filled and thus a steeper change in the flux, with $\beta>1$. 

 %To decide which physical process (hydrodynamic interactions or armoring) governs the erosion threshold, new theoretical predictions must be made and put to experimental test. In this letter, we achieve the first step of this goal while resolving the apparent contradictions of deposition models. 
 
 To settle these problems in previous views, we start from a dynamical model capturing the microscopic details missed before, 
 in particular, the interplay between the disorder that leads to channelization and the particle interaction. %We are curious to see whether the critical behavior of the erosion can be reproduced in the model, what is the universality class that the erosion falls in, and what phenomena to expect from the prediction of the universality class.
 We then build a theoretical framework based on the model to reveal the microscopic cause for the critical flux-drive relation, Eq.(\ref{intro_order}). Our theory also provides a testable description on the spatial organization of the flux.

 \section{Critical dynamics}
 \label{intro_4}
 %SOC models, marginal stability
 %SOC what, why important glass ageing, avalanche, and examples in different systems, features, explained with sand pile models and force chain models. Similar idea: marginal stability, what and why, critical dynamics in different materials. Newton deterministic dynamics and stochastic dynamics, critical depinning as an example. 
 In the erosion problem, the granular bed self-organizes into an ``armored'' state where a minimal number of particles are mobile. 
 Athermal glassy systems with long-range interactions can also self-organize into microscopic states that have rich dynamics and sensitive responses~\cite{Bak87,Paczuski96,Bak13}, characterized by diverging length and time scales. Their dynamics are in some sense critical: a tiny local perturbation ends up with a response, called an ``avalanche'', which can spread over the system and last very long. 
 The frequency distributions of the lifespans and sizes of the response obey a power law, 
 \be
 P(A)\sim A^{-\tau}p(A/N^{\sigma}),
 \label{intro_critical}
 \ee
 where $A$ can be either the avalanche duration or size, $N$ is the system size, and $N^{\sigma}$ sets a cutoff that diverges in the thermodynamic limit. 
 Common examples of these ``scale-free'' avalanches are Barkhausen noises and earthquakes. 
 
\begin{figure}[h!]
\centering
\includegraphics[width=.9\columnwidth]{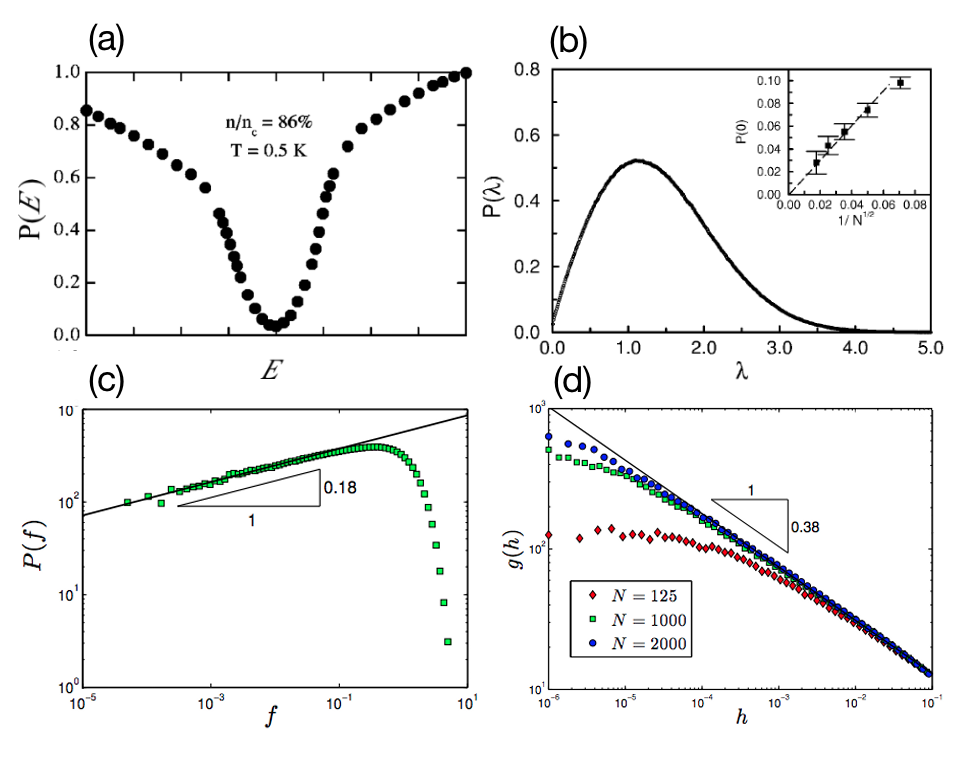}
\caption{\small{Pseudogap in distribution of local stabilities in various glassy systems. (a) Density of states $P(E)$ in Coulomb glasses, where the gap is centered at the Fermi level; (b) Distribution of local fields $P(\lambda)$ in spin glasses; (c) Distribution of contact forces $p(f)$ and (d) distribution of contact gaps $g(h)$ in packings of hard spheres. The plots are reproduced from References~\cite{Massey95,Pazmandi99,Lerner13a}.
}}\label{intro_merge}
\end{figure}

 An important feature of such glassy systems is that the densities of soft excitations are singular~\cite{Muller14}, corresponding to the so-called pseudogap: 
 \be
 \rho(\lambda)\sim\lambda^{\theta}
 \label{intro_rho}
 \ee
 where the local stability $\lambda$ quantifies the external field needed to cause an elementary excitation in the system. As shown in Fig.~\ref{intro_merge}, such singular distributions are observed in various glassy systems. In Coulomb glasses~\cite{Muller14,Efros75,Pollak13}, the energy to excite an electron is determined by the electron energy $E$ and its distance to the Fermi level $E_f$. Experiments find the density of states of electrons is gapped at $E_f$, shown in Fig.~\ref{intro_merge}(a). In spin glasses~\cite{Muller14,Pazmandi99,Pankov06,Horner07,Doussal10,Le-Doussal12}, the magnitude of the local magnetic field defines the energy to flip the corresponding spin. Numerics show that the density of local fields $\lambda$ vanishes linearly at $\lambda=0$ in mean-field spin glasses, as shown in Fig.~\ref{intro_merge}(b). In random packings of hard particles~\cite{Muller14,Wyart12,Lerner13a,DeGiuli14b,Lerner12,Lerner13a,Charbonneau15}, the contact force $f$ characterizes the difficulty to break a contact and the depth $h$ of a gap between two particles features the difficulty to close it and to form a new contact. The densities of both contacts with force $f$ and gaps separated by $h$ follow the singular distribution, Eq.(\ref{intro_rho}), shown in Fig.~\ref{intro_merge} (c) and (d). 
  
 The pseudogap exponent $\theta$ is in general bounded by a stability requirement~\cite{Muller14}, first recognized by Effros and Shklovskii~\cite{Efros75}. %For instance, in hard sphere packings, the distribution of forces between particles in contact must vanish analogously while the distribution of gaps must diverges, preventing that collective motions of particles from breaking soft contacts without making new contacts. 
 For instance, in the example of a mean-field spin glass, the Sherrington-Kirkpatrick (SK) model, Ising spins are randomly coupled with each other, defined by the Hamiltonian,
 \be
 \mh = -\frac{1}{2}\sum_{i,j}J_{ij}s_is_j-h\sum_is_i,
 \label{intro_sk}
 \ee
 where $J_{ij}$ are independent random variables obeying a Gaussian distribution with zero mean and variance $1/N$ ($N$ is the system size.). 
 The energy cost (or gain, if the sign is negative) of flipping a set of spins $\mf$ from an initial state is,
 \be
 \Delta\mh=2\sum_{i\in\mf}\lambda_i-2\sum_{i,j\in\mf}J_{ij}s_is_j
 \label{intro_cost}
 \ee
 where $\lambda_i\equiv h_is_i=(h+\sum_{j\neq i} J_{ij}s_j)s_i$ defines the local stability, the energy cost to flip one spin. A trivial stability requirement demands that every single-spin flip costs energy, i.e., $\lambda_i>0$ for $\forall i$. For two-spin flips, from Eq.(\ref{intro_cost}), the stability meets when $\lambda_i+\lambda_j-2J_{ij}s_is_j\geq0$ for $\forall i,j$. The worst case to ensure this two-spin flip stability is to flip the two of the lowest $\lambda$ which are correlated with a positive $J_{ij}s_is_j$. In this case, $\lambda\sim N^{-1/(1+\theta)}$, presuming the pseudogap distribution, Eq.(\ref{intro_rho}), and the typical magnitude of the correlation is determined by the variance of the Gaussian distribution $J_{ij}s_is_j\sim1/\sqrt{N}$. Therefore, the pseudogap exponent $\theta$ is bounded by $\theta\geq1$ in the SK model~\cite{Palmer79,Anderson79}..

\begin{figure}[h!]
\centering
\includegraphics[width=.8\columnwidth]{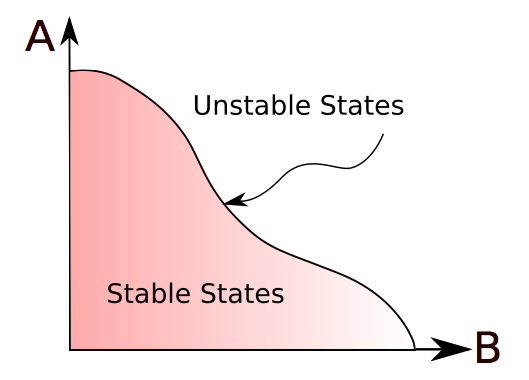}
\caption{\small{Illustration of the evolution of a dynamical system from unstable states to the margin of the stable states in the phase space. A and B are arbitrary coordinates of the phase space. The plot is reproduced from Reference~\cite{Muller14}.
}}\label{intro_marginal}
\end{figure}

 Very often, these stability bounds are saturated, so that the exponent $\theta$ is the minimum one guaranteeing stability, $\theta=1$, as shown in Fig.~\ref{intro_merge}(b) for the SK spin glass.  Such marginal stability has been proven for dynamical, out-of-equilibrium situations under a quasi-static driving at zero temperature in the glassy systems with sufficiently long-range effective interactions~\cite{Muller14}. The emerging scenario is described in Fig.~\ref{intro_marginal}, underlying that the dynamics can only probe the boundary between the stable states and the unstable states.
 
 %Despite the great success of the self-organized criticality theory, what set of general characteristics that guarantee a system to spontaneously display critical dynamics is yet unknown. Recent studies on the marginal stability of jammed hard spheres and spin glasses inspire some hints to the question. Marginal stability assumption~\cite{Muller14,Wyart12} argues that the states visited in the critical dynamics are marginally stable, that is to say, the states are stable but vulnerable to infinitesimal perturbation in driving force.
  
 The presence of a pseudogap can be shown to be intrinsically related to the presence of power law avalanches~\cite{Muller14}. %Scale-free transient processes are triggered by changing adiabatically the external field in spin glasses, or the charge density in Coulomb glasses, or the shear strain in amorphous packings. 
 %where local excitations are strongly coupled. As a result, a local dynamical event, like flipping a spin, may destabilize many distant components, which cascades the excitations and ends up into a large dynamical response. Consider a typical change in the driving force, perturbing the least stable sites, 
 %\be
 %\Delta\lambda\sim\lambda_{\rm min}\sim N^{-1/(1+\theta)}
 %\label{intro_lm}
 %\ee
 %where $N$ is the system size. It triggers an avalanche characterized by a change in the coupled dynamical quantity $A$, whose density usually changes linearly with the driving force $\rd A/N\rd\lambda\sim o(1)$. Therefore, the average avalanche size in terms of change in the dynamical quantity scales as, $\overline{\Delta A}=\frac{\rd A}{\rd\lambda}\Delta\lambda\sim N^{\theta/(\theta+1)}$, a power law to system size if $\theta>0$. 
 %The pseudogap distributions in these glassy systems are commonly deduced from a marginal stability argument -- equating the total stability cost and the energy gain from the interactions in a set of local dynamical events~\cite{Palmer79,Anderson79}. 
 %Although we know the pseudo gaps play key roles in the critical dynamics in these glassy systems, and their existence can be justified by stability, 
 However, how such pseudogaps emerge dynamically is not understood. %You will understand that in the simple system of spin glasses. Say what you have shown/understood, and how generic it may be. 
 Currently, thermodynamic calculation of the distribution of local stabilities has been worked out for the ground state of specific glassy systems~\cite{Doussal10,Le-Doussal12,Charbonneau14,Charbonneau15}, but these arguments do not apply in the relevant context of driven athermal systems. In this dissertation, we will explain how the pseudogap appears dynamically in the mean-field spin glass. This work leads to the novel idea that soft excitations are singularly anti-correlated, which we believe will apply broadly to other glassy systems.

 \section{Structure of the dissertation}
 \label{intro_5}
 %In this dissertation, I integrate all my graduate studies to the above questions. 
 The dissertation is organized in the following order. 
 In {\it Chapter 2}, we develop an elastic network model of network glasses to unveil the microscopic mechanism that determines the dynamic and thermodynamic properties of glasses, and we derive a thermodynamic theory to predict the non-monotonic dependence of the specific heat on the coordination number. The chapter is a reproduction of the work published in Reference~\cite{Yan13}. 
 In {\it Chapter 3}, we study how the network topology evolves under cooling with a network model with a topology adaptive to the temperature, and we show that the mean-field rigidity transition rather than the rigidity window would apply  to the real network glasses due to the existence of Van der Waals interactions. We reproduce the chapter essentially from the Reference~\cite{Yan14}.
 In {\it Chapter 4}, we concentrate on the same question as in {\it Chapter 2}, but including the fact that the network topology evolves under cooling with the model studied in {\it Chapter 3}. We derive the thermodynamics of the model with features not captured in mean-field rigidity transition. The chapter is a duplication of our work in the Reference~\cite{Yan15b}.
 In {\it Chapter 5}, we investigate the dynamics of driven particles on random surface, which models the erosion of granular river beds. We show model recovers the critical flux-drive relation observed in experiments, as it captures a mechanism missed in the literature. We derive testable predictions on the spatial organization of the erosion flux. The equivalent work is published as the Reference~\cite{Yan15a}.
 In {\it Chapter 6}, we reveal the dynamical emergence of the pseudogap in the mean-field spin-glass with a stochastic description of the critical dynamics. A non-trivial singular correlation among soft excitations arising in the dynamics is the key to the pseudogap. This mechanism can be generalized to other glassy systems. We have published the main content of the chapter in the Reference~\cite{Yan15}.
 In {\it Chapter 7}, we list the open questions following up with our published works and we discuss briefly the possible methods and  preliminary results to tackle them. 
 We conclude the questions and the results on the dynamics of glassy systems in the end.

%% file: thermodyn/thermodyn.tex
\chapter{Why glass elasticity affects the thermodynamics and  fragility of super-cooled liquids}
\label{2_elasticity}
%\begin{abstract}

%When a liquid is cooled, its viscosity increases up to the glass transition where the material becomes solid. 
Super-cooled liquids are characterized by their fragility:  the slowing down of the dynamics under cooling is more sudden and the jump of specific heat at the glass transition is generally larger in fragile liquids than in strong ones. Despite the importance of this quantity in classifying liquids, explaining what aspects of the microscopic structure controls fragility remains a challenge. 
Surprisingly, experiments indicate that the linear elasticity of the glass -- a purely local property of the free energy landscape -- is a good predictor of fragility.  In particular, materials presenting a large excess of  soft elastic modes, the so-called boson peak, are strong. This is also the case for  network liquids near the rigidity percolation, known to affect elasticity. Here we introduce a model of the glass transition based on the assumption that particles can organize locally into distinct configurations,  which are coupled  spatially via elasticity. The model captures the mentioned observations connecting elasticity and fragility. We find that materials presenting an abundance of soft elastic modes have little elastic frustration: energy is insensitive to most  directions in phase space, leading to a small jump of specific heat. In this framework  strong liquids turn out to lie the closest to a critical point associated with a rigidity or jamming transition, and their thermodynamic properties are related to the problem of number partitioning and to Hopfield nets in the limit of small memory.

%\end{abstract}

%\keywords{Glass transition | Elasticity | Fragility | Rigidity percolation | Boson peak}

%\abbreviations{DOS, density of states; MCS, Monte Carlo step; CLT, central limit theorem; REM, random energy model; ROM, random orthogonal model}

%\pagebreak

\section{Introduction}

When a liquid is cooled rapidly to avoid crystallization, its viscosity increases up to the glass transition where the material becomes solid. Although this phenomenon was already  used in ancient times to mold artifacts, the nature of the glass transition and the microscopic cause for the slowing down of the dynamics 
remain controversial. Glass-forming liquids are characterized by their fragility \cite{Ediger96,Debenedetti01}: the least fragile liquids are called strong, and their characteristic time scale $\tau$
follows approximatively an Arrhenius law $\tau(T)\sim \exp(E_a/k_BT)$, where the activation energy $E_a$ is independent of temperature. Instead in fragile liquids the activation energy 
grows as the temperature decreases, leading to a sudden slowing-down of the dynamics. The fragility of liquids strongly correlates with their thermodynamic properties \cite{Wang06,Martinez01}: the jump in the specific heat that characterizes the glass transition is large in fragile liquids and moderate in strong ones. Various theoretical works \cite{Adam65,Lubchenko07,Bouchaud04}, starting with Adam and Gibbs, have proposed explanations for such correlations. By contrast few propositions, see e.g. \cite{Hall03,Shintani08}, have been made to understand which aspects of the microscopic structure of a liquid determines its fragility and the amplitude of the jump in the specific heat at the transition. This  question is conceptually important, but also practically, as solving it would help engineering materials with desired properties.  %Thus a framework is lacking to unify the various observations characterizing the relationships between these quantities.% have received no explanations. 
%This state of affair is surprising, since various empirical data characterize this relationship, and could be used to constrain our understanding of the glass transition. 

Observations indicate that the linear elasticity of the glass is a key factor determining fragility -- a fact a priori surprising since linear elasticity is a local property of the energy landscape, whereas fragility is a non-local property characterizing transition between  meta-stables states. In particular (i) glasses are known to present an excess of soft elastic modes with respect to Debye vibrations at low frequencies, %their crystalline counterpart, 
the so-called boson peak  that appears in scattering measurements \cite{Phillips81}. The amplitude of the boson peak is strongly anti-correlated with fragility, both in network and molecular liquids: structures presenting an abundance of soft elastic modes tend to be strong \cite{Novikov05,Ngai97}.
(ii) In network glasses, where particles interact via covalent bonds and via the much weaker Van der Waals interactions, the microscopic structure and the elasticity can be monitored  by changing continuously the composition of compounds \cite{Tatsumisago90,Ito99,Boolchand05,Kamitakahara91}. %, 
As the average valence $r$ is increased toward some threshold $r_c$, the covalent networks display a rigidity transition \cite{Phillips79,Phillips85} where the number of covalent bonds is just sufficient to guarantee mechanical stability.  Rigidity percolation has striking effects on the thermal properties of super-cooled liquids: in its vicinity, liquids are strong \cite{Boolchand05,Bohmer92} and the jump of specific heat  is  small \cite{Tatsumisago90}; whereas they become %more 
fragile with a large jump in specific heat  {\it both} when the valence is increased,  and %or 
decreased \cite{Boolchand05,Tatsumisago90,Bohmer92}. There are currently no explanations to why increasing the valence affects the glass transition properties in a non-monotonic way, and why such properties are extremal when the covalent network acquires rigidity.

Recently it has been shown that the presence of soft modes in various amorphous materials, including granular media \cite{OHern03,Brito10,Wyart05,Wyart05b}, Lennard-Jones glasses \cite{Wyart05b,Xu07}, colloidal suspensions \cite{Brito09,Ghosh10,Chen10} and silica glass \cite{Trachenko00,Wyart05b} was controlled by the proximity of a jamming transition\cite{Liu10}, a sort of rigidity transition  that occurs for example when purely repulsive particles are decompressed toward vanishing pressure \cite{OHern03}. Near the jamming transition spatial fluctuations play a limited role and simple theoretical arguments \cite{Wyart05,Wyart05b} capture the connection between elasticity and structure. They imply that soft modes must be abundant near the transition,  suggesting a link between observations (i) and (ii). However these results apply to linear elasticity and cannot explain intrinsically non-linear phenomena such as those governing fragility or the jump of specific heat. 
In this article we propose to bridge that gap by introducing a  model for the structural relaxation in super-cooled liquids. Our starting assumption is that  particles can organize locally into distinct configurations,  which are coupled at different points 
in space via elasticity. We study what is perhaps the simplest model  realizing this idea, and show numerically that it captures qualitatively the relationships between elasticity, rigidity, thermodynamics and fragility. 
The thermodynamic properties of this model can be treated theoretically within a good accuracy in the temperature range we explore. Our key result is the following physical picture:  when there is an abundance
of soft elastic modes, elastic frustration %what is this jargon mean? 
vanishes, in the sense that 
a limited number of directions in phase space cost energy. Only those directions contribute to the specific heat, which is thus small. Away from the critical point, elastic frustration increases: 
more degrees of freedom contribute to the jump of specific heat, which increases while the boson peak is reduced.
%The framework we propose makes a new connection between elasticity and energy landscape, in which strong liquids turn out to lie the closest to a critical point governing fragility. In this view strong liquids are closely related to the problem of number partitioning, and to Hopfield neuron networks in the limit of small memory. %Should we mention this here? 
%

\section{Random Elastic Network Model}

Our main assumption is 
that in a super-cooled liquid, nearby particles can organize themselves into a few distinct configurations. Consider for example covalent networks sketched in Fig.~1, where we use the label $\langle ij\rangle$ to indicate the existence of a covalent bond between particles $i$ and $j$. If two  covalent bonds $\langle 12\rangle$ and $\langle34\rangle$  are adjacent, there exists locally another configuration for which these bonds are broken, and where the bonds $\langle13\rangle$ and $\langle24\rangle$ are formed instead. These two configurations do not have the same energy in general. Moreover going from one configuration to the other generates a local strain, which creates an elastic stress  that propagates in space. In turn, this stress changes the energy difference between local configurations elsewhere in the system. This process leads to an effective interaction between local configurations at different locations. %

\begin{figure}[h!]
   \begin{center}   
   {\includegraphics[width=0.8\columnwidth]{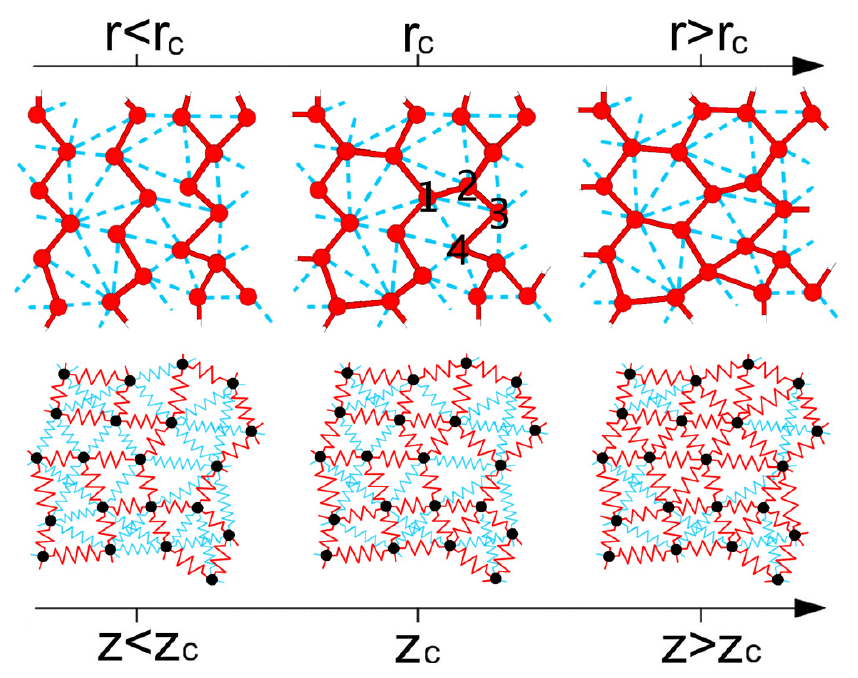}}     
   \caption{Top row:  sketches of covalent networks with different mean valence $r$ around the valence $r_c$: red solid lines represent covalent bonds; cyan dash lines represent van der Waals interactions. Bottom:  sketch of our elastic network model with varying coordination number $z$ (defined as the average number of strong springs in red) around Maxwell threshold $z_c$; cyan springs have a much weaker stiffness, and model weak interactions.}
  \label{2_fig1}
   \end{center}
\end{figure}

Our contention is that even a simple description of the  local configurations  -- in our case we will consider two-level systems, and we will make the approximation that the elastic properties do not depend on the levels -- can capture several unexplained aspects of super-cooled liquids, as long as the salient features of the elasticity of amorphous materials are taken into account. 
To incorporate in particular the presence of soft modes in the vibrational spectrum  we  consider random elastic networks. The elasticity of three types of networks have been studied extensively: networks of springs randomly deposited on a lattice \cite{Garboczi85}, on-lattice self-organized networks \cite{Boolchand05} and off-lattice random networks with small spatial fluctuations of coordination \cite{Wyart08,During13,Ellenbroek09}. We shall  consider the third class of networks, which  are known to 
capture correctly the scaling properties of elasticity near jamming, and can be treated analytically \cite{Wyart05,During13,Wyart10a}.  %by contrast with purely repulsive.
 In our model two kinds of springs connect the $N$ nodes of the network: strong ones, of stiffness $k$ and coordination $z$, and weak ones,  of stiffness $k_{\rm w}$ and coordination $z_{\rm w}$.  These networks undergo a rigidity %why not rigidity? With soft interactions like springs, I don't think a jamming transition happens, right?
transition as $z$ crosses $z_c=2d$, where $d$ is the spatial dimension. For $z<z_c$ elastic stability is guaranteed by the presence of the weak springs. As indicated in Fig.~1, this situation is similar to covalent networks, where the weak Van der Waals interactions are required to insure stability when the valence $r$ is smaller than its critical value $r_c$.

Initially when our network is built, every spring $\langle ij\rangle$ is at rest: the rest length follows $l^0_{\langle ij\rangle}=||{\bf R}_i^0-{\bf R}_j^0||$, where ${\bf R}_i^0$ is the initial position of the node $i$. To allow for local changes of configurations we shall consider that any strong spring $\langle ij\rangle$ can switch between two rest lengths: $l_{\langle ij\rangle}=l^0_{\langle ij\rangle}+\epsilon \sigma_{\langle ij\rangle}$, where  $\sigma_{\langle ij\rangle}=\pm1$ is a spin variable. % that can take the value $+1$ or $-1$. 
There are thus two types of variables: the $N_s\equiv zN/2$ spin variables $\{\sigma_{\langle ij\rangle}\}$, which we shall denote using the ket notation $|\sigma\rangle$, and the $Nd$ coordinates of the particles denoted by $|{\bf R}\rangle$. % Note that the rest length of the $N_{\rm w}\equiv z_{\rm w} N/2$ weak spring remain unchanged.  
The elastic energy ${\cal E}( |{\bf R}\rangle,|\sigma\rangle)$ is a function of both types of variables.
%\ba
%\label{2_1}
%{\cal E}( |{\bf R}\rangle,|\sigma\rangle)=\sum_{strong \langle ij\rangle}\frac{k}{2}(||{\bf R}_i-{\bf R}_j||-||{\bf R}_i^0-{\bf R}_j^0||-\epsilon \sigma_{\langle ij\rangle})^2 \notag\\
%+\sum_{weak \langle ij\rangle}\frac{k_{\rm w}}{2}(||{\bf R}_i-{\bf R}_j||-||{\bf R}_i^0-{\bf R}_j^0||)^2
%\ea
The inherent structure energy  ${\cal \tilde H}(|\sigma\rangle)$ associated with any configuration $|\sigma\rangle$ is defined as:
\be
\label{2_2}
{\cal \tilde H}(|\sigma\rangle)\equiv \hbox{ min}_{|{\bf R}\rangle}{\cal E}(|{\bf R}\rangle,|\sigma\rangle)\equiv k\epsilon^2 {\cal H}(|\sigma\rangle) 
\ee
where we have introduced the dimensionless Hamiltonian ${\cal H}$. 
We shall consider the limit of small $\epsilon$, where the vibrational energy is simply that of harmonic oscillators. In this limit all the relevant information is contained in the inherent structures energy, since including the vibrational energy would increase the specific heat by a constant, which does not contribute to the jump of that quantity at the glass transition.
In this limit,  linear elasticity implies the form:
\be
\label{2_3}
%{\cal H}(|\sigma\rangle)= \frac{k \epsilon^2}{2} \sum_{\gamma \neq \beta} {\cal G}_{\gamma,\beta}\sigma_\gamma\sigma_\beta + o(\epsilon^2)\equiv \frac{k \epsilon^2}{2}\langle \sigma|{\cal G}|\sigma\rangle +o(\epsilon^2)
{\cal H}(|\sigma\rangle)= \frac{1}{2} \sum_{\lambda, \beta} {\cal G}_{\lambda,\beta}\sigma_\lambda\sigma_\beta + o(\epsilon^2)\equiv \frac{1}{2}\langle \sigma|{\cal G}|\sigma\rangle +o(\epsilon^2)
\ee
%I feel the diagonal terms are unnecessarily zero. so the sum should include gamma=beta
where $\lambda$ and $\beta$ label strong springs, ${\cal G}_{\lambda,\beta}$ is the Green function describing how a dipole of force applied on the contact $\lambda$ changes the amplitude of the force in the contact $\beta$. ${\cal G}_{\lambda,\beta}$ is computed in  Appendix Sec.~\ref{app_A1} and reads: 

\be
\label{2_green}
{\cal G}= {\cal I}-{\cal S}_{\rm s}{\cal M}^{-1}{\cal S}^t_{\rm s}
\ee
where ${\cal I}$ is the identity matrix, and   ${\cal S}_{\rm s}$ and the dimensionless stiffness matrix  ${\cal M}$ are standard linear operators connecting forces and displacements in elastic networks \cite{Calladine78}. They can be formally written as:
\begin{eqnarray}
\label{2_stiffM}
{\cal M}&=&{\cal S}^t_{\rm s}{\cal S}_{\rm s}+\frac{k_{\rm w}}{k}\,{\cal S}^t_{\rm w}{\cal S}_{\rm w},\\
{\cal S}_{\bullet}&=&\sum_{\langle ij\rangle_\bullet\equiv \gamma}| \gamma \rangle {\bf n}_{ij} (\langle i | -\langle j |)\nonumber
\end{eqnarray}
where $\langle i| {\bf R}\rangle\equiv {\bf R}_i$, $\langle ij\rangle_{\rm \bullet}$ indicate a summation over the strong springs ($\bullet=\rm s$) or the weak springs ($\bullet=\rm w$). 
 ${\cal S}_{\bullet}$ is a $N_\bullet\times dN$ matrix which projects any displacement field   onto the contact space of strong  or weak  springs. The components of this linear operator   are uniquely determined by  the unit vectors ${\bf n}_{ij}$   directed  along the contacts $\langle ij\rangle$ and point toward the node $i$.  

Finally note that the topology of the elastic network is frozen in our model. This addition of frozen disorder is obviously an approximation, as the topology itself should evolve as local configurations change. 
Building models which incorporate this possibility, while still  tractable numerically and  theoretically, remains a challenge. 
%Our simplified model can be understood theoretically, at least its thermodynamic properties,
%and as we shall see it appears to capture essential aspects of the relationship between elasticity, thermodynamics and dynamics. 

\section{Numerical Results of the Model}

\subsection{Network structure} Random networks with weak spatial fluctuations of coordination can be generated from random packings  of compressed soft particles \cite{Wyart08,During13,Ellenbroek09}. We consider packings with periodic boundary conditions. 
The centers of the particles correspond to the nodes of the network,  of unit mass $m=1$, and un-stretched springs of  stiffness $k=1$ %(whose unit is such that the energy scale $k\epsilon^2$ is unity)
 are put between particles in contact. 
 %Their average length $a$ defines our unit of length. 
 Then springs are removed,
preferably where the local coordination is high, so as to achieve the desired coordination $z$. 
%{\bf Removing the springs preferably where the local coordination is high, the networks achieve the desired coordinations $z$ with average spring length $a$ as unit of length.} %the unit of length is calculated for network after $z$ reached, and this $a$ is not even necessary here. What we are using latter is just $\epsilon$
In a second phase, $N_{\rm w}$ weak springs are added between the closest unconnected pairs of nodes.
The relative effect of those weak springs is best characterized by $\alpha\equiv (z_{\rm w}/d) (k_{\rm w}/k)$, which we modulate by fixing $z_{\rm w}=6$ and changing $(k_{\rm w}/k)$.
Note that an order of magnitude estimate of $\alpha$ in covalent glasses can be obtained by comparing  the behavior of the shear modulus $\ G$ in the elastic networks \cite{Wyart08} and in network glasses near the rigidity transition. As shown In Fig.~\ref{app_AA} of Appendix Sec.~\ref{app_A2}, this comparison yields the estimate that $\alpha \in [0.01,0.05]$.

\subsection{Thermodynamics} We introduce the rescaled temperature $T={\tilde T}k_B/(k\epsilon^2)$ where ${\tilde T}$ is the temperature. %, $\epsilon\ll a$. 
To equilibrate the system, we perform a one spin-flip Monte Carlo algorithm. The energy ${\cal H}$ of configurations are computed using Eq.(\ref{2_3}). 
%An alternative computation uses Eq.(\ref{2_2}) together with a conjugate gradient method, which  gives identical results when $\epsilon$ is small but is much slower. 
We  use 5 networks of $N=256$ nodes in two dimensions and $N=216$ in three dimensions, each run  with 10 different initial configurations. Thus our results are averaged on these 50 realizations. We perform $10^9$ Monte Carlo steps  at each $T$.  The time-average inherent structure energy $E(T)$ is calculated, together with the specific heat $C_v=\partial E/\partial T$.  The intensive quantity  $c(T)\equiv C_v/N_s$ is represented in Fig.~2 for various excess coordination $\delta z=z-z_c$ and $\alpha=3\times10^{-4}$. We observe that the specific heat increases under cooling, until the glass transition temperature $T_g$ where $c(T)$ rapidly vanishes, indicating that the system falls out of equilibrium. %that a glass is formed. 

The amplitude of $c(T)$ just above $T_g$ thus corresponds to the jump of specific heat $\Delta c$, and is shown in Fig.~3. Our key finding is that as the coordination increases,  $\Delta c(z)$ varies non-monotonically and is  minimal in the vicinity of the rigidity transition for all values of $\alpha$ investigated, as observed experimentally \cite{Tatsumisago90,Boolchand05}. This behavior appears to result from a sharp asymmetric transition at $\alpha\to 0$. For $z>z_c$ we observe that $\Delta c(z)\propto \delta z$. The jump in specific heat thus vanishes as $\delta z\rightarrow 0^+$ where the system can be called  ``perfectly strong".  For $z<z_c$, $\Delta c$ is very rapidly of order one. %It's not obvious here.
When $\alpha$ increases, this sharp transition becomes a cross-over, marked by a minimum of $\Delta c(z)$ at some coordination larger but close to $z_c$.

\begin{figure}[h!]
   \begin{center}
     {\includegraphics[width=0.8\columnwidth]{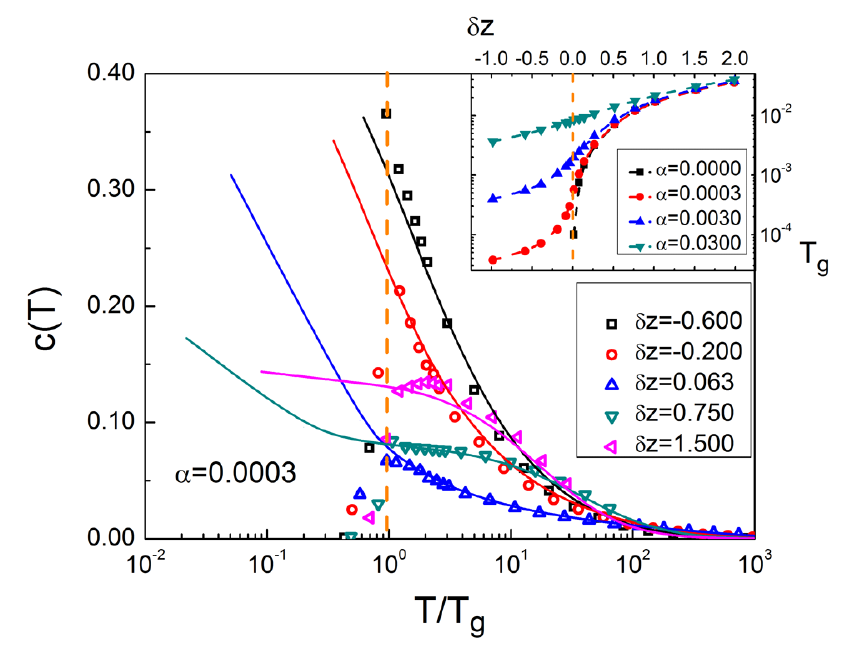}}
     \caption{Specific heat $c(T)$ {\it v.s.} rescaled temperature $T/T_g$ for various excess coordination $\delta z\equiv z-z_c$ as indicated in legend, for $\alpha=3\times10^{-4}$ and $d=2$.  $c(T)$ displays a jump at the glass transition. Solid lines are theoretical predictions, deprived of any fitting parameters, of our mean-field approximation. They terminate at the Kautzman temperature $T_K$. Inset: glass transition temperature $T_g$ {\it vs} $\delta z$ for several amplitude of weak interactions $\alpha$, as indicated in legend. }
     \label{2_f2A}
   \end{center}
\end{figure}

%\vspace{5cm}
\begin{figure}[h!]
   \begin{center}
     {\includegraphics[width=0.8\columnwidth]{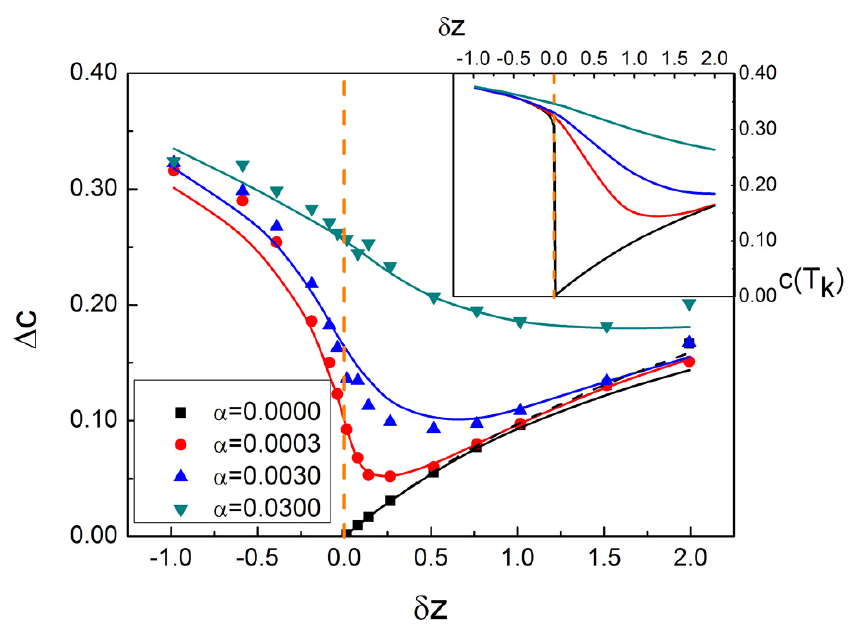}}
     \caption{Jump of specific heat $\Delta c$ versus excess coordination $\delta z$ in $d=2$ for different strength of weak springs $\alpha$, as indicated in legend. Solid lines are mean-field predictions not enforcing the orthogonality of the $|\delta r_p\rangle$, dashed-line corresponds to the ROM where orthogonality is enforced. In both cases the specific heat is computed at the numerically obtained temperature $T_g$. Inset: theoretical predictions for $\Delta c$ {\it vs} $\delta z$ computed at the theoretical temperature $T_K$.}
     \label{2_f2B}
   \end{center}
\end{figure}

\subsection{Dynamics} To characterize the dynamics  we compute the correlation function $C(t)=\langle\sigma(t)|\sigma(0)\rangle$, which decays to zero at long time in the liquid phase. We define the relaxation time $\tau$ as $C(\tau)=1/2$, and the glass transition temperature $T_g$ as $\tau(T_g)/\tau(\infty)=10^5$. Finite size effects on $\tau$ appear to be weak, as shown in Appendix Sec.~\ref{app_A3}.
The Angell plot representing the dependence of $\tau$ with inverse rescaled temperature is shown in the inset of Fig.~4. It is found that the dynamics follows an Arrhenius behavior for $\alpha\rightarrow 0$ and $z\approx z_c$. Away from the rigidity transition, the slowing down of the dynamics is faster than Arrhenius. To quantify this effect we compute the fragility $m\equiv\frac{\partial\log\tau}{\partial(T_g/T)}|_{T=T_g}$, whose variation with coordination is presented in Fig.~4. Our key finding is that for all weak interaction amplitudes $\alpha$ studied, the fragility depends non-monotonically on coordination and is minimal near the rigidity transition, again as observed empirically in covalent liquids \cite{Bohmer92}. As was the case for the thermodynamic properties, the fragility appears to be controlled by a critical point present at $\alpha=0$ and $z=z_c$ where the liquid is strong, and the dynamics is simply Arrhenius. As the coordination changes and $|\delta z|$ increases, the liquid becomes more fragile. The rapid change of fragility near the rigidity transition is smoothed over  when the amplitude of the weak interaction $\alpha$ is increased. 

\subsection{Correlating boson peak and fragility} The presence of soft elastic modes in glasses is traditionally analyzed by considering the maximum of  $Z(\omega)\equiv \frac{D(\omega)}{\omega^2}$ \cite{Phillips81}, where $D(\omega)$ is the vibrational density of states. $Z(\omega)$  quantifies the departure from Debye behavior. The maximum of $Z(\omega)$ defines the boson peak frequency $\omega_{BP}$ \cite{Phillips81}. To characterize the amplitude of the peak, Sokolov and coworkers \cite{Novikov05,Ngai97} have introduced a dimensionless quantity $\tilde R_1\equiv \frac{Z(\omega_{min})}{Z(\omega_{BP})}$, where $\omega_{min}$ is the minimum of $Z(\omega)$ for $\omega\in[0,\omega_{BP}]$. $\tilde R_1$ characterizes the inverse amplitude of the boson peak, and was shown to strongly correlate with  fragility \cite{Novikov05,Ngai97} both in molecular liquids and covalent networks. %I remember they also tested polymer glasses.

To test if our model can capture this behavior we compute the density of states via a direct diagonalization of the stiffness matrix, see Eq.(\ref{2_stiffM}). Then we extract the maximum $Z(\omega_{BP})$ of $D(\omega)/\omega^2$. We find that below this maximum, $Z(\omega)$ is monotonic, implying that $\omega_{min}=0$.  For all coordinations if $\alpha>0$ the density of states follows a Debye behavior at low frequency in such networks \cite{During13,Wyart10a}, and in three dimensions $D(\omega)\sim \omega^2/G^{3/2}$ where $G$ is the shear modulus. Thus $\tilde R_1\sim 1/(G^{3/2} Z(\omega_{BP}))\equiv R_1$. The dependence of $R_1$ is represented in Fig. 5 and shows a minimum near the rigidity transition, and even a cusp in the limit $\alpha\rightarrow 0$.  % Thus $R_1$ is a good indicator of the proximity of the rigidity transition in three dimensions. 
This behavior can be explained in terms of previous theoretical results on the density of states near the rigidity transition, that supports that $R_1\sim |\delta z|^{1/2}$ when $\alpha\rightarrow 0$ \footnote[1]{ When $\alpha \rightarrow 0$ and $\delta z>0$, $\omega_{BP}\sim \delta z$ and $D(\omega_{BP})\sim 1 $ \cite{Wyart05}, whereas $G\sim \delta z$ \cite{Wyart05b}, leading to $R_1\sim \sqrt{\delta z}$. For $\delta z<0$, $G\sim -\alpha/\delta z$ \cite{Wyart08}. On the other hand the boson peak is governed by the fraction $\sim \delta  z$ of floppy modes, which gain a finite frequency $\sim \sqrt{\alpha}$ \cite{During13} thus we expect $D(\omega_{BP})\sim -\delta z/\sqrt{\alpha}$ and $R_1\sim \sqrt{-\delta z}$.}.

\begin{figure}[hb!]
   \begin{center}
        {\includegraphics[width=0.8\columnwidth]{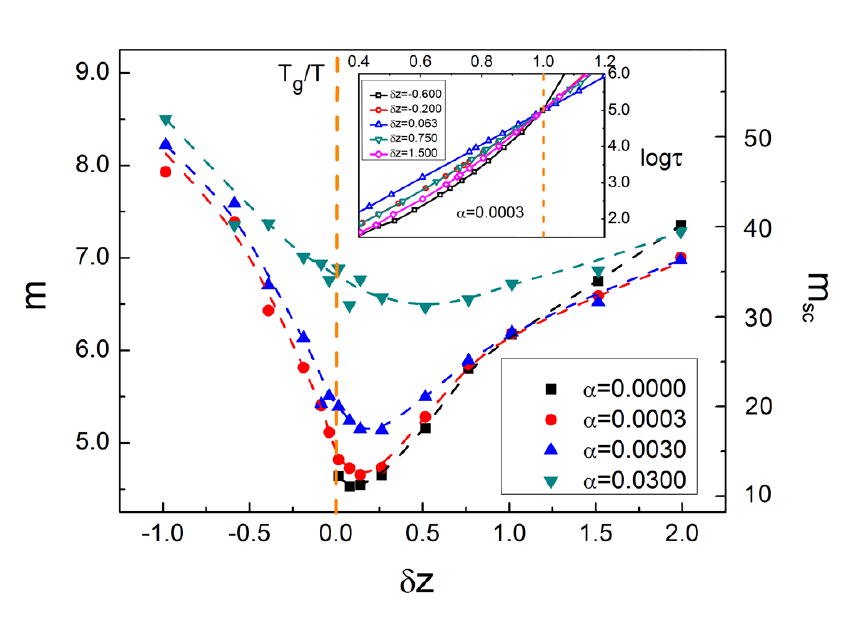}}
        \vspace{-0cm}
     \caption{Fragility $m$ versus excess coordination $\delta z$ for different strength of weak interactions $\alpha$ as indicated in legend, in $d=2$. Dash dot lines are guide to the eyes, and reveal the non-monotonic behavior of $m$ near the rigidity transition. Inset: Angell plot representing $\log\tau$ {\it v.s.} inverse temperature $T_g/T$ for different $\delta z$ and $\alpha=3\times10^{-4}$. }
     \label{2_f3}
   \end{center}
\end{figure}

%If $R_1$ indicates the proximity of the rigidity transition, it should correlate in our model with the jump of specific heat and with fragility, since  both are strongly affected by the transition. %the latter infers the rigidity transition?

Fig.~5 shows that $R_1$ and the liquid fragility $m$ are correlated in our model, thus capturing observations in molecular liquids. The model also predicts  that $R_1$ and the jump of specific heat are correlated. %  (data not shown).
 Note that the correlation between fragility and $R_1$ is not perfect, and that two branches, for glasses with low and with high coordinations, are clearly distinguishable. In general we expect physical properties  to depend on the full structure of the density of states, as will be made clear for the thermodynamics of our model below. The variable $R_1$, which is a single number, cannot capture fully this relationship. In our framework it is a useful quantity however, as it characterizes well the proximity of the jamming transition. 

\section{Theory on Thermodynamics of the Model}

\subsection{Thermodynamics in the absence of weak interactions ($\alpha=0$)} 
In the absence of weak springs the thermodynamics is non-trivial if $z\geq z_c$, otherwise the inherent structure energies are all zero. Then Eq.(\ref{2_stiffM}) implies ${\cal M}={\cal S}^t_{\rm s} {\cal S}_{\rm s}$, and Eq.(\ref{2_green}) leads to ${\cal G}={\cal I}-{\cal S}_{\rm s}({\cal S}^t_{\rm s}{\cal S}_{\rm s})^{-1}{\cal S}^t_{\rm s}$. Inspection of this expression indicates that ${\cal G}$ is  a projector on the kernel of ${\cal S}^t_{\rm s}$, 
which is generically of dimension $N_s-Nd\equiv \delta z N/2$. This kernel corresponds to all the sets of contact forces that balance forces on each node \cite{Wyart05b}.  We denote by $|\delta r_p\rangle, p=1,...,\delta z N/2$ an orthonormal basis of this space. We may then rewrite ${\cal G}=\sum_p |\delta r_p\rangle\langle \delta r_p|$ and Eq.(\ref{2_3}) as:
\be
\label{2_7}
{\cal H}(|\sigma\rangle)=\frac{1}{2}\sum_{p=1}^{\delta zN/2} \langle \sigma|\delta r_p\rangle^2
\ee
Eq.(\ref{2_7}) is a key result, as it implies that near the rigidity transition the number $\delta z N/2$ of directions of phase space that cost energy vanishes. Only those directions can contribute to the specific heat, which must thus vanish linearly in $\delta z$ as the rigidity transition is approached from above.

Eq.(\ref{2_7}) also makes a connection between strong liquids in our framework and well-know problems in statistical mechanics. In particular Eq.(\ref{2_7}) is similar to that describing Hopfield nets \cite{Hopfield82} used to store $\delta z N/2$ memories consisting of the spin states $|\delta r_p\rangle$. The key difference is the sign: in Hopfield nets memories correspond to  meta-stables states, whereas in our model the vectors $|\delta r_p\rangle$ corresponds to maxima of the energy. A particularly interesting case is $\delta z N/2=1$, the closest point to the jamming transition which is non-trivial.  In this situation  the sum in Eq.(\ref{2_7}) contains only one term: ${\cal H}(|\sigma\rangle)= \frac{1}{2} \langle \sigma|\delta r_1\rangle^2= \frac{1}{2} (\sum_{\alpha=1}^{N_s} \delta r_{1,\alpha} \sigma_\alpha)^2$. %The vector  $|\delta r_1\rangle$ must be  random, since  the typical length scale become of the size of the system. 
This Hamiltonian corresponds to the  NP complete partitioning problem \cite{Hayes02}, where given a list of numbers (the $\delta r_{1,\alpha}$) one must partition this list into two groups 
whose sums are as identical as possible. Thermodynamically this problem is known \cite{Mertens01} to map into the random energy model \cite{Derrida81} where energy levels are randomly distributed.

It is in general very difficult  to compute the thermodynamic functions
of the problem defined by Eq.(\ref{2_7})  because the vectors $|\delta r_p\rangle$ present spatial correlations,
as must be the case since the amplitude of the interaction kernel ${\cal G}_{\gamma,\beta}$ must decay with distance. However this effect is expected to be mild near the rigidity transition. Indeed there exists a diverging  length scale at the transition, see  \cite{During13} for a recent discussion, below which ${\cal G}_{\gamma,\beta}$ is dominated by fluctuations and decays mildly with distance. Beyond this length scale ${\cal G}_{\gamma,\beta}$ presents a dipolar structure, as in a standard continuous elastic medium.  We shall thus assume that $|\delta r_p\rangle$ are random unitary  vectors, an approximation of mean-field character expected to be good near the rigidity transition.

\begin{figure}[ht!]
   \begin{center}
     {\includegraphics[width=1.\columnwidth]{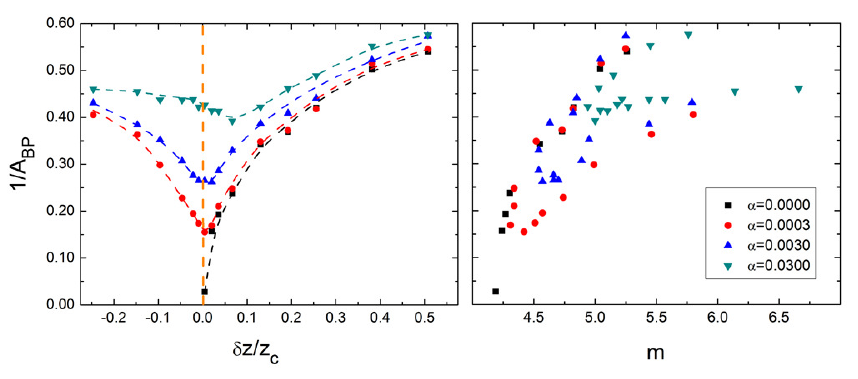}}
%     \rotatebox{+0}{\resizebox{4.2cm}{!}{\includegraphics{R1m}}
     \caption{Left:  Inverse boson peak  amplitude $R_1$ versus excess coordination $\delta z$ in our $d=3$ elastic network model, for different weak interaction strenghts as indicated in legend. Dash dot lines are drawn to guide one's eyes. Right: Inverse boson peak  amplitude  $R_1$ versus fragility $m$ for different weak springs $\alpha$.}
     \label{2_f4}
   \end{center}
\end{figure}

 Within this approximation, the thermodynamic properties can be derived for any spectrum of ${\cal G}$ \cite{Cherrier03}. If the orthogonality of the vectors $|\delta r_p\rangle$ is preserved,  the Hamiltonian of Eq.(\ref{2_7}) corresponds to the Random Orthogonal Model (ROM) whose thermodynamic properties have been derived \cite{Cherrier03} as well as some aspects of the dynamics \cite{Caltagirone12}. Comparison of the specific heat of our model and the ROM predictions of \cite{Cherrier03} is shown in Fig. 3 and are found to be very similar. For sake of simplicity, in what follows we shall also relax  the orthogonal condition on the vectors $|\delta r_p\rangle$. This approximation allows for a straightforward analytical treatment in the general case $\alpha\neq 0$, and is also very accurate near the rigidity transition since the number of vectors $\delta z N/2$  is significantly smaller than the dimension of the space $dN$ they live in, making random vectors effectively orthogonal. Under these assumptions we recover the random Hopfield model with negative temperature.

In the parameter range of interest, the Hopfield free energy ${\cal F}={\overline{\ln({\cal Z})}}$ (here $\overline{(...)}$ represents the disorder average on the $|\delta r_p\rangle$) is approximated very precisely by the annealed free energy $\ln({\overline {\cal Z}})$  (this is obviously true for the number partitioning problem that maps to the Random Energy Model), which can be easily calculated. Indeed in our approximations the quantities $X_p\equiv\langle \sigma|\delta r_p\rangle$ are independent gaussian random variables  of variance one, and:
\begin{equation}
\label{2_8}
{\cal {\overline Z}}\propto \int\left(\prod_{p=1}^{N\delta z/2}dX_p\frac{1}{\sqrt{2\pi}}e^{-X_p^2/2}\right)e^{-\frac{1}{2T}\sum_pX_p^2}
\end{equation}
Performing the Gaussian integrals we find:
\be
%\varepsilon(T)&=&\frac{\delta z}{2z}\frac{T}{1+T}\\
\label{2_9} c(T)=\frac{\delta z}{2z}\frac{1}{(1+T)^2}
%s(T)&=&\ln2-\frac{\delta z}{2z}\left[\ln(1+1/T)-\frac{1}{1+T}\right]
\ee
%where $e(T)$ ans $s(T)$ are the energy and entropy per spring  respectively. 
The Kautzman temperature defined as $s(T_K)=0$ is found to follow $T_K\approx\frac{2}{e}2^{-2z/\delta z}$.
 Eq.(\ref{2_9}) evaluated at $T_g$ is tested against the numerics in Fig.~3 and performs remarkably well for the range of coordination probed. %{\bf We find the heat capacity from inherent structure increases when cooling down the temperature, which mainly }

\subsection{General case ($\alpha\neq 0$)} 
To solve our model analytically in the presence of weak interactions, we make the additional approximation that the associated coordination $z_{\rm w}\rightarrow \infty$, while keeping $\alpha\equiv z_{\rm w} k_{\rm w}/(kd)$ constant. % the alpha defined here is different from the one in simulation, network structure
In this limit weak springs lead to an effective interaction between each node and  the center of mass of the system, that is motionless. Thus the restoring force stemming from  weak interactions $|{\bf F}_{\rm w}\rangle$ follows $|{\bf F}_{\rm w}\rangle=-\alpha |\delta{\bf R}\rangle$, leading to a simple expression in  the stiffness matrix Eq.(\ref{2_stiffM})  for the weak spring contribution $\frac{k_{\rm w}}{k}{\cal S}_{\rm w}^t{\cal S}_{\rm w}=\alpha  {\cal I} $. It is useful to perform the eigenvalue decomposition:
\be
\label{2_99}
{\cal S}_{\rm s}^t{\cal S}_{\rm s}=\sum_\omega \omega^2 |\delta {\bf R}_\omega\rangle\langle\delta {\bf R}_\omega|
\ee
where $|\delta {\bf R}_\omega\rangle$ is the vibrational mode of frequency $\omega$ in the elastic network without weak interactions. We introduce the orthonormal eigenvectors in contact space $|\delta r_\omega\rangle\equiv {\cal S}_{\rm s} |\delta {\bf R}_\omega\rangle/\omega$ defined for $\omega>0$. For $\delta z<0$ these vectors form a complete basis of that space, of dimension $N_s$. When $\delta z>0$ however, this set is of dimension $Nd<N_s$, and it  must be completed by the kernel of ${\cal S}^t_{\rm s}$, i.e. the set of the $|\delta r_p\rangle, p=1,...,\delta z N/2$ previously introduced. Using this decomposition in Eq.(\ref{2_green},\ref{2_stiffM}) we find:
\be
\label{2_10}
{\cal H}(|\sigma\rangle)=\frac{1}{2}\sum_{p=1}^{\delta zN/2} \langle \sigma|\delta r_p\rangle^2+\frac{1}{2}\sum_{\omega>0} \frac{\alpha}{\alpha+\omega^2}\langle \sigma|\delta r_\omega\rangle^2
\ee
where the first term exists only for $\delta z>0$. Using the mean field approximation that the set of  $|\delta r_p\rangle$ and $|\delta r_\omega\rangle$ are random gaussian vectors,
the annealed free energy is readily computed, as shown in Appendix Sec.~\ref{app_A4}. We find in particular for the specific heat: 
\begin{equation}
\label{2_11}
c(T)=\frac{\delta z }{2z}\frac{\theta(\delta z)}{(1+T)^2}+ \frac{1}{2 N_s}\sum_{\omega>0}\left(\frac{\alpha}{\alpha+(\omega^2+\alpha)T}\right)^2
\end{equation}
where $\theta(x)$ is the unitary step function. To compare this prediction with our numerics without fitting parameters, we compute numerically the vibrational frequencies for each value of the coordination.
Our results  are again in excellent agreement with our observations, as appears in Figs.~2,~3.

To obtain the asymptotic behavior near jamming, we  replace  the summation over frequencies in Eq.(\ref{2_11})  by an integral $\sum_{\omega>0}\rightarrow N_s\int \rm{d}\omega D(\omega)$. 
The associated density of vibrational modes $D(\omega)$ in such networks has been computed theoretically \cite{Wyart10a,During13,Wyart05}. 
%On each side of the transition, there is  a frequency scale $\omega^*\sim |\delta z|$ above which $D(\omega)$ displays a plateau, up to some frequency of order one. For $\omega<\omega^*$, the density of states follows a Debye law for $\delta z>0$, but the associated contribution in Eq.(\ref{2_11}) is subdominant. For $\delta z<0$, $D(\omega)$ displays a gap for $0<\omega<\omega^*$. 
These results allows us to compute the scaling behavior of thermodynamic properties near the rigidity transition, see Appendix Sec.~\ref{app_A5}. We find that the specific heat increases monotonically with decreasing temperature.  Its value at the Kautzman temperature thus yields an  upper%lower 
bound on the jump of specific heat. In the limit $\alpha\rightarrow 0$, we find that a sudden discontinuity of the jump of specific heat occurs at the rigidity transition:
\bea
\label{2_12}c(T_K)&\sim& \frac{\delta z}{2z} \ \ \ \ \  \hbox{    for  }  \delta z>0\\
\label{2_13}\lim_{\delta z\rightarrow 0^-} c(T_K)&\sim&\frac{\pi z_c}{8z} \ \ \  \hbox{    for  }  \delta z<0
\eea
Eq.(\ref{2_12}) states that adding weak interactions is not a singular perturbation for $\delta z>0$,  and we recover Eq.(\ref{2_9}). 
On the other hand for $\delta z<0$, the energy of inherent structures is zero in the absence of weak springs, which thus have a singular effect.  The relevant scale of temperature is then a function of $\alpha$. In particular we find that the Kautzman temperature is sufficiently low that all the terms in the second sum of Eq.(\ref{2_11}) contribute significantly to the specific heat, which is therefore large  as Eq.(\ref{2_13}) implies.
Thus as the coordination decreases below the rigidity transition, one goes discontinuously from a regime where at the relevant temperature scale the energy landscape consists of a vanishing number of costly directions in phase space, whose cost is governed by the strong interaction $k$, to a regime where the weak interaction $\alpha$ is the relevant one, and where at the relevant temperature scale all directions in phase space contribute to the specific heat.

Note that although the sharp change of thermodynamic behavior that occurs at the rigidity transition is important conceptually, empirically a smooth cross-over will always be observed. This is the case because (i) $\alpha$ is small but finite. As $\alpha$ increases this sharp discontinuity is replaced by a cross-over at a coordination $\delta z\sim \ln(1/\alpha)^{-1}$ (see Appendix Sec.~\ref{app_A5}) where $c(T_K,z)$ is minimal, as indicated in the inset of Fig.~3. (ii) The Kautzman temperature range is not accessible dynamically, i.e. $T_g>>T_K$ near the rigidity transition. Comparing Fig.~3 with its inset, our theory predicts that the minimum of $c(T_g)$ is closer to $z_c$ and more pronounced than at $T_K$.

\section{Discussion}
 
Previous work \cite{Liu10} has shown that well-coordinated glasses must have a small boson peak, which increases as the coordination (or valence for network glasses) is decreased toward the jamming (or rigidity) transition.
Here we have argued that as this process occurs,  elastic frustration vanishes:  thanks to the abundance of soft modes, any configuration (conceived here as a set of local arrangements of the particles) can relax more and more of its energy as jamming is approached from above. As a result, the effective number of degrees of freedom that cost energy and contribute to the jump of specific heat at the glass transition vanishes. 
As the coordination is decreased further below the rigidity transition, the scale of energy becomes governed by the weak interactions (such as Van der Waals) responsible for the finite elasticity of the glass. At that scale, all direction in phase space have a significant cost and the specific heat increases. This view potentially explains why linear elasticity strongly correlates to key aspects of the energy landscape in network and molecular glasses \cite{Tatsumisago90,Ito99,Boolchand05,Novikov05,Ngai97}. 
This connection we propose between structure and dynamics can also be tested  numerically. For example, the amplitude of weak interactions can be increased  by adding long-range forces to the interaction potential \cite{Wyart05b,Xu07}. According to our analysis, doing so should increase fragility, in agreement with existing observations \cite{Berthier09b}. 

The model of the glass transition we introduced turns out to be a spin glass model,  with the specificity that (i) the interaction is dipolar in the far field, and that  (ii) the sign of the interaction is approximatively  random below some length scale $l_c$ that diverges near jamming, where the coupling matrix has a vanishingly small rank. Applying spin glass models to structural glasses have a long history. In particular the Random First Order Theory (RFOT) \cite{Lubchenko07} is based on mean-field spin glass models that display a thermodynamic transition at some $T_K$  where the entropy vanishes.
A phenomenological description of relaxation in liquids near $T_K$  based on the nucleation of random configurations leads to a diverging time scale and length scale $\xi$ at $T_K$ \cite{Lubchenko07,Bouchaud04}. One limitation of this approach is that no finite dimensional spin models have been shown to follow this scenario so far \cite{Cammarota12}, and it would thus be important to know if our model does display a critical point at finite temperature.  Our model  will also allow one to investigate the generally neglected role of the action at a distance allowed by elasticity, characterized by a scale $l_c$. In super-cooled liquids heterogeneities of elasticity (that correlates to irreversible rearrangements) can be rather extended \cite{Widmer-Cooper08} suggesting that $l_c$  is large. This length scale may thus play an important role in a description of relaxation in liquids, and in deciphering the relationship between elastic and dynamical heterogeneities. 

\iffalse
\begin{acknowledgments} We thank A. Grosberg, P. Hohenberg, E. Lerner, D. Levine, D. Pine, E. Vanden-Eijnden, M. Vucelja., A. Lef\`evre for discussions, and E.Lerner for discussions leading to Eq.(\ref{app_5}). This work has been supported primarily by the  National Science Foundation  CBET-1236378, and partially by the Sloan Fellowship, the NSF DMR-1105387, and the Petroleum Research Fund 52031-DNI9.
\end{acknowledgments}

%\bibliographystyle{pnas.bst}
%\bibliography{reference.bib}

\fi

%% file: rigidity/rigidity.tex
\chapter{Evolution of Covalent Networks under Cooling: Contrasting the Rigidity Window and
Jamming Scenarios}
\label{3_rigidity}

%\begin{abstract}
We study the evolution of structural disorder under cooling in supercooled liquids, focusing on  covalent networks. We introduce a model 
for the energy of networks that incorporates weak non-covalent interactions. We show  that at low-temperature, these interactions
considerably affect the network topology near the rigidity transition that occurs as the coordination increases. As a result, this transition  becomes mean-field  and does not present a line of critical points
previously argued for, the  ``rigidity window".   Vibrational modes are then not fractons, but instead are similar to the anomalous modes observed in packings of particles near jamming. 
These results suggest an alternative interpretation for  the intermediate phase observed in chalcogenides.

%We study a self-organizing model to reveal how network topology of covalent glass evolves under cooling. The model shows that the weaker interactions like Van der Waals forces play an important role on low temperature covalent network topology. The topology goes from the rigidity percolation universality at high temperature to the jamming universality at low temperature when weak constraints exist. We do finite size scaling and measure the density of states, where we also verify the fractons with measuring fracton dimension, to show this. We give an intuitive argument on the origin of the jamming transition scenario in the low temperature limit and disprove the topological rigidity window in covalent glass based on the numeric results and the argument.
%\end{abstract}

%\keywords{rigidity percolation, jamming, vibration modes, weak constraints}

%\maketitle
\section{Introduction}
The physics of amorphous materials is complicated by the presence of structural disorder, which depends on temperature in supercooled liquids,
and on system preparation in glasses. As a result, various properties of amorphous solids are much less understood than in their crystalline counterparts, such as the non-linear phenomena that control plasticity under stress \cite{Argon79,Falk98} or  the glass transition \cite{Ediger96}, or even linear properties like elasticity.  
Concerning the latter, glasses present a large excess of soft elastic modes,  the so-called boson peak \cite{Phillips81}, and their response to a point perturbation can be heterogeneous on  a scale $l_c$ larger than the particle size \cite{Jaeger96,Leonforte06,Lerner14,Ellenbroek09a}. Recent progress has been made on these questions for short-ranged particles with radial interactions \cite{Liu10}. A central aspect  of these systems is the {\it contact network} made by interacting particles, and its associated average coordination $z$. Scaling behaviors \cite{Liu10} are observed at the unjamming transition where $z\rightarrow z_c$, where $z_c=2d$ is the minimal coordination required for stability \cite{Maxwell64} in spatial dimension $d$. As this bound is approach most of the vibrational spectrum consists of strongly-scattered but extended modes \cite{OHern03,Silbert05} coined {\it anomalous modes} \cite{Wyart05b}, whose characteristic onset frequency $\omega^*$ vanishes \cite{Silbert05,Wyart05b} and length scale $l_c$ diverges \cite{Silbert05,Lerner14} at threshold. Surprisingly, these critical behaviors can be computed correctly by mean-field approximations, which essentially assume that the spatial fluctuations of coordination are small \cite{Wyart05,Wyart10,DeGiuli14}. Likewise, some detailed aspects of the structure of random close packing are well captured by infinite dimensional calculations  \cite{Charbonneau14,Kurchan13}. However, it is unclear if these results, which assume that structural fluctuations are mild, apply generically to glasses.

%However, it is presently unclear if these technics and results apply specifically to repulsive particles, or more generically to glasses. %These observations suggest that the amplitude of disorder is mild at relatively low temperature, and raise the hope that other mean-field approaches, sufch as those used for the glass transition \cite{Kirkpatrick89,Berthier11b} may fair well too. 

\begin{figure}[h!]
\centering
\includegraphics[width=0.50\columnwidth]{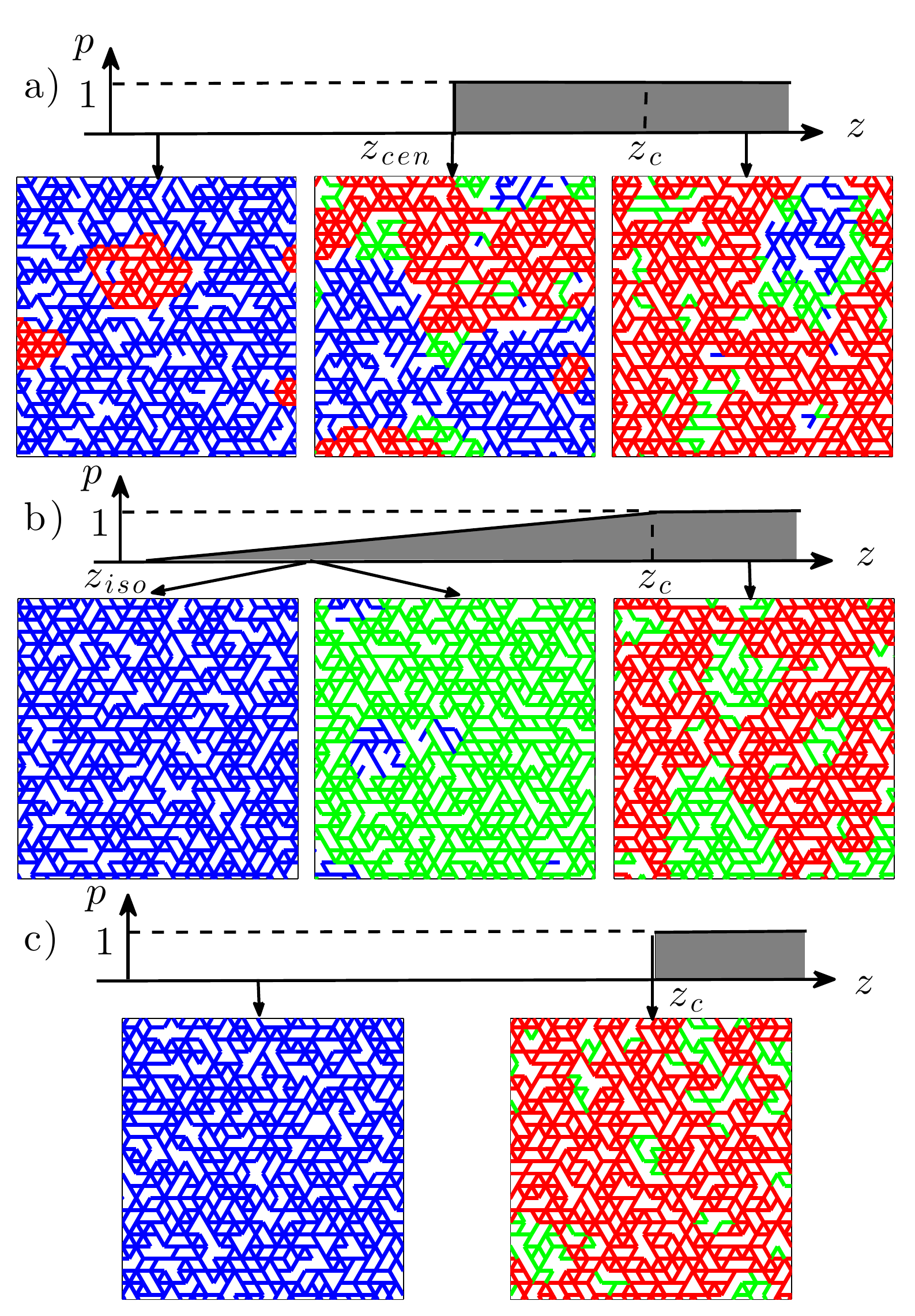}
\caption{\small{Three distinct scenarios for the rigidity transition in chalcogenide glasses. Bonds in blue, green, red corresponds respectively to floppy (under-constrained), isostatic (marginally-constrained)  and over-constrained regions.  $p(z)$ is the probability that a rigid cluster (made of green and red bonds) percolates, as a function of the valence $z$. (a) Rigidity percolation model where bonds are randomly deposited on a lattice. Percolation occurs suddenly and $p(z)$ jumps  from 0 to 1 at $z_{cen}<z_c$. At $z_{cen}$, the rigid network is fractal. (b) The self-organizing network model at  zero temperature. Over-constrained regions are penalized energetically and are absent for $z<z_z$. For $z\in[z_{iso},z_c]$,  $0<p(z)<1$ even in the thermodynamic limit. (c) Mean-field scenario, where $p(z)$ jumps from 0 to 1 at $z_c$, and where the rigid cluster at $z_c$ is not fractal.}}\label{3_f1}
\end{figure}

In particular, it is generally believed that fluctuations in the structure are fundamental  in covalent glasses such as chalcogenides. 
 In these systems the degree of bonding $z$ plays a role analogous to coordination, and can be changed continuously in compounds such as $Se_xAs_yGe_{1-x-y}$, allowing to go from a polymeric, under-coordinated glass $(x=1,y=0)$ to well-connected structures. Around a mean valence $z_c=2.4$ one expects the network to become rigid \cite{Phillips79,Thorpe85}. Near $z_c$ there is a range of valence, called the intermediate phase \cite{Selvanathan99,Wang00,Georgiev00,Chakravarty05,Wang05}, where the supercooled liquid is strong and the jump of specific heat is small \cite{Tatsumisago90,Bohmer92}, and where the glass almost does not age at all \cite{Selvanathan99,Wang00,Georgiev00,Chakravarty05,Wang05,Bauchy13,Micoulaut13}. Theoretically, at least three distinct scenarios were proposed (but see \cite{Moukarzel13} for a recent fourth proposition) to describe this rigidity transition, see Fig.~\ref{3_f1}. Fluctuations are important in the first two. The {\it rigidity percolation} model \cite{Jacobs98,Duxbury99,Feng84,Jacobs95} assumes that bonds are randomly deposited on a lattice. This leads to a second order transition at some $z_{cen}$ where a rigid cluster (a subset of particles with no floppy modes) percolates. Near $z_{cen}$ vibrational modes are fractons \cite{Feng85, Nakayama94}. This model does not take into account that rigid regions cost energy, and thus corresponds to infinite temperature.  To include these effects {\it self-organizing network models} were introduced \cite{Thorpe00,Chubynsky06,Briere07,Micoulaut03}, where rigid regions are penalized. A surprising outcome of these models is the emergence of a rigidity window: a range of valence for which rigidity occurs with a probability $0<p(z)<1$, even in the thermodynamic limit. This rigidity window was proposed to correspond to the intermediate phase observed experimentally \cite{Thorpe00}. 
Finally, in the {\it mean-field  or jamming scenario}, fluctuations of coordinations are limited, and $p(z)$ jumps from 0 to 1 at $z_c$. The rigid cluster at $z_c$ is not fractal, and is similar to that of packings of repulsive particles. Specific protocols to generate such networks were used to study elasticity \cite{Wyart08,Ellenbroek09} as well as the thermodynamics and fragility of chalcogenides \cite{Yan13}.

In this Letter we introduce an on-lattice model of networks, and study how  structure and vibrational modes evolve under cooling. Unlike  previous models supporting the existence of a rigidity window \cite{Chubynsky06,Briere07}, our model  includes  weak interactions (such as Van der Waals), always present in addition to covalent bonds. We show numerically and justify theoretically that the rigidity window is not robust: it disappears at low temperature as soon as weak interactions are added. At zero temperature the rigidity transition is then well described by the mean field scenario, and the vibrational modes consist of anomalous modes and not fractons.

\section{Adaptive Elastic Network Model}

\begin{figure}[h!]
\centering
\includegraphics[width=1.0\columnwidth]{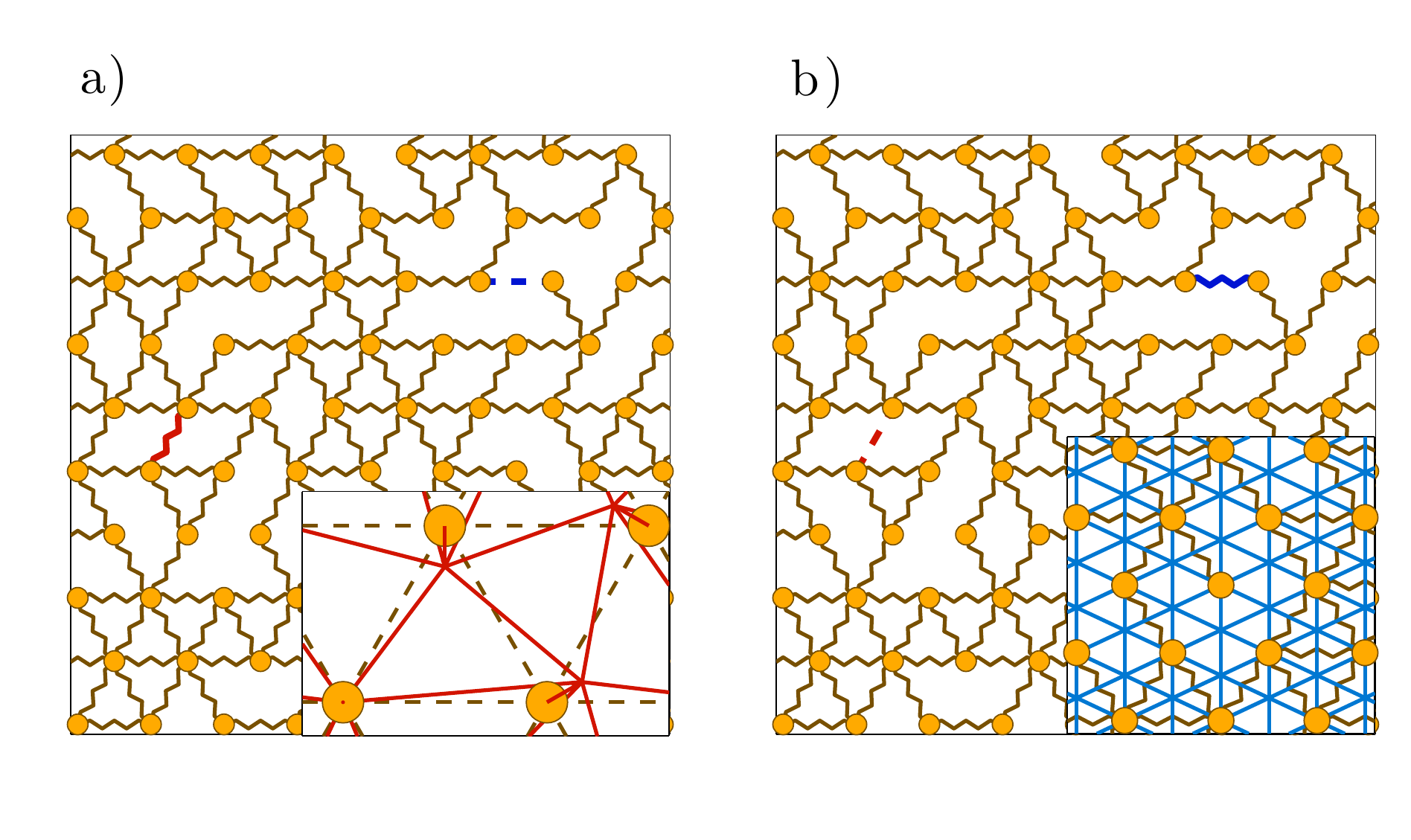}
\caption{\small{Illustration of our model. The triangular lattice is slightly distorted as shown in the inset of (a), and  weak springs connecting all second neighbors are present, as shown in blue in the inset of (b). Our Monte-Carlo considers the motion of strong springs such as that leading from (a) to (b). }}\label{3_model}
\end{figure}

Our model shares similarity to glasses of polydisperse particles, but it is on-lattice, and  particles are replaced by springs. 
Specifically, in the spirit of \cite{Jacobs95} we consider  a triangular lattice with slight  periodic distortion   to avoid straight lines (non-generic in disordered solids), as shown in Fig.~\ref{3_model}.
The lattice spacing between neighboring nodes $i$ and $j$ is $r_{\langle i,j\rangle}= 1+\delta_{\langle i,j\rangle}$ where the periodic distortion $\delta_{\langle i,j\rangle}$ is specified in Appendix Sec.~\ref{app_B1}.
Springs of identical stiffness $k$ can jump from an occupied to an unoccupied link, as shown in Fig.~\ref{3_model}. Their number is controlled by fixing the coordination $z$. The rest length $l_\gamma$ of the spring $\gamma$ positioned on the link $\langle ij\rangle$ is  $l_\gamma=r_{\langle i,j\rangle}+\epsilon_\gamma$, where $\epsilon_\gamma$ is taken from a Gaussian distribution of zero mean and variance $\epsilon^2$ \footnote[1]{The dependence of $l_\gamma$ on link $ij$ is a trick to remove the effect of straight lines on vibrational modes, unphysical for amorphous solids. In an elastic network it could be implemented in two dimensions by forcing the spring to bend in the third dimension, with a position-dependent amount of bending.}.   $k\epsilon^2$ is set to unity as the energy scale. To mimic Van der Waals interactions, we add weak springs of stiffness $k_{\rm w}$ between  second neighbors, so that the coordination of weak springs is $z_{\rm w}=6$. For a given choice of spring location, indicated as  $\Gamma\equiv\{ \gamma\leftrightarrow \langle i,j\rangle\}$, forces are unbalanced if the positions of the nodes are fixed. Instead we allow the  nodes to relax to a minimum of elastic energy $H(\Gamma)$, which depends only on the location of the springs $\Gamma$:
\be
\label{3_e1}
H(\Gamma)=
\min_{\{ {\vec R}_i\}} \sum_\gamma \frac{k}{2} \left[||{\vec R}_i-{\vec R}_j||-l_\gamma\right]^2\\+\sum_{\langle i,j\rangle_2}\frac{k_{\rm w}}{2}\left[||{\vec R}_i-{\vec R}_j||-\sqrt{3}\right]^2
\ee
where ${\vec R}_i$ is the position of node $i$ and ${\langle i,j\rangle_2}$ labels second neighbors. How the minimization of Eq.(\ref{3_e1}) is performed in practice is described in Appendix Sec.~\ref{app_B2}. 
Having defined an energy functional on all possible network structures $\Gamma$, we perform a Monte Carlo simulation using Glauber dynamics (illustrated in Fig.~\ref{3_model}) at temperature $T$.

%Subtracting the trivial degrees of freedom, the Maxwell counting gives $z_c=4-3\frac{2}{N}$ for a network of $N$ particles. 
%Our model is more realistic that previous ones where energy was based on topology alone \cite{JacobsPRL1995,Thorpe2000,BarrePRL2005,ChubynskyPRE2006},
%which did not allow to model weak interactions.

 Our model has two parameters: the temperature $T$ and $\alpha\equiv(z_{\rm w}/d)(k_{\rm w}/k)$  characterizing the relative strength of the weak forces,
estimated from experiments to be of order $\alpha=0.03$ \cite{Yan13}.  We find that we can equilibrate networks in the vicinity of the rigidity transition for $T\geq \alpha$. As we shall see below, for $T\gg1$ we naturally recover rigidity percolation. When $\alpha=0$ and $T\ll1$, a rigidity window appears, as previously reported in \cite{Jacobs95,Thorpe00,Barre05,Chubynsky06}, 
which we exemplify below using $T=3\times10^{-4}$ and $\alpha=0$.  We refer to this condition as {\it strong-force regime}. Finally, our main contention is that for $\alpha>0$ and for $T\leq\alpha$,
the rigidity window disappears, and the rigidity transition is mean-field.  We show that this is already the case  for extremely weak additional interactions $\alpha=T=0.0003$, a condition we refer to as {\it weak-force regime}.
%From these results and estimations of parameters in chalcogenides, we will conclude that network models do not support  the rigidity window scenario.

\section{Numerical Proofs of the Mean-field Rigidity Transition}
\begin{figure}[h!]
\centering
\includegraphics[width=0.8\columnwidth]{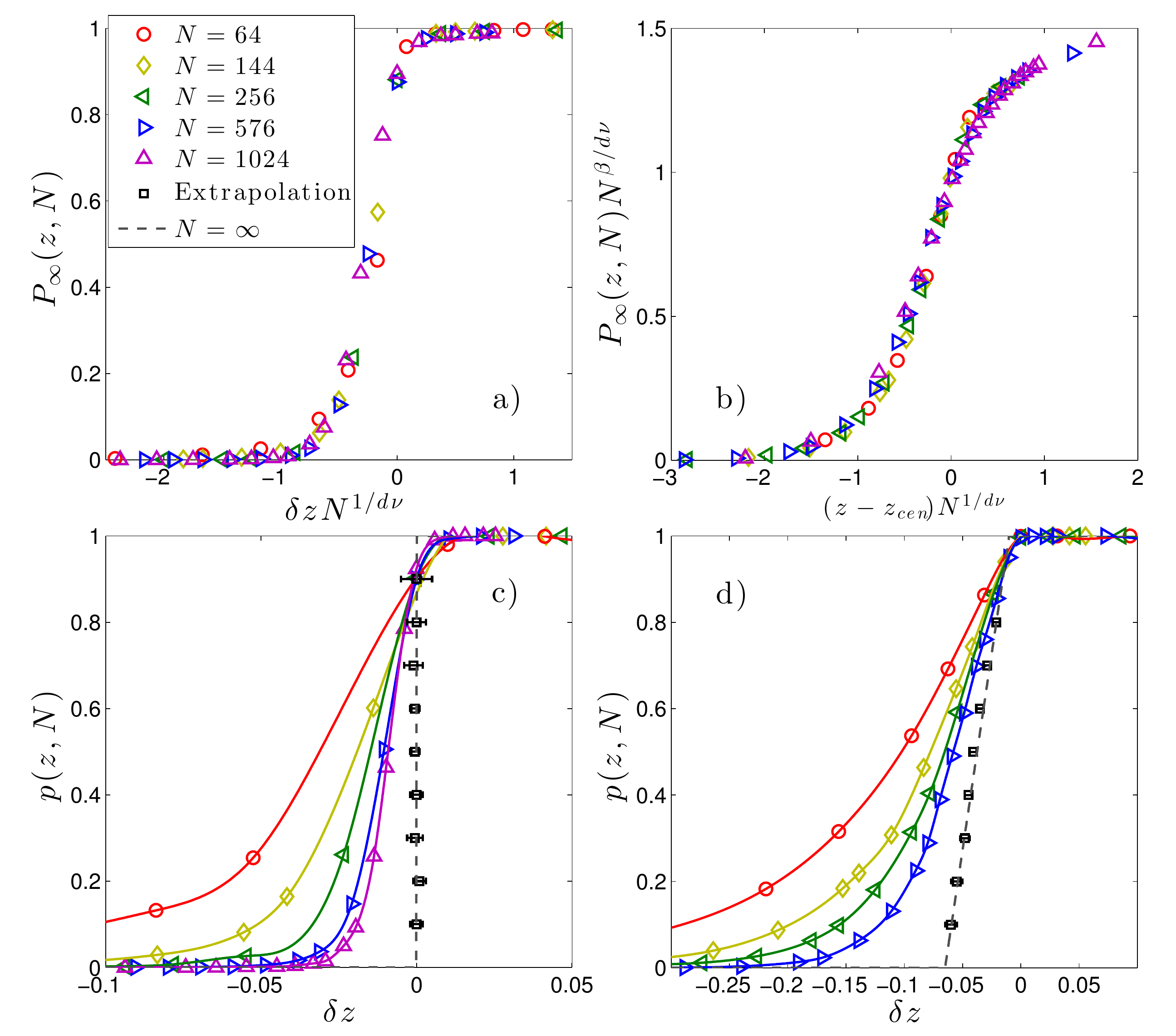}
\caption{\small{$P_{\infty}$ {\it vs} $(z-z_c)N^{1/d\nu}$ in the  {\it weak-force} condition (a) and for $T=\infty$ (b). $p$ {\it vs} $\delta z\equiv z-z_c$ in the {\it weak-force} (c) and {\it strong-force} (d) conditions.  The black squares are extrapolations of the finite $N$ spline curves,  as detailed in the main text.  In (c), the gray line is a step function at $z=z_c$, whereas in (d) it corresponds to the result of~\cite{Chubynsky06}.}}\label{3_pzsc}
\end{figure}

%\begin{enumerate}
{\it Percolation probabilities:} the probability $p(z)$ to have a rigid cluster spanning the system, and the  probability $P_{\infty}(z)$ for a bond to belong to this cluster are key quantities to distinguish scenarios. They can be computed for the network of strong springs using the Pebble Game algorithm~\cite{Jacobs97}. For rigidity percolation and infinite system size $N\rightarrow \infty$, $P_{\infty}(z)\sim (z-z_{cen})^{\beta}$. For finite $N$ one then expects \cite{JohnL.Cardy88}  $P_{\infty}(z,N)= N^{-\beta/d\nu}f_{RP}((z-z_{cen})N^{1/d\nu})$, where $f_{RP}$ is a scaling function and $\nu$ the length scale exponent. As shown in Fig.~\ref{3_pzsc}(b), we recover this result for $T=\infty$, with $z_{cen}=z_c-0.04$, $\beta=0.17$ and $\nu=1.3$, which perfectly matches previous works~\cite{Jacobs96}. Here the Maxwell threshold is set to $z_c=4-6/N$, as expected in two dimensions with periodic boundary conditions. In mean-field, the transition is discontinuous at $z_c$ and one therefore expects $P_{\infty}(z,N)= f_{J}((z-z_c)N^{1/d\nu})$. Our first key evidence that the  {\it weak-force regime} is mean-field is shown in Fig.~\ref{3_pzsc}(a), where this collapse is satisfied with $\nu=1.0$ - an exponent  consistent with the prediction of ~\cite{Wyart05}.

 Our second key evidence considers $p(z)$, which varies continuously ~\cite{Thorpe00, Chubynsky06,Briere07} in  the rigidity window scenario, but abruptly in mean-field, see Fig.~\ref{3_f1}. For finite size systems, it turns out to be easier to extract the inverse function $z(p)$, proceeding as follows. 
We first compute $p(z, N)$ for various $z$ and $N$. For each $N$ we use a spline interpolation to obtain continuous curves, as shown in Fig.~\ref{3_pzsc}(c,d).
We then extract $z(p)$ by fitting the following correction to scaling  $|z(p)-z(p,N)|\sim N^{-1/d\nu}$. Our central result is that for the {\it weak-force regime}, $p(z)$ discontinuously jumps from 0 to 1 at $z_c$ (which simply corresponds to the crossing of the spline lines) as shown in Fig.~\ref{3_pzsc}(c), again supporting that the mean-field scenario applies. By contrast, in the {\it strong force regime} this procedure predicts a rigidity window   for $z\in [z_c-0.06,z_c]$. This result is essentially identical to previous work using much larger $N$ ~\cite{Chubynsky06} (which is impossible in our model).

\begin{figure}[h!]
\centering
\includegraphics[width=0.7\columnwidth]{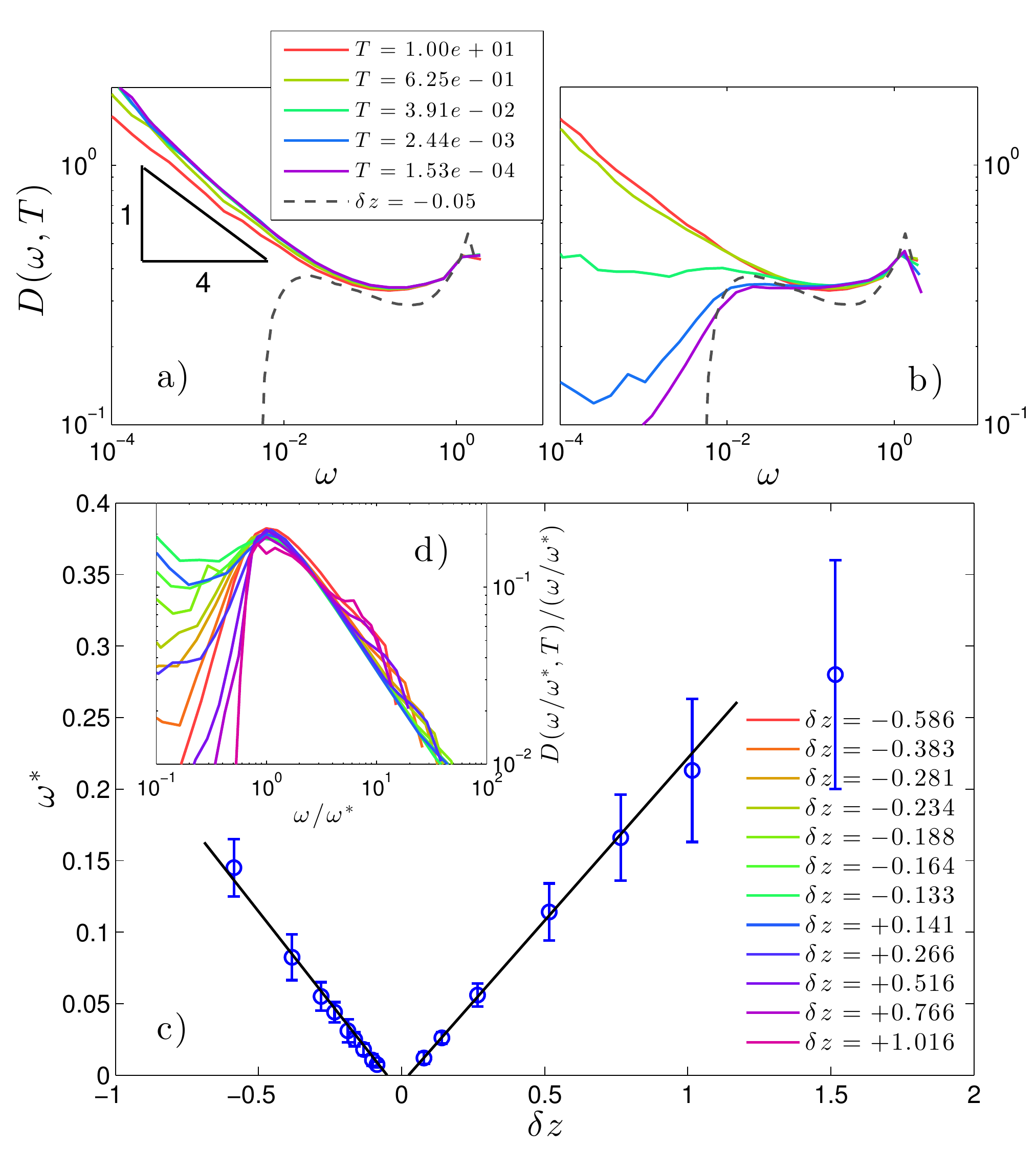}
\caption{\small{ $D(\omega)$ at $z-z_c=-0.05$ and various $T$ indicated in legend for (a) $\alpha=0$ and (b) $\alpha=0.0003$. Gray dashed lines are numerical solution of mean-field networks generated in \cite{Wyart08}. (c) Boson peak frequency $\omega^*$ {\it vs} coordination $z$ for the {\it weak-interaction } regime. $\omega^*$ is defined as the peak frequency of $D(\omega)/\omega^{d-1}$, a quantity shown in (d).}}\label{3_dos}
\end{figure}

{\it Density of vibrational modes (DOS) $D(\omega,T)$:} %and Boson peak frequency $\omega^*(z)$.
The DOS is a sensitive observable to characterize  network structure. In the mean-field scenario, anomalous modes appear above a frequency  $\omega^*\sim |z-z_c|$~\cite{Wyart05,During13}, above which the DOS displays a plateau: $D(\omega)\sim \omega^0$, as observed in packings \cite{Silbert05}.
By contrast, at rigidity percolation the rigid cluster is fractal and the spectrum consist of fractons, leading to  $D(\omega)\sim\omega^{\tilde{d}-1}$ ~\cite{Feng85, Nakayama94},  where $\tilde{d}$ is the fracton dimension. Numerically we compute the DOS associated with the  network of strong springs by diagonalization of the stiffness matrix. Within the rigidity window, we find that the DOS is insensitive to temperature for $\alpha=0$ as shown in Fig.~\ref{3_dos}(a), supporting that normal modes are fractons in the rigidity window, with $\textcolor{black}{\tilde{d}}\approx 0.75$~\cite{Nakayama94}. By contrast, already at small $\alpha=0.0003$, a key observation is that the DOS evolves under cooling  toward the mean-field prediction, as illustrated in Fig.~\ref{3_dos}(b). At low-temperature, one recovers a frequency scale $\omega^*\sim |z-z_c|$ as shown in Fig.~\ref{3_dos}(c,d), supporting further that the mean-field scenario applies. Note that there is a very narrow region around $z_c$ where the mean-field prediction does not work well and instead one finds $\omega^*\approx 0$ (see discussion  below).

\begin{figure}[h!]
\centering
\includegraphics[width=0.8\columnwidth]{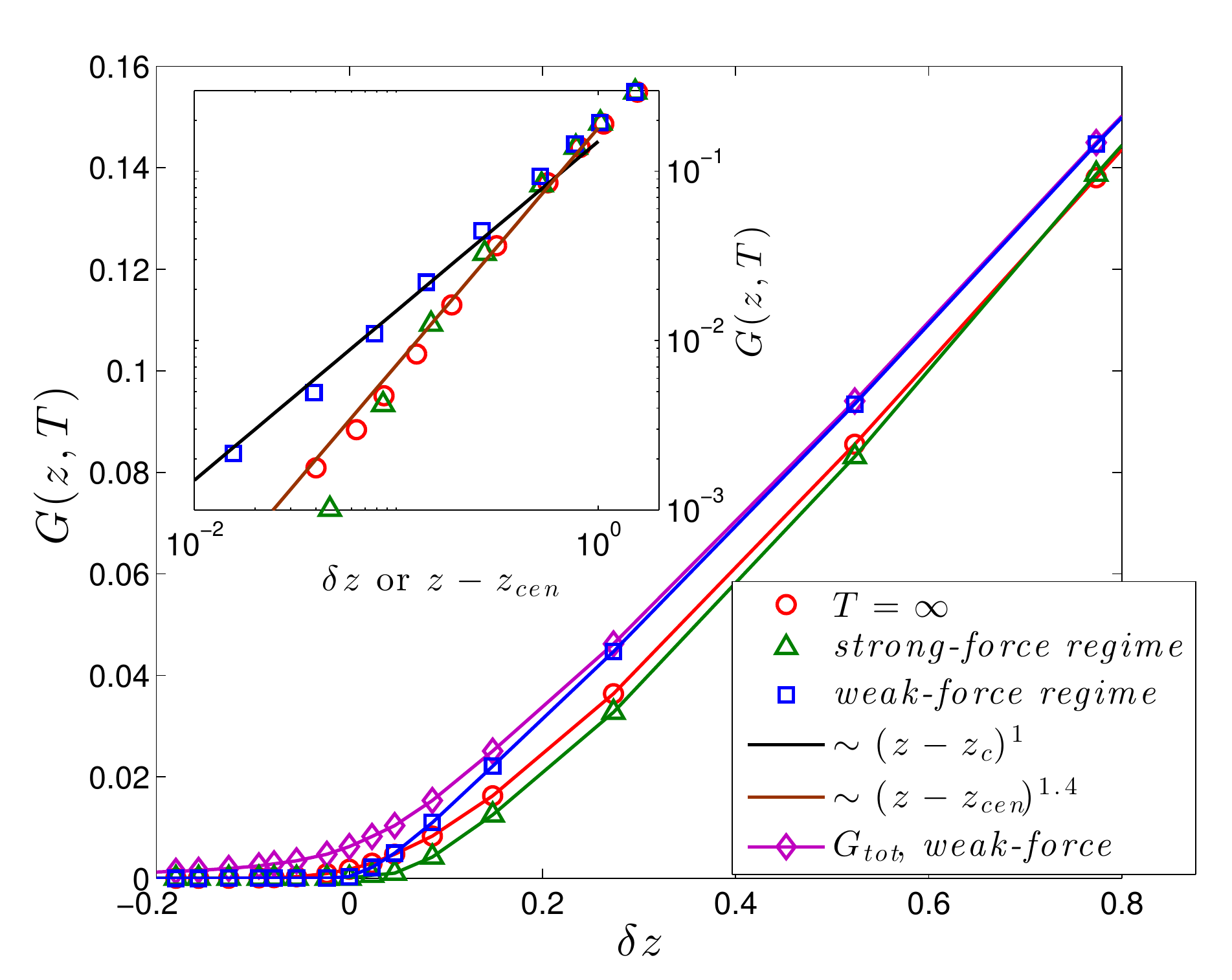}
\caption{\small{Shear modulus of the strong network $G$ {\it vs} $\delta z\equiv z-z_c$ for parameters indicated in legend. The total shear modulus $G_{tot}$ including the effect of weak springs is represented for the {\it weak-interaction regime}.  Inset: same plot in log-log scale, the horizontal axis is $z-z_c$ for low-temperatures conditions (blue and green), and $z-z_{cen}$ at  $T=\infty$ (red).}}\label{3_Gz}
\end{figure}

{\it Shear modulus $G(z)$:} Lastly, we compute the shear modulus for the strong network numerically, as shown in Fig.~\ref{3_Gz}. As expected, we find for $T=\infty$ the rigidity percolation result $G(z)\sim(z-z_{cen})^f$,  with $f\approx1.4$~\cite{Broedersz11}. In the {\it weak-interaction regime}, we find that the mean-field result \cite{Thorpe85} $G(z)\sim\delta z$ holds, supporting further our main claim. 
 In the {strong interaction regime}, we find that the shear modulus is zero up to $z_c$. However no power law scaling is found near $z_c$, and $G$ is much lower than in mean-field, in agreement once again with previous models that observed a window~\cite{Chubynsky03}.

\section{A Simple Argument for the Mean-field Scenario}

We have shown numerically that the {\it weak interaction regime} is well-described by the mean-field scenario. To explain this fact,
we argue that this scenario is stable  if $\alpha>0$, but unstable if $\alpha=0$. Consider  the elastic energy per unit volume $E_p$. Qualitatively this quantity is expected to behave as $E_p\sim G_{tot}\epsilon^2$,
where $G_{tot}$ is the total shear modulus  that includes weak interactions, also shown in Fig.~\ref{3_Gz}. 
The central point is that in mean-field, if $\alpha=0$ then $E_p(z)$ linearly grows for $z>z_c$ and is strictly 0 for $z<z_c$.
It implies that there is no penalty for increasing spatial fluctuations of coordination as long as $z<z_c$ locally. Thus large fluctuations of coordination are expected, 
the mean-field scenario is not stable and one finds a rigidity window instead. 
By contrast, $E_p(z)$ is strictly convex for all $z$ as soon as $\alpha>0$. Then spatial fluctuations of coordination are penalized energetically, and they disappear at low $T$.
In Appendix Sec.~\ref{app_B3}., Fig.~\ref{app_dist}, we find numerically that in our model, fluctuations of coordination indeed decay under cooling only if $\alpha>0$.

%\begin{figure}[h!]
%\includegraphics[width=0.8\columnwidth]{omegaz}
%\caption{\small{Boson peak frequency $\omega^*$ as  a function of excess coordination $\delta z=z-z_c$ for different $\alpha$  as indicated in legend, at temperatures $T=\alpha$. $\omega^*=0$ indicates that no maximum was observed in $D(\omega)/\omega^{d-1}$, consistent with the presence of fractons at very low frequency.}}\label{3_flu}
%\end{figure}

It is apparent from Fig.~S2 that this process of homogenization is already playing a role at temperatures of order $T\sim 10 \alpha$. 
In practice the glass transition $T_g$ is of order $\alpha$ (the typical covalent bond  energy is between 1 and 10ev, Van der Waals interactions are of between 0.01 and 0.1ev, and the glass transition temperature $T_g$ is of order 100 to 1000K, which is about 0.01 to 0.1ev), supporting that spatial fluctuations of coordination are strongly tamed due to the presence of weak interactions in real covalent glasses.

To conclude, we have argued that weak interactions induce a finite cost to spatial fluctuations of coordination,
which therefore vanish with temperature.   As a consequence, the rigidity transition is mean-field in character,
 if equilibrium can be achieved up to $T=0$. In this light, the mean-field scenario is a convenient starting point to 
 describe these materials. Note that although we focussed on $d=2$, our arguments go through unchanged in $d=3$,
 where the order of the rigidity percolation transition appears to be non-universal \cite{Chubynsky07}.
 
However, as $T$ increases  fluctuations must be included in the description, as appears in Fig.~\ref{3_dos}(b).
 Since covalent networks freeze at some  $T_g>0$,  one still expects a finite amount of fluctuations in the glass phase.
Indeed one must cross-over from rigidity percolation at $T=\infty$ to a mean-field scenario at $T=0$. We shall investigate this cross-over in detail elsewhere,
and instead speculate on its nature here. We expect
this cross-over to be continuous, implying  that at any finite temperature,
there is a narrow region around $z_c$ where fluctuations still play a role, and where the transition lies in 
the  rigidity percolation universality class. The size of this region vanishes with vanishing temperature but is finite at $T_g$. Inside this region, one expects  the boson peak to be dominated by fractons, whereas outside 
the mean-field approximation holds and anomalous modes dominate the spectrum. Fig.~\ref{3_dos}(c) supports this view since already at the very low-temperature considered, there is a narrow region for which $\omega^*\approx 0$, at odds with the mean-field prediction.  As expected, this effect is stronger as $T_g$ increases (as occurs in our model when $\alpha$ increases), as illustrated in Fig.~\ref{app_Hamiltonian} of Appendix Sec.~\ref{app_B4}.  This qualitative difference in elasticity 
\textcolor{black}{is} likely to affect thermodynamic and aging properties near the glass transition, since these properties are known to be strongly coupled \cite{Novikov05,Yan13}.
The region surrounding the rigidity transition where fluctuations are important is thus a plausible candidate for the intermediate phase observed in chalcogenides, which would then  result from a dynamical effect, namely the freezing of fluctuations at the glass transition.

%% file: adaptive/adaptive.tex
\chapter{Adaptive Elastic Networks as Models of Supercooled Liquids}
\label{4_adaptive}

The thermodynamics and dynamics of supercooled liquids correlate with their elasticity. In particular for covalent networks,   the jump of specific heat is  small and the liquid is {\it strong} near the threshold valence where the network acquires rigidity. By contrast, the jump of specific heat and the fragility are large away from this threshold valence. In a previous work [Proc. Natl. Acad. Sci. U.S.A., 110, 6307 (2013)], we could explain these behaviors by introducing a  model of supercooled liquids in which local rearrangements interact via elasticity. However, in that model the disorder characterizing elasticity was frozen, whereas it is itself a dynamic variable in supercooled liquids. Here we study numerically and theoretically adaptive elastic network models where polydisperse springs can move on a lattice, thus allowing for the geometry of the elastic network to fluctuate and evolve with temperature. We show numerically that our previous results on the relationship between structure and thermodynamics hold in these models. We introduce an approximation where redundant constraints (highly coordinated regions where the frustration is large) are treated as an ideal gas, leading to analytical predictions that are accurate in the range of parameters relevant for real materials. Overall, these results lead to a description of supercooled liquids, in which the distance to the rigidity transition controls the number of directions in phase space that cost energy and the specific heat.

%\end{abstract}

\section{Introduction}

Liquids undergo a glass transition toward an amorphous  solid state when cooled rapidly enough to avoid  crystallization ~\cite{Debenedetti01}. The glass lacks  structural order: it is a liquid ``frozen'' in a local minimum in the energy landscape, due to the slowing down of relaxation processes. It is very plausible that the thermodynamics and the dynamics in supercooled liquids strongly depend on the microscopic structure of these configurations -- hereafter referred to as ``inherent structures"~\cite{Stillinger84}.  However, a majority of glass theories~\cite{Adam65,Kirkpatrick89,Lubchenko07,Bouchaud04,Chamberlin99,Dyre06,Chandler10} have focused on explaining the correlations between macroscopic observables seen in experiments (such as the relationship between thermodynamics and dynamics ~\cite{Martinez01,Wang06}), while only a few~\cite{Hall03,Bevzenko09,Souza09,Rabochiy13} have investigated  the role of structure. %The question on how the evolution of the microscopic structures affects the glassy properties~\cite{Wyart10} is buried even deeper. 
%However, it is unclear how the liquid structure evolves under cooling and how much this evolution contributes to the macroscopic properties of glasses. 

Experiments reveal that elasticity plays a key role in both the thermodynamic and dynamical properties in supercooled liquids, such as the jump of specific heat and the fragility characterizing the glass transition. Specifically, it has been found that 
(I) glasses present an excess of low-frequency vibrational modes with respect to Debye modes. The number of these excess anomalous modes, quantified as the intensity of the boson peak~\cite{Phillips81}, shows a strong anti-correlation with the fragility~\cite{Ngai97,Novikov05}. (II) The rigidity of the inherent structures is tunable by changing the fraction of components with different valences in network glasses~\cite{Tatsumisago90,Kamitakahara91,Selvanathan99}, where atoms interact via covalent bonds and much weaker Van der Waals force. %that a network becomes rigid when the number of constraints exceeds its degrees of freedom (degrees of freedom), 
The covalent network becomes rigid~\cite{Maxwell64,Phillips79,Phillips85}, when the average valence $r$ exceeds a threshold $r_c$, determined by the balance between the number of covalent constraints and the degrees of freedom of the system. %And a critical coordination number $z_c=2d$ defines the on-set point of the rigidity of covalent network. 
Both the fragility and the jump of specific heat depend nonmonotonically on $r$, and their minima coincide with $r_c$~\cite{Tatsumisago90,Bohmer92}. { Interesting works using density functional theory~\cite{Hall03,Micoulaut03a} investigated the relationship between structure and fragility, but they do not capture this nonmonotonicity}. 
% context: non-montonic
%We rationalized the correlation by connecting elasticity of the structure to thermodynamics. However, the topology not evolves.

Recent observations \cite{Trachenko00,OHern03,Chen08,Chen10,Ghosh10} and theory \cite{Wyart05a,Wyart05b,Xu07,Brito09,Souza09,Souza09a,DeGiuli14, DeGiuli14b,DeGiuli15,Franz15b} indicate that  in various amorphous materials, the presence of soft elastic modes is regulated by the proximity of the rigidity transition, linking  evidence (I) and (II). To rationalize this connection,  we have introduced a frozen elastic network model that bridges the gap between network elasticity and geometry on one hand, elasticity and the thermodynamics and dynamics of liquids on the other~\cite{Yan13}. This model incorporated the following aspects of supercooled liquids: ({\it i}) particles interact with each other with interactions that can greatly differ in strength, such as the covalent bonds and the much weaker Van der Waals interaction found in network glasses. ({\it ii}) Neighboring particles can organize into a few distinct local configurations.  ({\it iii}) The choices of local configurations are coupled at different location in space via elasticity. These features were modeled using a random elastic network whose topology was frozen, as illustrated in Fig.~\ref{4_model}. The possibility for local configurations to change was incorporated by letting each spring switch between two possible rest lengths. 
Despite its simplicity, this model recovered (I) and (II). In particular, it reproduced the nonmonotonic variance of the jump of specific heat and the fragility with the coordination  $z$ of the network: they are extremal at $z_c=2d$ ($d$ is the spatial dimension),  where a rigidity transition occurs. This model could be solved analytically, and it led to the view that near the rigidity transition, the jump of specific heat is small because frustration vanishes: most directions in phase space do not cost energy, and thus do not contribute to the specific heat.

\begin{figure}[h!]
\centering
\includegraphics[width=1.0\columnwidth]{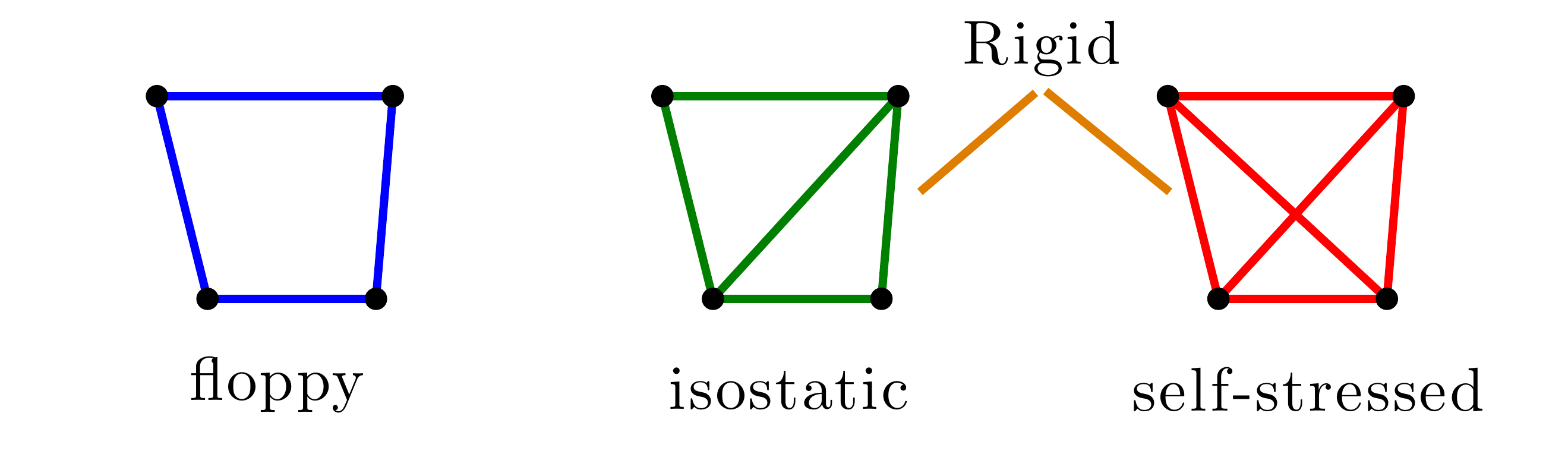}
\caption{\small{Illustration of rigidity transition. Blue, green, and red color the floppy, isostatic, and stressed clusters, respectively.}}\label{4_isostatic}
\end{figure}

This is a novel explanation for a long-standing problem, and it is important to confirm that this view is robust
when more realism is brought into the model. In particular, the model used frozen disorder to describe elasticity,
whereas it is itself a dynamical property in liquids, where there cannot be any frozen disorder. 
The thermal evolution of the topology of the contact network and its effects on rigidity transition were also not addressed. 
 A network is rigid when an imposed global strain induces stress, and the rigidity can be achieved topologically by adding  constraints~\cite{Maxwell64}, see Fig.~\ref{4_isostatic} for an illustration in a small network. The network is said to be self-stressed if some of the constraints are redundant, removing those leaves  the network rigid. 
Three scenarios of rigidity transition have been extensively studied in the literature~\cite{Yan14,Ellenbroek15} (but see Ref. \cite{Moukarzel13} for a recent fourth proposition). 
Spatial fluctuations of coordination  are important in the first two. The {\it rigidity percolation} model \cite{Jacobs98,Duxbury99,Feng84,Jacobs95} assumes that bonds are randomly deposited on a lattice. Fluctuations  lead to over-constrained (self-stressed) clusters  even when the average coordination number is not sufficient to make the whole network rigid.
This  model corresponds to the infinite temperature limit.  To include these effects, self-organized network models were introduced \cite{Thorpe00,Chubynsky06,Briere07,Micoulaut03}, where overconstrained regions are penalized. A surprising outcome of these models is the emergence of a rigidity window:  rigidity emerges at a small coordination number before the self-stress appears 
(even in the thermodynamic limit). 
Finally, in the {\it mean-field  or jamming scenario}, fluctuations of coordinations are limited.  Similar to the simple picture in Fig.~\ref{4_isostatic}, the rigidity, and the stress appear at the same $z_c$ in the thermodynamic limit. The rigid cluster at $z_c$ is not fractal and is similar to that of packings of repulsive particles. The model of Ref. \cite{Yan13} assumed that networks were of this last type.

Recently, we have introduced adaptive elastic network models \cite{Yan14}, where the topology of the network is free to evolve to lower its elastic energy as the system is cooled. We found that as soon as weak interactions are present, the network of strong interactions becomes mean-field like at low temperature.  However, the thermodynamic properties  were not studied to test the robustness of the thermodynamic predictions of Ref. \cite{Yan13} relating structure to the jump of specific heat. In this work, we directly show numerically and theoretically that the prediction for the jump of specific heat is essentially identical in adaptive and frozen elastic network models. Section II describes the adaptive network models. Section III presents the numerical results of the model, while Section IV gives the explicit derivation of the thermodynamic properties, developing an approximation scheme to deal with the temperature-dependence of the number of over-constraints in the system, treating them as an ideal gas.

\section{Adaptive Network Model}

\begin{figure}[h!]
\centering
\includegraphics[width=.65\columnwidth]{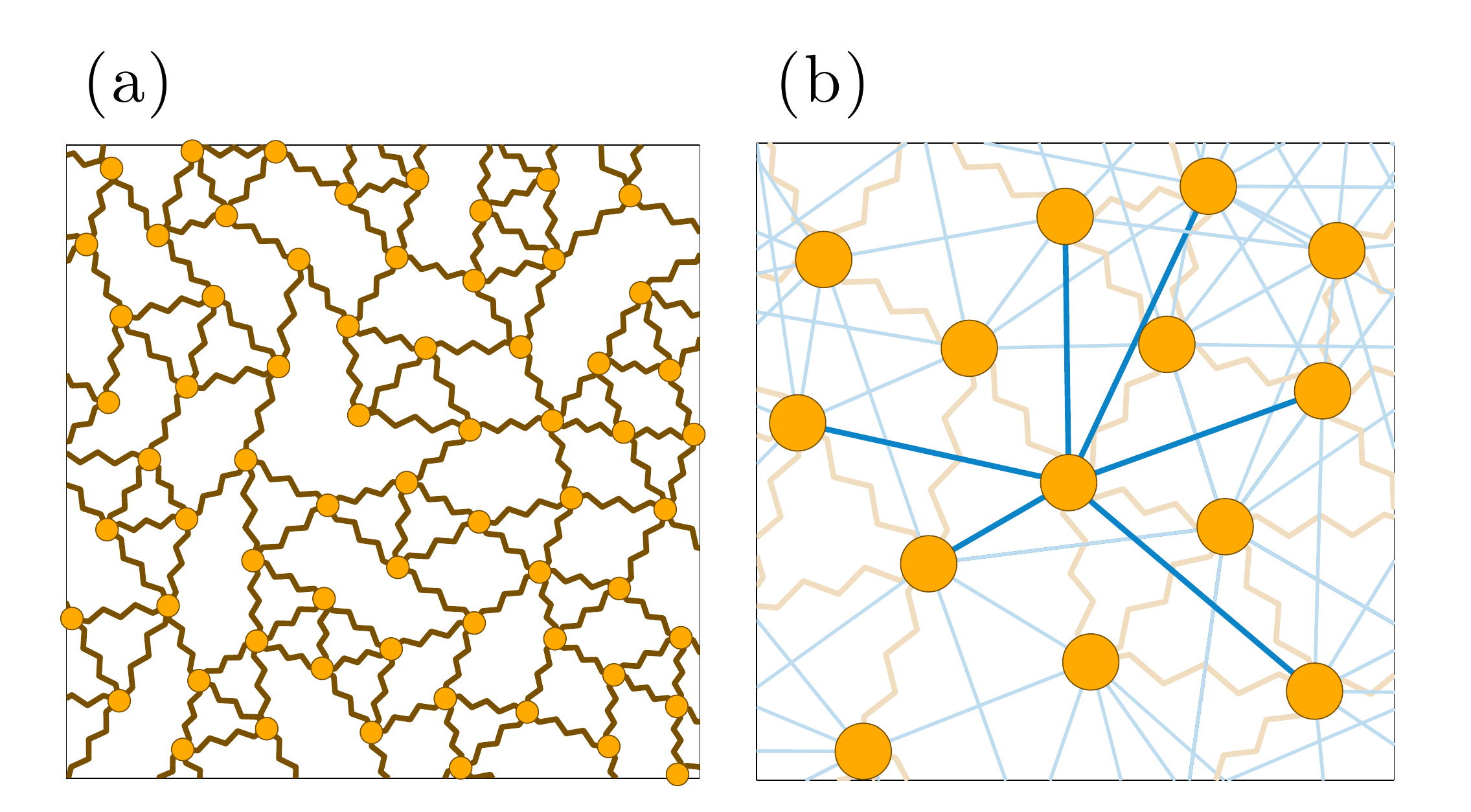}\\
\includegraphics[width=.65\columnwidth]{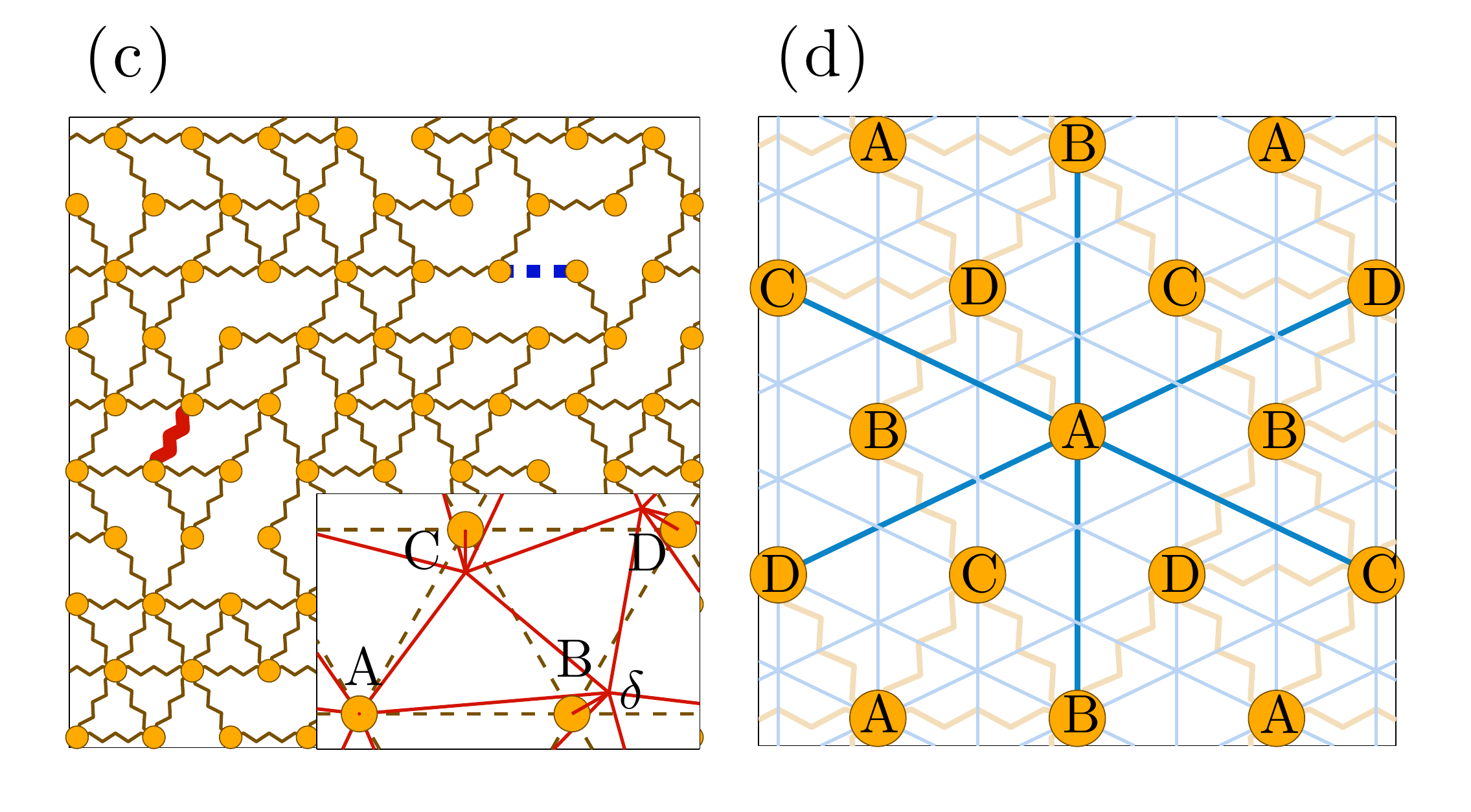}
\caption{\small{(Color online) (a) and (b) Illustration of the frozen network model~\cite{Yan13}; (c) and (d) illustrate the adaptive network model~\cite{Yan14}. In the latter case, the triangular lattice is systematically distorted in a unit cell of four nodes shown in the inset of (c). We group nodes by four, labeled as, A, B, C, and D in Fig.~\ref{4_model}. One group forms the unit cell of the crystalline lattice. Each cell is distorted identically in the following way: node A stays,  while nodes B, C, and D move by a distance $\delta$, B along the direction perpendicular to BC, C along the direction perpendicular to CD, and D along the direction perpendicular to DB. $\delta$ is set to $0.2$ with the lattice constant as unity. Weak springs connecting (b) six nearest neighbors without strong springs and (d) six next-nearest-neighbors are indicated in straight cyan lines, emphasized for the central node.  (c) Illustration of an allowed step, where the strong spring in red relocates to a vacant edge indicated  by a dashed  blue line.}}\label{4_model}
\end{figure}

% detailed description (distortion, mismatch, coordination, weak, energy)
In our  model degrees of freedom are springs, which are poly-disperse and can move  on a lattice. The lattice is built using a triangular lattice with periodic boundary conditions, see Fig.~\ref{4_model}(c), with a slight regular distortion to { minimize the non-generic presence of zero modes that occurs when straight lines are present}, as illustrated in the inset of Fig.~\ref{4_model}(c).    
Polydisperse and mobile ``strong" springs of identical stiffness $k$ connect the nearest neighbors on the lattice and model the covalent constraints. 
 % among atoms solidify the glass when the covalent network itself is floppy, and play important roles in self-organization of the covalent network~\cite{Yan14}. W
We model weak Van der Waals interactions with ``weak" and stationary  springs of stiffness $k_{\rw}\ll k$ adding to all next-nearest-neighbors on the triangular lattice, illustrated in Fig.~\ref{4_model}(b). We introduce a control parameter $\alpha\equiv(z_{\rw}/d)(k_{\rw}/k)$ to characterize the relative strength of the weak interactions, where the spatial dimension is $d=2$ and the  number of weak constraints per node is chosen $z_{\rw}=6$.

 The number of ``covalent'' springs $N_s$, equivalent to the coordination number $z\equiv2N_s/N$ ($N$ is the number of nodes in the lattice), is also a dimensionless control parameter. 
For a given { $\delta z\equiv z-z_c$}, the valid configurations are defined by the locations of the $N_s$ springs, indicated as $\Gamma\equiv\{\gamma\leftrightarrow\langle i,j\rangle\}$, where the Greek index $\gamma$ labels springs and the Roman indices $\langle i,j\rangle$ label the edges on triangular lattice between nodes $i$ and $j$. We introduce the { occupation of an edge:} $\sigma_{\langle i,j\rangle}=0$ if there is no strong spring on the edge $ij$, and $\sigma_{\langle i,j\rangle}=1$ if there is one. %And the associated local strain equals the quench random mismatches assigned to springs, which frustrate crystallization. %and avoid local fields of the distorted lattice by ignoring the link lengths. 
If $r_{\langle i,j\rangle}$ denotes the geometric length between nodes $i$ and $j$ on the lattice, we assume that the spring $\gamma$ has a rest length $l_{\gamma}=r_{\langle i,j\rangle}+\epsilon_{\gamma}$, where the mismatch $\epsilon_{\gamma}$ is a feature of a given spring. $\epsilon_{\gamma}$ are sampled independently from a Gaussian distribution with mean zero and variance $\epsilon^2$, which thus characterizes the polydispersity of the model. $k\epsilon^2$ is set to unity as the natural energy scale.

\begin{figure}[h!]
\centering
\includegraphics[width=1.0\columnwidth]{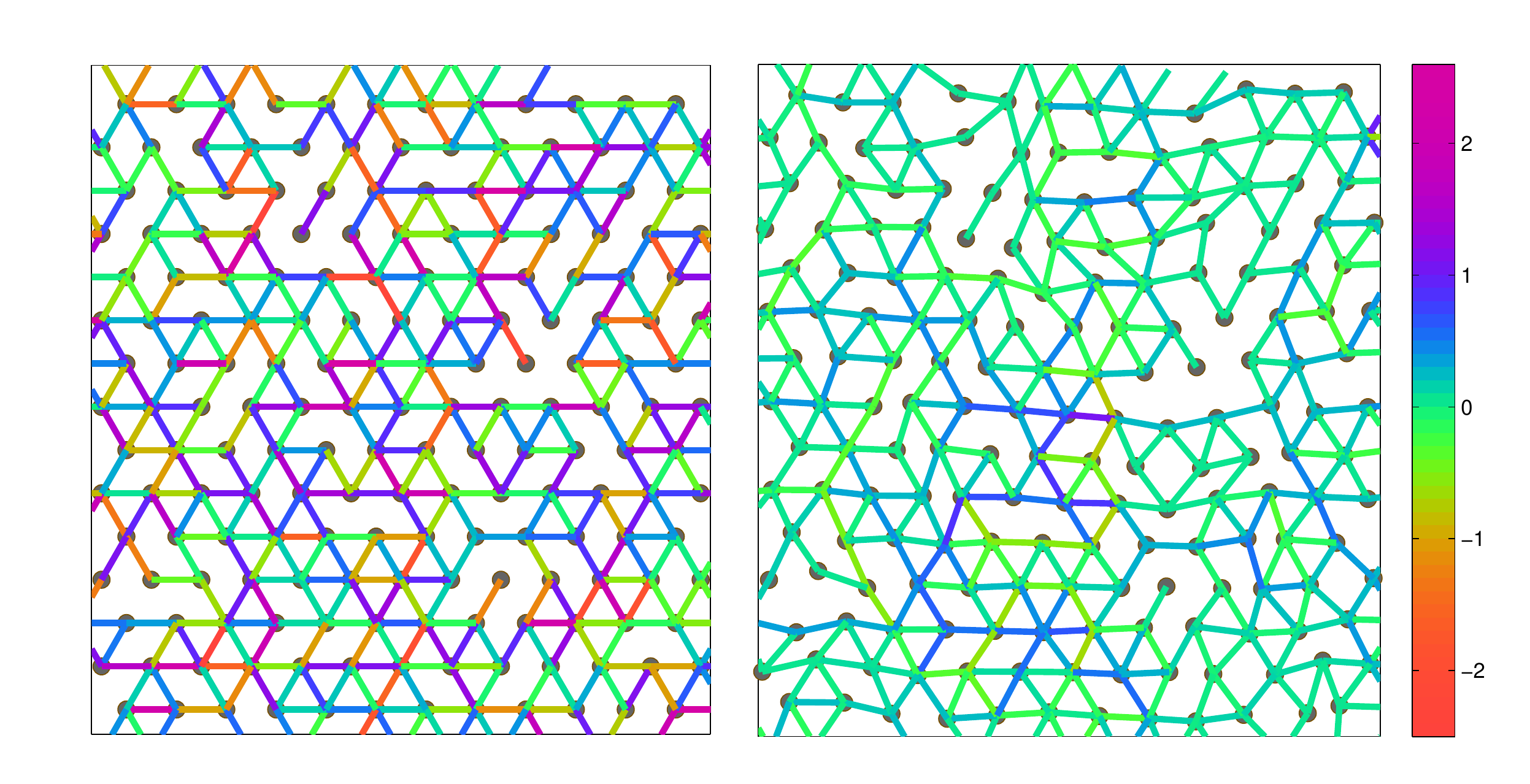}
\caption{\small{(Color online) Illustration of configuration energy of the adaptive network model ($\delta z=0.27$). { Solid lines are springs, colored according to their extensions: from red to purple, the springs go from being  stretched to being compressed,  with spring extensions shown in the unit of $\epsilon$. Left: Nodes sit at lattice sites, so the color shows the rest length mismatches of the springs $\{\epsilon_{\gamma}\}$. Right: Nodes are relaxed to mechanical equilibrium. Most links appear in green, indicating that most of the elastic energy is released. The configuration energy is defined by the residual energy.}}}\label{4_relax}
\end{figure}

The energy of an inherent structure is denoted $\mh(\Gamma)$. The configuration $\Gamma$ is sampled with probability proportional to $\exp(-\mh(\Gamma)/T)$ in the liquid phase, with $k_B=1$. Temperature $T$ serves as a third dimensionless control parameter. $\mh(\Gamma)$ is defined as the remaining energy once the nodes of the network are allowed to relax to mechanical equilibrium:
\be
\label{4_e1}
\mh(\Gamma)=
\min_{\{ {\vec R}_i\}} \left\{\sum_\gamma \frac{k}{2} \left[||{\vec R}_i-{\vec R}_j||-l_\gamma\right]^2\right.\\\left.+\sum_{\langle i,j\rangle_2}\frac{k_{\rm w}}{2}\left[||{\vec R}_i-{\vec R}_j||-r_{\langle i, j\rangle_2}\right]^2\right\}
\ee
where $\vec{R}_i$ is the position of particle $i$ and $\langle i, j\rangle_2$ labels the next-nearest neighbors. The minimal energy can be calculated by steepest decent as illustrated in Fig.~\ref{4_relax}, but this is computationally expensive. Instead, we approximate the elastic energy in the linear response range, setting that $\epsilon^2\ll1$~\footnote{We have tested the validity of the linear approximation: the energy difference from the steepest decent results keeps below 3\% for $\epsilon<0.02$.}. The above minimization expression Eq.(\ref{4_e1}) could then be written as,
\be
\mh(\Gamma)=\frac{k}{2}\sum_{\Gamma}\epsilon_{\langle i,j\rangle}\mg_{\langle i,j\rangle,\langle l,m\rangle}\epsilon_{\langle l,m\rangle}+o(\epsilon^3)
\label{4_hamiltonian}
\ee
where $\epsilon_{\langle i,j\rangle}=\epsilon_{\gamma}$ when spring $\gamma$ connects $i$ and $j$. 
The coupling matrix $\mg=\mpp-\ms(\ms^t\ms+\frac{k_{\rw}}{k}\ms_{\rw}^t\ms_{\rw})^{-1}\ms^t$, derived in our previous works~\cite{Yan13,Yan14} (or see Appendix Sec.~\ref{app_C1}), is a product of the structure matrix $\ms$ and its transpose $\ms^t$, the structure matrix of the weak spring network $\ms_{\rw}$, and $\mpp$ the projection operator of the triangular lattice onto occupied edges. The structure matrices $\ms$ and $\ms_{\rw}$ describe the topology of the networks of strong and weak springs: if neighbor nodes $i$ and $j$ are connected, the change of the distance between $i$ and $j$, $\delta r_{\langle i,j\rangle}=\ms_{\langle i,j\rangle,i}\cdot\delta\vec{R}_i+\ms_{\langle i,j\rangle,j}\cdot\delta\vec{R}_j+o(\delta\vec{R}^2)$, due to displacements of nodes $\delta\vec{R}$. We point out that as the weak network is fixed, $\ms$ and thus $\mg$ depend only on the network topology of strong springs, but not on the mismatches $\epsilon_{\gamma}$. 
% distinct from previous models

% extended discussions (what is not included)
Our model is a generalization of on-lattice network models: setting the interaction strength control parameter $\alpha=0$, it naturally recovers the randomly diluted lattice model~\cite{Jacobs95} when $T=\infty$. It is also related to the self-organized lattice model~\cite{Thorpe00,Chubynsky06}, which postulates that elastic energy is linearly proportional to the number of redundant constraints~\cite{Thorpe00,Barre09}. We will find that this assumption holds true for $\alpha=0$ and $T\ll1$.
However, the existence of weak interactions among sites means that in real physical systems $\alpha>0$. This turns out to completely change the physics, an effect that our model can incorporate.

\section{Numerical Results of the Model}
%We concentrate on the thermodynamics of the model, especially the specific heat's dependence on control parameters: temperature $T$, coordination number $z$, and strength of weak interactions $\alpha$.
We implement a Monte Carlo simulation to sample the configuration space of the model, with $10^6$ Monte Carlo steps at each $T$. At each step, a potential configuration is generated by a Glauber dynamics - moving one randomly chosen spring to a vacant edge, as illustrated in Fig.~\ref{4_model}(c). 
We numerically compute the elastic energy of the proposed configuration using Eq.(\ref{4_hamiltonian}): calculating the structure matrix $\ms$ and then the corresponding $\mg$. On computing $\mg$, the matrix { inversion, $(\ms^t\ms+\frac{k_{\rw}}{k}\ms_{\rw}^t\ms_{\rw})^{-1}$,} is singular when the network contains floppy structures, which do not appear except when $k_{\rw}=0$. When $\alpha=0$, we implement the ``pebble game'' algorithm~\cite{Jacobs97} to identify the over-constrained sub-networks, and then do matrix division in the subspace, as the isostatic and floppy regions store no elastic energy after relaxation. 
We have found little finite size effect by varying the system size from $N=64$ to $N=1024$ nodes in the triangular lattice. In the following, we present our numerical results of networks with $N=256$ nodes, averaged over 50 realizations of random mismatches if not specified.

\subsection{Dynamics}
\begin{figure}[h!]
\centering
\includegraphics[width=.8\columnwidth]{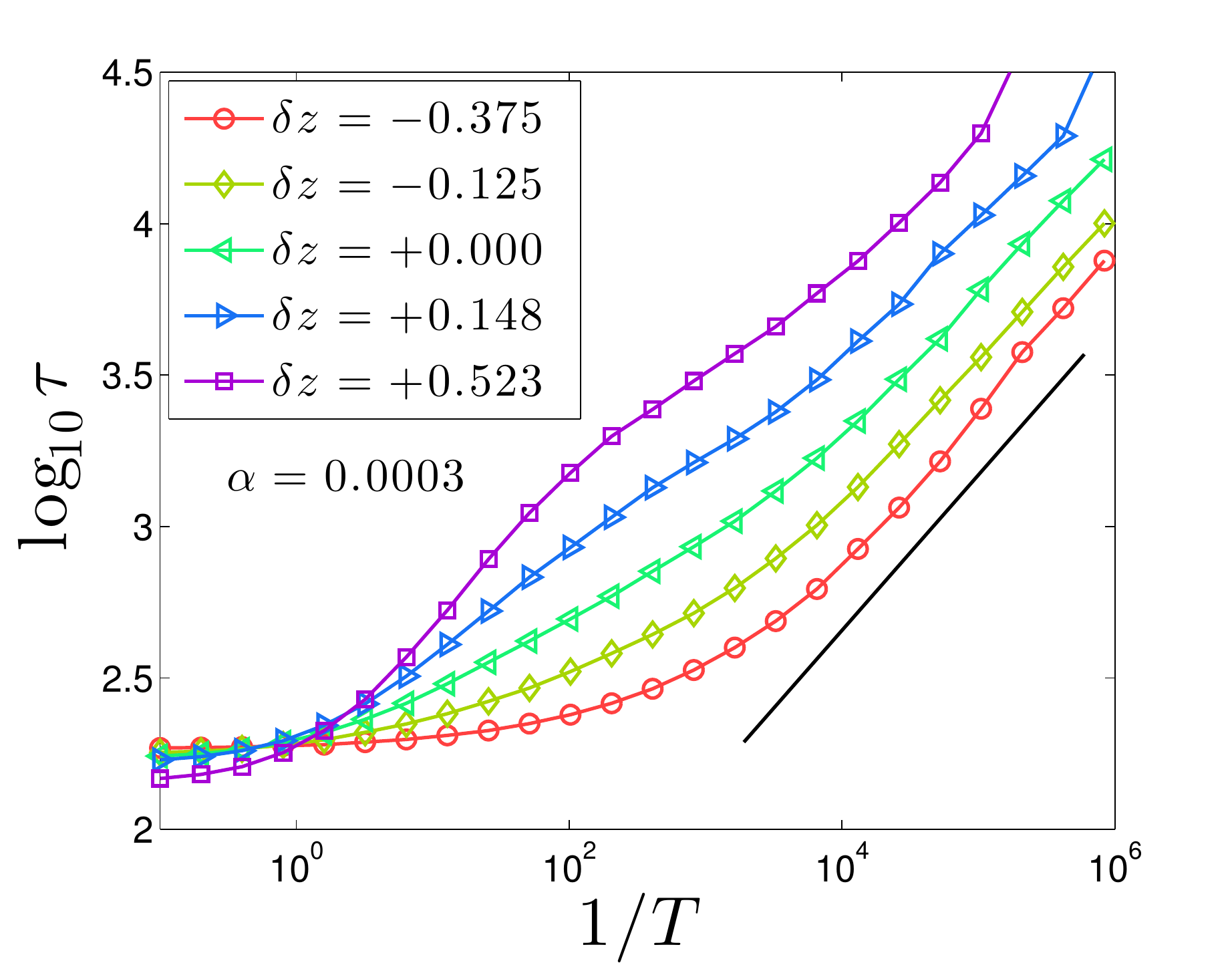}
\caption{\small{(Color online) Relaxation time $\tau$ in log-scale versus inverse temperature $1/T$ for different coordination numbers $\delta z$ and $\alpha=0.0003$. The solid black line indicates a power law relation between $\tau$ and $T$: $\tau\sim T^{-1/2}$.}}\label{4_logt}
\end{figure}

We investigate the dynamics by computing the correlation function $C(t)=\frac{1}{N_{s}(1-N_s/3N)}(\langle\sigma(t)|\sigma(0)\rangle-N_s^2/3N)$, { where $|\sigma(t)\rangle$ is the vector indicating the occupation  of all edges at time $t$. The correlation $C(t)$} decays from one to zero at long time scales.  We define the relaxation time $\tau$ as the time $C(\tau)=1/2$, and the numerical results of $\tau$ as a function of temperature $T$ for several different coordination numbers are shown in the Fig.~\ref{4_logt}. 

We find that the implemented dynamics is not glassy. The relaxation time increases as a power law of the temperature $T^{-0.5}$, even much slower than a strong glass that would display an Arrhenius behavior $\log_{10}\tau\propto1/T$.  This result is very surprising because the frozen elastic network model we studied earlier was glassy (its fragility was similar to that of network liquids). Despite being dynamically very different, these two models are almost identical as far as thermodynamics is concerned, as we will see below. It could be that the lack of glassiness comes from our choice of Monte-Carlo where springs can try other locations anywhere in the system~\cite{Grigera01a}. %(it would be interesting to see if similar Monte-Carlo methods used in real models of polydisperse particles can achieve rapid equilibration). 

%Locally, the energy barrier among configurations is small: relaxation can be achieved by locomotions of springs with small mismatches. Globally, the polydispersity of springs makes the entropy diverge logarithmically as the system size: so the system is far from any possible ideal glass transition~\cite{Lubchenko07} at any finite temperature. Understanding this non-glassy dynamics of the model may hint on the question that which of dynamic processes dominate the slow dynamics, the nucleation~\cite{Kirkpatrick89,Bouchaud04} or the elastic activation~\cite{Dyre06} or the mixture of both~\cite{Bevzenko09,Rabochiy13},  and is left as an open question. 
%A benefit of the non-glassy dynamics is that it enables us to equilibrate the network configurations at a very low temperature~\cite{Yan14}. 
%Whereas, we have to seek for an alternative to define the glass transition temperature and discuss the specific heat jump around the transition temperature. 
%
%\subsection{B. Definition of $T_g$}

\begin{figure}[t!]
\centering
\includegraphics[width=1.0\columnwidth]{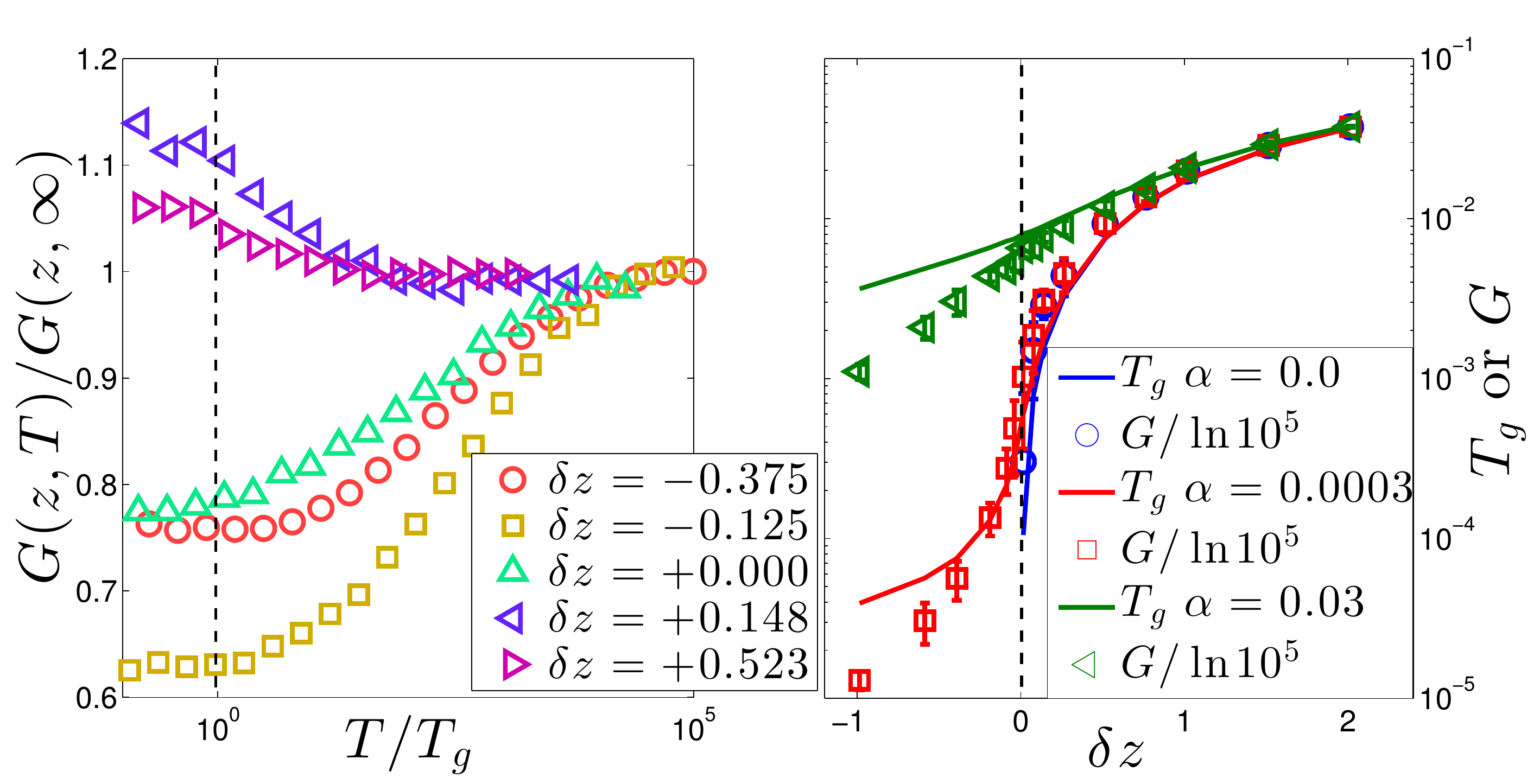}
\caption{\small{(Color online) Left: Shear modulus of adaptive networks at temperature $T$ rescaled by $G$ at $T=\infty$ $G(z,T)/G(z,\infty)$, $\alpha=0.0003$. The temperature $T$ is rescaled by $T_g$. Right: Correlation between transition temperature $T_g$ and shear modulus $G$ in the frozen network model~\cite{Yan13}. %shown for different $z$ and $\alpha$.
}}\label{4_GT}
\end{figure}

To compare the thermodynamics of these models we now need to define an effective glass temperature $T_g$ (even if we do not see a real glass transition). { We do that by using the  empirical Lindemann criterion~\cite{Lindemann10} according to which an amorphous solid melts when the standard deviation $\langle \delta R^2\rangle^{1/2}$ of particles' displacements is greater than a fraction $c_L$ of the particle size $a$. The coefficient $c_L$ must depends on the quench rate $q$, since this is also the case for $T_g$. This dependence is logarithmic, because the dependence of relaxation time on temperature in experimental glass formers is  at least exponential (for typical experimental quench rate in supercooled liquids, $c_L\approx 0.15$   ~\cite{Nelson02})}.  We can estimate this standard deviation via the elastic modulus if we treat the glass as a continuum $\langle \delta R^2\rangle\sim T/G a$ where $G$ is the instantaneous shear modulus of the structure~\cite{Dyre06}, we thus get $T_g\propto Ga^3/\ln(1/q)$. We set the lattice length $a$ in our model to unity. 

We measure the shear modulus averaging over configurations at given temperatures, shown in the left panel of Fig.~\ref{4_GT}. 
Practically, we choose { $T_g=\langle G\rangle_{T_g}/\ln(1/10^3 q)$, where the cooling rate $q$ is defined as the inverse of  the number of Monte Carlo steps performed at each temperature in the model}. $\langle\bullet\rangle_{T_g}$ is the mean value at temperature $T_g$. The prefactor in this definition of $T_g$ does not affect qualitatively { our conclusions, but for this pre-factor the definition of $T_g$ in the frozen model \cite{Yan13} is essentially identical to the dynamical definition used in \cite{Yan13}, as shown in the right panel of Fig.~\ref{4_GT} by lining up $G$ and $T_g$.  The specific values of $T_g$ following that definition are shown in the inset of the bottom panel of Fig.~\ref{4_cp}, they correspond to $T_g=\langle G\rangle_{T_g}/\ln(10^3 )$ in the present model, and $T_g=\langle G\rangle_{T_g}/\ln(10^5 )$ in the frozen network model~\cite{Yan13}, which is simpler to simulate and can thus be equilibrated longer.} 
%Note that this definition of $T_g$ is consistent with the dynamics in the frozen model, as shown in the right panel of Fig.~\ref{4_GT} by lining up $G$ and $T_g$ (which can be defined from the dynamics in that case \cite{Yan13}). 

\begin{figure}[ht]
\centering
\includegraphics[width=1.0\columnwidth]{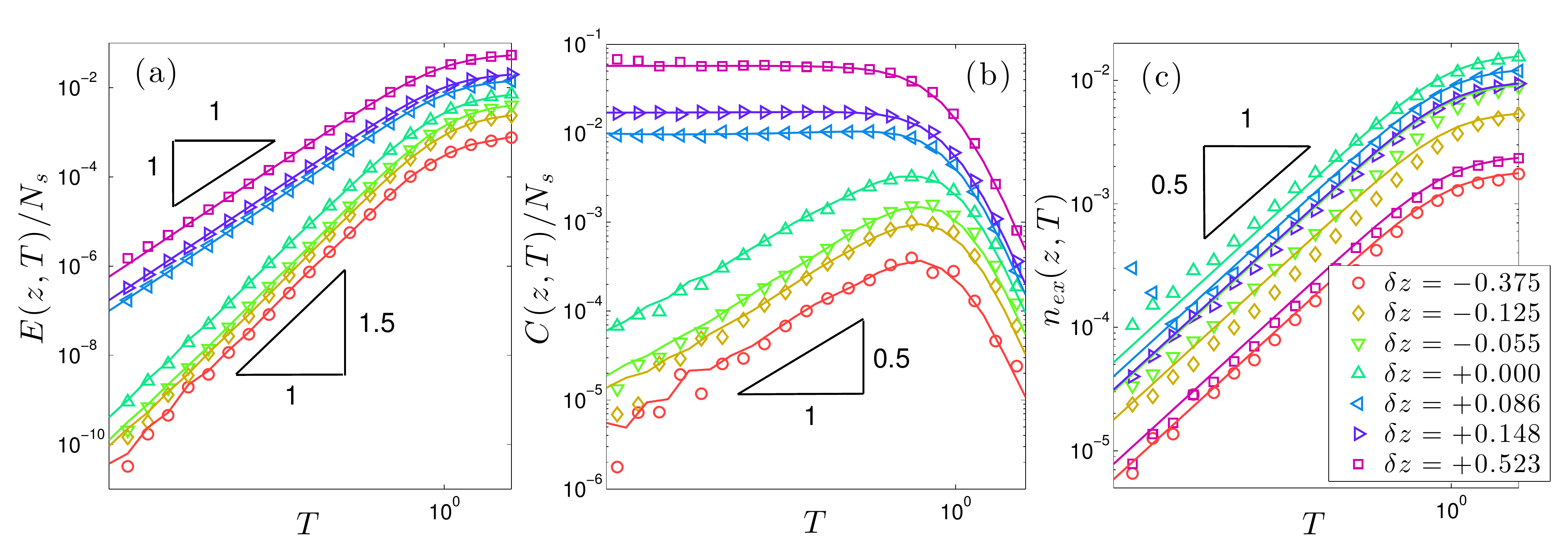}
\caption{\small{(Color online) Thermodynamics of the adaptive network model without weak constraints $\alpha=0$. (a) Energy $E/N_s$ vs. temperature; (b) specific heat $C/N_s$ vs. temperature; (c) excess number density of redundant constraints $n_{\rm ex}$ extracted using the pebble game algorithm vs. temperature. Symbols are numerical data; solid lines are theoretic predictions.
}}\label{4_thermo0}
\end{figure}

\subsection{Specific heat}

\begin{figure}[h!]
\centering
\includegraphics[width=.48\columnwidth]{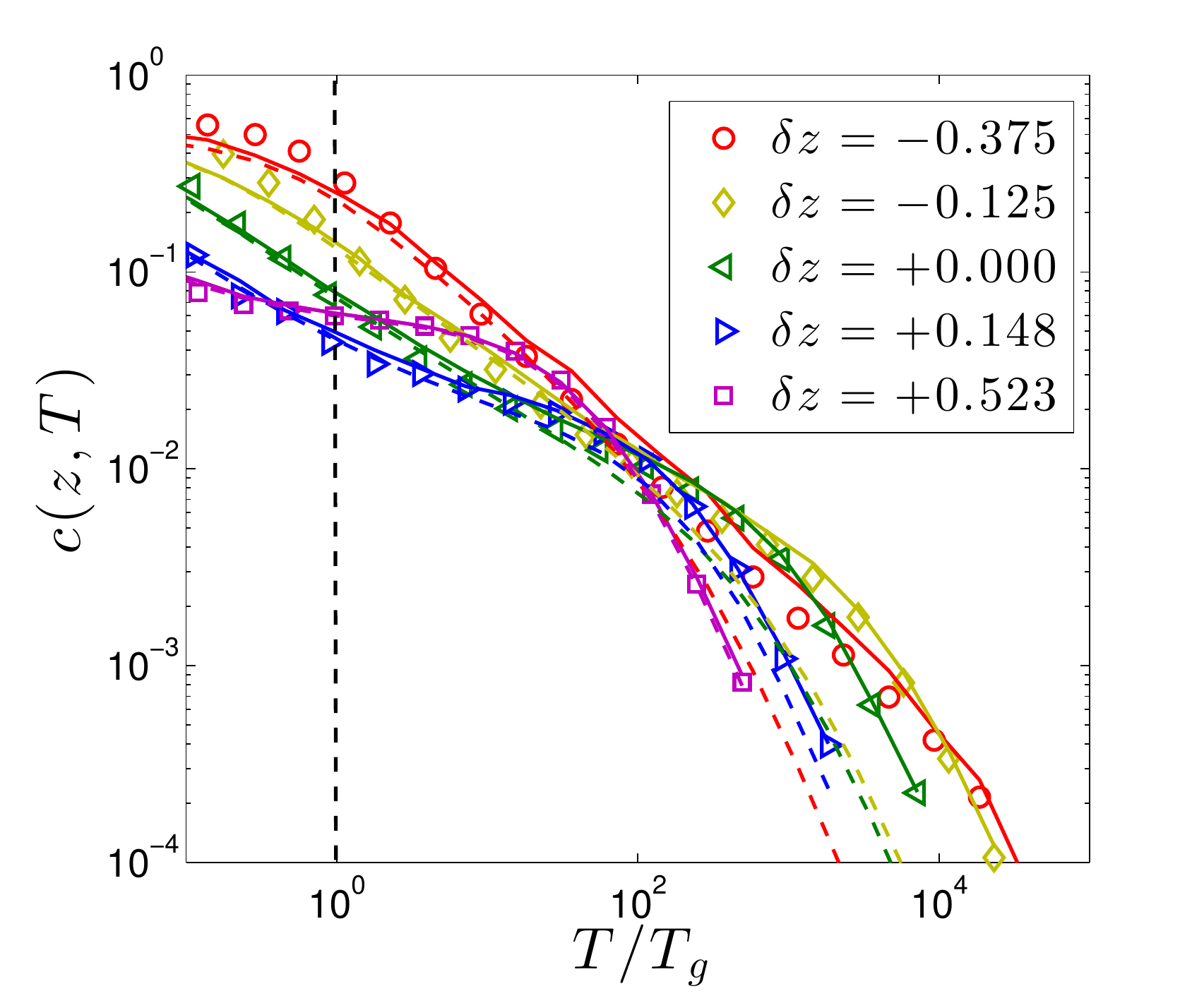}
\includegraphics[width=.48\columnwidth]{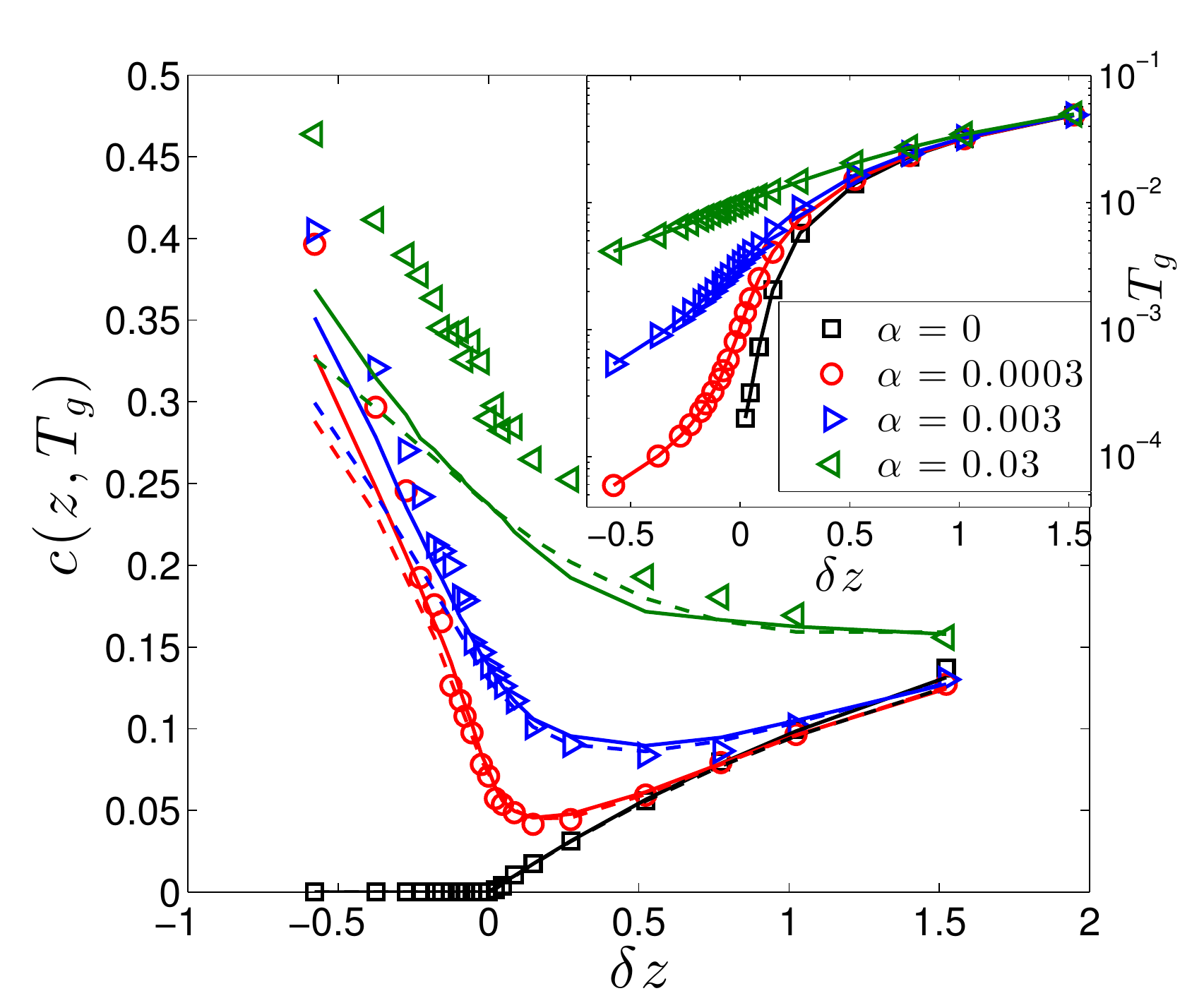}
\caption{\small{(Color online) Top: Specific heat $c(z,T)$ {\it vs} scaled temperature $T/T_g$ for networks with average coordination numbers near and away from the isostatic on both floppy and rigid sides. The strength of the weak constraints is given by $\alpha=0.0003$. Bottom: Specific heat at temperature $T_g$, $c(z,T_g)$, {\it vs} coordination number $\delta z$ for $\alpha=0,\ 0.0003,\ 0.003,\ 0.03.$  The inset shows the transition temperature $T_g$ for different $z$ and $\alpha$. Symbols are numerical results, and lines are theoretical predictions: dashed lines are for frozen network model and solid lines are for the new model derived in section IV.}}\label{4_cp}
\end{figure}

% depends on T, depends on z  at Tg, justify Tg
The specific heat data shown in Figs.~\ref{4_thermo0} and \ref{4_cp} are our central numerical results. 
The energy $E=\langle \mh\rangle$ is  obtained using a time-average over Monte Carlo steps, and is shown in Fig.~\ref{4_thermo0}(a). The specific heat is calculated as its derivative $c\equiv\frac{1}{N_s}\rd E/\rd T$, and is shown   versus $T$ for several coordination numbers when $\alpha=0$ in Fig.~\ref{4_thermo0}(b)  and $\alpha=0.0003$ in the top panel of Fig.~\ref{4_cp}. When $\alpha=0$, the specific heat increases as temperature decreases for networks with $\delta z>0$ while it meets a maximum at $T_a\sim1$ and decreases under cooling when $T<T_a$ if $\delta z\leq0$. By contrast, the specific heat increases under cooling close to the transition temperature for all coordination numbers when $\alpha>0$. In addition, when $T\lesssim\alpha$, $c\to0.5$.%, almost every strong springs stores energy and contributes a fraction of $0.5$ to the total heat capacity. %, as indicated in the equipartition theorem. 
All these results are qualitatively identical to our previous frozen model. 

%However, due to the non-glassy dynamics of the model, no obvious specific heat jump signals the glass transition. 
%We define the jump of specific heat  $\Delta c$ in an alternative way. %from directly measuring the jump in $c$-$T$ plot. 
%We define the ``jump'' by the specific heat contribution from the configurational entropy (vibrational entropy contribution discussed in Sec. V-A) at the ``glass transition'' temperature as if the network was frozen in one of inherent structures and the contribution jumped to zero below $T_g$. 
To define the jump of the specific heat at the glass transition, we simply measure the specific heat at our glass transition $T_g$ defined above.  This definition is natural, since in a real glassy system, below $T_{g}$ the liquid is  essentially frozen in an inherent structure, and the contribution to the specific heat from configurational entropy (i.e. the bottom energy of inherent structures) vanishes.  %The specific values of $T_g$ are shown in the inset of the bottom panel of Fig.~\ref{4_cp}.

Our central numerical result is shown in the bottom panel of Fig.~\ref{4_cp}:  $c(T_g)$ varies nonmonotonically with the coordination number $z$ when $\alpha>0$. When the network of strong springs is poorly coordinated $\delta z\lesssim0$, $c(T_g)$ decreases as $z$ increases; When the strong network gets better coordinated $\delta z\gtrsim0$, $c$ gradually changes to increase with $z$; $c$ is minimal at the proximity of the rigidity transition $z_c$ for finite $\alpha$. These numerical results are very similar to empirical observations, see Point (II) in the introduction.
Our data are in fact very similar to that of the frozen model, which essentially follows the dotted lines in Fig.~\ref{4_cp}. 

\subsection{Number of redundant constraints $R$}
When $\alpha=0$ and $T\to0$, the specific heat is simply proportional to $R$, as shown in Fig.~\ref{4_thermo0}(b). This number is fixed, $R=N\delta z/2$, in the frozen network models. It varies in the adaptive network model and depends on the temperature. As the Maxwell counting gives the minimal number of redundant constraints of a network, we can define an excess number of redundant constraints
\be
\label{4_nex}
n_{\rm ex}\equiv \frac{1}{N_s}\left(R-\frac{N\delta z}{2}\Theta(\delta z)\right),
\ee
where $\Theta(x)$ is the Heaviside step function. $n_{\rm ex}$ counts the average number of redundant constraints, additional to the Maxwell counting. 
This excess number of redundant constraints decreases monotonically to zero under cooling. When $\alpha=0$, $n_{\rm ex}$ is proportional to $\sqrt{T}$ in the adaptive network model at low temperature, shown in Fig.~\ref{4_thermo0}(c).

\begin{figure}[h!]
\centering
\includegraphics[width=.9\columnwidth]{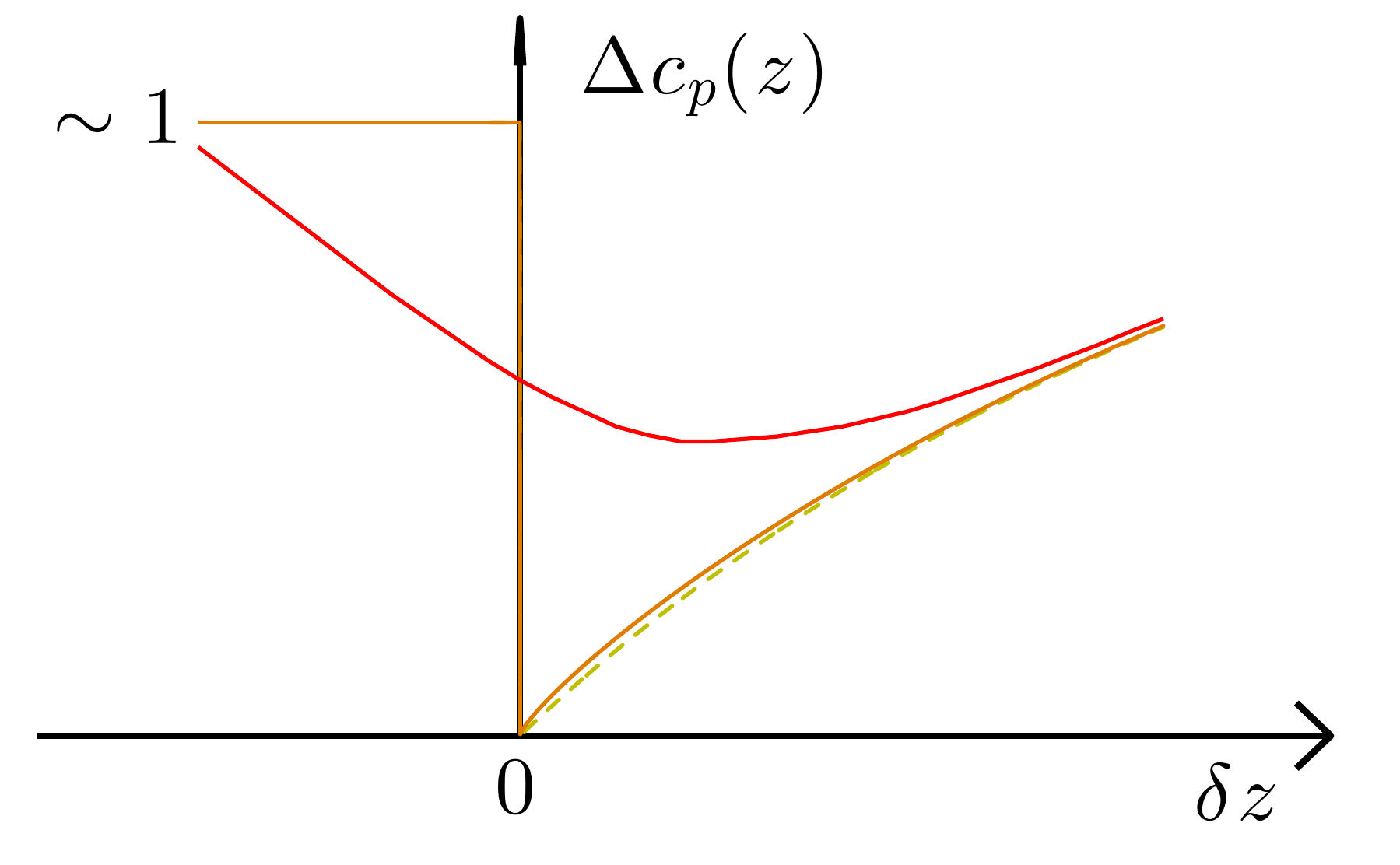}
\caption{\small{(Color online) Theoretical predictions for the jump of specific heat. For vanishingly weak springs $\alpha\rightarrow 0$, it is predicted that the jump is essentially constant for $z<z_c$ and then drops to zero a $z_c$. For larger $z$, it behaves as $z-z_c$. As $\alpha$ grows this sharp curve becomes smooth, but a  minimum is still present near $z=z_c$.  }}\label{4_theory}
\end{figure}

\section{Theory of Thermodynamics}
%In the previous section, we have numerically shown that the specific heat of the adaptive network model recovers the one of the frozen network model. Moreover, we have numerically confirmed that the number of redundant constraints converges to the Maxwell counting. We have also shown that the density of states $D(\omega)$ converges to the one predicted by the mean-field theory~\cite{Wyart08,During13,DeGiuli14,Mao13} due to a homogenizing adaptation under cooling of the model~\cite{Yan14}. We are now going to show these two facts guarantee a qualitative consistence of thermodynamic behaviors between the two models. 

As illustrated in Fig.~\ref{4_theory}, in the frozen elastic model we found that as $\alpha\rightarrow 0$, $c$ converges to a constant if $z<z_c$,
whereas it behaves as $z-z_c$ for $z>z_c$. As $\alpha$ is increased, the discontinuous behavior becomes smooth and looks similar to experimental data. We seek to derive these same features in the adaptive network models.

%Our central picture of the nonmonotonic behavior of the specific heat bases on networks with non-zero but vanishing weak forces $\alpha\to0^+$ as follows: for rigid covalent networks $\delta z>0$, the number of directions contributing elastic energy, equal to the number of redundant constraints, vanishes as $\delta z\to0$, and thus the specific heat vanishes linearly with $\delta z$; for floppy networks $\delta z<0$, weak constraints stabilize the structure, every local configuration stores an energy portion comparable to temperature near the transition, and the specific heat becomes a constant in the zero temperature limit. This discontinuous step in the jump of specific heat shown as the yellow dashed line in Fig.~\ref{4_theory} becomes a continuous crossover with a minimal near the rigidity threshold when $\alpha$ is finite, shown as the red line in Fig.~\ref{4_theory}. 
%
%In this section, we first derive that the thermodynamics of the adaptive network model. Then we give a quantitative prediction on the number of redundant constraints, rooting in a simple ideal gas picture of redundant constraints. In the end, we point to the similarity in formalism between the adaptive network model and the frozen network model.

% general theoretic formalism
\subsection{Partition function}
For simplicity, we consider the annealed free energy $\mf_{\rm ann}=-T\ln\overline{\mz}$. It is exact in the random energy model~\cite{Derrida81} above the ideal glass transition~\cite{Mezard09} and we find it to be a good approximation of $\overline{\mf}$ in our models~\cite{Yan13}. 
%We start from considering the free energy of the system 
%as quenched randomness $\epsilon$ are independent Gaussian variables. 
The over-line implies an average over disorder $\epsilon$, 
\be
\overline{\mz}=\overline{\sum_{\{\sigma\}}\sum_{\text{perm}[\gamma]}\exp[-\mh(\Gamma)/T]}
\label{4_partfunc}
\ee
where a given configuration $\Gamma$ is characterized by $\{\sigma\}$  indicating which  edges are occupied on  the triangular lattice, %$\sigma_{\langle i,j\rangle}=1$ for the link connected by a spring or $0$ for none, 
and $\text{perm}[\gamma]$ labels the possible permutations of springs' rest lengths. %The Hamiltonian is defined in Eq.(\ref{4_e1}). %Hamiltonian is defined in the model for a given configuration shown as Eq.(\ref{4_hamiltonian}).

We first average over the quenched randomnesses. Using the linear approximation Eq.(\ref{4_hamiltonian}) and the Gaussian distribution $\rho(\epsilon_{\gamma})=\frac{1}{\sqrt{2\pi\epsilon^2}}e^{-\epsilon_{\gamma}^2/2\epsilon^2}$,
\be
\overline{\mz}=\sum_{\{\sigma\}}\left(\frac{Nz}{2}\right)!\exp\left[-\frac{1}{2}\tr\ln\left(\mi+\frac{\mg(\{\sigma\})}{T}\right)\right]
\label{4_partz}
\ee
The factorial comes from $N_s!=\sum_{\text{perm}[\gamma]}{\bf 1}$ as $\mg$ is independent of the permutation. $\mi$ is a $3N\times 3N$ identity matrix; each component corresponds to an edge on the lattice. 
To compute the trace in the exponent, we first make the approximation that the  weak springs are weak and numerous $\ms_{\rw}^t\ms_{\rw}\approx\frac{z_{\rw}}{d}\mi_{Nd\times Nd}$, which corresponds to the highly connected limit $z_{\rw}\to\infty$ and finite $\alpha$. We can then decompose the coupling matrix $\mg\approx\mpp-\ms(\ms^t\ms+\alpha\mi)^{-1}\ms^t$ as ~\cite{Yan13}:
\be
\mg(\{\sigma\})=\sum_{p(\{\sigma\})}|\psi_{p}\rangle\langle\psi_p|+\sum_{\omega(\{\sigma\})>0}\frac{\alpha}{\omega^2+\alpha}|\psi_{\omega}\rangle\langle\psi_{\omega}|
\label{4_coupmat}
\ee
where $p$ labels the vectors $|\psi_p\rangle$ satisfying $\ms^t|\psi_p\rangle=0$ (i.e. a basis for the kernel of $\ms^t$), and where the $|\psi_{\omega}\rangle$ satisfy $\ms\ms^t|\psi_{\omega}\rangle=\omega^2|\psi_{\omega}\rangle$. The number of redundant directions is $\sum_{p}{\bf 1}=N_s-(Nd-F)\equiv R$. Note that  $\tr\mpp=N_s$, $Nd-F$ gives the number of frequencies $\omega$, and $F$ counts the number of floppy modes. The modes  $|\psi_p\rangle$, $|\psi_{\omega}\rangle$, $R$, and $\omega$ depend on occupation $\{\sigma\}$. As the $|\psi\rangle$'s are orthonormal, the trace in Eq.(\ref{4_partz}) gives %Because all different directions $|\psi\rangle$ are orthonormal, the trace in Eq.~\ref{4_partz} is still easy to work out,
%\begin{widetext}
\begin{multline}
\overline{\mz}=\left(\frac{Nz}{2}\right)!\sum_{n_r,D(\omega)}\exp\left[N_s\left(s(n_r,D(\omega))-\frac{n_r}{2}\ln(1+\frac{1}{T})\right.\right.\\
\left.\left.-\frac{1-n_r}{2}\int\rd\omega D(\omega)\ln(1+\frac{1}{T}\frac{\alpha}{\omega^2+\alpha})\right)\right],
\label{4_part}
\end{multline}
%\end{widetext}
where $s(n_r,D(\omega))\equiv\frac{1}{N_s}\ln\sum_{\{\sigma\}}{\bf 1}_{R,D(\omega)}$ is configurational entropy density with given number of redundant constraints $n_r\equiv R/N_s$ and density of vibrational modes, $D(\omega)$, satisfies $(1-n_r)\int\rd\omega D(\omega)\equiv\lim_{N\to\infty}\frac{1}{N_s}\sum_{\omega>0}$.

\subsection{No weak interactions}

\begin{figure}[h!]
\centering
\includegraphics[width=.8\columnwidth]{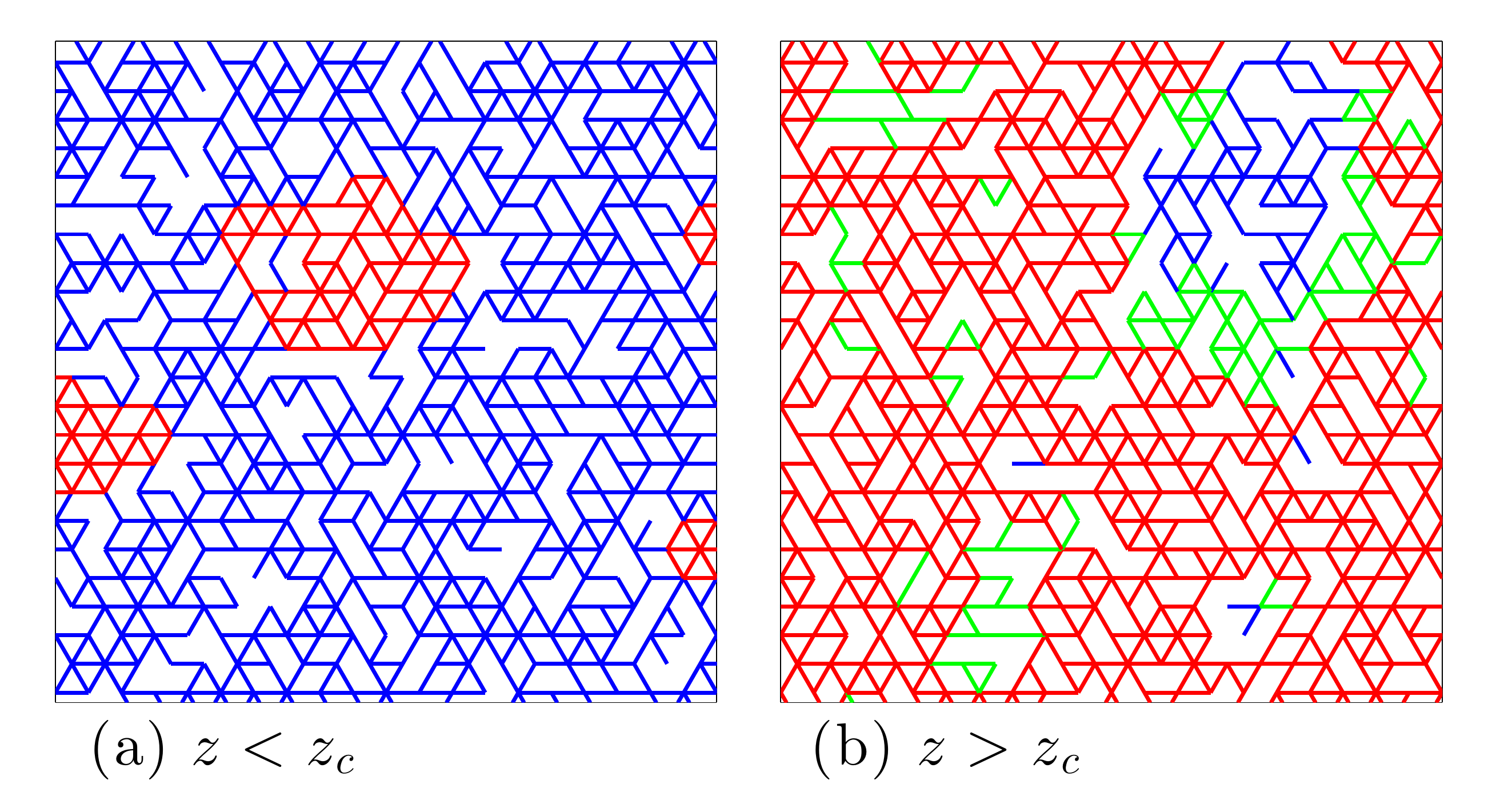}
\caption{\small{(Color online) (a) $z<z_c$, localized redundant constraints (red) in a floppy sea (blue); (b) $z>z_c$ localized floppy modes (blue) in a rigid sea (red and green).}}\label{4_defect}
\end{figure}

%The coupling matrix $\mg$ is a projection operator, $\mg^2=\mg$ and $\tr\mg=R$.
Neglecting the weak constraints $\alpha=0$, the last term in the exponential vanishes and the summation over states with given density of states can be absorbed into the entropy, which then depends only on the number of redundant constraints. 
\be
\overline{\mz}=\left(\frac{Nz}{2}\right)!\sum_{n_r}e^{N_s[s(n_r)-\frac{n_r}{2}\ln(1+\frac{1}{T})]}
\label{4_part0}
\ee

We propose an ideal-gas picture of ``defects''  to find an approximation form of the entropy $s(n_r)$. 
When the coordination number is very small $z<z_c$ and the network is mostly floppy, redundant constraints are defects localized in rigid islands. Similarly, when the coordination number is very large $z>z_c$ with most regions of the network rigid, there are localized floppy modes in regions where there are negative fluctuations of coordination number, which we again described as defects, see illustration in   Fig.~\ref{4_defect}. The number of such floppy modes is equal to the number of additional over-constrained in the rigid cluster. The entropy gains from having these defects. Assuming that such defects are independent, we approximate the entropy by that of an ideal  gas: 
\be
s(n_{\rm ex})\approx s(0)-n_{\rm ex}\ln\frac{n_{\rm ex}}{en_0(z)}
\label{4_s}
\ee
where $n_{\rm ex}$ is the excess number of redundant constraints defined in Eq.(\ref{4_nex}) and is thus counting the number of defects. $s(0)$ is the entropy density of the states with a minimal number of redundant constraints (i.e. they satisfy the Maxwell counting); and $n_0(z)$ is the excess number of redundant constraints at $T=\infty$. Both $s(0)$ and $n_0$ depend only on $z$ and the lattice structure. This form of Eq.(\ref{4_s}) fails when the assumption of independent ``defects'' breaks down, as must occur near the rigidity transition. However,  our numerical results indicate that this approximation is very accurate, we see deviations only for $|\delta z|\lesssim0.1$. % and small $\alpha$, a parameter range  presumably not achievable  in experiments (where we estimated $\alpha\approx 0.05$ \cite{Yan13}). % on the triangular lattice. %The idea is that when $\delta z<0$, the entropy characterizes the independent redundant constraints in floppy media, while $\delta z>0$, the entropy features the ``holes'' in rigid media. 

\begin{figure}[h!]
\centering
\includegraphics[width=1.0\columnwidth]{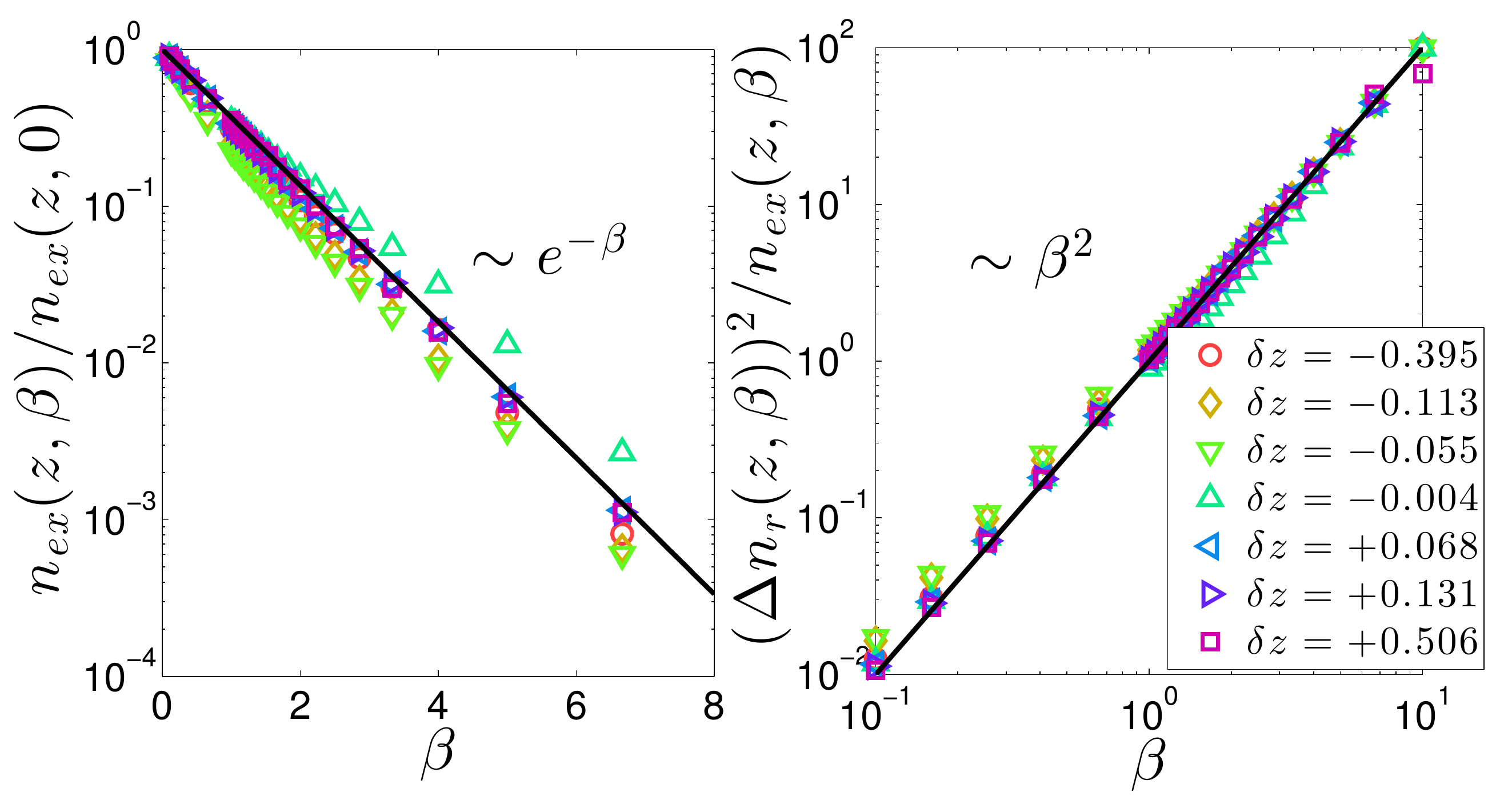}\\
\caption{\small{(Color online) Left: Excess number density of redundant constraints  $n_{\rm ex}(z,\beta)$.  Right: Fluctuation of the number density of redundant constraints  $(\Delta n_r)^2$. The solid black lines show the predictions from the approximate entropy Eq.(\ref{4_s}). %Bottom Left: Entropy $s$ for given excess number of redundant constraints density $n_{\rm ex}$, the black solid line rescaled from Eq.(\ref{4_s}) collapse the data for different $z$. Bottom Right: Correlation between two rescaling quantities $s(0)$ and $n_0$ in the left plot for $z$ and $N$, the black solid line corresponds to the ideal gas entropy, Eq.(\ref{4_s}).
}}\label{4_snr}
\end{figure}

% numeric test of the formula
%From above analysis, we expect the above approximation works well for $|z-z_c|\gtrsim0.03$. 
We numerically test the formula Eq.(\ref{4_s}) for a triangular lattice. % by implementing the same Glauber dynamics described before to sample the configurations with given number of redundant constraints. 
The configurations with $R$ redundant constraints are weighted by $e^{-\beta R}$ for different values of the parameter $\beta$. From Eq.(\ref{4_s}), the mean and variance of the excess number density of redundant constraints, $n_{\rm ex}$, satisfy the following formulas: % $n_r\equiv R/N_s$ is equivalent to probing configurations with different $n_{\rm ex}$. Numerically, we can directly measure the mean and variance of the excess number of redundant bond with given parameter $\beta$ in Monte Carlo dynamics.
\begin{subequations}
\begin{equation}
\beta\equiv\frac{\partial s}{\partial n_{\rm ex}}\ \Rightarrow n_{\rm ex}(z,\beta)=n_0(z)e^{-\beta}
\label{4_nbeta}
\end{equation}
\begin{equation}
\Delta n_{\rm ex}^2(z,\beta)=-\beta^2\frac{\partial}{\partial\beta} n_{\rm ex}(z,\beta)=\beta^2 n_{\rm ex}(z,\beta)
\label{4_cbeta}
\end{equation}
\end{subequations}
%Eqs.~\ref{4_nbeta},\ref{4_cbeta} are tested for different coordination numbers, even close to $z_c$. 
Our numerical results coincide with Eqs.(\ref{4_nbeta}) and (\ref{4_cbeta}) remarkably well, with minor deviations for $|\delta z|\lesssim0.1${, as shown in Fig.~\ref{4_snr}}.% (see Fig.~\ref{4_snr}). 

%The trace of this logarithm is in fact not as hard as it seems to be. 
Applying Eq.(\ref{4_s}), we derive the thermodynamics of our model when $\alpha=0$.  %Averaged over frozen randomness, the partition function depends only on number of redundant constraints,
%where $s(R)\equiv\frac{1}{N_s}\ln\sum_{\{\sigma\}(R)}$ is exactly the entropy density given $R$ redundant constraints. Inserting Eq.(\ref{4_s}) and taking the saddle point value:
%\be
%\frac{1}{2}\ln\left(1+\frac{1}{T}\right)=\frac{\partial}{\partial n_{\rm ex}}s(n_{\rm ex})=-\ln\frac{n_{\rm ex}}{n_0}
%\label{4_saddle0}
%\ee
\begin{subequations}
Solving the saddle point of Eq.(\ref{4_part0}), we obtain the average energy density:
\be
\frac{1}{N_s}E(z,T)=\frac{r_0+n_{\rm ex}(z,T)}{2}\frac{T}{1+T}
\label{4_e0}
\ee
the specific heat:
\be
\frac{1}{N_s}C(z,T)=\frac{r_0+\frac{3}{2}n_{\rm ex}(z,T)}{2}\frac{1}{(1+T)^2}
\label{4_c0}
\ee
and the excess number density of redundant constraints:
\be
n_{\rm ex}(z,T)=n_0(z)\left(1+\frac{1}{T}\right)^{-1/2}
\label{4_n0}
\ee
\label{4_noweak}
\end{subequations}
where $r_0\equiv \frac{\delta z}{z}\Theta(\delta z)$. 

%When $T\ll1$, $n_{\rm ex}(z,T)\sim T^{1/2}$ and thus $E/N_s\sim T^{3/2}$ leading to $C/N_s\sim T^{1/2}$ for $\delta z\leq0$.
%
%For $\delta z>0$ we recover our central result  $C/N_s\sim \delta zT^{0}+o(T^1/2)$.

 As $n_0(z)$ is expected to be an analytic function of $z$, Eqs.(\ref{4_noweak}) indicate that $c$ converges to the one found in frozen network model in the limit $T\to0$:  $c=0$ when $\delta z<0$ and $c=\delta z/2z$ when $\delta z>0$ - the dashed yellow line in Fig.~\ref{4_theory}. This is our first central result,
which shows that our previous results hold even when the network is adaptive. 

Eqs.(\ref{4_noweak}) predict the energy, specific heat, and the number density of redundant constraints at an arbitrary temperature without any fitting parameter. The solid lines, shown in Fig.~\ref{4_thermo0}(a) and (b), are predictions of Eqs.(\ref{4_e0}) and (\ref{4_c0}), respectively, with $n_{\rm ex}$ as the numerical input. They are closely consistent with the data points, which confirms the annealed free energy approximation when $\alpha=0$. A $T^{1/2}$ power-law with numerical prefactor $n_0(z)=n_{\rm ex}(z,\infty)$ predicted by Eq.(\ref{4_n0}) coincides well with data points in Fig.~\ref{4_thermo0}(c).  %The coincidence in Fig.~\ref{4_thermo0}(c) again justifies our ideal gas picture of redundant constraints and the approximate entropy Eq.(\ref{4_s}). %In fact, $n_0(z)$ is the only necessary input to theory when $\alpha=0$. %It depends in general on embedding structure. %, or practically the specific geometry of the atoms forming the covalent network. % bring less error than the idea gas assumption of the redundant constraints, at least for $z\in[-0.15,0.03]$, in our model.

Extending to finite glass transition $T_g$ at $\alpha=0$, we find a correction vanishing as $\sqrt{\delta z}$ in addition to $c\approx \delta z/2z$, assuming $T_g\sim G\sim\delta z$ for $z>z_c$. But this correction is quantitatively unimportant as $n_0\leq0.03$ and does not change qualitatively the linear growth of the specific heat when $\delta z> 0$, as illustrated by the solid orange  line in Fig.~\ref{4_theory}. %, the additional contribution induces a finite correction near the rigidity threshold $\delta z=0$, and $c$ vanishes as $\sqrt{\delta z}$ rather than $\delta z$ in the self-organized network model, indicated by the orange solid line in Fig.~\ref{4_theory}. But this small correction does not change qualitatively the results at $\alpha>0$. %, and suggests that our model should reproduce the rigidity window at low temperature when $\alpha=0$~\cite{Yan14}.

Our theoretic prediction that $n_{\rm ex}\to0$ when $T\to0$ validates the assumptions of \cite{Thorpe00,Chubynsky06,Barre09} that the energy of redundant bonds is proportional to their number, and that this number is  $R_0$ at $T=0$. 

\subsection{General case}
% density of states and fractons I
In the thermodynamic limit, $N_s\to\infty$, we take the saddle point of Eq.(\ref{4_part}),
\begin{subequations}
\be
\frac{2\partial s}{\partial n_r}=\ln\left(1+\frac{1}{T}\right)-\int\rd\omega D(\omega)\ln\left(1+\frac{1}{T}\frac{\alpha}{\omega^2+\alpha}\right)
\label{4_saddlen}
\ee
and
\be
\frac{2\delta s}{\delta D(\omega)}=(1-n_r)\ln\left(1+\frac{1}{T}\frac{\alpha}{\omega^2+\alpha}\right)
\label{4_saddleD}
\ee
\label{4_saddle}
\end{subequations}
and solve for energy,
\begin{multline}
\frac{1}{N_s}E(z,T,\alpha)=\frac{n_r(T)}{2}\frac{T}{1+{T}}\\+\frac{1-n_r(T)}{2}\int\rd\omega D(\omega,T)\frac{\alpha T}{\alpha+(\omega^2+\alpha){T}}
\label{4_e}
\end{multline}
The specific heat predictions from differentiating Eq.(\ref{4_e}) with numerical inputs $n_r(z,T,\alpha)$ and $D_{z,T,\alpha}(\omega)$ are plotted as solid lines in Fig.~\ref{4_cp}. (See Appendix Secs.~\ref{app_C2}, \ref{app_C3}, and \ref{app_C4} for the temperature dependence of $D(\omega)$.) %Consistency between the theory and the numerics supports the annealed free energy approximation. 
Notice that replacing $n_r(T)$ by $\delta z/z$ and $D(\omega, T)$ by its low-temperature limit $D(\omega)$ studied in \cite{Wyart08,During13,Yan13}, Eq.(\ref{4_e}) recovers exactly the one obtained in the frozen network model, whose predictions are plotted as dashed lines in Fig.~\ref{4_cp}. The dashed lines converge to the solid lines despite differences at high temperatures for weakly coordinated networks. %Note that the two terms in Eq.(\ref{4_e}) correspond to the elastic energy from the redundant strong springs and the contribution from the eigen-modes in the effective weak field, respectively. Our previous explanation on nonmonotonic jump of specific heat~\cite{Yan13}, based on the dominance of two contributions on different sides of the rigidity transition, would apply if the temperature derivatives of $n_r$ and $D(\omega)$ contribute little to the specific heat near the glass transition.

In the limit $\alpha\to0$ and $T\ll\alpha$, Eq.(\ref{4_e}) converges to $E/N_s=T/2$, which indicates a constant specific heat $c=0.5$ when $\delta z<0$ independent of the models. This is shown by the solid orange line and the dashed yellow line in Fig.~\ref{4_theory}, and is our second key theoretical result showing the robustness of our conclusions for adaptive networks. %The additional contribution to the specific heat from the temperature dependence of density of states vanishes, and almost all local configurations carry a small energy of the same order as the temperature. Together, 
%a specific heat $c\approx\frac{1}{2}\int\rd\omega D(\omega)\frac{1}{(1+(\omega^2+\alpha)T/\alpha)^2}$ of order one is expected as $\alpha\to0$.

\section{Conclusions}
% main result, implication, prospectives
In this work, we have studied the correlation between the elasticity of inherent structures and the thermodynamics in covalent glass-forming liquids using adaptive network models. We found numerically and explained theoretically why these models have a thermodynamic behavior similar to frozen network models \cite{Yan13} which captures nicely experimental facts. 

{ The main prediction conclusion of \cite{Yan13} is thus robust: as the coordination number approaches $z_c$ from above, elastic frustration vanishes. This leads both to an abundance of soft elastic modes, as well as a diminution of the number of directions in phase space that cost energy, which is directly proportional to the jump of specific heat. 
Below the rigidity transition, the elasticity of strong force network vanishes, thus the energy landscape is governed by the weak Van der Waals interactions. At these energy scale, all directions in contact space have a cost,  and thus the specific heat increases. Thus thermodynamic properties are governed by a critical point at $\delta z=0$, $\alpha=0$ where the jump of specific heat is zero. }
This prediction focuses on the configurational part of the jump of specific heat, since we considered only the energy minima in the metastable states.  In Appendix Sec.~\ref{app_C5}, we argue that the vibrational  contribution to this jump is so small in our models. { Thus the main prediction of the specific heat still holds, even when including the vibrational part}. 

%Restate, compare to HW and SH, application to other systems.
{ Beyond network glasses, our main result potentially explains the correlation between   elasticity and the key aspects of the energy landscape in molecular glasses~\cite{Tatsumisago90,Bohmer92,Boolchand05}. Indeed  according to our work we expect glasses with a strong Boson peak to display less elastic frustration, so that they have a limited number of directions in phase space costing energy, see discussion in \cite{Yan13}. }

%% file: erosion/erosion.tex
\chapter{A Model for the Erosion Onset of a Granular Bed Sheared by a Viscous Fluid}
\label{5_erosion}
%\begin{abstract}
We study theoretically the erosion threshold of a granular bed forced by a viscous fluid. We first introduce a novel model
of  interacting particles driven on a rough substrate. It predicts a continuous transition at some threshold forcing $\theta_c$,
beyond which the particle current grows linearly $J\sim \theta-\theta_c$. The stationary state is reached after a transient time $t_{\rm conv}$ which diverges near the transition as $t_{\rm conv}\sim |\theta-\theta_c|^{-z}$ with $z\approx 2.5$. { Both features  agree with experiments.} 
The model also makes quantitative testable predictions for the drainage pattern: the distribution $P(\sigma)$ of local current is found to be extremely broad with $P(\sigma)\sim J/\sigma$, spatial correlations for the current are negligible in the direction transverse to forcing, but long-range parallel to it. We explain some of these features using a scaling argument and  a mean-field approximation that builds an analogy with $q$-models. We discuss the relationship between our erosion model and models for the plastic depinning transition of vortex lattices in dirty superconductors, where our results may also apply. 

%Under a weak external drive, sediments are eroded on a solid bed when the drive is above a critical value. We propose an experimentally testable model of the erosion. The model shows a linear relation between the erosion rate and the external drive above the critical threshold. The linear relation is nontrivially rooted in the fact that the erosion covers the whole landscape, due to a splitting effect of the interaction among eroding particles in dense particle drains. We emphasize this splitting effect with a theoretic analysis on the probability distribution in the local erosion flux.
%\end{abstract}

\section{Introduction}

Erosion shapes Earth's landscape, and occurs when a fluid  exerts a sufficient shear stress on a sedimented layer. It is controlled by the dimensionless Shields number $\theta\equiv\Sigma/(\rho_p-\rho)gd$, where $d$ and $\rho_p$ are the particle diameter  and density, and $\rho$ and $\Sigma$ are the fluid density and the shear stress.  Sustained sediment transport can take place above some critical value $\theta_c$ \cite{Shields36,White70,Lobkovsky08}, in the vicinity of which motion is localized on a thin layer of  order of the particle size, while deeper particles are  static or very slowly creeping \cite{Charru04,Aussillous13,Houssais15}. This situation is relevant in gravel rivers, where erosion occurs until the fluid stress approaches threshold \cite{Parker07}. In that case, predicting the flux $J$ of particles as a function of $\theta$ is difficult, both for turbulent and laminar flows \cite{Bagnold66,Charru04,Charru13,Chiodi14}. We focus  on the latter, where  experiments show that: (i)  in a stationary state, $J\propto (\theta-\theta_c)^\beta$ with $\beta\approx 1$ \cite{Charru04,Ouriemi09,Lajeunesse10,Houssais15}, although other exponents are sometimes reported \cite{Lobkovsky08}, (ii) transient effects occur on a time scale that appears to diverge as  $\theta\rightarrow\theta_c$ \cite{Charru04,Houssais15} and (iii) as $\theta\rightarrow\theta_c$ the { density  $m$} of moving particles vanishes, but not their %characteristic 
speed \cite{Charru04,Lajeunesse10,Duran14}.

%Although a continuous description of erosion appears successful for $\theta\gg\theta_c$ \cite{Leighton86,Ouriemi09,Aussillous13}, it is not expected to apply for $\theta\sim\theta_c$. 

{ Two distinct views have been proposed to describe erosion near threshold. For Bagnold~\cite{Bagnold66} and followers \cite{Chiodi14}, hydrodynamics is key:  moving particles carry a fraction of the  stress  proportional to their density $m$, such that the  bed of static particles effectively remains at the critical Shields number. This argument  implies $m\sim \theta-\theta_c$, in agreement with (i,iii). However, it treats the hydrodynamic effect of a moving particle on the static bed in an average (mean-field)  way, and its application when  moving particles are far apart  (i.e. $m\ll1$) may thus not be warranted. 
By contrast, erosion/deposition models \cite{Charru04} emphasize the  slow ``armoring" or ``leveling" of the particle bed. One assumes that a $\theta$-dependent fraction of initially mobile particles evolve over a frozen static background, which contain  holes. In this view, $\theta_c$ occurs when the number of holes matches the number of initially moving particles. This phenomenological model also leads to $m\sim \theta-\theta_c$ and captures (i,ii,iii) qualitatively well.
However, the implicit assumption that the moving particles visit  the static bed entirely (thus filling up all holes) is highly non-trivial.  Indeed, due to the disorder of the static bed one expects mobile particles to follow favored paths and to eventually flow  in a few channels, thus exploring a tiny region of space. Such disorder-induced coarsening dynamics occurs for example in  river networks models \cite{Dhar06}, as well as plastic-depinning models of vortex lattices in dirty superconductors \cite{Watson96, Watson97} which also display a transition, but with $\beta\approx 1.5$, at odd with (i). 

To decide which physical process (hydrodynamic interactions or armoring) governs  the erosion threshold,
new theoretical predictions must be made and put to experimental test. In this letter, we achieve the first step of this goal while resolving the apparent contradictions of  deposition models. }
Specifically,  we introduce a model of interacting particles  forced along one direction on a disordered substrate. Particle interactions based on mechanical considerations are incorporated. Such model recovers (i,ii,iii) with $\beta=1$ and an equilibration time $t_{\rm conv}\sim |\theta-\theta_c|^{- 2.5}${, which agrees quantitatively with experiments \cite{Houssais15}}. { Our most striking predictions concerns the  spatial organization of the flux near threshold, which emerges from the interplay between disorder and particle interaction: }(a) the  distribution of local flux $\sigma$ is extremely broad, 
and follows $P(\sigma)\sim 1/\sigma$ and (b) spatial correlations of flux are short-range and very small in the lateral direction, but are power-law in the mean flow direction. We derive $\beta=1$ and explain why $P(\sigma)$ is broad using a mean-field description of our model, leading to  an analogy with $q$-models \cite{Liu95,Coppersmith96} used to study force propagation in grains.

 \section{Erosion Model}

We consider a density $n$ of particles  on a frozen background. $n$ should be chosen to be of order one, but its exact value does not affect our conclusions. The background is modeled via a square lattice, whose diagonal indicates the direction of  forcing, referred to as ``downhill". The lattice is bi-periodic, of dimension $L\times W$, where $L$ is the length in downhill direction and $W$ the transverse width.  Each node $i$ of the lattice is ascribed a height $h_i\in [0,1]$, chosen randomly with a uniform distribution. Lattice bonds $i\to j$ are directed in the downhill direction, and characterized by an inclination $\theta_{i\to j} = h_i-h_j$. We denote by $\theta$ the amplitude of the forcing. For an isolated particle on site $i$, motion will occur along the steepest of the two outlets (downhill bonds) \cite{Rinaldo14}, if it satisfies $\theta+\theta_{i\to j}>0$. Otherwise, the particle is trapped.

\begin{figure}[!ht]
\centering
\includegraphics[width=.8\columnwidth]{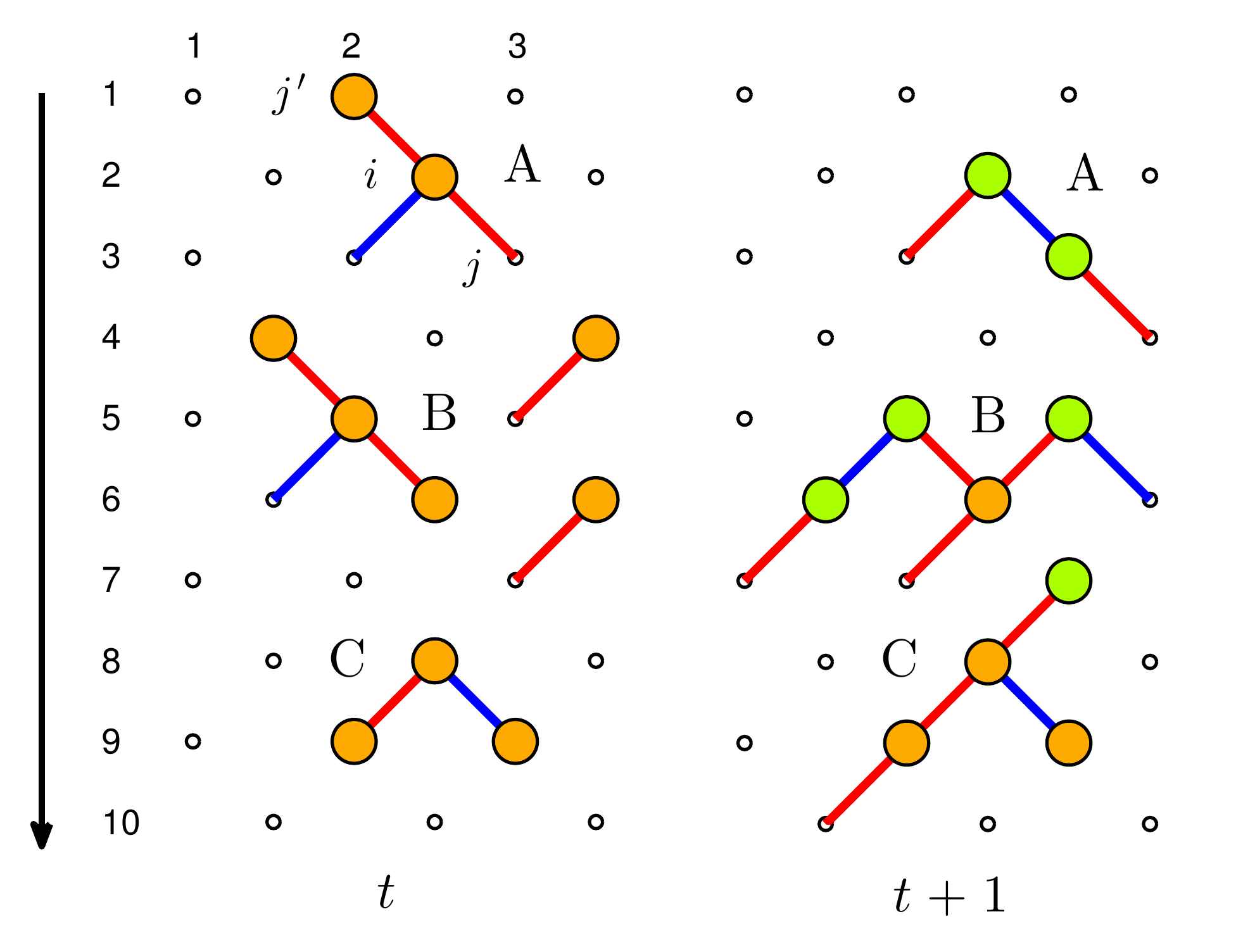}
\caption{\small{Illustration of the model. Small circles indicate lattice sites, particles are represented by discs in yellow, or green if motion occurred between $t$ (left) and $t+1$ (right). The black arrow is in the downhill direction. Solid lines indicate outlet with positive forces. If a particle has two outlets with positive forces, the larger (smaller) one is colored in red (blue). %The red ones show the primary directions and the blue ones are the secondary directions if exist. Particles move according to rules i-iv at each time step. For the specific configuration on the left at time $t$, the particles moved are shown in green at time $t+1$ on the right.
}}\label{5_model}
\end{figure}

However, if particles are adjacent, interaction takes place. First, particles cannot overlap, so they will only move toward unoccupied sites.  Moreover, particles can push  particles below them, potentially un-trapping these or affecting their direction of motion.  
To model these effects, we introduce scalar forces $f_{i\to j}$ on each outlet of occupied sites, which satisfy:
\be
\label{5_force}
f_{i\to j} = \max(f_{j'\to i}+\theta_{i\to j}+\theta,0)
\ee
where  $f_{j'\to i}$ is the force on the input bond $j'\to i$ along the same direction as $i\to j$, as depicted in Fig.~\ref{5_model}. Eq.(\ref{5_force}) captures that forces are positive for repulsive particles, and that particle $i$ exerts a larger force on toward site $j$ if the bond inclination $\theta_{i\to j}$ is large, or if other particles above $i$ are pushing  it in that direction. From our analysis below, we expect that the details of the interactions  (contacts, lubrication forces, etc...) are not relevant, as long as the  direction of motion of one particle can depend on the presence of particles above it- an ingredient not present in \cite{Watson96, Watson97}.

% When there is no particle at site $j'$, $f_{j'\to i}=0$. 
%
%The particles move on the landscape according to the following dynamical rules:

We update the position of the particles as follows, see Fig.~\ref{5_model} for illustration. We first compute all the  forces  in the system.
Next we consider one row of $W$ sites, and consider the motion of its  particles. Priority is set by considering  first outlets with the largest $f_{i\to j}$ and unoccupied downhill site $j$.   Once all possible moves ( $f_{i\to j}>0$, $j$ empty) have been made, forces are computed again in the system, 
and the next uphill row of particles is updated. When the $L$ rows forming the periodic system have all been updated, time increases by one.

For  given parameters $\theta,n$ we prepare the system via  two protocols.  In the ``quenched'' protocol, one considers a given frozen background, and launch the numerics with a large $\theta$ and randomly placed particles - parameters are such that the system  is well within the flowing phase. Next, $\theta$ is lowered slowly so that stationarity is always achieved. We also consider the ``Equilibrated'' protocol: for {\it any} $\theta$,  particles initial positions are  random. Dynamical properties are measured after the memory of the random initial condition is lost. We find that using different protocols does not change critical exponents, but that the quenched protocol appears to converge more slowly with system size. % $\theta_c$. % In what follows we make use of the ``quenched'' protocol for  explaining some of the results. 
Below we present most of our results obtained from the ``equilibrated'' protocol with $W=4\sqrt{L}$~\cite{Watson97}, and $n=0.25$ unless specified.

\begin{figure}[!ht]
\centering
\includegraphics[width=1.\columnwidth]{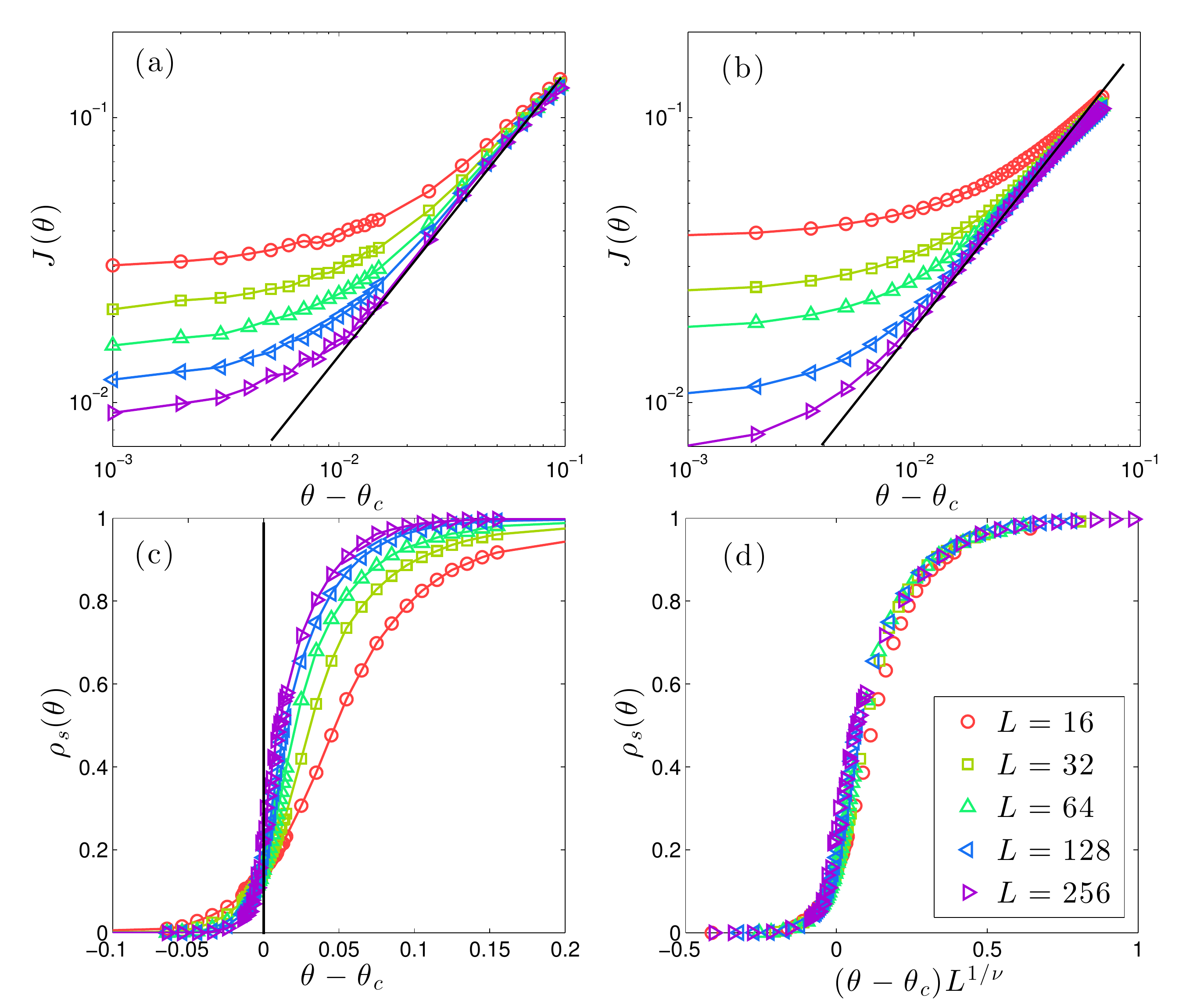}
\caption{\small{%Properties of the eroding phase of the model for different system sizes.
Average current $J$ versus $\theta-\theta_c$ in log-log scale for the (a) ``equilibrated'' and (b) ``quenched'' protocols, for which $\theta_c=0.164\pm0.002$ and $\theta_c=0.172\pm0.002$ respectively-  a difference plausibly due to finite size effects. The black solid lines with slope one indicate the linear relation  $J\propto \theta-\theta_c$. (c) Density of conducting sites  $\rho_s$ versus $\theta-\theta_c$ for the ``equilibrated'' protocol. (d) $\rho_s$ curves collapsed by rescaling $\theta-\theta_c$ with $L^{1/\nu}$, where $\nu=3.0\pm0.2$. %(e) Correlation in transversal direction $C_T$ of the conduction $\sigma$ of the sites in the steady state, the transversal distance $x$ is rescaled by the system with $W$.
}}\label{5_J}
\end{figure}

\section{Numerical Results on Dynamics of the Model}

Once the steady state is reached, we measure the average current of particles $J$ and the number density of sites carrying a finite current $\rho_s$.  Measurements of both quantities indicate a  sharp dynamical transition at some $\theta_c$ below which $J=0$ and $\rho_s=0$ as $L\rightarrow \infty$, see Fig.~\ref{5_model}. $\theta_c$ can be accurately extracted by considering the crossing point of the curves $\rho_s(\theta)$ as $L$ is varied, yielding $\theta_c=0.164\pm0.002$  for the equilibrated protocol. In the limit $L\rightarrow \infty$ our data extrapolates to:
\bea
\label{5_Jthe}
J(\theta)&\sim& \theta-\theta_c \ \  \hbox{ for} \ \theta>\theta_c\\
\label{5_rhot}
\rho_s(\theta)&=&\Theta(\theta-\theta_c),
\eea
where $\Theta$ is the Heaviside function. Eq.(\ref{5_Jthe}) corresponds to $\beta=1$, whereas Eq.(\ref{5_rhot}) indicates that all sites are visited by particles in the flowing phase.
Introducing the exponent $\rho_s(\theta)\sim (\theta-\theta_c)^\gamma$, this corresponds to $\gamma=0$.  The collapse of Fig.~\ref{5_J}(d) shows how convergence to Eq.(\ref{5_rhot}) takes place as $L\rightarrow \infty$, from which a finite size scaling length $\xi \sim (\theta-\theta_c)^{-\nu}$ with $\nu\approx 3$ can be extracted.

\begin{figure}[!ht]
\centering
\includegraphics[width=1.\columnwidth]{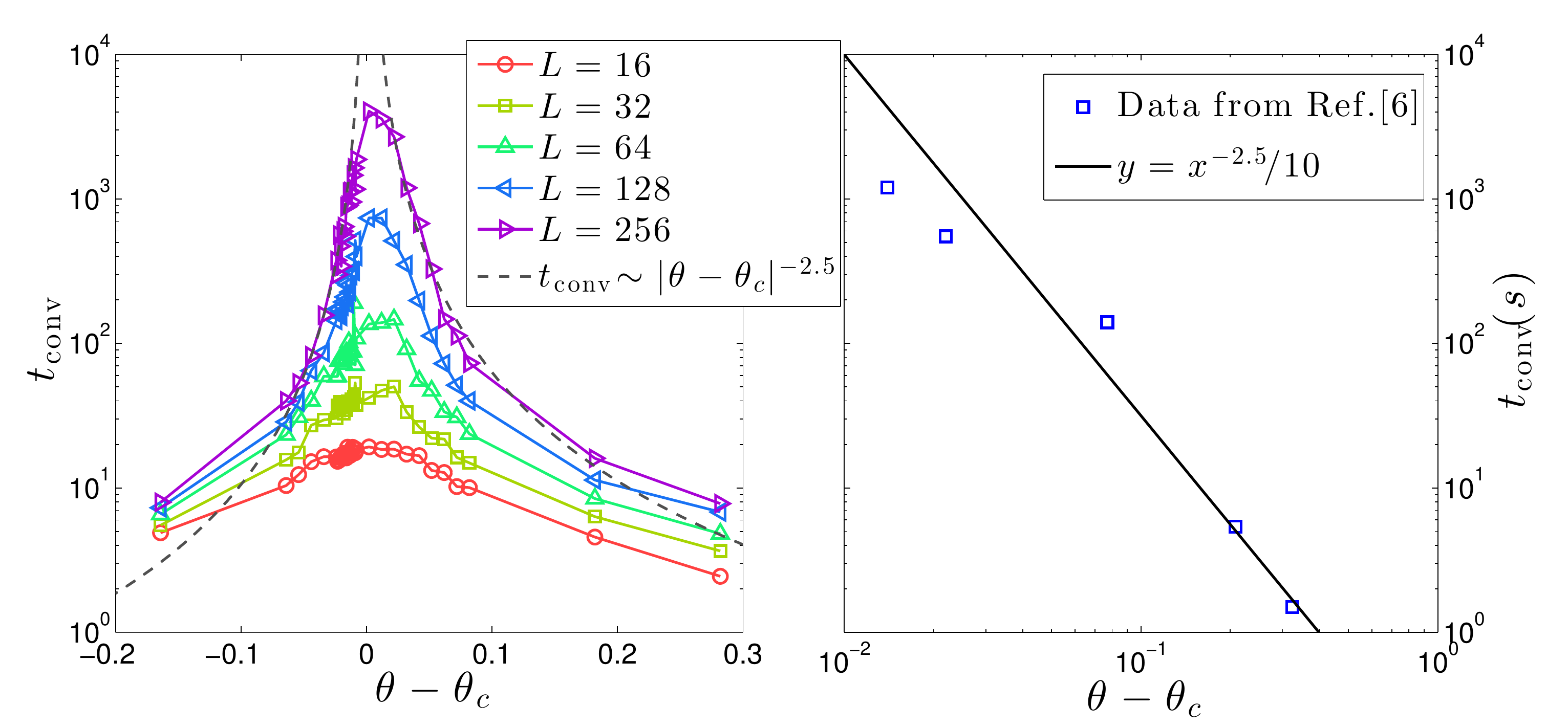}
\caption{\small{
Left: Transient time $t_{\rm conv}$ {\it v.s.}  $\theta$. For a given realization, $t_{\rm conv}$ is defined as the smallest time for which $J(t)-J \leq \sqrt{Var(J)}$ where $Var(J)=\lim_{T\to\infty}\frac{1}{T}\sum_{t=1}^{T}(J(t)-J)^2$. The gray dashed lines correspond to $t_{\rm conv}\sim|\theta-\theta_c|^{-2.5}$. Right: The obtained exponent fits well the observations of \cite{Houssais15}.
}}\label{5_tau}
\end{figure}

Criticality is also observed in the transient time $t_{\rm conv}$ needed for the current to reach its stationary value.
Fig.~\ref{5_tau} reports that $t_{\rm conv}\sim |\theta-\theta_c|^{-z}$ with $z\approx 2.5$ on both sides of the transition, which captures accurately the experiments of \cite{Houssais15} and the numerics of \cite{Clark15}. %We also find that the dynamics becomes more intermittent as $\theta\rightarrow \theta_c$, as indicated by the rescaled variance of the flux in stationary state shown in Fig.\ref{5_tau}.a.

\begin{figure}[!ht]
\centering
\includegraphics[width=.8\columnwidth]{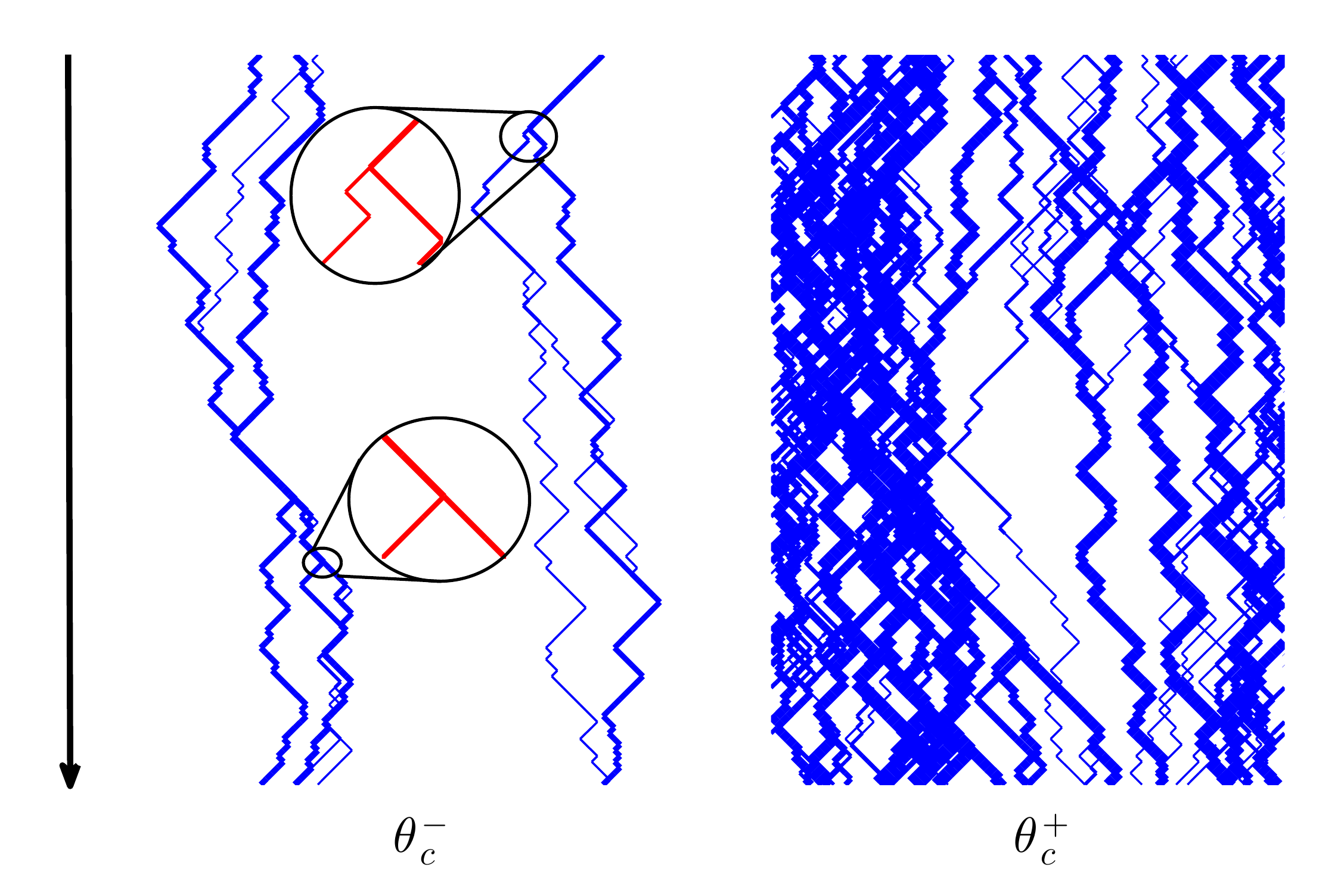}
\caption{\small{ Examples of drainage pattern  just below $\theta_c$ (Left) and above (Right). The black arrow shows the downhill direction. The thickness of the lines represents $\sigma_{i\to j}$ in logarithmic scale. A few examples showing  splitting events are magnified on the left. Here $W=45$ and $L=128$, and $J>0$ even below  $\theta_c$ due to finite size effects.}}\label{5_drain}
\end{figure}

The spatial organization of the current in steady state can be studied by considering the time-averaged local current $\sigma_i$ on site $i$, or the time-averaged  outlet current $\sigma_{i\to j}$. %(\textcolor{red}{$\sigma_i$ is the sum of the two outlets.}) 
The spatial average of each quantity is $J$.  Fig.~\ref{5_drain}  shows  an example of drainage pattern, i.e.  one realization of the map of the $\sigma_{i\to j}$.

%as the number of eroding particles passing through the site in a unit time, which is a static quantity in steady state for a given landscape and a certain drive $\theta$. Typical drainage patterns below and above threshold $\theta_c$ are shown in Fig.~\ref{5_drain}.The mean of the conductions is the current.

\begin{figure}[!ht]
\centering
\includegraphics[width=1.\columnwidth]{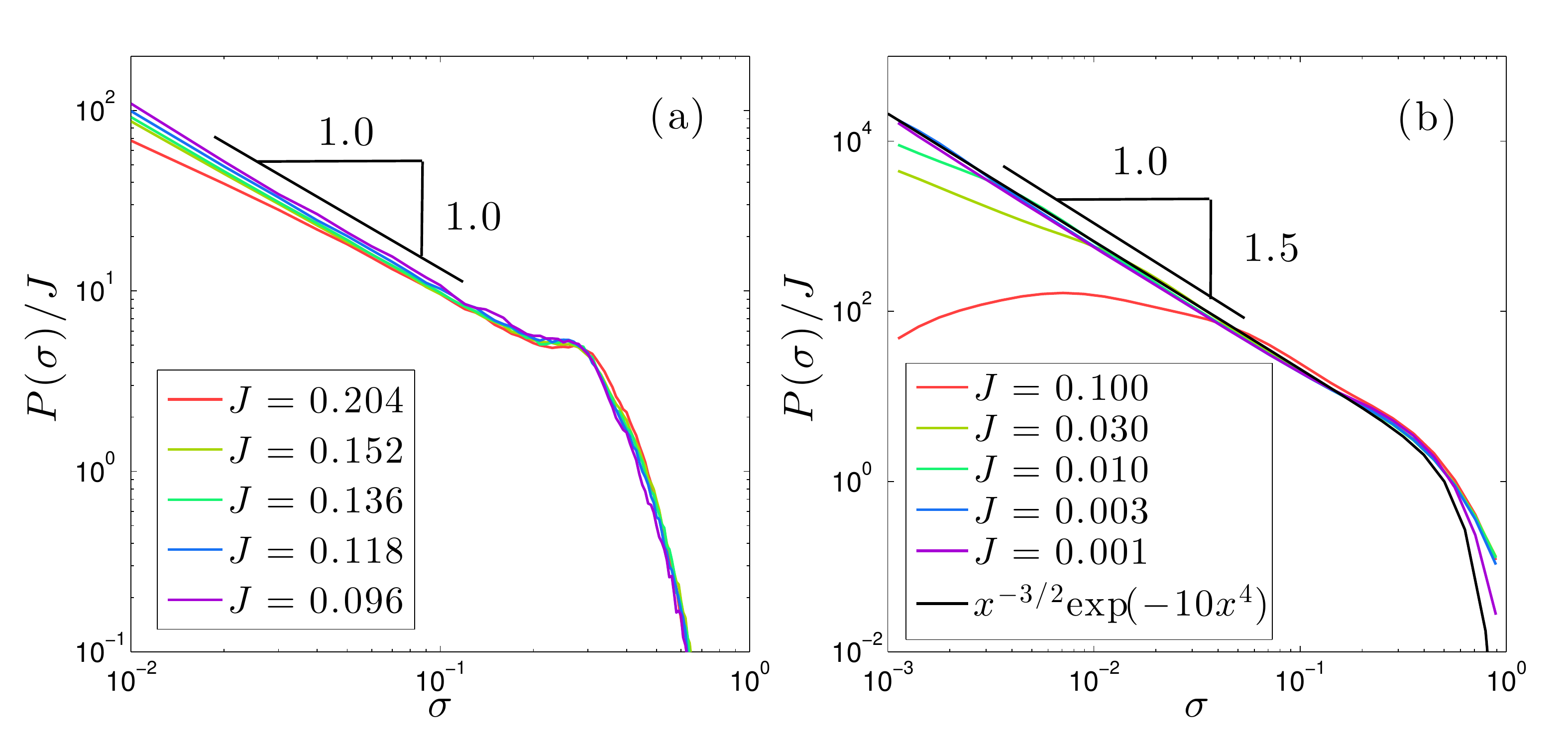}
\caption{\small{Distribution of the site current $P(\sigma)$ in steady state for given average currents $J$ of (a) the erosion model ($L=256$, $W=64$) and (b) our mean-field model ($W=1600$).}}\label{5_dist}
\end{figure}

To quantify such  patterns, we compute in Fig.~\ref{5_dist}(a) the distribution $P(\sigma)$ of the local current $\sigma_i$ for various mean current $J$.
We observed that:
\be
\label{5_psig}
P(\sigma)= J\sigma^{-\tau}f(\sigma)
\ee
where $\tau\approx1$ and $f$ is a cut-off function, expected since in our model $\sigma_i<1$. Eq.(\ref{5_psig}) indicates that $P(\sigma)$  is remarkably broad.
In fact, the divergence at small $\sigma$ is so pronounced that a cut-off $\sigma_{\min}$ must be present in Eq.(\ref{5_psig}) to guarantee a proper normalization of the distribution $P(\sigma)$, although we cannot detect it numerically. 

\begin{figure}[!ht]
\centering
\includegraphics[width=1.0\columnwidth]{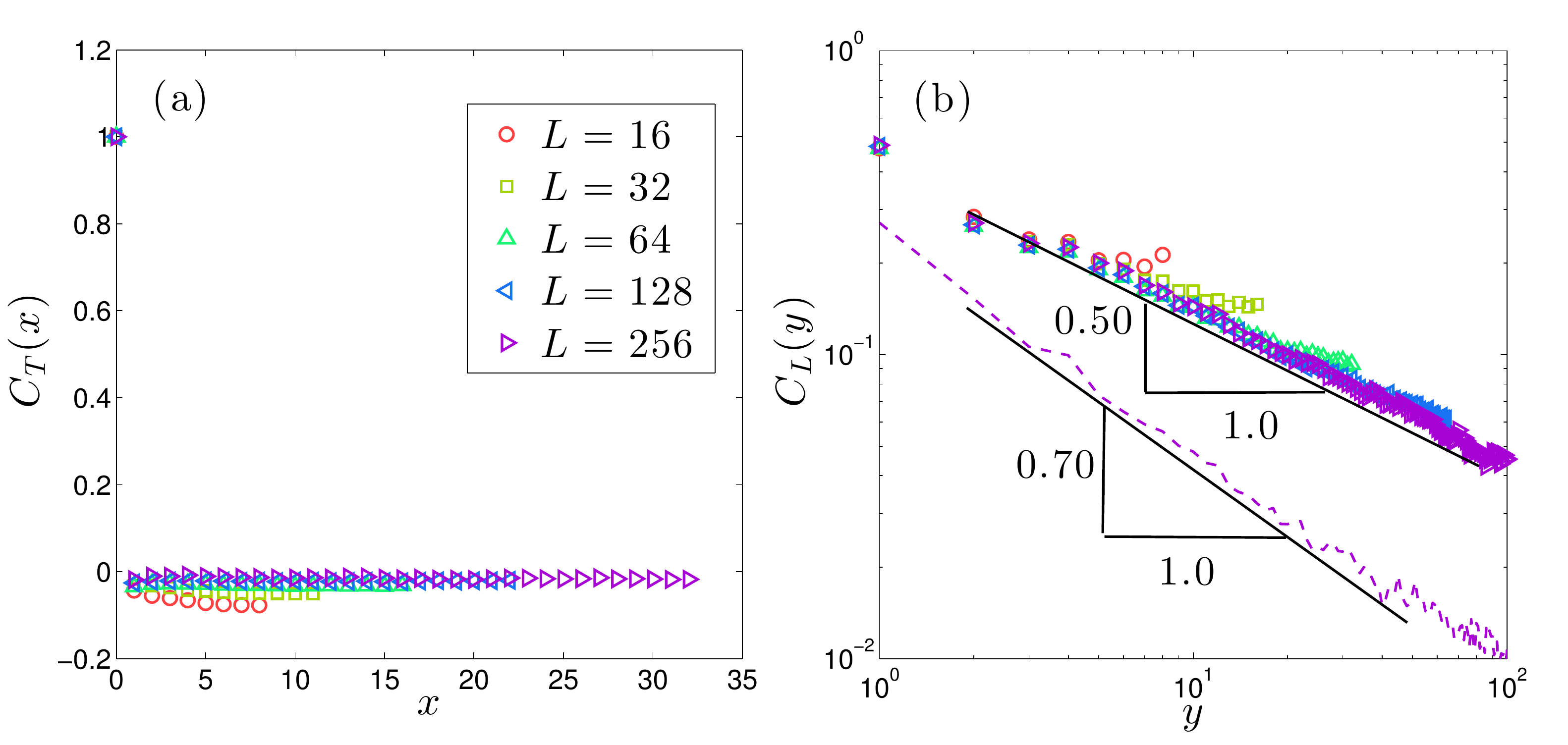}
\caption{\small{(a) Transverse current correlations $C_T$ at $\theta_c$  and (b) longitudinal current correlation   $C_L$ at $\theta_c$ and at $\theta-\theta_c=0.25$ for $L=256$ (dashed line).}}\label{5_core}
\end{figure}

Next, we compute the spatial correlation of the local current  in the transverse direction $C_T(x)$, defined as:
\be
\label{5_corr}
C_T(x)=\overline{(\langle\sigma_i\sigma_{i+x}\rangle-J^2)/(\langle\sigma_i^2\rangle-J^2)}
\ee 
where the site $i$ and $i+x$ are on the same row, but at a distance $x$ of each other. Here the brackets denote the spatial average, whereas the overline indicates averaging over the quenched randomness (the $h_i$'s).  Fig.~\ref{5_core}(a) shows that no transverse correlations exist for distances larger that  one site. However, long-range, power-law correlations are observed in the longitudinal direction, as can be seen by defining a longitudinal correlation function  $C_L(y)$, where $y$ is the vertical distance between two sites belonging to the same column. We find that $C_L(y)\sim 1/\sqrt{y}$ at $\theta_c$, but that $C_L(y)$ decays  somewhat faster deeper in the flowing phase, as shown in Fig.~\ref{5_core}(b).

\section{An Argument on the Scaling Relation}
 We now derive a relationship between the exponents $\beta$ characterizing $J$ and $\gamma$  characterizing $\rho_s$. It holds true for both protocols, but is presented here in 
the ``quenched" case. Near threshold, at any instant of time the density of moving particles is $J\ll n<1$, thus most of the particles are trapped and will move only when a mobile particle passes by. As $\theta$ is decreased by some $\delta\theta$, a finite density of new traps $\delta m\sim \delta \theta$ is created. If these traps appear on the region of size $\rho_s$ where mobile particles flow, they will reduce the fraction of mobile particle by $\delta J= \rho_s \delta m\sim \rho_s \delta\theta$, which implies:
\be
\label{5_screlation}
\beta=\gamma+1
\ee
%For the ``equilibrated'' protocol, particles need to fill all traps in the drainage basin~\cite{Narayan94,Watson97}, $\phi$, the landscape explored by mobile particles in transient. $\phi\geq\rho_s=1$ when $J>0$, and $\phi=0$ if $J=0$.  
Eq.(\ref{5_screlation}) shows that the result $\beta=1$ is a direct consequence of the fact that in our model, all sites are explored by mobile particles for $\theta>\theta_c$, a result which is not obvious. In the dirty superconductor models of  \cite{Narayan94,Watson97}, this is not the case and for the ``equilibrated" protocol  $\beta>1$ was found. We argue that this difference comes from the dynamical rules chosen in \cite{Narayan94,Watson97}, according to which ``rivers" forming the drainage pattern never split: their current grows in amplitude in the downhill direction, until it reaches unity. In these models the drainage pattern thus consists of rivers of unit current, avoiding each other, and separated by a typical distance of order $1/J$. Our model behaves completely differently because rivers can split, as emphasized in Fig.~\ref{5_drain}. This comes about because the direction taken by a particle can depend on the presence of a particle right above it, as illustrated in case A of Fig.~\ref{5_model}. This effect is expected to occur in the erosion problem due to hydrodynamic interactions or direct contact between particles, and may also be relevant for superconductors.

%a. Theoretically and numerically solving the q-model with different choice of the splitting distribution $\eta(q)$. 
\section{Mean-field Model on the Distribution of Local Currents}
We now seek to quantify the effect of splitting. Its relevance is not obvious a priori, as splitting stems from particle interactions, 
and may thus become less important as the fraction of moving particles vanishes as $J\rightarrow 0$. To model this effect we consider that the current $\sigma_i$  on a site $i$ is decomposed in its two outlets as $\sigma_i=q \sigma_i +(1-q)\sigma_i$, where $q$ is a random variable of distribution $\eta(q)$. If there were no splitting then $\eta(q)=\frac{1}{2}\delta(q)+\frac{1}{2}\delta(1-q)$. Here instead, we assume that $\eta(q)=\frac{1}{2}\delta(q-J)+\frac{1}{2}\delta(1-J-q)$. %(\textcolor{red}{I tried to explain why the split is proportional to $J$ in my initial version with both case A and B. Do you think it's too complicated with little information?}) 
This choice captures that the probability of splitting is increased if more moving particles are present, and can occur for example if two particles flow behind each other, as exemplified in case A of Fig.~\ref{5_model}. Next, we make the mean field assumption that two adjacent sites $i$ and $j$ on the same row are uncorrelated, $P(\sigma_i,\sigma_j)=P(\sigma_i)P(\sigma_j)$. We then obtain the self-consistent equation  that $P(\sigma)$ must be equal to:
\be
\label{5_selfconsist}
%P(\sigma)=
\int\rd q_1\rd q_2\rd\sigma_1\rd\sigma_2\eta(q_1)\eta(q_2)
P(\sigma_1)P(\sigma_2)\delta(q_1\sigma_1+q_2\sigma_2-q) 
\ee
This mean-field model belongs to the class of $q$-models introduced to study force propagation \cite{Liu95, Coppersmith96}. It is easy to simulate, and some aspects of the solution can be computed. Numerical results are shown in Fig.~\ref{5_dist}(b). The result obtained for $P(\sigma)$ is very similar to Eq.(\ref{5_psig}) that describes our erosion model: $P(\sigma)$ is found to be power-law distributed (although $\tau=3/2$ instead of $\tau=1$) where with an upper cutoff at $\sigma_{\max}\sim 1$, and $P(\sigma) \propto J$. 

These results are of interest, as they explain why $P(\sigma)$ is very broad,  and is not dominated by sites displaying no current at all (which would correspond to a delta function at zero) even as $J\rightarrow 0$, thus confirming that $\gamma=0$. They can be explained by taking the Laplace transform $\tilde P$ of Eq.(\ref{5_selfconsist}). One then obtains a non-linear differential equation for $\tilde P$, from which it can be argued generically that $\tau=3/2$ \cite{Coppersmith96}. We have performed a Taylor expansion of $\tilde P$ around zero, which leads to relationship between the different moments of the distribution $P(\sigma)$. From it,
we can show that $P(\sigma)\propto J$ and $\sigma_{\max}\sim 1$. We also find that the cut-off  of the divergence of $P(\sigma)$ at small argument follows $\sigma_{\min}\sim J^{1/(\tau-1)}$.

\section{Potential Experimental Tests}
We have introduced a novel model for over-damped interacting particles driven on a disordered substrate.
It predicts a dynamical phase transition at some threshold forcing $\theta_c$, and makes quantitative predictions  for various quantities including  the particle current  and the drainage pattern, testable by tracking particles on the surface \cite{Charru04}. Our model includes the possibility that channels carrying most of the flow split, which may also be well-suited to describe plastic depinning phenomena including the pinning of vortices in dirty superconductors \cite{Kolton99,Watson96} or driven colloidal systems  \cite{Reichhardt02, Pertsinidis08}, which have never been received a proper analytical treatment \footnote[2]{ In plastic depinning an exponent $\beta\geq 1.5$ is often reported, larger than our predicted $\beta=1$. However we observe that large systems are required to measure $\beta$ accurately, and that in smaller systems  $\beta$ can appear significantly larger.}.
%Our prediction that the value of $\theta_c$ depends on the protocol, and is large if $\theta$ is decreased continuously, would also be interesting to test.

Note that our model assumes that particles are over-damped, and that their inertia is negligible. 
We expect inertia to lead to hysteresis and make the transition first order, as observed on inertial granular flows down an inclined plane \cite{Andreotti13}, although this effect may be small in practice  \cite{Ouriemi07}. We did not consider non-laminar flows, nor temperature (that can be relevant for colloids). Both effects should smooth  the transition, and lead to creep even below $\theta_c$.% \cite{Houssais15}.

%Finally, it has been proposed that the erosion threshold is a dynamical transition very similar to the jamming transition that occurs when a bulk amorphous material is sheared \cite{Houssais15}. If our model holds, this is not the case: due to the presence of the free interface, long-range elastic interactions between mobile particles are absent. In recent theoretical descriptions of the jamming transition such  interactions are central   both for soft \cite{Lin14} and hard particles \cite{Lerner12a,DeGiuli14d}. 

%% file: marginal/marginal.tex
\chapter{Dynamics and Correlations among Soft Excitations in Marginally Stable Glasses}
\label{6_marginal}

Marginal stability is the notion that stability is achieved, but only barely so. This property constrains the ensemble of configurations explored at low temperature in a variety of systems, including spin, electron and structural glasses. A key feature of marginal states is a (saturated) pseudo-gap in the distribution of soft excitations. { We examine how such pseudo-gaps appear dynamically by studying the Sherrington-Kirkpatrick (SK) spin glass.} After revisiting and correcting the multi-spin-flip criterion for local stability, we show that stationarity along the hysteresis loop requires soft spins to be frustrated among each other, with a correlation diverging as $C(\lambda)\sim 1/\lambda$, where $\lambda$ is the stability of the more stable spin. We explain how this arises spontaneously in a marginal system and develop an analogy between the spin dynamics in the SK model and random walks in two dimensions. { We discuss analogous frustrations among soft excitations in short range glasses and how to detect them experimentally.} We also show how these findings apply to hard sphere packings. %Finally we show numerically that a simple dynamical Markov model where these correlations are enforced generate the crackling noise observed in the SK model, with the same (currently unexplained) exponents. Some of our results are expected to hold in other glassy systems, such as hard sphere packings. % is the most important feature so that a dynamical model capturing the correlation sufficiently reproduces all the scaling relations of the pseudo-gap and the critical dynamics in the SK model. 
%\end{abstract}

\section{Introduction}
In glassy materials with sufficiently long-range interactions, stability at low temperature imposes an upper bound on the density of soft excitations~\cite{Muller14}. %{\bf only cite works that we want to cite later on, too} 
In electron glasses~\cite{Efros75,Pollak13,%BenChorin93,Goethe09,Massey95,Pankov05,
Muller04,Muller07,Palassini12,Andresen13a} stability towards hops of individual localized electrons requires that the density of states vanishes at the Fermi level, exhibiting a so-called Coulomb gap. Likewise, in mean-field spin glasses ~\cite{Thouless77,Pazmandi99,Pankov06,Eastham06,Horner07,Doussal10,Le-Doussal12,Andresen13%,Sharma14
} stability towards flipping several ``soft" spins implies that the distribution of local fields vanishes at least linearly. In hard sphere packings the distribution of forces between particles in contact must  vanish analogously, preventing that collective motions of particles lead to denser packings~\cite{Wyart12,Lerner13a,DeGiuli14b%,Kallus13,Kallus14
}. Often, these stability bounds appear to be saturated~\cite{Pazmandi99,Palassini12,Andresen13,Lerner13a,Lerner12,Charbonneau15}.  Such {\em marginal stability} can  be proven for dynamical, out-of-equilibrium situations  under slow driving at zero temperature \cite{Muller14} if the effective interactions do not decay with distance. This situation occurs in the {Sherrington-Kirkpatrick (SK) model (see Eq.~(\ref{6_h}) below)}, but also in finite-dimensional hard sphere glasses, where elasticity induces non-decaying interactions~\cite{Wyart05b}.  { Marginality} is also found for the ground state or for slow thermal quenches by replica calculations for spin glass \cite{Mezard87,Pankov06} and hard sphere systems~\cite{Charbonneau14,Charbonneau14a}, assuming infinite dimension.%, for which the result was extended to states reached by slow thermal quenches.

The presence of pseudo-gaps strongly affects the physical properties of these glasses. The Coulomb gap alters transport properties in disordered insulators \cite{Efros75,Pollak13}, while its cousin in spin glasses suppresses the specific heat and susceptibility. It was recently proposed that the singular rheological properties of dense granular and suspension flows near jamming are controlled by the pseudo-gap exponents in these systems \cite{DeGiuli14d}. More generally, an argument of Ref.~\cite{Muller14} shows
that a pseudo-gap implies avalanche-type response to a slow external driving force, so-called crackling~\cite{Sethna01}, { for a range of applied forcing. Such behavior is indeed observed in these systems \cite{Combe00,Palassini12,Pazmandi99} and in the plasticity of crystals \cite{Ispanovity14}, and contrasts with depinning or random field Ising models where crackling occurs only at one specific value of forcing~\cite{Perkovi95,Dhar97,Sabhapandit02}}.  Despite the central role of pseudo-gaps, it has not been understood how they emerge dynamically, even though some important elements of the athermal dynamics of the SK spin glass have been pointed out in earlier works~\cite{Eastham06,Horner07}. 

In this Letter we identify a crucial ingredient that was neglected in previous dynamical approaches, and also in considerations of multi-spin stability: Soft spins are strongly frustrated among each other, a correlation that becomes nearly maximal  for spins in the weakest fields. { We expect analogous correlations in short range spin glasses, which can be probed experimentally.}
These correlations require revisiting earlier multi-spin stability arguments that assumed opposite correlations. We then argue, assuming stationarity along the hysteresis loop, that the correlation $C(\lambda)$ between the softest spins and spins in local fields of magnitude $\lambda$ must follow $C(\lambda)\sim 1/\lambda^\gamma$, with $\gamma=1$. %\MW: removed , {\bf similarly as predicted in Ref.~\cite{Horner07}}.
%(equivalent to the hypothesis  that each spin flips a number of times that diverges with $N$ along this loop) . 
Using this in a Fokker-Planck description of the dynamics we predict the statistics of the number of times a given spin flips  in an avalanche. %Finally we show that a simple dynamical Markov model where these correlations are enforced generate the crackling noise observed in the SK model, with the same (currently unexplained) exponents.

%\textcolor{red}{...}
%To specify the concept of marginal stability and unveil the underlying mechanism of emergence of the nontrivial correlation and the pseudo-gap, we study the athermal dynamics of a spin glass model with long-range interactions, the Sherrington-Kirkpatrick (SK) model~\cite{Sherrington}, %where each 
\section{Dynamics of Sherrington-Kirkpatrick Model}
We consider the SK model with $N$ Ising spins ($s_i=\pm1$) in an external field $h$:
%Consider Sherrington-Kirkpatrick model of Ising spins $s_i=\pm1$, each spin is randomly coupled with every other spin in the system and the external field $h$,
\be
\label{6_h}
\mh=-\frac{1}{2}\sum_{i\neq j}J_{ij}s_is_j-h\sum_{i=1}^Ns_i.
\ee
 All spins are coupled to each other by a symmetric matrix $J_{ij}$, whose elements are {\it i.i.d.} {Gaussian random variables} with zero mean and variance $1/N$. %In the following discussion, if not specified, we set $J_0=0$ and $\tilde{J}=1$. , where 
The total magnetization is $M\equiv\sum_is_i$. We define the local field $h_i$ and the local stability $\lambda_i$ of spin $i$ by
\be
\label{6_lambda}
h_i\equiv -\frac{\partial \mh}{\partial s_i} = \sum_{j\neq i}J_{ij}s_j+h,\quad\lambda_i=h_is_i.
\ee
The spin $s_i$ is called stable when it aligns with the local field, i.e. if $\lambda_i>0$, and unstable otherwise. The energy to  flip the spin $s_i\to-s_i$ (and hence $\lambda_i\to-\lambda_i$) is:
\be
\label{6_e1}
\Delta\mh_1(i)\equiv\mh(-s_i)-\mh\\=2s_i(\sum_{j\neq i}J_{ij}s_j+h)=2\lambda_i. 
\ee
%which has the same sign as the local stability. So the system energy is lower when flipping an unstable spin.

%The athermal dynamics of the model~

\begin{figure}[h!]
\centering
\includegraphics[width=.9\columnwidth]{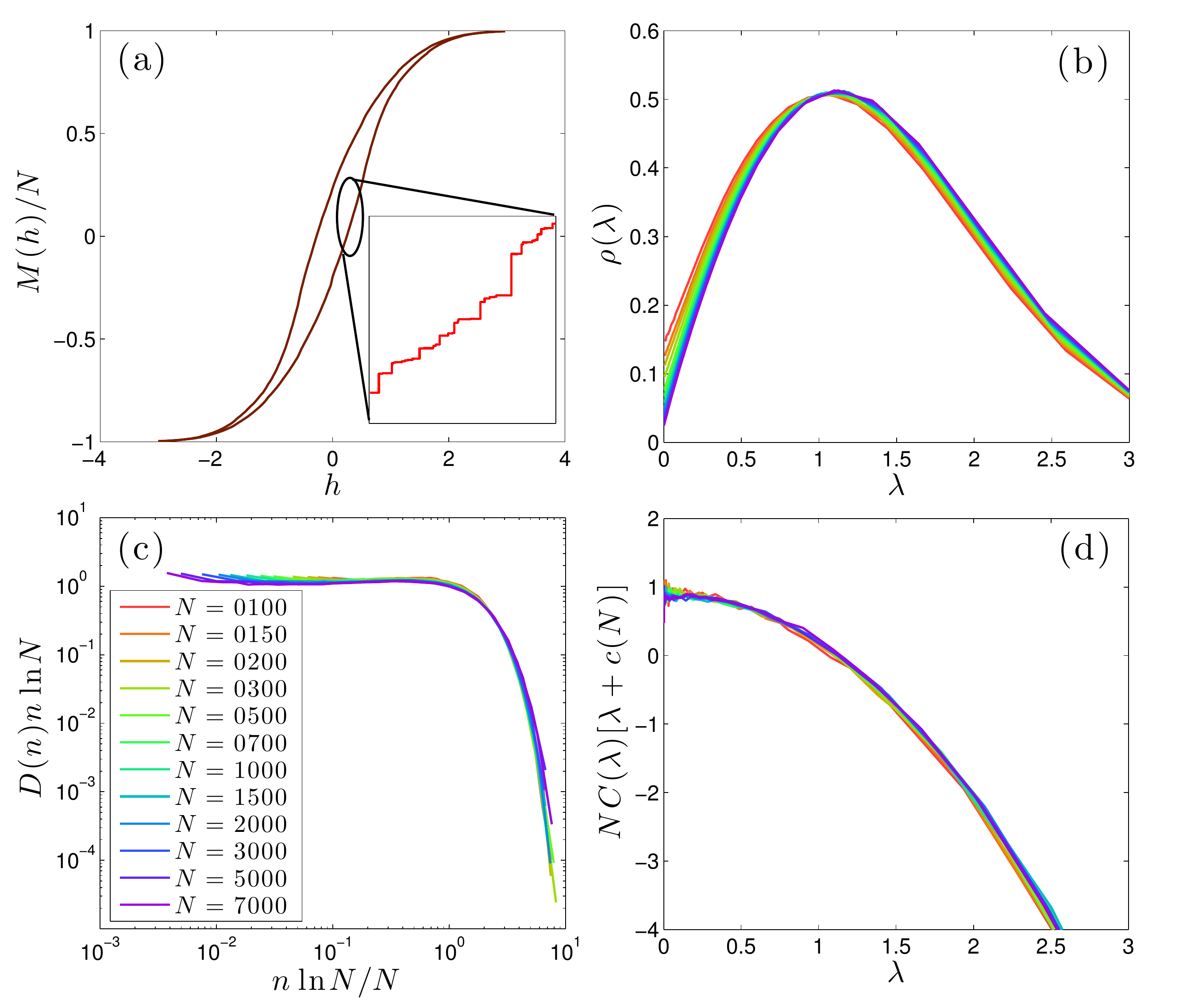}
\caption{\small{(a) Hysteresis loop: Magnetization $M$ under a periodic quasi-static driving of the external field $h$. Inset: magnified segment of the hysteresis loop of a finite size system. { (b)} Distribution of local stabilities, $\rho(\lambda)$, in locally stable states along the hysteresis loops for different system sizes $N$.
{ (c)} Finite size scaling of the avalanche size distribution $D(n)$ confirms $\tau=\sigma=1$ {up to logarithmic corrections}. %{\bf size legend here, or exchange with Fig c!} 
 (d) Correlation { $C(\lambda)$} between the least stable spin and spins { of} stability $\lambda$ in locally stable states along the hysteresis loop. 
The data for different system sizes collapses{, implying $C(\lambda \ll 1)\sim1/\lambda$ in the thermodynamic limit}.
}}\label{6_skmodel}
\end{figure}

As in Ref.~\cite{Pazmandi99}, we consider the hysteresis loop at zero temperature obtained by quasi-statically increasing the field, as shown in Fig.~\ref{6_skmodel}(a).  
When a spin turns unstable, we apply a greedy Glauber dynamics that relaxes the system in an avalanche-like process towards a new one-spin-flip stable state by sequentially flipping the most unstable spin. { Such hysteretic field ramping has also been used to find approximate ground states~\cite{Boettcher05, Pal06}.}
Those states empirically exhibit a pseudo-gap in the distribution of the $\lambda_i$ ~\cite{Pazmandi99,Eastham06,Andresen13},
\be
\label{6_pseudogap}
\rho(\lambda)=A\lambda^{\theta}+O(N^{-\theta/(1+\theta)}),
\ee
with $\theta=1$ for $\lambda\ll1$, as shown in Fig.~\ref{6_skmodel}(b), {but with a slope $A$ significantly larger than in equilibrium~\cite{Parisi03,Pankov06,Horner07}}. 
The avalanche size is power-law distributed~\cite{Pazmandi99}:
\be
\label{6_dn}
D(n)=n^{-\tau}d(n/N^\sigma)/\Xi(N),
\ee 
%We implement quasi-static driving to study the avalanches from one stable state to the other: we increase the external field with a minimal amount to flip one spin from a stable state to trigger the avalanche, $h$ is then fixed until the other stable state is reached. 
where $n$ is the number of flips in an avalanche. The scaling function $d(x)$ vanishes for $x\gg1$. $N^{\sigma}$ is the finite size cutoff, and $\Xi(N)$ is a size dependent normalization if $\tau\leq1$. 
% Finally, we introduce the characteristic local field $\lambda_0$ of the most unstable spins in an avalanche of size $n$:
Numerical studies of the dynamics of the SK model indicate that $\tau=\sigma=1$ and $\Xi=\ln N$~\cite{Pazmandi99,Andresen13}, as shown by the finite size collapse in Fig.~\ref{6_skmodel}(c). While one can argue  that $\theta=1$ along the hysteresis curve~\cite{Muller14}, the exponents $\tau$ and $\sigma$  have not been derived theoretically for the dynamics (unlike for ``equilibrium avalanches", for which $\tau=1$ has been obtained analytically~\cite{Doussal10,Le-Doussal12}). 

\begin{figure}[h!]
\centering
\includegraphics[width=.9\columnwidth]{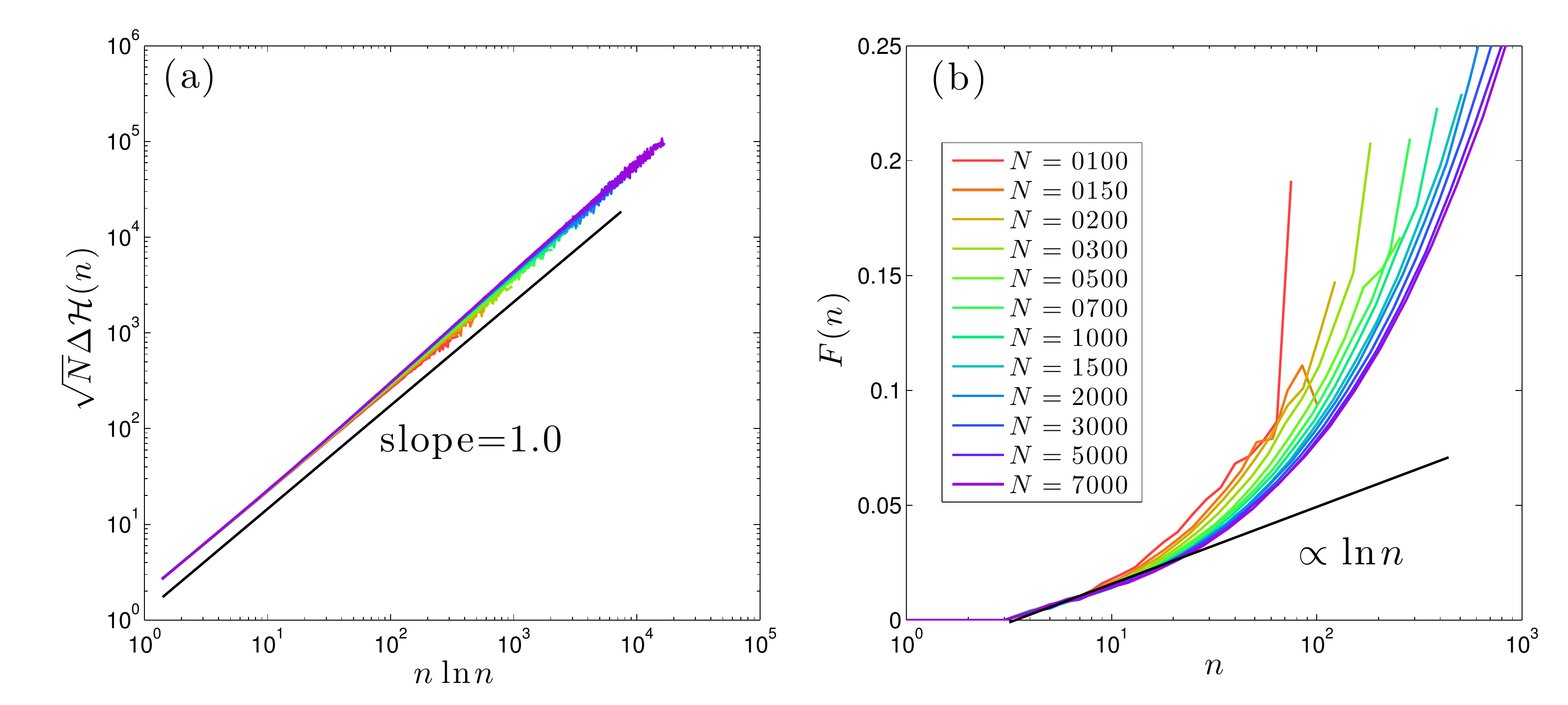}
\caption{\small{(a) The average dissipated energy $\Delta \mh$ in avalanches of size $n$ scales as $\Delta \mh\sim n\ln n/\sqrt{N}$. $-\Delta \mh/n$ is a measure of the typical value of the stability of most unstable spins, $\lambda_0(n)$. Thus, in the thermodynamic limit, { $\lambda_0\sim \ln n/\sqrt{N} \ll 1$ }even for very large avalanches. (b) The average number of times, $F(n)$, spins active in avalanches of size $n$ re-flip later on in the avalanche.}}\label{6_propt}
\end{figure}

Below we present a theoretical analysis of the dynamics. { We assume that the average number of times a spin flips along the hysteresis loop diverges with $N$ for any finite interval of applied field $[h, h+\Delta h]$ if $h=O(1)$. This assures that a stationary regime is reached rapidly. (For $\tau=1$ this condition simply reads $\sigma +1/(1+\theta)>1$)~\footnote[1]{The typical external field increment triggering an avalanche is $h_{\rm min}\sim\lambda_{\rm min}\sim N^{-1/(1+\theta)}$, so there are $N_{\rm av}\sim1/h_{\rm min}\sim N^{1/(1+\theta)}$ avalanches in a finite range of external field $dh$~\cite{Muller14}. Each avalanche contains on average $N_{\rm flip}\sim\int nD(n)\rd n\sim N^{(2-\tau)\sigma}$ flip events. The total number of flip events along the hysteresis curve is $N_{\rm av}N_{\rm flip}\sim N^{(2-\tau)\sigma+1/(1+\theta)}$, which we assume to be $\gg N$.}.
We further rely on  $\theta<\infty$. This implies a diverging number of avalanches in the hysteresis loop, each contributing a subextensive amount of dissipation \footnote[1].. The latter rules out avalanches running into strongly unstable configurations, with an extensive number of spins with negative stability $|\lambda|=O(1)$. Thus, the lowest local stability encountered in an avalanche, $\lambda_0$, must satisfy $\lambda_0\rightarrow 0$ as $N\rightarrow \infty$, as we confirm numerically in Fig.~\ref{6_propt}(a).} %{\bf put 3a bef 2??}

%{\it (i)} Contained Avalanches: Spins never become strongly unstable, that is, the lowest local stability encountered in an avalanche, $\lambda_0$, satisfies $\lambda_0\rightarrow 0$ as $N\rightarrow \infty$. This has been numerically verified in Fig.~\ref{6_propt}(a). % where $|\lambda_0|$ is estimated from the dissipated energy per spin. 

%{\it (ii)}  Stationarity: %\MEvery spin flips many times along the hysteresis loop, the first flip requiring an increase $dh$ in external field, which vanishes as $N\to \infty$. This ensures that a statistically stationary state is reached quickly along the hysteresis curve, and that its statistics can be defined for infinitesimal intervals $[h,h+dh]$. 
%the number of times each spin flips along the hysteresis loop diverges with $N$ for any finite interval of applied field $[h, h+\Delta h]$ if $h=O(1)$. This assures that a stationary regime is reached rapidly.
%The number of times each spin flips along the hysteresis loop diverges with $N$; {\bf thus a negligible fraction of the hysteresis loop consists of transients that precede a stationary regime.} 

\section{Multi-spin Stability Criterion}
A static bound for the pseudogap exponent $\theta$ is obtained by considering two of the softest spins $i,j$ (with stabilities $\lambda_{\rm min}\sim 1/N^{1/(1+\theta)}$) ~\cite{Palmer79,Anderson79,Muller14}. Their simultaneous flip costs an energy $2(\lambda_i+\lambda_j - 2 s_is_j J_{ij})$. %{\bf please check!}. 
The last term scales as $1/\sqrt{N}$ and is negative if the two spins are {\em unfrustrated}. If this occurs with finite probability, a strong enough pseudogap, $\theta\geq 1$, is necessary to prevent the last term from overwhelming the stabilizing terms. The extension of this argument to  multi-spin stability
%First recap the 2 spin-flip criterion, why it is powerful, and cite the literature on it. 
reveals its subtle nature.  Flipping a set $\mf$ of $m$ spins in a one-spin flip stable state costs 
\be
\label{6_de}
\Delta\mh(\mf)=2\sum_{i\in\mf}\lambda_i-2\sum_{i,j\in\mf}J_{ij}s_is_j.
\ee
The initial state is unstable to multi-flip excitations if $\Delta\mh<0$ for some $\mf$. Refs.~\cite{Palmer79, Anderson79}  considered just the set of the $m$ softest spins. Extremal statistics and the assumption of Eq.~(\ref{6_pseudogap}) implies the scaling of the maximal stabilities $\lambda(m)\sim (m/N)^{1/(1+\theta)}$, and thus $\sum_{i\leq m}\lambda_i\sim m \lambda(m)$.
%$\sim m^{(2+\theta)/(1+\theta)}/N^{1/(1+\theta)}$. 
The term $\sum_{i\leq m}J_{ij}s_is_j\sim m(m/N)^{1/2}$ was erroneously argued to be  positive {\em on average}, which yielded the bound $\theta\geq 1$ to guarantee $\Delta\mh_m>0$. 
However, numerically we find that on average $\sum_{i\leq m}J_{ij}s_is_j$ is negative for soft spins. More precisely, the correlation $C(\lambda)=-2\langle Jss\rangle$ between a spin of stability $\lambda$ and the softest spin in the system is positive for small $\lambda$, as shown in  Fig.~\ref{6_skmodel}(d). Postulating that:
\be
\label{6_correlation}
C(\lambda)\sim \lambda^{-\gamma} N^{-\delta},
\ee
it leads to $\langle -\sum_{i\leq m}J_{ij}s_is_j \rangle\sim m^2 C(\lambda(m))\sim m^{2-\gamma/(1+\theta)} N^{\gamma/(1+\theta)-\delta}$. { A more complete characterization of correlations is given in the Appendix Secs. \ref{app_D1}, \ref{app_D2}.}

It follows that the average r.h.s. of Eq.~(\ref{6_de}) is always positive. We argue that the stability condition nevertheless leads to a non-trivial constraint, because the last term of Eq.~(\ref{6_de}) can have large fluctuations. Indeed, consider all sets $\mf$ of $m$ spins belonging to the $m'>m$ softest spins, and for definiteness we choose $m'=2m$ here.
%and  use scaling arguments. % could be dropped: 
%{A more detailed derivation for a general ratio $m'/m$ is given in the Supplementary Materials Sec.~A. }
To determine the probability that the optimal set leads to a negative $\Delta \mh$ in Eq.~(\ref{6_de}), we use an { approximate estimate} akin to the random energy model~\cite{Derrida81}. %We assume that the  $\Delta\mh$ associated with different $\mf$ are independent, Gaussian distributed variables. 
The variance of the fluctuation $X\equiv \sum_{i,j\in\mf}J_{ij}s_is_j-\langle \sum_{i,j\in\mf}J_{ij}s_is_j\rangle$ is of order $m/\sqrt{N}$. Since there are $2^{2m}$ sets $\mf$, the number density having fluctuation $X$ follows ${\cal N}(X)\sim \exp[2m\ln(2)-X^2N/m^2]$. The most negative fluctuation $X_{\rm min}$ is determined by ${\cal N}(X_{\rm min})\sim 1$, leading to $X_{\rm min}\sim - m^{3/2}/\sqrt{N}$. { Correlations neglected by this argument should not affect the scaling.}  The associated energy change is thus, according to Eq.~(\ref{6_de}) and the subsequent estimates of each term:
\be
\label{6_sta}
\Delta\mh(\mf_{\rm min})=m^{(2+\theta)/(1+\theta)}/N^{1/(1+\theta)}+
m^{2-\gamma/(1+\theta)} N^{\gamma/(1+\theta)-\delta}- m^{3/2}/\sqrt{N}.
\ee
Multi-spin stability requires that for large $N$ and fixed $m$ this expression be positive. This yields the conditions:
\be
\label{6_sca}
\theta\geq 1, \ \ \hbox{ or}\ \ \gamma/(1+\theta)-\delta\geq-1/2.
\ee
However, the correlation in Eq.~(\ref{6_correlation}) cannot exceed the typical coupling among spins, $C\lesssim1/\sqrt{N}$, which requires $\gamma/(1+\theta)-\delta\leq -1/2$. Thus, if $\theta<1$, stability imposes
the equality $\gamma/(1+\theta)-\delta=-1/2$, while the scaling with $m\gg 1$ additionally requires $2-\gamma/(1+\theta)\geq 3/2$; or in other words, $\gamma\leq(1+\theta)/2\leq1$ and $\delta\leq1$. 
%Combining with the equality achieved from the second relation of Eq.(\ref{6_sca}), there are two other stability constraints,  . 
In the relevant  states, all three exponents $\theta$, $\gamma$, and $\delta$ turn out to equal $1$ and thus satisfy these constraints as exact equalities.
%which is not very constraining considering that 3 exponents were introduced. As we shall now see, dynamical considerations are more useful in constraining the form of the correlation $C(\lambda)$. 
We will now show how to understand this { emergent marginal stability} from a dynamical viewpoint.

%The pseudo-gap indicates a non-vanishing correlation among soft spins. 
\section{Fokker-Planck Description of the Dynamics}   
Consider an elementary spin flip event in the greedy relaxation dynamics, cf. Fig.~\ref{6_dynmod}. { The stability of the flipping spin $0$ (red) changes from $\lambda_0$ to $-\lambda_0$ as the spin flips from $s_0$ to $-s_0$. Due to the coupling $J_{0j}$, the stability of all other spins $j$ (green or blue) receives a kick,
% $-J_{0j}s_0s_j$,
%\be
%\label{6_dl}
$\lambda_j\to\lambda_j'=\lambda_j-2J_{0j}s_0s_j.$
%\ee
%These kicks have a random fluctuating part, as well as a mean value due to the correlation between the spins $0$ and $j$. 
Using an expansion in $1/N$, we can describe the dynamics of the distribution of local stabilities $\rho(\lambda,t)$ by a Fokker-Planck equation, similarly as in Refs.~\cite{Eastham06,Horner07}: 
%The Fokker Planck equation for 
\be
\label{6_fp}
\partial_t\rho(\lambda,t)=-\partial_{\lambda}\,\left[v(\lambda,t)-\partial_{\lambda}D(\lambda,t)\right]\rho(\lambda,t)-\delta(\lambda-\lambda_0(t))+\delta(\lambda+\lambda_0(t)),
\ee
%{\bf is it clear that $d_\lambda$ acts both on $[...]$ and $\rho$? introduce one more pair of brackets?}
where $t$ counts the number of {\em flips  per spin}.
%$t\equiv t'/N$ and $t'$ counts the number of flips that have taken place. 
The drift $v(\lambda,t) \equiv-2N\langle J_{0i}s_0s_i\rangle_{\lambda_i = \lambda}\equiv NC(\lambda,t)$ is the average positive kick received by a spin of stability $\lambda$.
The diffusion constant $D(\lambda,t)\equiv 2N \langle J_{0i}^2\rangle_{\lambda_i = \lambda}=2$ %-[v(\lambda,t)/N]^2\right)/2$ 
is the mean square of those kicks, where we have assumed that the random parts of successive kicks are uncorrelated, as our numerics support. %If we assume  the couplings to remain uncorrelated with the flipping spins, we have $D=2$.} %For $\lambda\gg 1/\sqrt{N}$,  $D=2$.} %{\bf A similar equation was obtained and studied in Ref.~\cite{Horner07}. Here we combine its analysis with an improved understanding of pseudogap exponents and spin correlations.}  
%{\bf where we neglected higher order spin correlations.} }
%The above equation is exact as $N\to \infty$. %However, a closed description requires an evolution equation for $v(\lambda,t)$ as well, which in turn requires knowledge about the evolution of higher order spin correlators. [So far we are not sure whether the tower of equations, that Le started deriving, terminates or not. I still suspect it might terminate at pairwise bond correlators.]}
For the dynamics to have a non-trivial thermodynamic limit the scaling $\langle J_{0i}s_0s_i\rangle\sim 1/N$ must hold, i.e., $\delta=1$ in Eq.~(\ref{6_correlation}).
We further recall that { $\lambda_0(t)\rightarrow 0$} as $N\to \infty$. We may thus replace the $\delta$-functions in Eq.~(\ref{6_fp}) by a reflecting boundary condition at $\lambda=0$,
% to obtain the simple equation: 
\be
\label{6_refb}
\left. \left[v(\lambda,t)-\partial_{\lambda}D(\lambda,t)\right]\rho(\lambda,t)\right|_{\lambda=0}=0.
\ee
%\be
%\label{6_fp2}
%\partial_t\rho(\lambda,t)=-\partial_{\lambda}\left[v(\lambda,t)-\partial_{\lambda}D(\lambda,t)\right]\rho(\lambda,t).
%\ee

{ Since we assume that spins flip many times along the hysteresis loop, finite intervals on the loop correspond to diverging times $\Delta t\rightarrow \infty$.} %\MW: this is implied by our FP model: Excluding entirely chaotic behavior or strange attractors, one expects that 
At those large times a dynamical steady state (ss) must be reached. In such a state the flux of spins must vanish everywhere:
\be
\label{6_steady}
v_{\rm ss}(\lambda)=D\partial_{\lambda}\rho_{\rm ss}(\lambda)/\rho_{\rm ss}(\lambda)\rightarrow 2\theta/\lambda\,,%=2/\lambda 
\ee
where we assumed that $\rho_{\rm ss}$ follows Eq.~(\ref{6_pseudogap}). { This result is tested in Fig. 1(d).} %and $D=2$.
A similar result was obtained in Ref.~\cite{Horner07} following a quench. 
\begin{figure}[h!]
 \centering
 \includegraphics[width=.9\columnwidth]{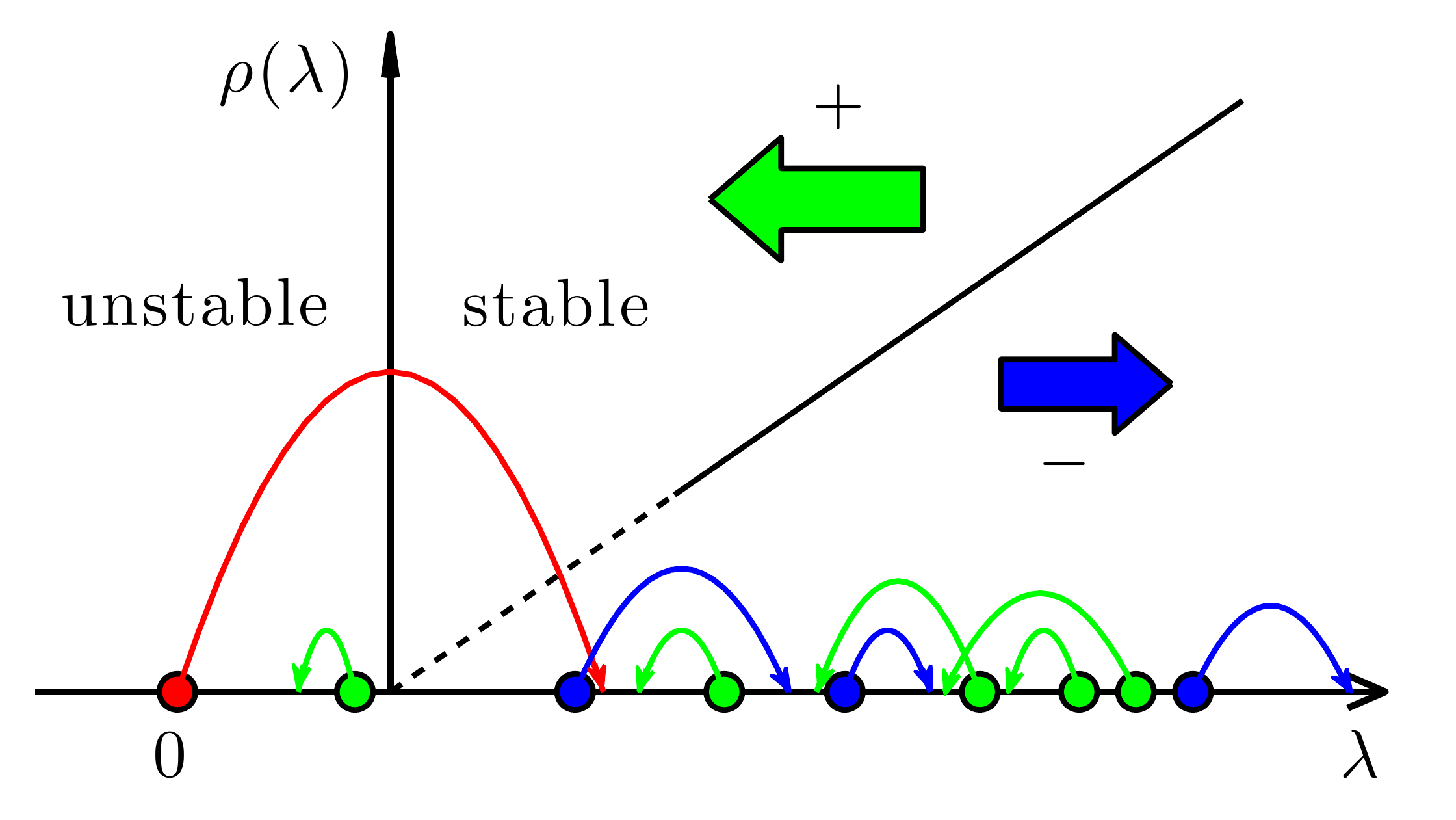}
  \caption{\small{Illustration of a step in the dynamics, in the SK model and the random walker model. Circles on the $\lambda$-axis represent the spins or walkers. At each step, the most unstable spin  (in red) is reflected to the stable side, while all others (in green or blue) receive a kick and move. The dashed and solid line outlines the density profile  $\rho(\lambda)\sim\lambda$ for $\lambda>1/\sqrt{N}$. %, the dashed part corresponds to the range where the linear profile breaks down with only of order one number of spins. 
 The blue spins  were initially frustrated with the flipping spin 0. They are stabilized and are now unfrustrated with 0. In contrast, green spins become frustrated with spin 0 and are softer now. Because of the  motion of spins depends on their frustration with spin 0, a correlation builds up at small $\lambda$, leading to an overall frustration of ``soft" spins among each other.% {\bf make big green/blue arrows same size !}
% {\bf maybe a red arrow for the flipping spin, to distinguish it from kick-arrows? Labelling the origin by O is not necessary} 
 }}\label{6_dynmod}
\end{figure}
%We understand a nontrivial correlation among soft spins must emerge to reinforce the pseudo-gap in stable state. 

\section{Emergence of Correlations}   

Equation~(\ref{6_steady}) implies that $\gamma=1$  in Eq.~(\ref{6_correlation}).
%,  assuming Eq.~(\ref{6_pseudogap}) for $\rho_{\rm ss}$. 
Such singular correlations are unexplained \footnote[0]{ The approximation Eq.~(21) in Horner yields an incorrect scaling behavior for $C(\lambda)$, assuming a pseudogap.}. We now argue that they naturally build up in the dynamics through the spin-flip induced motion of stabilities of frustrated and unfrustrated spins, as illustrated in Fig.~\ref{6_dynmod}.
To quantify this effect we define respectively $C_f(\lambda)$ and $C_f'(\lambda)$ as the correlation between the flipping spin 0 and the spins at $\lambda$ {\it before} and {\it after} a  flip event. %(A dynamical approach to the correlation between any two local stabilities is discussed in Supplementary Materials Section C.)  
As $s_0$ flips, the stability of spin $i$ increases by $x_i\equiv-2J_{0i}s_0s_i$, $\lambda'_i=\lambda_i+x_i$. 
The correlation $C_f'(\lambda)$ is  { an average} over all spins which migrated to $\lambda$ due to the flip:
 \[
\begin{aligned}
C_f'(\lambda)&=\frac{1}{\rho'(\lambda)}\int\rho(\lambda-x)(-x)f_{\lambda-x}(x)\rd x,\\
\rho'(\lambda)&=\int\rho(\lambda-x)f_{\lambda-x}(x)\rd x.
\end{aligned}
\]
$f_{\lambda}(x)$ is the Gaussian distribution of kicks $x$ given to spins of stability $\lambda$: $f_{\lambda}(x)=\exp\left[-\frac{(x-C_f(\lambda))^2}{4D/N}\right]/{\sqrt{4\pi D/N}}$. %$D=\frac{1}{2}\langle X^2\rangle=2$.
%
%To calculate the integral, notice that the Gaussian distribution $f_{\lambda}(x)$ is narrow, 
In the integrands we expand $\rho(\lambda-x)$ and $C_f(\lambda-x)$ for small $x$ and keep terms of order $1/N$, which yields %, $\rho(\lambda-x)=\rho(\lambda)-x\partial_{\lambda}\rho(\lambda)+o(x^2)$, 
%\[
%C_f(\lambda)=\frac{1}{\rho(\lambda)}\int\left(-x\rho(\lambda)+x^2\frac{\partial}{\partial\lambda}\rho(\lambda)\right)f_{\lambda-x}(x)\rd x
%\]
%$C_f(\lambda-x)=C_f(\lambda)-\partial_{\lambda}C_f(\lambda)x+o(x^2)$. Keep the terms to the order of $1/N$,  %but $C'\sim 1/N\lambda^2$ is negligible compare to 1 when $\lambda\gg1/\sqrt{N}$,
\begin{subequations}
\begin{align}
C_f'(\lambda)&=-C_f(\lambda)+2\frac{D}{N}\frac{\partial_{\lambda}\rho(\lambda)}{\rho(\lambda)},
\label{6_selfa}\\
\rho'(\lambda)&=\rho(\lambda)-\partial_{\lambda}\left[ C_f(\lambda)\rho(\lambda)-\frac{D}{N}\partial_{\lambda}\rho(\lambda)\right]. \label{6_selfb}
\end{align}
\end{subequations}
Thus,  even if correlations are initially absent, $C_f(\lambda)=0$,
they arise spontaneously, $C_f'(\lambda)=2D\partial_\lambda\rho(\lambda)/N\rho(\lambda)$. 

In  the steady state, $\rho'_{\rm ss}=\rho_{\rm ss}$, and Eq.~(\ref{6_selfb}) implies the vanishing of the spin flux, that is, Eq.~(\ref{6_steady}) with $v=NC_f$.
%$C_f(\lambda)=D\partial_{\lambda}\rho(\lambda)/N\rho(\lambda)$. 
Plugged into Eq.~(\ref{6_selfa}), we obtain that the correlations are steady, too, %$C'_f= C_f =v_{\rm ss}/N= 2/(N\lambda)$ 
\be
\label{6_cc}
C'_f(\lambda)=C_f(\lambda)=\frac{v_{\rm ss}(\lambda)}{N}
%\frac{D}{N}\frac{\partial_{\lambda}\rho(\lambda)}{\rho(\lambda)}=\frac{D\theta}{N\lambda}
=\frac{2\theta}{N\lambda}.
\ee 
These correlations are expected once the quasi-statically driven dynamics reaches a statistically steady regime, and thus should be present both
during avalanches and in the locally stable states reached at their end. 
%In a stable state obtained when an avalanche stop along the hysteresis curve, the correlation $C(\lambda)$ between the softest spins must therefore take the value prescribed by Eq.(\ref{6_cc}).

Interestingly, Eq.~(\ref{6_cc})  implies that all the bounds of Eq.~(\ref{6_sca}) are saturated if the first one is, i.e., if $\theta=1$. {The latter value was previously derived from dynamical considerations} in Ref.~\cite{Muller14}. It is intriguing that the present Fokker-Planck description of the dynamics  does not pin  $\theta$, as according to Eqs.~(\ref{6_steady},~\ref{6_cc}) any value of $\theta$ is acceptable for stationary states. However, additional considerations on the applicability of the Fokker-Planck description discard the cases $\theta>1$ and $\theta<1$, as discussed in the Appendix Sec.~\ref{app_D3}. 

%{\bf ADD: Dynamic considerations, discussed in Ref.~\cite{Muller14} and Supplementary Materials Section B, allow us to conclude that the only stable value for $\theta$ is $\theta=1$.} 

{ Those are related to the interesting fact that  
%It is interesting to observe 
that Eqs.~(\ref{6_fp},~\ref{6_refb},~\ref{6_steady}) with $\theta=1$ are equivalent to the Fokker-Planck equation for the radial component of unbiased diffusion in $d=2$} % - the marginal dimension with respect to the return to the origin} 
(as derived in Appendix Sec.~\ref{app_D4}), whose statistics is well known \cite{redner01,Bray13}. We can use this analogy to predict $F(n)$, the number of times an {initially soft} spin flips in an avalanche of size $n$. Indeed, a discrete random walker starting at the origin will visit that point $\ln(t)$ times after $t$ steps in two dimensions, and thus $F(n)\sim\ln(n)$, as supported by Fig.~\ref{6_propt}(b). Similarly we expect times between successive flips of a given spin to be distributed as $P(\delta t) \sim 1/( \delta t [\ln(\delta t)]^2)$.

{ {\it Short range systems and experiments:} In short range spin glasses we expect analogous frustrated correlations between pairs of directly interacting soft spins as in the SK model, except that the growth of correlations at small $\lambda$ is cut off at the typical coupling between spins. %Such correlations are a hallmark of frustrated systems, having opposite sign to those expected in unfrustrated magnets. 
This prediction can be tested in experiments akin to NMR protocols: First flip the spins of stability $\lambda$ by a $\pi$-pulse of appropriate frequency. Then flip those of stability $\lambda'$ and observe the resulting shift in the fluorescence spectrum around $\lambda$. From our findings we predict a systematic shift to higher frequencies.}

\section{Conclusion}
We have studied the quasi-static dynamics in a marginally stable glass at zero temperature, focusing on a fully-connected spin glass as a model system. Our central result is that the pseudo-gap appears dynamically due to a strong frustration among the softest spins, characterized by a correlation function $C(\lambda)$ which scales inversely with the stability $\lambda$. We provided a Fokker-Planck description of the dynamics that explains  the appearance of  both the pseudo-gap and the singular correlation, and suggests a fruitful analogy between spin glass dynamics and random walks in two dimensions.  

We expect our findings to apply to other marginally stable systems, in particular hard sphere packings that display a pseudo-gap with a non-trivial exponent: $P(f)\sim f^{\theta_e}$ \cite{Wyart12,Lerner13a,Charbonneau15,Charbonneau14} where $f$ is the contact force. Our analysis above suggests that a singular correlation function $C(f)\sim 1/f$ characterizes how { contacts are affected by the opening of a contact of very weak force,} %force, i.e., by pushing two weakly-interacting particles apart 
the relevant excitations in packings \cite{Wyart12,Lerner13a}. Contacts with small forces should on average be stabilized by $C(f)$ - a testable prediction. Our analysis also suggests a connection between sphere dynamics and random walks in dimension $1+\theta_e$, which is interesting to explore further.

%% file: future/future.tex
\chapter{Outlooks}

\section{Intermediate Phase}
%According to the ``phase diagram'' in last subsection, the real network glasses are far from the rigidity window scenario. 
In {\it Chapter 3}, we have shown that the ``rigidity window'' picture at zero temperature is not robust in the presence of weak interactions. Mapping the intermediate phase in chalcogenides~\cite{Chakravarty05,Boolchand05,Bhosle12,Bhosle12a,Georgiev00,Wang05,Thorpe01} to a ``rigidity window'' is thus unsound, as weak non-covalent interactions cannot be excluded in practical systems. Here, we propose an alternative idea to understand the intermediate phase. 

\begin{figure}[h!]
\centering
\includegraphics[width=.8\columnwidth]{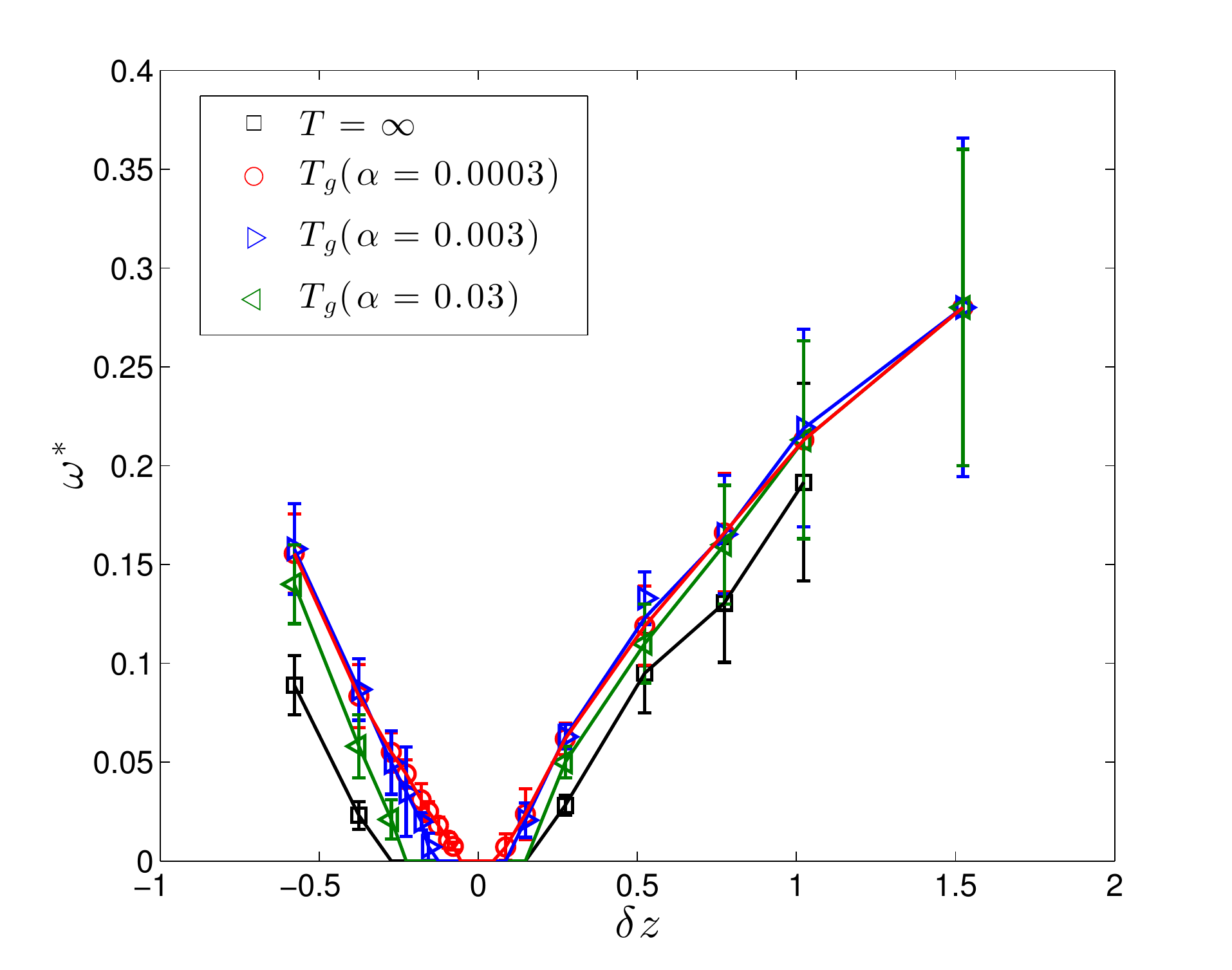}
\caption{\small{Boson peak frequency $\omega^*$ obtained from equilibrated configurations near $T_g$ {\it vs} coordination number $\delta z$ for different $\alpha$. %The inset shows reduced DOS $D(\omega)/\omega$ with boson peak, gives the systematic way to measure $\omega^*$.
}}\label{7_omegadz}
\end{figure}

% ``fractons'' why
We propose the coordination range near $z_c$, where the boson peak is ill-defined, as the intermediate phase. The boson peak becomes impossible to define as the density of states is filled at the low-frequency end in this range. % of ``fractons'', which characterize the elastic properties of large nearly isostatic regions. %Fractons fill up the low-frequency gap in DOS. 
We can rationalize experimental features of the intermediate phase from this consideration. 
In one set of the experiments on the elasticity of chalcogenides~\cite{Wang05}, experimentalists tested the pressure responses of the microscopic structures by measuring the phonon spectra, especially on the phonon line of corner-sharing tetrahedral units. The corner-sharing units in Ge-Se compounds are made of two Germanium atoms, which share two Selenium atoms connected by covalent bonds. The corresponding frequency in the density of states shifts due to nonlinear responses to the stresses on the units in the bulk. The pressure responses indicate an intermediate phase, in which the frequency begins to shift as soon as the pressure acts on, while out of which the frequency shifts only when the pressure is above a threshold. The intermediate phase defined in this way is consistent with the intermediate phase defined by non-reversal heat~\cite{Wang05,Boolchand05}
Indicated by Eq.(\ref{4_coupmat}), the elasticity of the covalent network is contributed mostly by the redundant constraints, i.e., self-stressed regions, and by the coupling with the network of weak forces through the low-frequency modes, $\omega\lesssim\sqrt{\alpha}$. When a stress applies on the material, the structure responses are limited mostly to the corresponding modes. The self-stressed modes are very anisotropic, distortion along which thus does not effect the structure of most corner-sharing units. By contrast, the low-frequency modes are more extended, response of which extends in the bulk. In the vicinity of the rigidity transition, the structures are abundant in these extended low-frequency modes, the pressure easily causes the distortion homogeneously in the bulk and shift the corner-sharing spectral line~\cite{Wang05,Briere07}. % frequency quantifies the elastic energy of the corresponding mode, thus a structure with a gapless density of states cannot sustain any external pressure. Therefore, 
%in the intermediate phase, an arbitrarily small external pressure penetrates the sample, distorts the corner-sharing tetrahedral units in the bulk and causes the Raman shift of the corner-sharing spectrum. Out of the intermediate phase, the lack of low-frequency modes blocks the low energy collective strains, and hence, a small pressure cannot propagate deeply into the sample and induce the Raman shift. 

We define the range by probing the boson peak frequency $\omega^*$ for strong networks equilibrated at $T_g$ in our model. In $D(\omega)/\omega^{d-1}$-$\omega$ plot, see for example Fig.~\ref{3_dos}(d), a peak becomes ill-defined in %In our model, $d=2$, we plot $D(\omega)/\omega$ versus $\omega$ shown as the inset in Fig.~\ref{7_omegadz}. 
a finite range of coordination number near $z_c$ when $\alpha>0$, shown in Fig.~\ref{7_omegadz}. This range of boson peak frequency $\omega^*=0$ is proposed as the intermediate phase. The width of intermediate phase depends on the glass transition temperature and thus on $\alpha$, as shown in Fig.~\ref{7_omegadz}. For $\alpha=0.03$ closest to the real strength of van der Waals interactions, the width of intermediate phase is $\Delta z\approx0.4$, which is significantly larger than $0.06$ obtained in the rigidity window~\cite{Chubynsky06,Briere07}, and much closer to the experimental values (approximately $0.1z_c$~\cite{Boolchand01}). Moreover, this definition of the intermediate phase covers both the floppy and rigid side on the coordination number as empirical observations~\cite{Boolchand01}, while the rigidity window appears only on the floppy side~\cite{Thorpe00}.

% Compare to rigidity window, experimental test
The importance of large homogeneous isostatic clusters to the intermediate phase is similar in both our boson peak definition and the rigidity window picture. %The main differences between the two is that: 
However, the rigidity window picture roots on a percolating isostatic cluster, which is fragile under the addition of weak forces and finite-temperature fluctuations. In contrast, our boson peak picture bases on a mean-field jamming transition, where large isostatic clusters near $z_c$ are induced by finite-temperature fluctuations, which is thus a more robust concept applicable to real glasses. Though isostatic clusters cannot be directly verified, some consequences can be empirically tested. The density of states in the intermediate phase may be measured through a Raman scattering experiment to check the abundance of the low-frequency vibrations; The width dependence on $\alpha$ of the intermediate phase may be studied by network glasses with different electronegativity or even patchy particle glasses~\cite{Wang12}, where the stiffness of the strong interactions is tunable. 

\section{Dynamical Transition at Finite Temperature}
The dynamical phase transition studied in {\it Chapter 5} is not limited to the erosion. Similar models and simulations have been extensively studied in the context of Type-II superconductors~\cite{Watson96,Watson97,Abrikosov04,Gronbech-Jensen96,Dominguez99,Kolton99,Kolton03,Olive06}, where vortices, carrying magnetic flux quanta, play the roles of driven particles. In the system, vortices repelling each other are driven by the flow of current in a randomly pinning substrate of impurities and defects. When the current exceeds the critical value, the flow of vortices induces an electric field and an electric resistance that breaks the superconducting state~\cite{Koshelev94}. Critical behaviors and spatial organizations, predicted in our erosion model, are under test in molecular dynamics simulations of these systems~\cite{Kolton99,Kolton03}. %Watson and Fisher~\cite{Watson96,Watson97} modeled this dynamical transition on a lattice with both a time-like direction and spatial directions perpendicular to it. Similar to many directed percolation models, the steady-state flow occurs when a structure free of pinning sites percolates in the time-like direction. %However, two significant features make it different from the directed percolation universality. One is that the dynamics is featured by the moving particles on the percolating structure. More importantly, particles, vortices in this context, may interact along the time-like coordinate: the particles that get pinned pave a pin-free way for later particles, while sites interact only at the same time level in directed percolation. This complex non-monotonic interaction between moving particles may also lead to a different universality class from depinning transition.  %The models are inspired by the Type-II Superconductors, where magnetic fields can penetrate the superconductor but confined into quantized vortices. The vortices repel each other and can be pinned by the defects in the materials. When the superconductor conducts a current, the ampere force will drive the vortices move in the perpendicular direction, induce a voltage difference in the direction along the moving vortices, and breaks down the superconducting state~\cite{}. The automaton of these vortex systems is equivalent to the model in {\it Chapter 5} of repelling particles driven on a random rough landscape. The most relevant difference is the form of interactions, which results into two other dynamical phases, a smectic flow phase and a transversal solid phase when driving force is large. However, the universality of phase transition indicates that the specific form of interactions would be irrelevant near the threshold of vortex current. 
In general, any system that consists of repulsive particles driven by a directed force on a glassy rough landscape can be modeled in the similar way, and its dynamical transition should fall in the same universality class. %, for instance, a system of charged colloidal particles or emulsions driven by an electric field on a rough surface or a region of random pins set by optical tweezers. 

However, in these microscopic systems, when the energy barrier becomes comparable to the temperature or Planck constant, thermal or quantum fluctuations are relevant. We need to include these fluctuations in the model to apply our theoretical framework to the dynamics in these systems. 
To incorporate the stochastic dynamics into the model, we introduce a control parameter probability $p$. At each time step, a particle may move to one of its two downhill sites with probability $p$, independent of the local heights if the target site is not occupied. If this stochastic movement is not executed, the particle will just follow the deterministic rule. The preliminary results from this simplest stochastic rule indicate that the sharp transition becomes a crossover from the force-driving dynamic state above $\theta_c$ to a fluctuation-induced creeping state below the threshold $\theta_c$~\cite{Barizien14}. 

To further capture the temperature excitation or quantum tunneling in the model, a Monte Carlo dynamics can be introduced to replace the deterministic dynamics. At each time step, a particle randomly moves to any of its four neighbor sites, according to a probability proportional to $\exp[\frac{1}{T}(f_{i\to j})]$, where $f_{i\to j}$ is the force at site $i$ to the direction on site $j$. In the creeping phase, the mean particle current is expected to be proportional to $\exp[-\frac{U_0}{T}(\theta_c/\theta)^{\mu}]$, where $U_0$ is a fitting parameter and $\mu$ is a critical exponent, in analogy to the creeping dynamics of elastic manifolds~\cite{Kolton09,Houssais15}. Further numerical works should be done to check this creeping behavior and the universal power law exponent $\mu$. 

%Applying the results from our erosion model, we expect that near the threshold, the voltage increase linearly with the current above the critical current, $V\sim(I-I_c)^{\beta}$, $\beta=1$. The fluctuation of current is large at microscopic level, due to the broad distribution of local currents. However, this fluctuation should be averaged out at macroscopic level, as the correlation of current in the transversal direction vanishes. 

%As both thermal and quantum fluctuations would be significant in vortex superconductors, we can extend our model by including some stochastic factors in the dynamics. Similar to 

\section{Universality of Critical Dynamics}
\begin{figure}[!ht]
\centering
\includegraphics[width=.9\columnwidth]{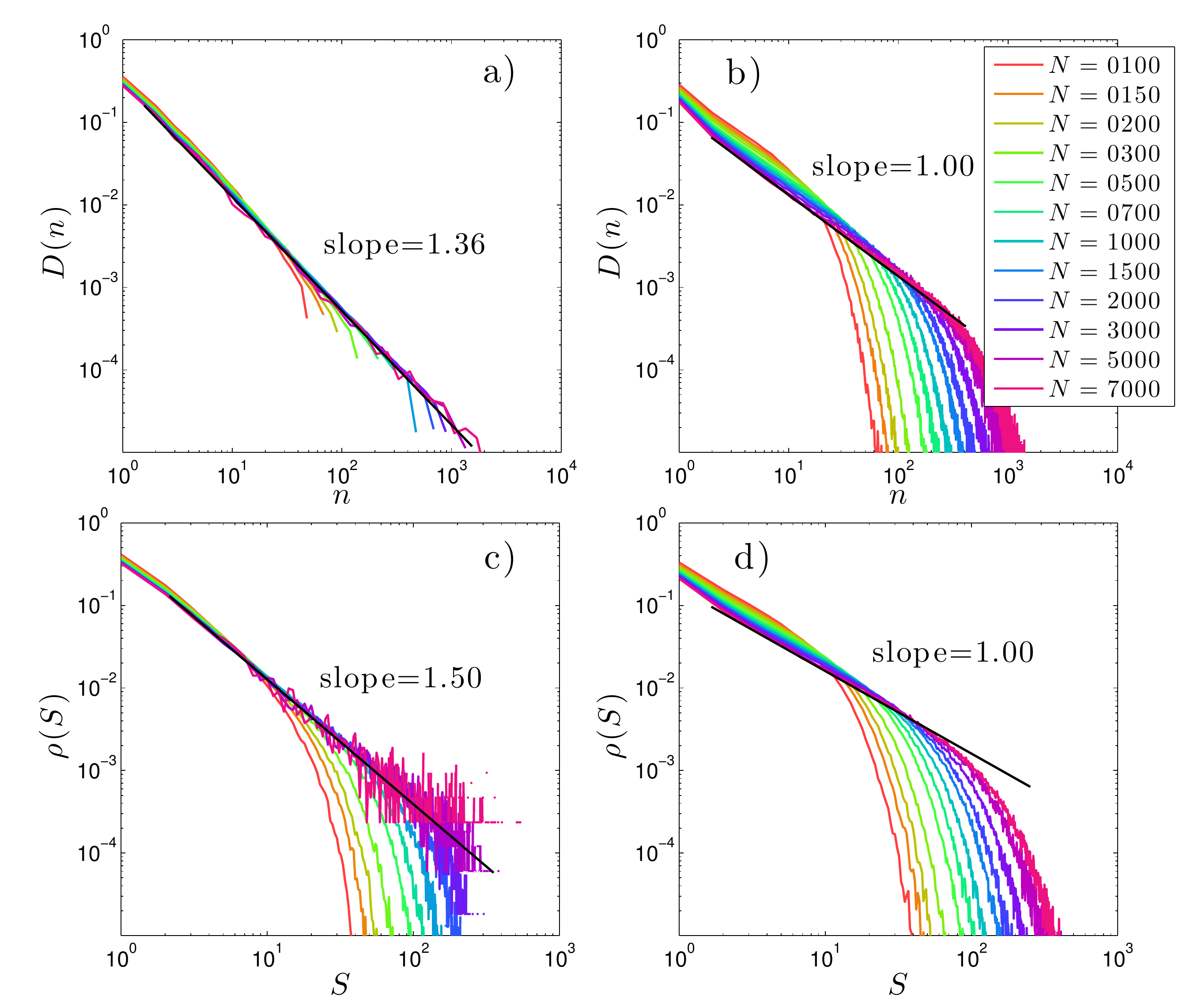}
\caption{\small{(a) (b) Avalanche size distribution $\md(n)$; (c) (d) Magnetization jump distribution $\mpp(S)$. (a) (c) Reluctant dynamics; (b) (d) Greedy dynamics.}}\label{7_dist}
\end{figure}

\begin{figure}[!ht]
\centering
\includegraphics[width=.9\columnwidth]{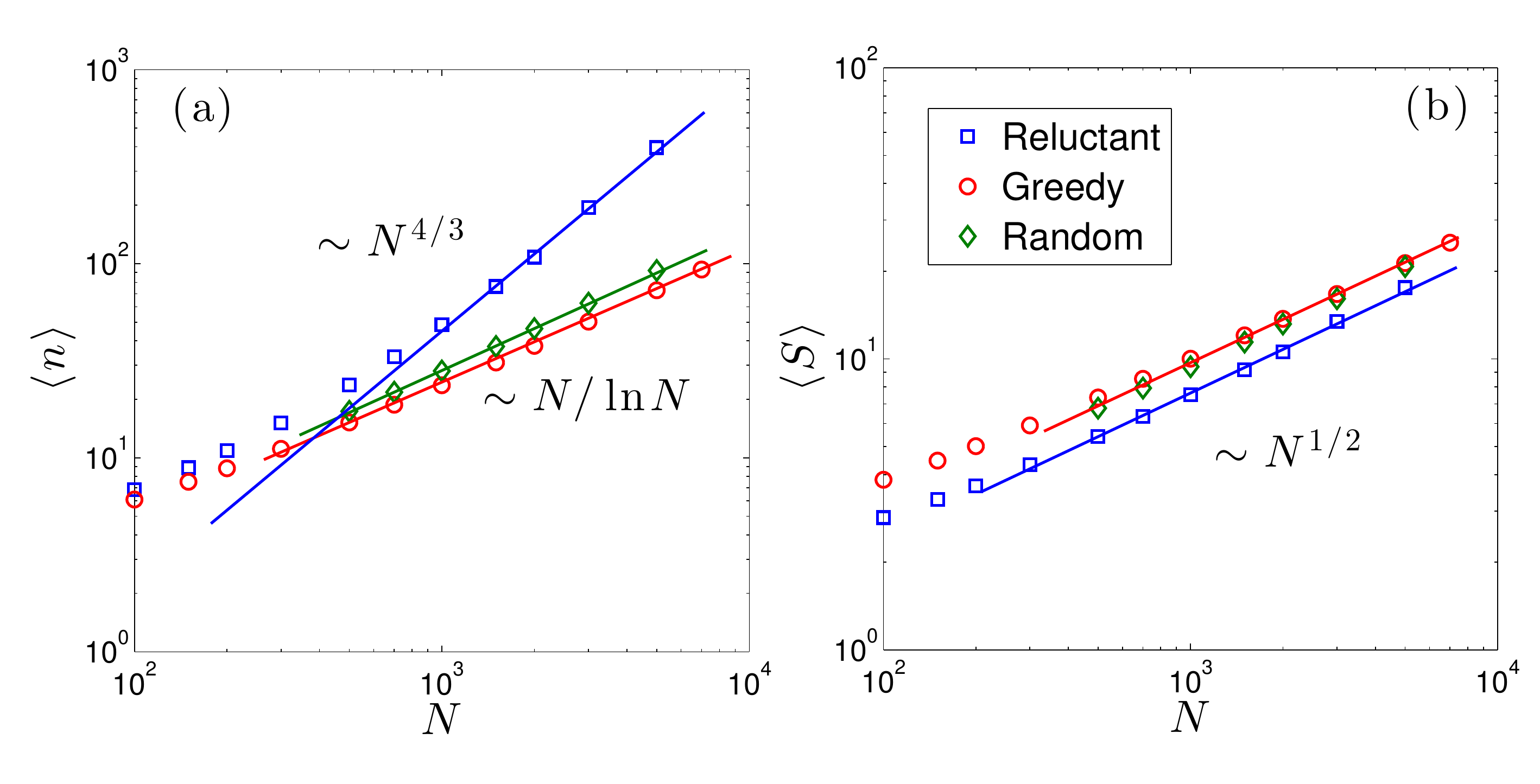}
\caption{\small{(a) Average avalanche size $\langle n\rangle $ (b) Average magnetization jump $\langle S\rangle$ of avalanches.}}\label{7_exp}
\end{figure}

The universal exponents of the critical dynamics in Sherrington-Kirkpatrick model studied in {\it Chapter 6} appear to be exact fractions. The exponents are defined in the power law distributions of the avalanche size and magnetization change, %The power-law exponents of the avalanche size and magnetization change distributions are exact fractions, 
\be
D(n) = n^{-\tau}d(n/N^{\sigma})/\Xi(N);
P(S) = S^{-\rho}p(S/N^{\beta})/Z(N).
\ee
where $\tau=\rho=1$, $\sigma=1$, $\beta=1/2$, and $\Xi(N)=Z(N)=\ln N$ for greedy dynamics described in {\it Chapter 6}~\cite{Pazmandi99,Eastham06}, as well as random dynamics~\cite{Eastham06} where a random unstable spin is flipped at a step. We have also investigated a peculiar dynamics -- reluctant dynamics~\cite{Contucci05}: at each step, the least unstable spin is chosen to flip. We have found $\tau=4/3$, $\rho=3/2$, $\beta=1$, and $\sigma=2$ for this dynamics, shown in Figs.~\ref{7_dist} and~\ref{7_exp}. 

We have argued in {\it Chapter 6}, dynamically, the pseudogap exponent $\theta=1$. The typical external field increment to trigger an avalanche (to destabilize the least stable spin) is thus $\Delta h\sim\lambda_{\rm min}\sim N^{-1/(1+\theta)}$. Indicated by the continuous hysteresis loop, Fig.~\ref{6_skmodel}, the average magnetization changes by an amount proportional to the system size under a finite change of the external field. Therefore, the average magnetization jump in an avalanche is $\langle S\rangle\sim N/\Delta h\sim N^{\theta/(1+\theta)}=N^{1/2}$, which leads to a scaling relation, 
\be
\beta(2-\rho)=\frac{1}{2}.
\ee
Due to the random couplings of spins, a spin does not necessarily align with the external field after flip. Assuming the sign is purely random, we find that the typical magnetization jump $S$ for an avalanche of size $n$ then scales as $S\sim n^{1/2}$, which indicates another scaling relation, 
\be
\beta=\sigma/2.
\ee
With the two scaling relations, four critical exponents are reduced to two independent ones, $\tau$ and $\sigma$. 

\begin{figure}[h!]
 \centering
   \def\svgwidth{.9\columnwidth}
   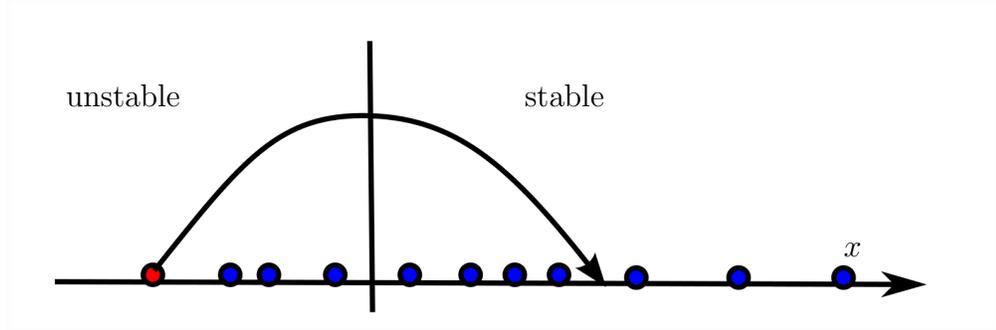
  \caption{\small{Illustration of the dynamic model. Circles on the axis represent the walkers. The leaping walker (in red) jumps to the stable side, and other walkers (in blue) move random steps according to Eq.(\ref{7_randw}).}}\label{7_dynmod}
\end{figure}

%Argument fixes some scaling relation,
To determine the two independent exponents, we assume that $\tau$ and $\sigma$ are purely determined by the avalanche dynamics. We introduce the following dynamic model, based on the stochastic description of the dynamics developed in {\it Chapter 6}. $N$ random walkers in one dimension are characterized by a random variable $X$. A boundary $B$ separates the axis into an unstable side $x<B$ and a stable side $x>B$. The system stops when unstable random walkers disappear. The boundary plays the role of the quasi-static external field: we retrigger avalanches by setting $B$ to the place between the two least stable walkers after the dynamics stop. 
At each time step during an avalanche, an unstable walker at $x_0<B$ leaps to the stable side at $(2B-x_0)$, mimicking the spin flip in SK model. All other walkers take random steps independently, as illustrated in Fig.~\ref{7_dynmod}, 
%\begin{widetext}
\be
\label{7_randw}
x_i(t+1)-x_i(t)=\frac{1}{N}\left[\frac{2\alpha}{\max(x_i(t)-x_0(t),\epsilon/\sqrt{N})}-\beta x_i(t)+2\sqrt{N}Z(x_i)-\gamma(t)\right]
\ee
%\end{widetext}
for the walker $i$ at $x_i$, where $x_0(t)$ is the position of the ``flipping'' walker at $t$, $Z(x)\sim\mathcal{N}(0,1)$. The drift terms resemble the nontrivial correlations, Fig.~\ref{6_skmodel}(d), between the flipping spin and other spins. The term proportional to $\alpha$ corresponds to the singular drift at the boundary, which is bounded by $\sqrt{N}$, and we set $\alpha=\theta=1$~\cite{Yan15}. The $\beta$ term limits the walkers from diffusing away %The similar trend exists in SK model, see Fig.~\ref{7_jss}, but the form is not justified as the singular term. 
but is a less relevant entry at small $x$. 
$\gamma(t)$ is a global constraint so that $\sum_{i\neq0}(x_i(t+1)-x_i(t))=0$. %, the importance  of which is discussed at the end. 
We present the results of ``flipping'' the most unstable walker at each step with $\alpha=1$, $\beta=1.5$, and $\epsilon=1$. 

\begin{figure}[h!]
\centering
\includegraphics[width=.9\columnwidth]{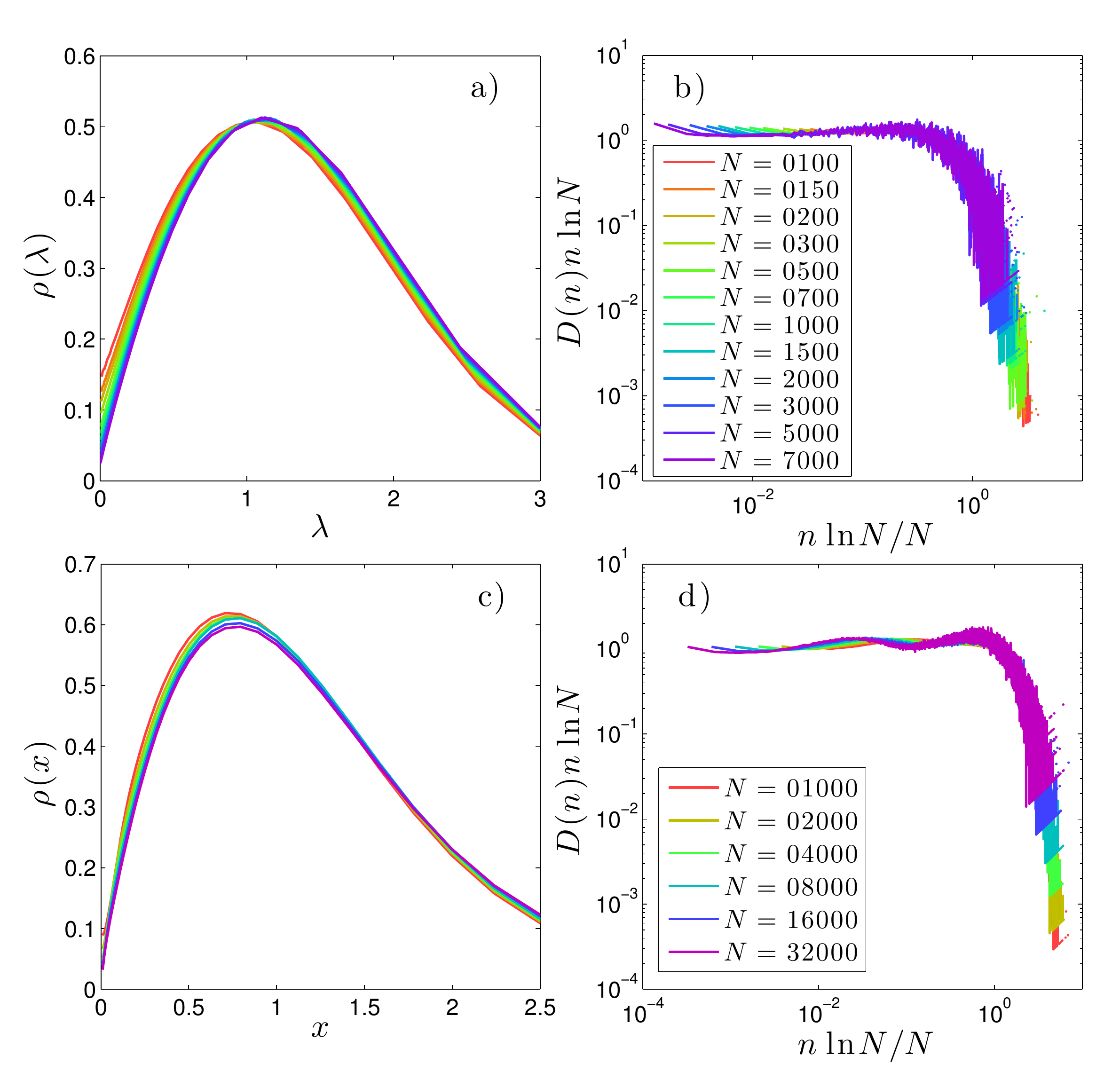}
\caption{\small{(a) (c) Distribution of local stability $\rho(\lambda)$ in stable states and (b) (d) Distribution of avalanche size $D(n)$ scaled by finite size $N$ for (a) (b) SK model and (c) (d) random walker model. In the pseudo-gap near $\lambda=0$, $\rho(\lambda)\propto\lambda$ in both (a) and (c). The collapses in (b) and (d) indicate the power law exponent $\tau=1.0$, and $\sigma=1$ with a logarithmic correction.}}\label{7_rdn}
\end{figure}

The stochastic model without any magnetization details reproduces the greedy dynamics where the unstable spins stick close to the boundary. Moreover, the numerical results of the stochastic model indicate avalanches with critical exponents, $\tau=1$ and $\sigma=1$, the same as in the greedy dynamics, shown in Fig.~\ref{7_rdn}. 
 
The model opens up a possibility of solving for the exponents from a well defined stochastic problem~\cite{Jagla15}. 
In addition, pointed out in {\it Chapter 6}, the stochastic description with a non-trivial correlation among soft excitations applies to jammed packings and other glassy systems. It is then possible to be generalized to solve for the critical exponents of avalanches in different glassy systems by considering the corresponding correlations.

%% file: future/dynmodel.eps_tex
%% Creator: Inkscape inkscape 0.48.2, www.inkscape.org
%% Accompanies image file 'dynmodel.eps' (pdf, eps, ps)
\begingroup%
  \makeatletter%
  \providecommand\color[2][]{%
    \errmessage{(Inkscape) Color is used for the text in Inkscape, but the package 'color.sty' is not loaded}%
    \renewcommand\color[2][]{}%
  }%
  \providecommand\transparent[1]{%
    \errmessage{(Inkscape) Transparency is used (non-zero) for the text in Inkscape, but the package 'transparent.sty' is not loaded}%
    \renewcommand\transparent[1]{}%
  }%
  \providecommand\rotatebox[2]{#2}%
  \ifx\svgwidth\undefined%
    \setlength{\unitlength}{381.2bp}%
    \ifx\svgscale\undefined%
      \relax%
    \else%
      \setlength{\unitlength}{\unitlength * \real{\svgscale}}%
    \fi%
  \else%
    \setlength{\unitlength}{\svgwidth}%
  \fi%
  \global\let\svgwidth\undefined%
  \global\let\svgscale\undefined%
  \makeatother%
  \begin{picture}(1,0.33774921)%
    \put(0,0){\includegraphics[width=\unitlength]{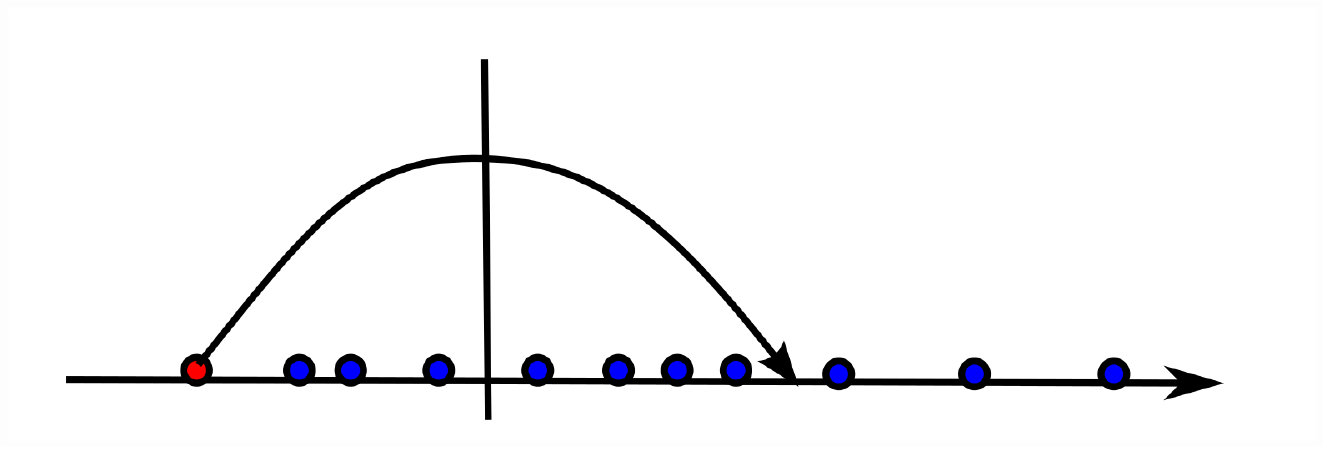}}%
    \put(0.06,0.23){\color[rgb]{0,0,0}\makebox(0,0)[lb]{\smash{unstable}}}%
    \put(0.52,0.23){\color[rgb]{0,0,0}\makebox(0,0)[lb]{\smash{stable}}}%
    \put(0.84,0.08){\color[rgb]{0,0,0}\makebox(0,0)[lb]{\smash{$x$}}}%
  \end{picture}%
\endgroup%

%% file: conclusion/conclusion.tex
\chapter{Conclusion}

% Correlation to the rigidity transition of microscopic structures of glass properties.
% Rigidity transition to mean field Jamming
% Non-monotonic specific heat with flexible topology
% Critical dynamics in logarithmic potential with reflective boundary
% Erosion transition

%\section{Summary}
In this dissertation, we have investigated several topics on the dynamics of glassy systems. The glass transition comes about as the dynamics rapidly slow down under cooling~\cite{Ediger96,Debenedetti01}. We have proposed an explanation for the correlation between the rapidity of slowing-down of glasses and the elasticity of their microscopic structures, by introducing novel network-glass models. Next, at zero temperature, a frozen glassy system becomes dynamic under a strong-enough driving force. We have proposed a model incorporating the interplay between disorder and particle interaction to describe the erosion of the river bed, which shows a dynamical phase transition with testable predictions associated to a new universality class. Finally, many athermal glassy systems self-organize to show critical dynamics under qausi-static drives~\cite{Parker07,Bak13}. The critical dynamics are results of a pseudogap of soft excitations in the systems with long-range effective interactions~\cite{Muller14}. We have done a case study of a spin glass system and developed a general stochastic description of the dynamics where the correlation and the pseudogap emerge spontaneously. %from equilibrium -- how the relaxation time of glasses depends on the elasticity of the microscopic structure, to the critical point -- how the criticality of marginally stable systems is self-reinforced dynamically, to non-equilibrium -- in which universality the erosion phase transition is of short-range interacting particles under an external drive on a random landscape. 

Self-organization plays a key role in all three topics. Network glasses self-organize at low temperature due to the existence of the weak constraints, which reduces the spatial fluctuations of covalent bonds, leading to mean-field like networks. As the result, the predictions of the thermodynamics and dynamics based on the frozen elastic networks obeying mean-field rigidity transition hold true near the glass transition even though the interaction networks are adaptive in real liquids. 
This result raises questions on the validity of mean-field approaches in other supercooled liquids at low temperature, which appear to apply in particular near the jamming transition. %What kind of microscopic structures will the self organization select out in supercooled liquids like polymer glasses? If yes, then our results can be extended to all liquids well. 

In the erosion of river beds, self-organization appears in form of ``armoring'' of the surface, for which, the ``holes'' are filled up, %. The feedback effect of the particle flow and the dynamical basin gives birth to a special universality class of the critical dynamics. The ``armoring'' effect releases the stress driving the system, which drives the system directly the critical point. Consequently, the flux reparation on the emerging landscape is 
leading to a subtle power-law distribution of the spatial organization of flux near the threshold. 
The results can be well generalized to other systems of short-range repelling particles driven by a external forcing on a random pinning substrate. Type-II superconductors are relevant examples, where magnetic quanta are driven by the electric current and pinned by impurities or defects in the substrate. Similar to the erosion, the ``armoring'' of the random pinning spots will end up with a spatial organization of flux of the magnetic quanta, which has also been overlooked in the relevant literature. 
Experimental works~\cite{Zou15} are currently testing our views. 
%It would be interesting to check the avalanche-like flux reparations, to see whether the critical exponents obeys the ones predicted in the model. Moreover, a kinetic description may also be developed to understand the self-organizing processes in these glassy systems.

Finally, the other set of athermal glassy systems with long-range, frustrated interactions also self-organizes under a quasi-static drive, which leads to marginal states, with pseudogaps and crackling responses. We have explained these phenomena dynamically with a stochastic-kinetic description in the mean-field spin glass. A challenge for the future is to generalize the description to other self-organizing systems, where the interactions are more practical ones, for example, power-law decaying, instead of fully-connected. 

%% file: appendices/appd_thermodyn.tex
\SkipTocEntry
\chapter{Why glass elasticity affects the thermodynamics and fragility of super-cooled liquids?}
\appcaption{\thechapter \space Why glass elasticity affects the thermodynamics and fragility of super-cooled liquids?} %this needs to be here to show up in the ''List of appendices'' which is required if you have more than one appendix
\label{app_thermodyn}
%In this appendix we explain a lot about elephants
%\keywords{rigidity percolation, jamming, vibration modes, weak constraints}

\SkipTocEntry
\section{Stiffness and Coupling matrices}
\label{app_A1}
Consider  a  network of $N$ nodes connected by $N_c$ springs. 
If an infinitesimal displacement field $|\delta {\bf R}\rangle $ is imposed on the nodes, the change of length of the springs  can be written as a vector $|\delta r\rangle$ of dimension $N_c$. For small displacements this relation is approximately linear: $|\delta r\rangle ={\cal S} |\delta {\bf R}\rangle $, where ${\cal S}$ is an $N_c\times Nd$ matrix.  To simplify the notation, we write ${\cal  S}$ as an $N_c\times N$ matrix of components of dimensions $d$, which gives ${\cal  S}_{\gamma,i}\equiv \partial r_\gamma/\partial {\bf R}_i=\delta_{\gamma,i} {\bf n}_\gamma$, where $\delta_{\gamma,i}$ is non-zero only if the contact $\gamma$ includes the particle $i$, and  ${\bf n}_\gamma$ is the unit vector in the direction of the contact $\gamma$, pointing toward the node $i$.  Using the bra-ket notation, we can rewrite   ${\cal  S}=\sum_{\langle ij\rangle\equiv \gamma}| \gamma \rangle {\bf n}_{\gamma} (\langle i | -\langle j |)\nonumber$, where the sum  is over all the springs of the network.  Note that the transpose ${\cal S}^t$ of ${\cal  S}$ relates the set of contact forces $|f\rangle$  to the set $|{\bf F}\rangle$ of unbalanced forces on the nodes:  $| {\bf F}\rangle={\cal  S}^t |f\rangle$, which simply follows from the fact that ${\bf F}_i=\sum_{\gamma} \delta _{\gamma,i} f_\gamma {\bf n}_\gamma=\sum_{\gamma} f_\gamma {\cal S}_{\gamma,i}$ \cite{Calladine78}. 

The stiffness matrix ${\cal \tilde M}$ is a linear operator connecting external forces to the displacements: ${\cal \tilde M}|\delta {\bf R}\rangle=|{\bf F}\rangle$. Introducing the $N_c\times N_c$ diagonal matrix ${\cal K}$, whose components are the spring stiffnesses ${\cal K}_{\gamma\gamma}=k_{\gamma}$, we have for harmonic springs $|f\rangle={\cal K} |\delta r\rangle$. Applying ${\cal  S}^t$ on each side of this equation, we get $|{\bf F}\rangle={\cal  S}^t |f\rangle={\cal  S}^t 
{\cal K} {\cal S} |\delta {\bf R}\rangle$, which thus implies \cite{Calladine78}:
\be
\label{app_4}
{\cal \tilde M}={\cal  S}^t {\cal K}{\cal  S}.
\nonumber
\ee
Let us assume that starting from a configuration where all springs are at rest, the rest lengths of the springs are changed by some amount $|y\rangle$. This will generate an unbalanced force  field $|{\bf F}\rangle ={\cal  S}^t{\cal K}|y\rangle$ on the nodes, leading to a displacement $|\delta {\bf R}\rangle={\cal \tilde M}^{-1}{\cal  S}^t{\cal K}|y\rangle$. The elastic energy ${\cal E}=\frac{1}{2}\langle y-\delta r|\mathcal{K}|y-\delta r\rangle$ is minimal for this displacement and the corresponding energy ${\cal \tilde H}$ is:
\be
\label{app_5}
{\cal \tilde H}(|y\rangle)=\frac{1}{2}\langle y|{\cal K}-{\cal K}{\cal  S}{\cal \tilde M}^{-1}{\cal  S}^t{\cal K}|y\rangle.
%\tag{SI-1}
\ee
In our model,  $y_\gamma=0$ for weak springs and $y_\gamma=\epsilon \sigma_\gamma$ for strong springs of stiffness $k$, implying that ${\cal K} |y\rangle=k |y\rangle$. Introducing the dimensionless stiffness matrix ${\cal M}\equiv {\cal \tilde M}/k$ and the restriction ${\cal  S}^t_{\rm s}$ of the operator ${\cal  S}^t$ on the subspace of strong contacts of dimension $N_s$, i.e. ${\cal S}^t_{\rm s} |\sigma\rangle\equiv {\cal  S}^t|y\rangle$, Eq.(\ref{app_5}) yields:
\be
\label{app_6}
{\cal H}(|\sigma\rangle)=\frac{1}{2}\langle\sigma|{\cal G}|\sigma\rangle \hbox{ where } {\cal G}={\cal I}-{\cal S}_{\rm s}{\cal M}^{-1}{\cal S}^t_{\rm s},
%{\cal H}(|\sigma\rangle)=\frac{k\epsilon^2}{2}\langle\sigma|{\cal G}|\sigma\rangle \hbox{ where } {\cal G}={\cal I}-{\cal S}{\cal M}^{-1}{\cal S}^t
\nonumber
\ee
where ${\cal I}$ is the identity matrix, and ${\cal G}$ is the coupling matrix used in {\it Chapter 2}. Note that in our model the diagonal matrix ${\cal K}$ contains only two types of coefficients $k_{\rm w}$ and $k$, corresponding to the stiffnesses of weak springs and stiff springs respectively. Then the dimensionless  stiffness matrix can be written as ${\cal M}={\cal S}^t_{\rm s}{\cal S}_{\rm s}+\frac{k_{\rm w}}{k}\,{\cal S}^t_{\rm w}{\cal S}_{\rm w}$, where ${\cal S}^t_{\rm w}$ is the projection of the operator ${\cal S}^t$ on the subspace of weak contacts.
\begin{figure}[h!]
   \begin{center}
   {\includegraphics[width=0.8\columnwidth]{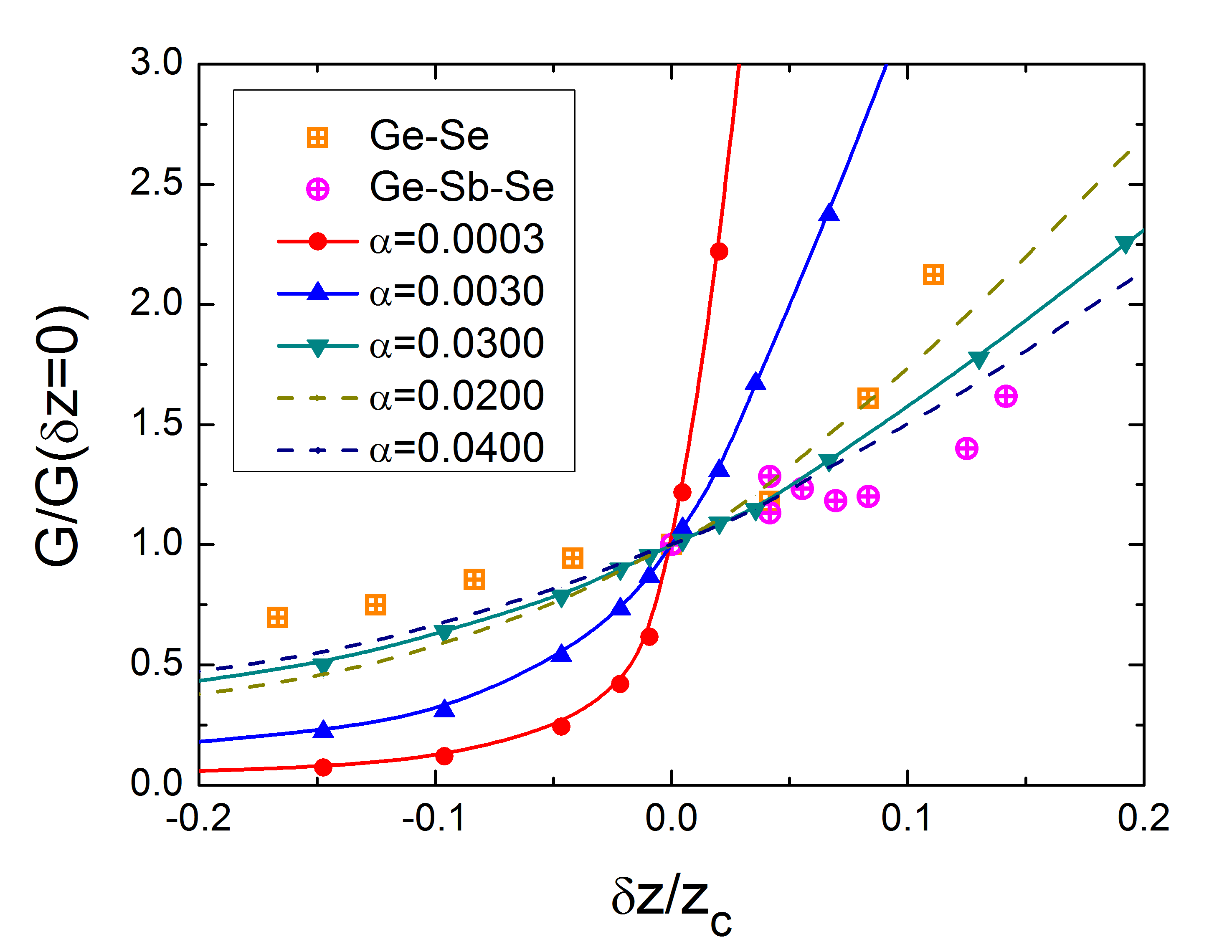}}
   \caption{Squares show the shear modulus $G$ normalized by its value at the rigidity threshold for Ge-Se, taken from Ref.~\cite{Ota78}. Circles show $G$ for Ge-Sb-Se, taken from Ref.~\cite{Mahadevan83}. Lines display the shear modulus $G$ for network models in $d=3$  using  different $\alpha$, as indicated in the legend.}
     \label{app_AA}
   \end{center}
%\tag{SI-1}
\end{figure}

\SkipTocEntry
\section{Shear modulus of the random elastic networks}
\label{app_A2}
An explicit  expression for the shear modulus ${G}$  of an elastic network can be found using linear response theory \cite{Lutsko89,Karmakar10b}. In particular, let us consider  the shear on the $(x, y)$ plane.  In the contact vector space, a shear strain can be written as  $\delta|y^{sh}\rangle$ with $\delta\ll1$ which represents the amplitude of the strain and $|y^{sh}\rangle$ corresponds to a unit shear strain. The components of  $|y^{sh}\rangle$  are given by  $y^{sh}_\gamma=\frac{\Delta x_\gamma\Delta y_\gamma}{l_\gamma}$, where  $l_{\gamma}$ is the rest length of the spring $\gamma$, and  $\Delta x_{\gamma}$ and $\Delta y_{\gamma}$ are its projections along the x and y directions. From the last section (A), we can obtain  the total energy induced by a shear strain  ${\cal \tilde H}(|y^{sh}\rangle)\delta^2$; hence the shear modulus $G=2 {\cal \tilde H}(|y^{sh}\rangle)/ V$.

To estimate the value of $\alpha$, we consider the dependence of the shear modulus with coordination or valence in the vicinity of the rigidity transition, which is smooth for large $\alpha$ and sudden for small $\alpha$ in our networks, see Fig.~\ref{app_A1}.  Comparing networks and  real chalcogenide glasses we find that the cross-over in the elastic modulus is   qualitatively reproduced for  $\alpha\in [0.01,0.05]$.

\SkipTocEntry
\section{Finite size effects on fragility}
\label{app_A3}

To estimate the role of finite size effects on the dynamics, we use  two different system sizes $N=64$ and $N=256$. As shown in Fig.~\ref{2_f2A}, the Angell plot for the relaxation time, and therefore our estimation of the fragility,  appears to be nearly independent of the system size. Note, however, that the correlation function $C(t)$  shows some finite size effects very close to the isostatic point $(z=z_c,\alpha=0)$, but that it does not affect our measure of $\tau$ significantly. 
In particular we find that near isostaticity, the distribution of relaxation time is broad for small systems, and becomes less and less so when the system size increases. 
%This effect can be understood in the limiting case of number partitioning where the slow spins corresponds to those for which the coefficient $\delta r_{1,\alpha}$ defined in Eq.(5) is large. 
We noticed that this effect also disappears if a two-spin flips Monte-Carlo is used, instead of the one-spin flip algorithm we perform. 

\SkipTocEntry
\section{Fragility in experimental dynamical range}
\label{app_A4}

The value of fragility depends on the definition of glass transition, in particular on the dynamical range. In super-cooled liquids the glass transition occurs when the relaxation time is about $10^{16}$ larger than the relaxation time at high temperature. Thus the dynamical range in experiments (which corresponds to the fragility of a perfectly Arrhenius liquid)  is $R=16$. In our simulation, the same quantity is $R=5$. It is possible however to rescale our values of fragility  to compare with experimental data, if we extrapolate the dynamics.
We shall assume a Vogel-Fulcher-Tammann (VFT)  relation at low temperature,
\[
\log_{10}\frac{\tau(T)}{\tau_0}=\frac{A}{T-T_0},
\]
We define the dynamical range as:
\[
R=\log_{10}\frac{\tau(T_g)}{\tau_0}=\frac{A}{T_g-T_0}.
\]
Thus we can express the fragility as: 
\be
\label{app_111}
m_R=\left.\frac{\partial\log_{10}\tau(T)/\tau_0}{\partial{T_g/T}}\right|_{T=T_g}=R+R^2\frac{T_0}{A}.
\ee
$T_0$ and $A$ are assumed to be independent of dynamical range.
Using the notation $m=m_5$ and $m_{sc}=m_{16}$ we get from Eq.(\ref{app_111}):
\[
m_{sc}=16+\frac{16^2}{5^{2}}(m-5)=10.24m-35.2
\]
The amplitude of fragility we find turn out to be comparable to experiments when $\alpha=0.03$, in particular for $z
\geq z_c$. For the smallest coordination explored our results underestimate somewhat the fragility, slightly above 50 in our model and about 80 experimentally.
This is not surprising considering that our model is phenomenological, and the extrapolation we made to compare different dynamical ranges.

\SkipTocEntry
\section{Theory in appearance of weak springs}
\label{app_A5}

In the case where $\alpha\neq0$, the annealed free energy can be easily calculated under the assumption that  $|\delta r_p\rangle$ and $|\delta r_{\omega}\rangle$ are random Gaussian vectors. The Hamiltonian in Eq.(\ref{2_11}) can be rewritten as:
$$ {\cal H}=\frac{1}{2}\sum_{p=1...\delta zN/2} X_p^2+\frac{1}{2}\sum_{\omega>0} \frac{\alpha}{\alpha+\omega^2}X_\omega^2,$$
where  $X_p=\langle\delta r_{p}|\sigma\rangle$ and $X_{\omega}=\langle\delta r_{\omega}|\sigma\rangle$ represent  independent  random variables for each configuration $|\sigma\rangle$.  In the thermodynamic limit the random variables $X_p$ and $X_\omega$ are Gaussian distributed  with zero mean and unit variance. The averaged partition function is  given by: 

\be
\label{app_a1}
\begin{split}
\overline{\mathcal{Z}}&=2^{N_s}\int e^{-{\cal H}/T}\,\prod_{p}\frac{e^{-\frac{X_{p}^2}{2}}}{\sqrt{2\pi}}\mathrm{d}X_{p}\prod_{\omega}\frac{e^{-\frac{X_{\omega}^2}{2}}}{\sqrt{2\pi}}\mathrm{d}X_{\omega}\\
&=2^{N_s}\prod_{p=1}^{\delta zN/2}\left(1+\frac{1}{T}\right)^{-1/2}\prod_{\omega>0}\left(1+\frac{\alpha/T}{\alpha+\omega^2}\right)^{-1/2}.
\end{split}
\nonumber%\tag{SI-1}
\ee

\begin{figure}[h!]
%\vspace{1cm}
  \label{app_BB}
   \begin{center}
   {\includegraphics[width=0.8\columnwidth]{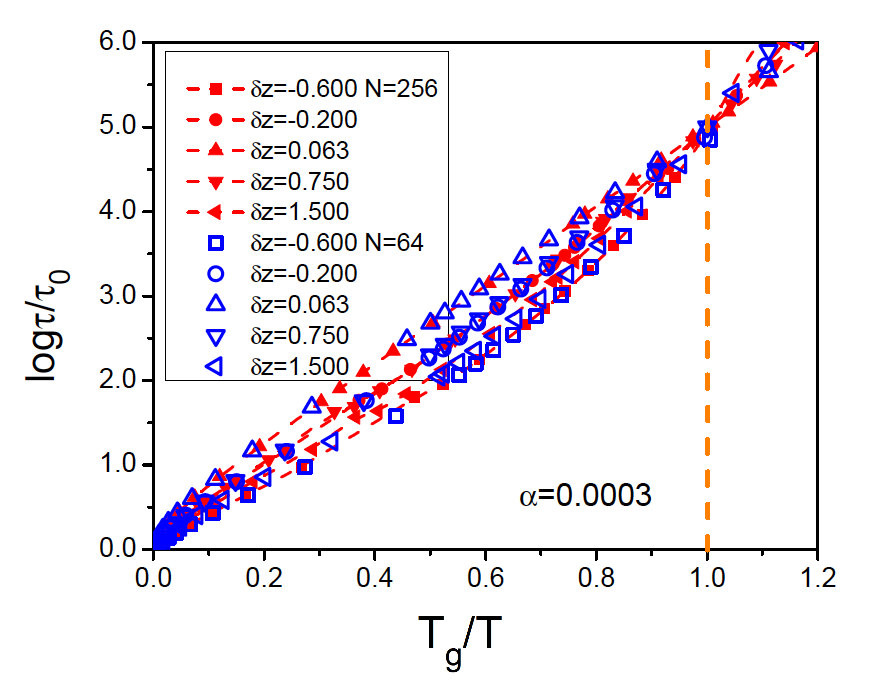}}
   \caption{Angell plot representing $\log \tau$ {\it v.s.} inverse temperature $T_g/T$ for different $\delta z$ and two system sizes $N=64$ and $N=256$, $\alpha=0.0003$.}
      \end{center}
%\tag{SI-1}
\end{figure}

From the average partition function the density of free energy per spring $f(T)$  and any other thermodynamic quantities are readily computed. In particular,  the energy density $ \varepsilon(T)$, the specific heat $c(T)$ and the entropy density $s(T)$ write:
 
\begin{equation}
\label{app_a2}
f(T)=\frac{T}{2N_s}\left[\sum_{p=1}^{\delta zN/2}\ln\left(1+\frac{1}{T}\right)+\sum_{\omega>0}\ln\left(1+\frac{\alpha/T}{\alpha+\omega^2}\right)\right]-T\ln2
\nonumber %\tag{SI-2}
\end{equation}
\begin{equation}
\label{app_a3}
\varepsilon(T)=\frac{1}{2N_s}\left[\sum_{p=1}^{\delta zN/2}\frac{T}{1+T}+\sum_{\omega>0}\frac{\alpha T}{\alpha+(\omega^2+\alpha)T}\right]
\nonumber%\tag{SI-3}
\end{equation}
\begin{equation}
\label{app_a4}
c(T)=\frac{1}{2N_s}\left[\sum_{p=1}^{\delta zN/2}\frac{1}{\left(1+T\right)^2}+\sum_{\omega>0}\left(\frac{\alpha}{\alpha+(\omega^2+\alpha)T}\right)^2\right]
%\tag{SI-2}
\end{equation}
\begin{equation}
\label{app_a5}
\begin{split}
s(T)=&\ln2-\frac{1}{2N_s}\sum_{p=1}^{\delta zN/2}\left[\ln(1+\frac{1}{T})-\frac{1}{1+T}\right]\\
&-\frac{1}{2N_s}\sum_{\omega>0}\left[\ln\left(1+\frac{\alpha/T}{\alpha+\omega^2}\right)-\frac{\alpha}{\alpha+(\omega^2+\alpha)T}\right].
\end{split}
\nonumber%\tag{SI-5}
\end{equation}

In the limit $\alpha\to0$, for any finite temperature $T$, the  sum over the  vibration modes ($\omega>0$) vanishes, and we recover the expressions in the  absence of weak springs  for pure rigid networks. Note that  Eq.(\ref{app_a4}) corresponds to Eq.(\ref{2_12}) in the article.

\SkipTocEntry
\section{Continuous density of states limit: Analytical results}
\label{app_A6}

In the thermodynamic limit $N\to\infty$, we can replace the sum over frequencies by an integral: $\sum_{\omega>0}\to N_s\int\mathrm{d}\omega D(\omega)$ for $\delta z\leq0$, and $\sum_{\omega>0}\to Nd\int\mathrm{d}\omega D(\omega)$ for $\delta z>0$. The density of states $D(\omega)$ is the distribution of vibrational modes of random elastic networks,  which    has been  computed theoretically~\cite{Wyart05,During13,Wyart10a}.
There are two frequency scales in the random network : $\omega^*\sim|\delta z|$  above which a plateau of soft modes exist,  and a cut-off  frequency $\omega_c\sim1$. Below $\omega^*$, rigid networks show plane wave modes \cite{Wyart08,Wyart05,Wyart10a} with a characteristic Debye regime $D(\omega)\sim \omega^2$, unlike  floppy networks, which show no modes in this gap \cite{During13}. 

It turns out that the Debye  regime contribution to the integrals is negligible near the jamming threshold.
To capture the scaling behavior near jamming, we approximate  $D(\omega)$ by a square function. This simplified  description allows  further analytical progress while preserving the same qualitative behavior. Since the Debye regime can be neglected,  we choose:
\[
D(\omega)=\left\{
\begin{array}{l l}
\frac{1}{\omega_c-\omega^*} & \quad \omega^*\leq\omega\leq\omega_c\quad \delta z\leq0\\
\frac{1}{\omega_c} & \quad 0\leq\omega\leq\omega_c\quad \delta z>0.\\
\end{array} \right.
\]
Considering $\omega^*=\frac{\vert\delta z\vert}{z_c}\omega_c$, the cut-off frequency $\omega_c\sim1$ is  the only fitting parameter of the simplified continuum model. Rescaling as $\alpha\rightarrow\alpha \omega^2_c$, we obtain that all the thermodynamic functions depend uniquely on $\alpha=\frac{z_{\rm w}k_{\rm w}}{d k \omega_c^2}$,  $T$ and $\delta z$. In particular, the specific heat is:
 
%Substituting the approximating DOS, and do integrations.
\begin{equation}
c(T,\delta z,\alpha)=\left\{
\begin{array}{l l}
\frac{z_c}{4z}\frac{\sqrt{\alpha(1+1/T)}}{(1+T)^2}\left[\arctan(\frac{1}{\sqrt{\alpha(1+1/T)}})\right. &\\
\quad+\frac{\sqrt{\alpha(1+1/T)}}{1+\alpha(1+1/T)}+\arctan(\frac{\delta z/z_c}{\sqrt{\alpha(1+1/T)}})\\
\quad+\left.\frac{\delta z}{z_c}\frac{\sqrt{\alpha(1+1/T)}}{\delta z^2/z_c^2+\alpha(1+1/T)}\right] & \delta z\leq0\\
\\
\frac{z_c}{4z}\frac{\sqrt{\alpha(1+1/T)}}{(1+T)^2}\left[\arctan(\frac{1}{\sqrt{\alpha(1+1/T)}})\right. &\\
\quad+\left.\frac{\sqrt{\alpha(1+1/T)}}{1+\alpha(1+1/T)}\right]+\frac{\delta z}{2z}\frac{1}{(1+T)^2} & \delta z>0.\\
\end{array} \right.
%\tag{SI-6}
\nonumber
\end{equation}

We compute the jump of specific heat at the Kautzman temperature, where the entropy vanishes $s(T_K,\delta z,\alpha)=0$.  In the continuous limit, the equations for $T_K$ can be approximated by:

\[
\ln2\approx \frac{z_c}{2z}\left[\ln(1+\frac{\alpha}{T_K})+2\sqrt{\frac{\alpha}{T_K}}\arctan\sqrt{\frac{T_K}{\alpha}}\right] -\frac{\delta z}{2z}\theta(\delta z)\ln{T_K},
\]
where  the conditions  $\delta z\ll z_c$ and $\alpha\sim k_{\rm w}/k\ll1$ have been used. There is no simple analytical expression for $T_K$, however,  one can  observe the existence of two asymptotic regimes: $T_K\sim\alpha$ for $\delta z\ll \vert1/\ln{\alpha}\vert$ and $T_K\sim2^{-2z/\delta z}$ for $\delta z\gg \vert1/\ln{\alpha}\vert$. Then the specific heat at the transition temperature is given by:

\[
c(T_K,\delta z,\alpha)\sim\left\{
\begin{array}{l l}
\frac{z_c}{4z}\frac{\pi}{2} & \delta z\ll \vert1/\ln{\alpha}\vert\\
\\
\frac{\delta z}{2z} & \delta z\gg \vert1/\ln{\alpha}\vert .\\ 
\end{array} \right.
\]
From these asymptotic  behaviors one gets that the  specific heat display a non-monotonous behavior with coordination, with a minimum whose position scales as $\delta z\sim \vert1/\ln{\alpha}\vert$.

%% file: appendices/appd_rigidity.tex
\SkipTocEntry
\chapter{\textcolor{black}{Evolution of covalent networks under cooling}}%Going from rigidity percolation to jamming under cooling}
\appcaption{\thechapter \space Evolution of covalent networks under cooling} %this needs to be here to show up in the ''List of appendices'' which is required if you have more than one appendix
\label{app_rigidity}
%In this appendix we explain a lot about elephants
%\keywords{rigidity percolation, jamming, vibration modes, weak constraints}

\SkipTocEntry
\section{Periodic distortion of triangular lattice.}
\label{app_B1}
\begin{figure}[h!]
   {\def\svgwidth{0.48\columnwidth}
   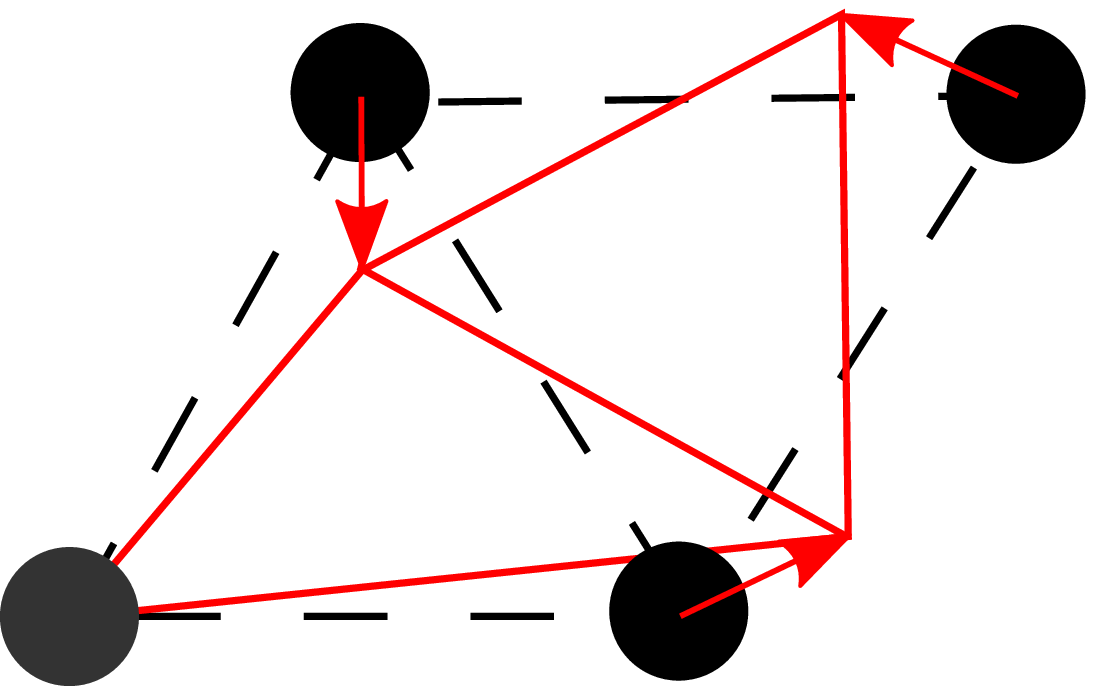}\hfill
   {\includegraphics[width=0.48\columnwidth]{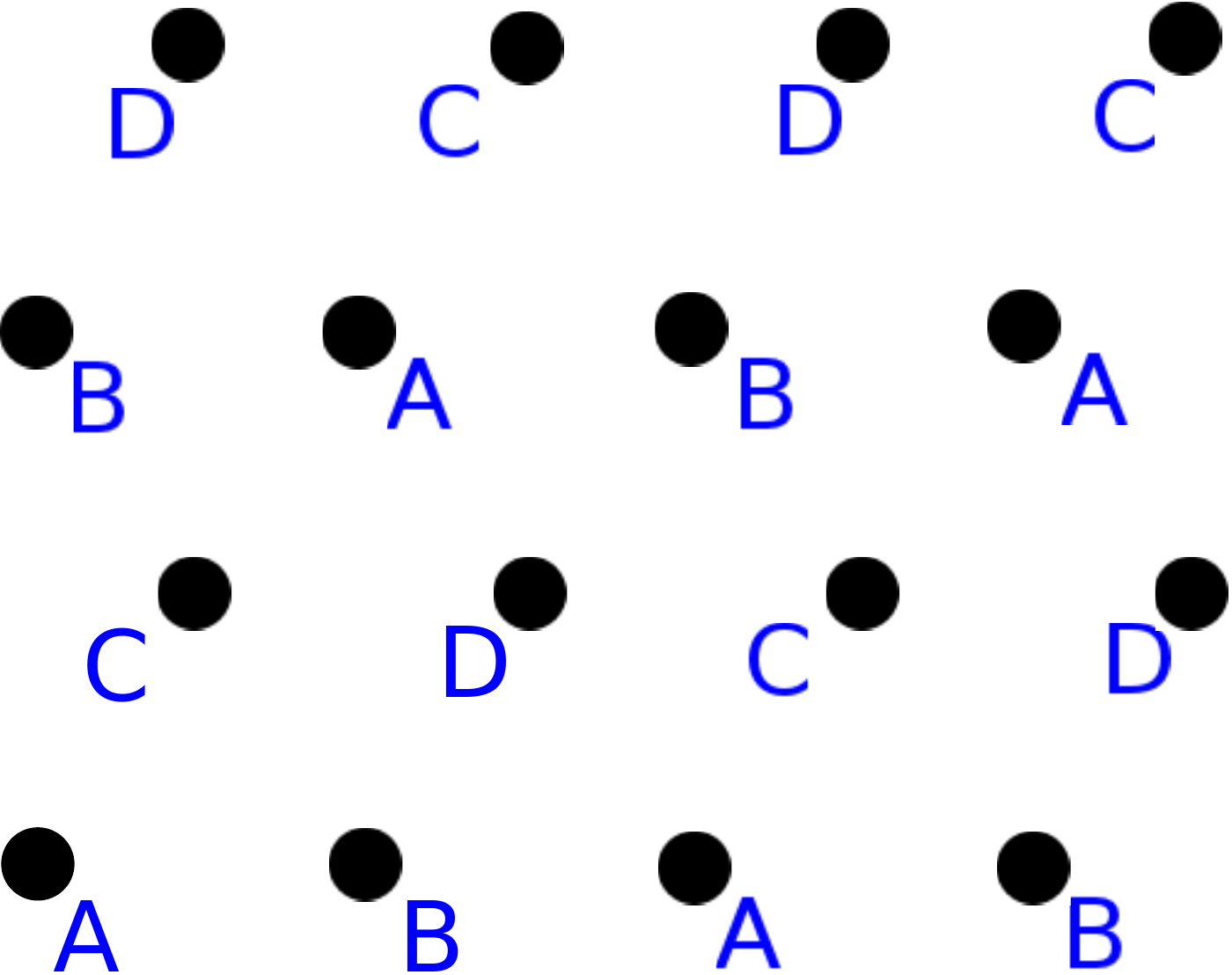}}\\
 \caption{Illustration of distortion of the triangular lattice, performed to remove straight lines.}
 \label{app_crystalline}
\end{figure}

In our model, we introduce a slight distortion of the lattice to remove the straight lines that occur in a triangular lattice, in the  spirit of \cite{Jacobs95}. Such straight lines would lead to unphysical localized floppy modes orthogonal to the lines.  In ~\cite{Jacobs95} random disorder is introduced to achieve this goal. 
Instead, we seek to distort the lines while avoiding frozen disorder (the only disorder we use corresponds to the polydispersity of the spring rest length, but it does not break translational symmetry  because springs can move). We group nodes by four, labeled as A B C D in Fig.~\ref{app_crystalline}. \textcolor{black}{One group forms a cell }%These groups form cell 
of our crystalline lattice. Each cell is distorted identically as follows: node A stays in place,  while nodes B, C, and D move by some distance $\delta$: B along the direction perpendicular to BC, C along the direction perpendicular to CD, and D along the direction perpendicular to DB, as illustrated in the figure.  $\delta$ is set to  $0.2$. %The new crystalline is assembled through aligning the cells along the original AB and AC directions.

\SkipTocEntry
\section{Numerical computation of the elastic energy.}
\label{app_B2}
The energy $H(\Gamma)$ of a given spring configuration $\Gamma\equiv\{\gamma\leftrightarrow\langle i,j\rangle\}$ is defined in Eq.(\ref{3_e1}) in {\it Chapter 3} as a minimization on the position\textcolor{black}{s} of the nodes. This minimum can be calculated using conjugate gradient methods. However for small mismatches $\epsilon$, it is more efficient to use  linear algebra  ~\cite{Yan13}, as we now recall.
Consider a displacement field $\delta\vec{R}_i\equiv\vec{R}_i-\vec{R}_{i0}$, where $\vec{R}_{i0}$ is the position of the node $i$ in the crystal described in the previous section. We define the distance $||\vec{R}_{i0}-\vec{R}_{j0}||\equiv r_{\langle i,j\rangle}$. At first order in $\delta\vec{R}_i$, the distance among  neighboring nodes can be written as:
\be
||\vec{R}_i-\vec{R}_j||=r_{\langle i,j\rangle}+\sum_{k}\ms_{\langle i,j\rangle,k}\delta\vec{R}_k+o(\delta\vec{R}^2)
\label{app_distc}
\ee
Where $\ms$ is the structure matrix, which gives the linear relation between displacements and changes of  distances, as indicated in Eq.(\ref{app_distc}). Minimizing Eq.(\ref{3_e1}) in {\it Chapter 3} leads to \cite{Yan13}:
\be
H(\Gamma)=\frac{k}{2}\sum_{\gamma,\rho}\epsilon_{\gamma}\mg_{\gamma,\rho}\epsilon_{\rho}+o(\epsilon^3)
\label{app_Hamiltonian1}
\ee
where $\mg=\ms(\ms^t\ms+\alpha\mi)^{-1}\ms^t$, and $\bullet^t$ is our notation for the transpose of a matrix.
In practice, we solve Eq.(\ref{app_Hamiltonian1}) for every configuration $\Gamma$ our Monte Carlo considers. One issue with Eq.(\ref{app_Hamiltonian1}) is that the inverse in the expression for $\mg$ is ill-defined when $\alpha=0$  if floppy modes are present in the network. To study the  case $\alpha=0$, we implement the Pebble Game algorithm~\cite{Jacobs95,Jacobs97} to distinguish stressed, hyperstatic clusters from floppy or isostatic regions.  Since only the stressed regions can contribute to the energy, we  reduce the matrix $\ms$ to this associated subspace, and solve Eq.(\ref{app_Hamiltonian1}) in this subspace.  We have compared this method and a direct minimization via conjugate gradients; the two results coincide within $1\%$ as long as $\epsilon\lesssim0.01$. In {\it Chapter 3}, our results are based on Eq.(\ref{app_Hamiltonian1}), and thus hold as long as $\epsilon$ is small enough. In this case the choice of $\epsilon$ only affects the energy scale.

\begin{figure}[ht!]
\includegraphics[width=1.0\textwidth]{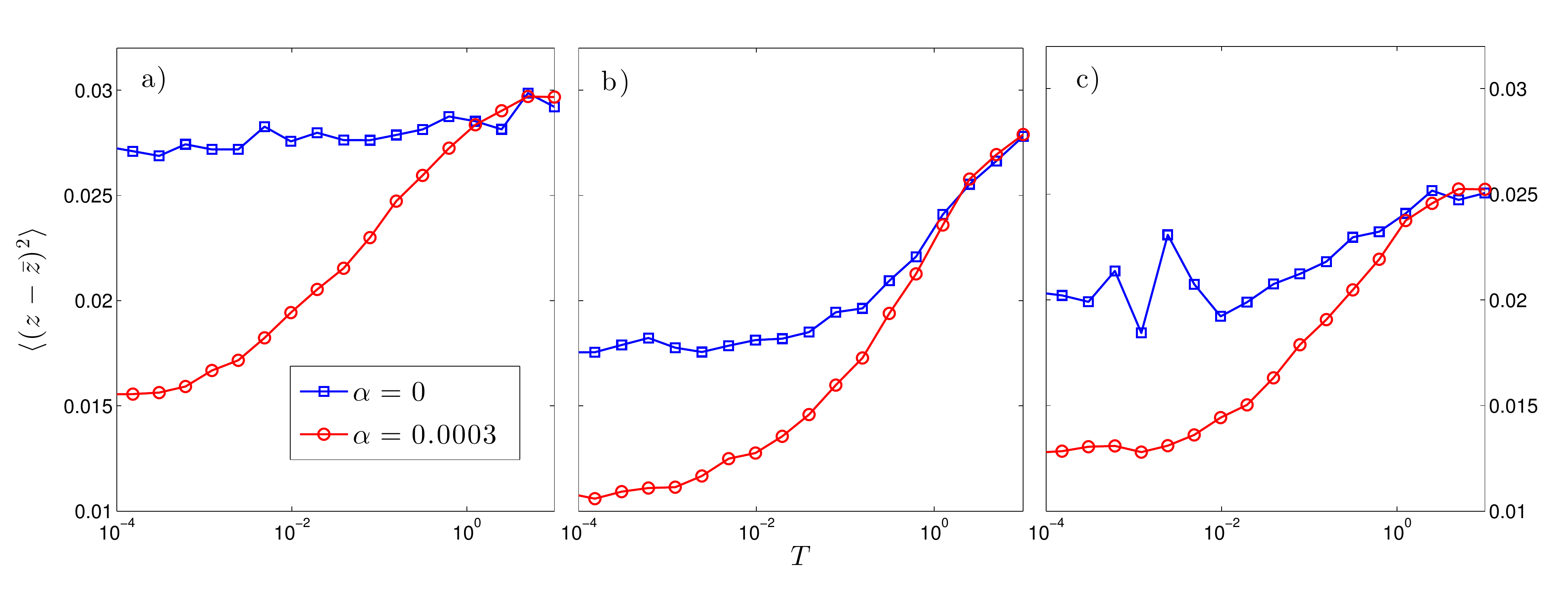}
\caption{\small{Fluctuations of coordination $\langle(z-\bar{z})^2\rangle$ {\it vs} temperature $T$ for different  $\alpha$ as indicated in legend. The network size is $N=256$ and the block size is $N^*=64$. Mean coordination number corresponds to a) $\bar{z}-z_c=-0.383$, b) $\bar{z}-z_c=-0.055$, c) $\bar{z}-z_c=0.523$.}}\label{app_zfT}
\end{figure}

\SkipTocEntry
\section{Removal of fluctuations under cooling: numerical evidence}
\label{app_B3}
%In the Letter, we argue that the weak constraints homogenize the network, so that the coordination number of a sub-network is getting closer to the mean value $\bar{z}$ under cooling when $\alpha>0$. In this supplementary section, we show some data on fluctuation of the coordination number of sub-networks to numerically support our argument in the Letter. 
The mean coordination number of the whole network is fixed in our model; in this section we denote it as $\bar{z}$. To characterize spatial fluctuations of coordination, we divide the network into four identical blocks of size $N^*=N/4$ sites. We then measure the coordination number $z$ in each  bloc\textcolor{black}{k}, and in many configurations equilibrated at some temperature $T$. We then compute the variance $\langle(z-\bar{z})^2\rangle$, where the average is over all blocks and configurations. Fig.~\ref{app_zfT} shows this quantity versus temperature for three choices of excess coordination $\delta z=\bar{z}-z_c$, corresponding to  a) below c) above  and b) near the rigidity transition. For all these choices we find that the amplitude of fluctuations does not vary for low temperatures when  $\alpha=0$. By contrast when $\alpha>0$, fluctuations of coordination are smaller  at low temperature, where they continue to decay under cooling.

\SkipTocEntry
\section{Effect  of frozen-in fluctuations of coordination at $T_g$}
\label{app_B4}

\begin{figure}[ht!]
\centering
\includegraphics[width=.8\columnwidth]{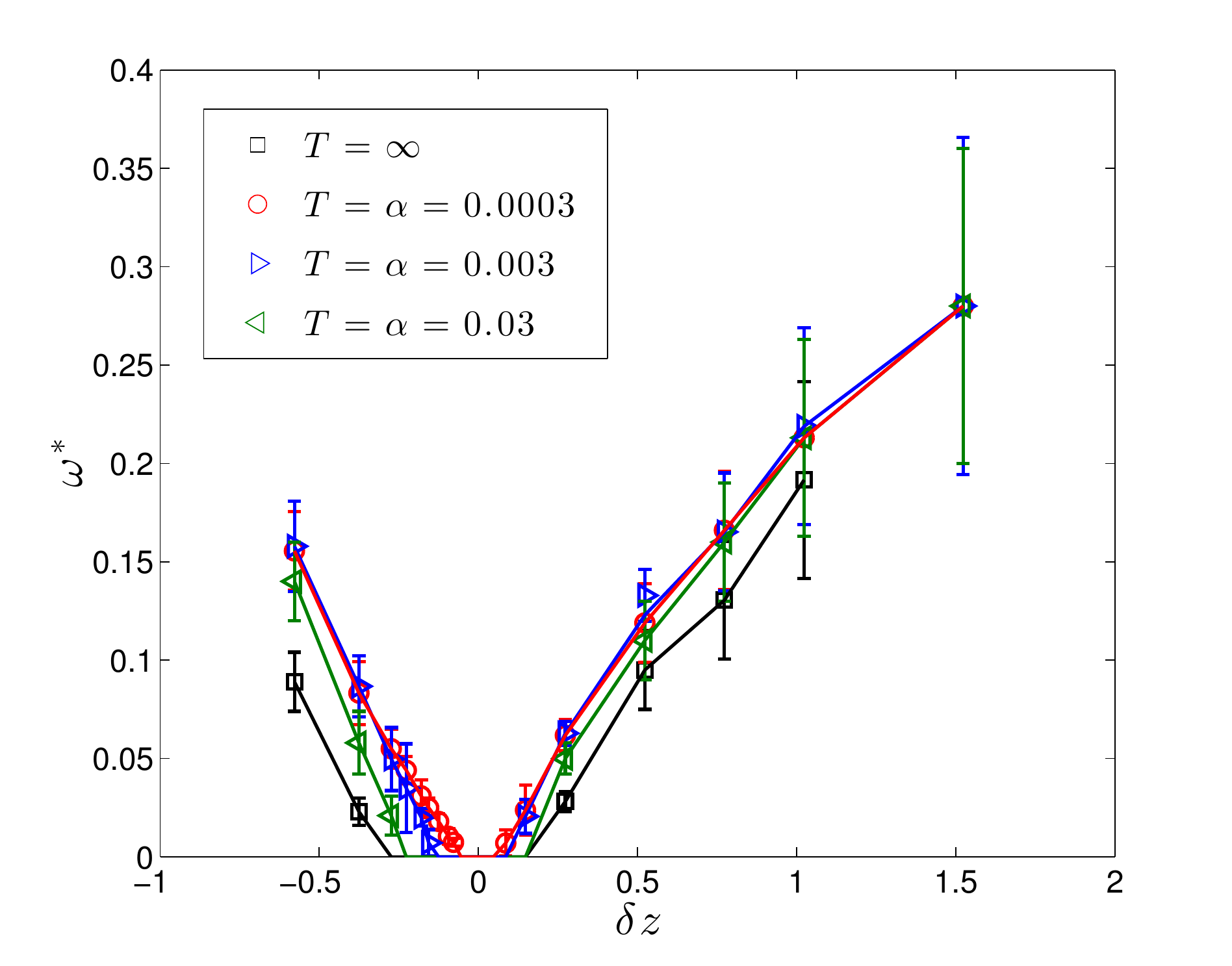}
\caption{\small{Boson peak frequency $\omega^*$ as  a function of excess coordination $\delta z=z-z_c$ for different $\alpha$  as indicated in legend, at temperatures $T=\alpha$. $\omega^*=0$ indicates that no maximum was observed in $D(\omega)/\omega^{d-1}$, consistent with the presence of fractons at very low frequency.}}\label{app_flu}
\end{figure}

To study the role of frozen-in spatial fluctuations of coordination, we increase $T_g$ in our model, which can be achieved  by increasing the strength of weak interactions $\alpha$. 
We can equilibrate our system up to temperatures of order $T= \alpha$, and in what follows we fix these two parameters to be equal. We then study the vibrational properties of the network of strong springs by computing the boson peak frequency $\omega^*$, defined as in {\it Chapter 3} as the maximum of $D(\omega)/\omega^{d-1}$. If we observe no maximum in this quantity we posit that $\omega^*=0$. Results are shown in Fig.~\ref{app_flu}. The key point is that as $T_g$ increases, a broader region appears in the vicinity of $z_c$ where mean-field predictions do not apply. Instead one finds that for a range of coordination, $\omega^*\approx 0$, consistent with the presence of fractons.  

%% file: appendices/rotate.eps_tex
%% Creator: Inkscape inkscape 0.48.2, www.inkscape.org
%% Accompanies image file 'rotate.eps' (pdf, eps, ps)
\begingroup%
  \makeatletter%
  \providecommand\color[2][]{%
    \errmessage{(Inkscape) Color is used for the text in Inkscape, but the package 'color.sty' is not loaded}%
    \renewcommand\color[2][]{}%
  }%
  \providecommand\transparent[1]{%
    \errmessage{(Inkscape) Transparency is used (non-zero) for the text in Inkscape, but the package 'transparent.sty' is not loaded}%
    \renewcommand\transparent[1]{}%
  }%
  \providecommand\rotatebox[2]{#2}%
  \ifx\svgwidth\undefined%
    \setlength{\unitlength}{344.2062563bp}%
    \ifx\svgscale\undefined%
      \relax%
    \else%
      \setlength{\unitlength}{\unitlength * \real{\svgscale}}%
    \fi%
  \else%
    \setlength{\unitlength}{\svgwidth}%
  \fi%
  \global\let\svgwidth\undefined%
  \global\let\svgscale\undefined%
  \makeatother%
  \begin{picture}(1,0.56664617)%
    \put(0,0){\includegraphics[width=\unitlength]{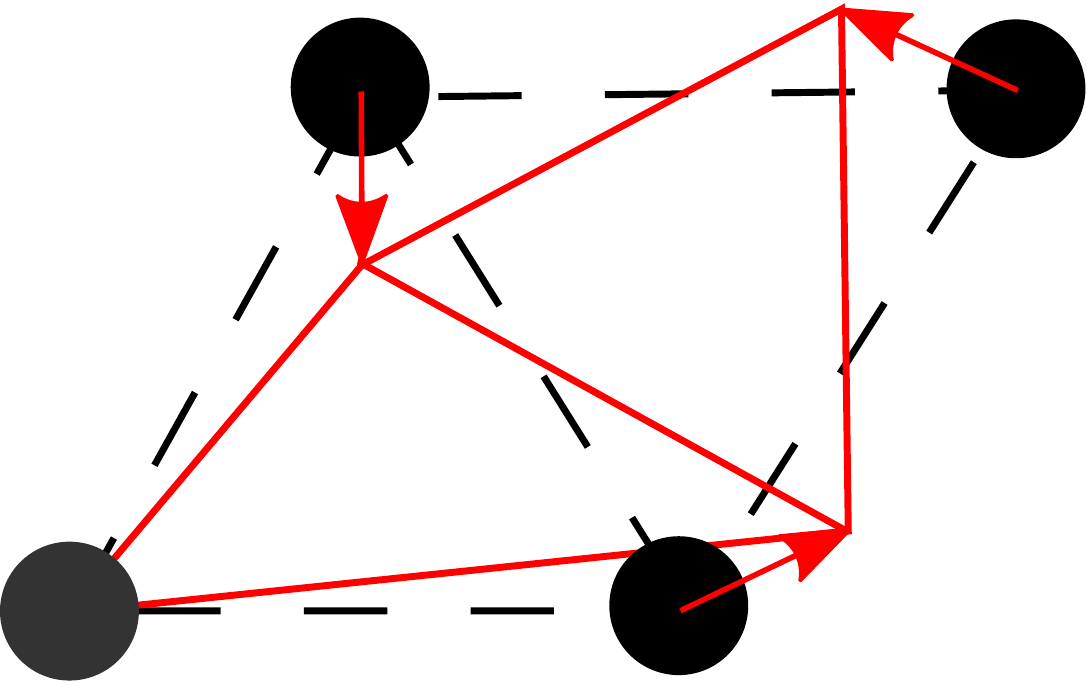}}%
    \put(0.80245566,0.10222848){\color{cyan}\makebox(0,0)[lb]{\smash{$\delta$}}}%
    \put(0.12033001,0.00212225){\color{blue}\makebox(0,0)[lb]{\Large\smash{A}}}%
    \put(0.70240303,0.002123246){\color{blue}\makebox(0,0)[lb]{\Large\smash{B}}}%
    \put(0.1841454,0.5144811){\color{blue}\makebox(0,0)[lb]{\Large\smash{C}}}%
    \put(0.78919474,0.51164523){\color{blue}\makebox(0,0)[lb]{\Large\smash{D}}}%
  \end{picture}%
\endgroup%

%% file: appendices/appd_adaptive.tex
\SkipTocEntry %these things have to be there for the appendices - or their sections not to show up in the table of contents (which according to NYU is baaad)
\chapter{Adaptive elastic networks as a model for supercooled liquids}
%\date{}                                           % Activate to display a given date or no date
\appcaption{\thechapter \space Adaptive elastic networks as a model for supercooled liquids} %this needs to be here to show up in the ''List of appendices'' which is required if you have more than one appendix
\label{app_adaptive}
%In this appendix we explain a lot about elephants

\SkipTocEntry %\tocless
\section{Formalism of elastic energy}
\label{app_C1}
The energy $H(\Gamma)$ of a given spring configuration $\Gamma\equiv\{\gamma\leftrightarrow\langle i,j\rangle\}$ is defined in Eq.(\ref{4_e1}) in {\it Chapter 4} as a minimization on the position\textcolor{black}{s} of the nodes. This minimum can be calculated using conjugate gradient methods. However for small mismatches $\epsilon$, it is more efficient to use  linear algebra~\cite{Yan13}, as we now recall.
Consider a displacement field $\delta\vec{R}_i\equiv\vec{R}_i-\vec{R}_{i0}$, where $\vec{R}_{i0}$ is the position of the node $i$ in the crystal described in the previous section. We define the distance $||\vec{R}_{i0}-\vec{R}_{j0}||\equiv r_{\langle i,j\rangle}$. At first order in $\delta\vec{R}_i$, the distance among  neighboring nodes can be written as:
\be
||\vec{R}_i-\vec{R}_j||=r_{\langle i,j\rangle}+\sum_{k}\ms_{\langle i,j\rangle,k}\delta\vec{R}_k+o(\delta\vec{R}^2)
\label{app_dist}
\ee
Where $\ms$ is the structure matrix, which gives the linear relation between displacements and changes of  distances, as indicated in Eq.(\ref{app_dist}). Minimizing Eq.(\ref{4_e1}) in {\it Chapter 4}, one gets:
\begin{multline}
H(\Gamma)=\min_{\{\delta\vec{R}_i\}}\left\{\frac{k}{2}\sum_{\gamma}(\sum_i\ms_{\gamma,i}\delta\vec{R}_i+\epsilon_{\gamma})^2
+\frac{k}{2}\sum_{\sigma} \frac{k_{\rm w}}{k}(\sum_{i}\ms_{\rm w{\ }\sigma,i}\delta\vec{R}_{i})^2+o(\delta\vec{R}^3)\right\}\\
=\min_{\{\delta\vec{R}_i\}}\frac{k}{2}\left[\langle\epsilon|\mpp|\epsilon\rangle+2\langle\epsilon|\ms|\delta\vec{R}\rangle+\langle\delta\vec{R}|\mm|\delta\vec{R}\rangle\right]
\label{app_linearexp}
\end{multline}
where we use bra-ket notations to indicate summation over edges or nodes, $\mpp$ projects the edge space to the subspace occupied by springs, $\mm\equiv\ms^t\ms+\frac{k_{\rm w}}{k}\ms_{\rm w}^t\ms_{\rm w}$ is the stiff matrix connecting the responding forces and displacements of nodes in an elastic network~\cite{Calladine78}, and $\bullet^t$ is our notation for the transpose of a matrix.
% One can show in the limit of connectivity of weak spring network $z_{\rm w}\to\infty$, $\ms^t_{\rm w}\ms_{\rm w}\to\frac{z_{\rm w}}{d}\mi$, all non-diagonal elements converges to zero as $1/z_{\rm w}$. 
Solving Eq.(\ref{app_linearexp}), one finds the linear response,
\be
|\delta\vec{R}\rangle=-\mm^{-1}\ms^t|\epsilon\rangle
\label{app_response}
\ee
which for a given mismatch field $|\epsilon\rangle$ minimizes the elastic energy in Eq.(\ref{4_e1}). Inserting Eq.(\ref{app_response}) back into the linear approximation Eq.(\ref{app_linearexp}), we have~\cite{Yan13}:
 %leads to \cite{Yan13}:
\be%gin{multline}
H(\Gamma)=\frac{k}{2}\langle\epsilon|\mpp-\ms\mm^{-1}\ms^t|\epsilon\rangle
=\frac{k}{2}\sum_{\Gamma}\epsilon_{\langle i,j\rangle}\mg_{\langle i,j\rangle,\langle l,m\rangle}\epsilon_{\langle l,m\rangle}
\label{app_Hamiltonian}
\ee%nd{multline}
with $\mg=\mpp-\ms(\ms^t\ms+\frac{k_{\rw}}{k}\ms_{\rw}^t\ms_{\rw})^{-1}\ms^t$, and $\epsilon_{\langle i,j\rangle}=\epsilon_{\gamma}$ for $\Gamma=\{\gamma\leftrightarrow\langle i,j\rangle\}$.

\SkipTocEntry %\tocless
\section{Density of states}
\label{app_C2}
\begin{figure}[h!]
\centering
\includegraphics[width=.9\columnwidth]{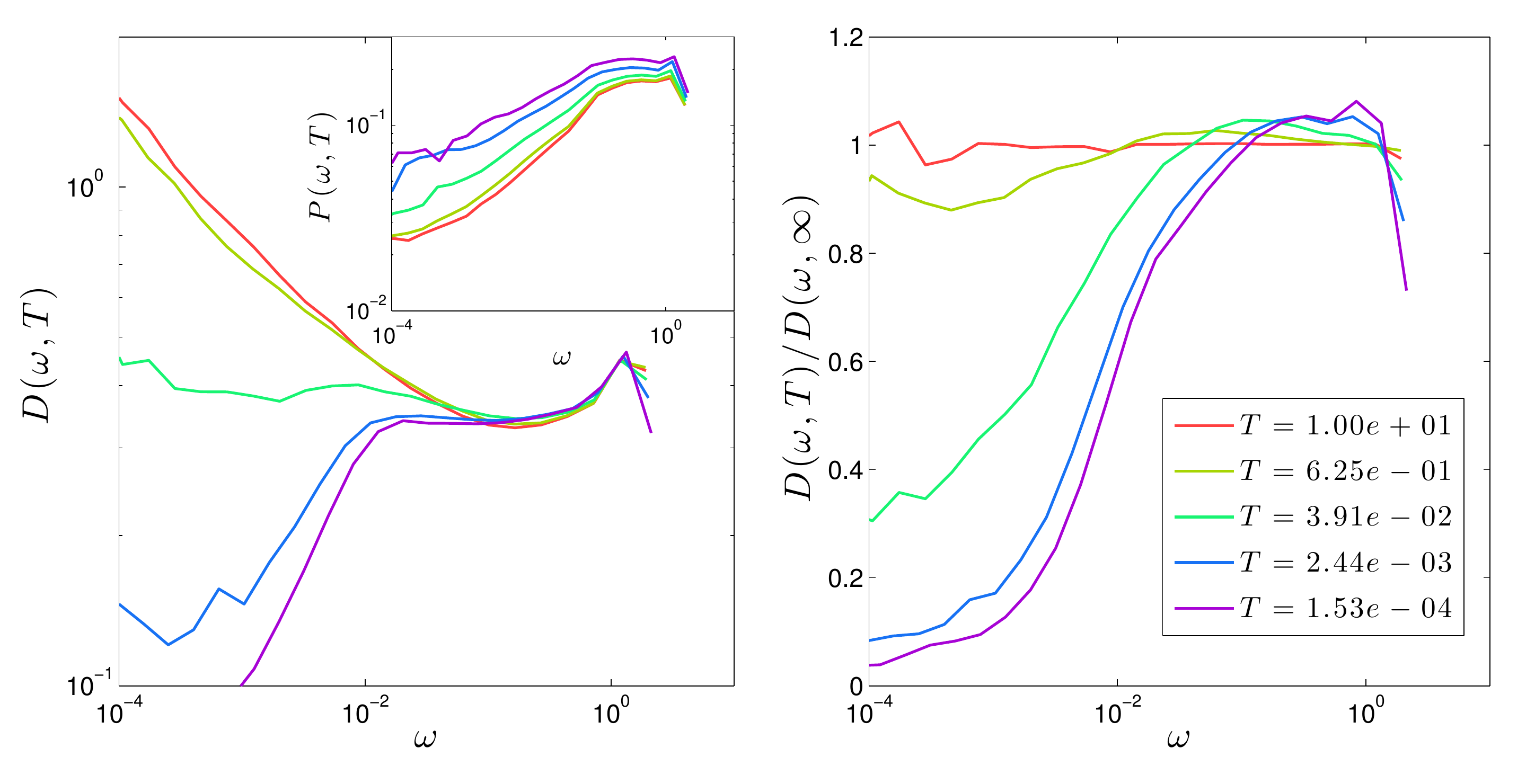}
\caption{\small{Variation of density of states $D(\omega,T)$ with temperature for the same $z=-0.055$, $\alpha=0.0003$. Left: density of states in log-log scale. Right: density of states normalized by its $T=\infty$ value, emphasizing its difference under cooling. Inset: participation ratio $P(\omega,T)$ variation under cooling. %Solid lines have the same color code in all plots for different temperatures.% Grey dashed line is the density of states of random network generated from random packing of soft particles~\cite{WyartPRL2008}.
}}\label{app_doscp}
\end{figure}

\begin{figure}[ht!]
\centering
\includegraphics[width=1.0\columnwidth]{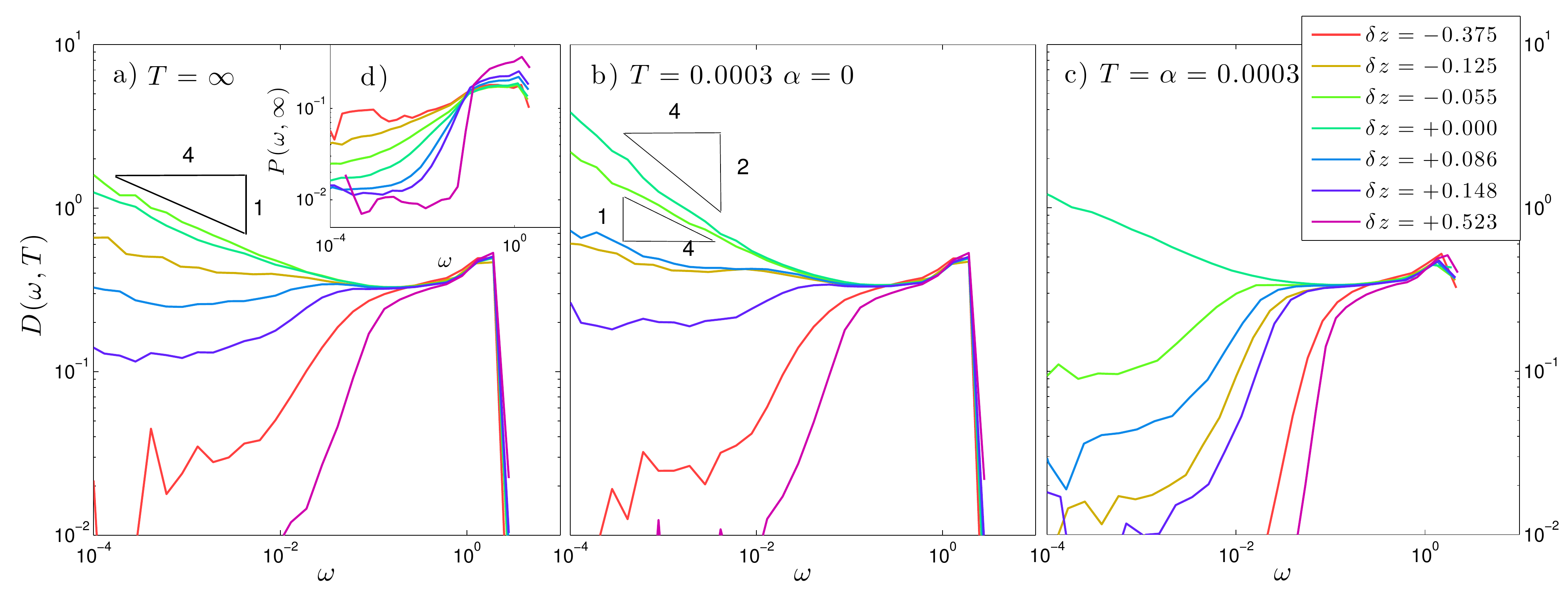}
\caption{\small{Density of states $D(\omega,T)$ for adaptive networks with different $z$. (a) Random diluted networks $T=\infty$; a power law $D(\omega)\sim\omega^{-0.25}$ is shown in low frequency range for networks near $z_{cen}$. (b) Adaptive networks without weak constraints ($\alpha=0$) at $T=0.0003$; power laws with different exponents are shown for networks in the rigidity window: $D(\omega)\sim\omega^{-0.25}$ for $\delta z=-0.055$, $D(\omega)\sim\omega^{-0.5}$ for $\delta z=0.0$. (c) Adaptive networks with weak constraints ($\alpha=0.0003$) at $T\approx\alpha$; away from isostatic, density of states are gapped between zero frequency and Boson peak, where $D(\omega)\sim\omega^0$. Inset (d) Participation ratio $P(\omega,T)$ at $T=\infty$, see text for definition.}}\label{app_dos}
\end{figure}

%The density of vibrational modes determines how quantitatively the energy not fully relaxed in these modes due to weak constraints contributes to the total energy and thus the heat capacity. 
We have shown the density of states converges to the one of mean-field networks~\cite{Yan14}. 
Cooling strongly suppresses low frequency vibrational modes, as seen in Fig.~\ref{app_doscp}. %Because the lower frequency modes store higher elastic energy when $\alpha>0$. 
This temperature effect on the density of states is primarily induced by the weak interactions: the density of states changes little under cooling when $\alpha=0$, as appeared in comparing (a) and (b) of Fig.~\ref{app_dos}. The slight change indicates that density of states depends on the presence of redundant constraints. However, when $\alpha>0$, the low temperature density of states strongly differs from its high temperature counterpart, as shown in Fig.~\ref{app_dos}(a) and (c).

The modes that rarefy under cooling are localized vibrations. The participation ratio, $P(\omega)\equiv\frac{1}{Nd}(\sum_i\Psi_{\omega i}^2)^2/\sum_i\Psi_{\omega i}^4$, quantifies the extensity of characteristic modes: $P\to0$ corresponds to a localized mode, while $P\to1$ means that the mode extends over the system. Both the low and high frequency ends of the density of states are reduced under cooling, but the modes in the middle are enhanced, as shown in the right panel of Fig.~\ref{app_doscp}. This agrees with the small participation ratio of modes with low and high frequencies, see Fig.~\ref{app_dos}(d). In fact, all modes become extended -- the participation ratio increases over the whole spectrum -- when the temperature decreases, as shown in the inset of Fig.~\ref{app_doscp}. 

In addition to localization, another prominent feature of reduced low frequency modes is the power law diverging density of states $D(\omega)\sim\omega^{\tilde{d}-1}$, see Fig.~\ref{app_dos}. The abundance of low frequency localized modes appearing with a power law density of states signals the ``fractons'' that appear near the rigidity percolation~\cite{Alexander82,Feng84,Nakayama94}. %, which are predicted in the random percolation model and we find them more generally near rigidity transition except the mean-field network in %More precisely, our model corresponds to the vector elastic potential model first studied by Feng~\cite{Feng85}. 
The exponent of the diverging tail, in Fig.~\ref{app_dos}(a), implies the fracton dimension $\tilde{d}\approx0.75$, which is consistent with $0.78$ observed for the rigidity percolation~\cite{Feng85,Nakayama94}. Different fracton dimensions $\tilde{d}$ are observed for different coordination number in the case of rigidity window shown in Fig.~\ref{app_dos}(b), although more work would be needed to establish this fact empirically. 

We discuss when the temperature affects the mode with frequency $\omega$ in Section C and show illustrations of ``fractons'' in Section D.

%In random networks at $T=\infty$, the large isostatic regions appear near $z_{cen}$, where a robust power law density of states characterizes fractons in Fig.~\ref{app_dos}(a). In the $\alpha=0$ self-organized networks, large isostatic clusters appear in a rigidity window due to the strong spatial fluctuations of the coordination number~\cite{Thorpe00,Chubynsky06,Yan14}, and the whole network is exactly isostatic at $z_c$. The ``fractons'' do not fade with lowering the temperature and the density of states shows power laws in the rigidity window range near $z_c$, see Fig.~\ref{app_dos}(b). When $\alpha>0$, the networks adapt in a homogeneous way under cooling~\cite{Yan14}. But large isostatic structures inevitably appear when $z=z_c$, and we only see abundant low frequency modes at $\delta z=0.0$ in Fig.~\ref{app_dos}(c). 

\SkipTocEntry %\tocless
\section{Adaptation effects on density of states}
\label{app_C3}
%We give a scaling argument in the following to quantify the deviation at high temperature. 
When $\alpha>0$, following Eq.(\ref{4_coupmat}), we see the typical elastic energy corresponding to a mode of frequency $\omega$ scales as ${\alpha}/({\omega^2+\alpha})$, which is proportional to $\alpha$ for $\omega\sim1$, while proportional to $1$ when $\omega\ll\sqrt{\alpha}$. This implies that the elastic energy in the degrees of freedom corresponding to the modes of low frequency is of the same magnitude as the one in the redundant constraints. Similar to the redundant constraints, these low frequency modes are reduced under cooling. 

From Eq.(\ref{4_saddleD}), 
$T^*(\omega,\alpha)\sim{\alpha}/({\omega^2+\alpha})$ gives an estimate on the temperature scale the mode $\omega$ begins to be reduced. The adaptation effect at this temperature scale can be seen in the right panel of Fig.~\ref{app_doscp}. For example, the green line at $T\approx0.04\ll1$ shows a density of states with frequencies $\omega\lesssim\sqrt{\alpha}\approx0.01$ strongly suppressed, while the shape of density of states with $\omega\approx0.1$ and above is almost unchanged. The purple line, $T\approx10^{-4}\sim\alpha$, shows a density of states whose highest frequency $\omega\sim1$ is also significantly reduced. %$\rd D(\omega,T)/\rd T\approx0$ except when $T\lesssim T^*(\omega)$ and $\omega<\omega^*$, the lower cutoff of the anomalous modes. %This is an important temperature scale when we study the temperature dependence of density of states. 

\SkipTocEntry %\tocless
\section{Fractons}
\label{app_C4}
\begin{figure}[h!]
\centering
\includegraphics[width=1.0\columnwidth]{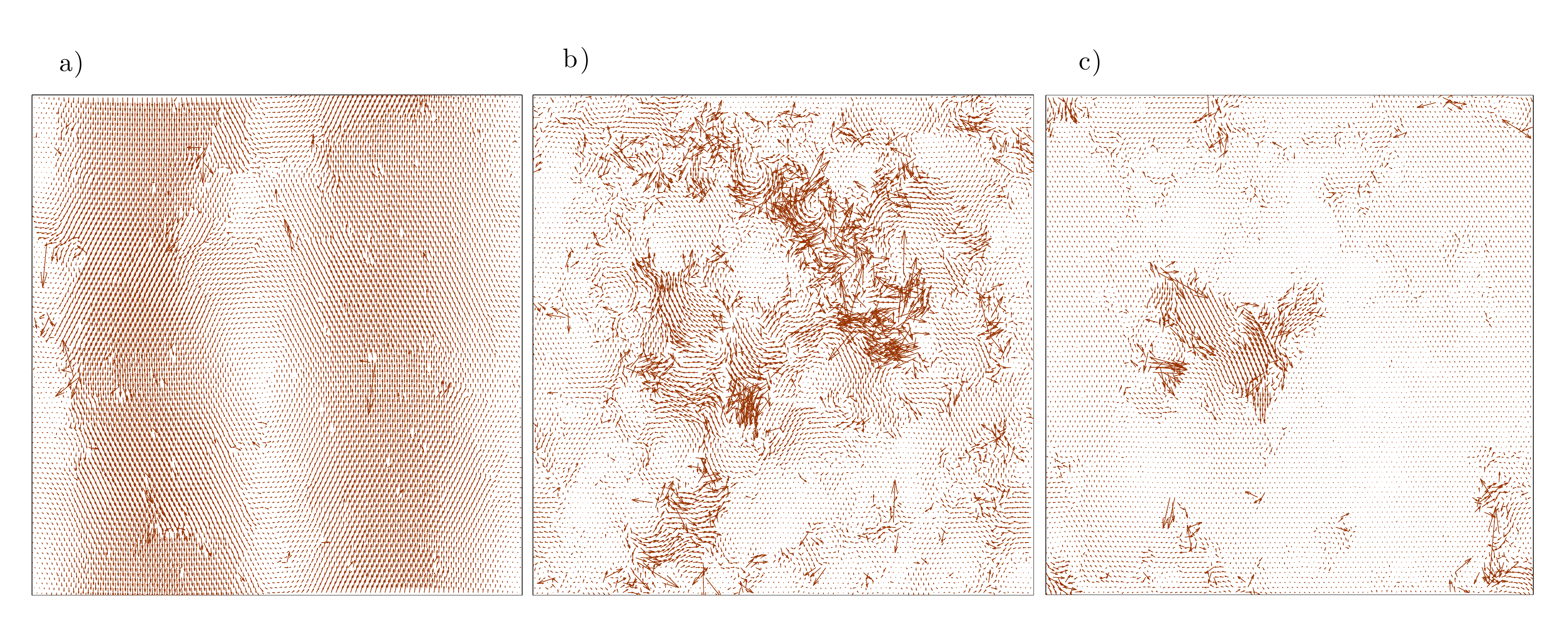}
\caption{\small{Vector plots of vibrational modes in randomly diluted networks, $N=100\times100$. (a) A typical Debye mode, $\delta z=0.501$, $\omega=0.017$. (b) A typical anomalous mode on boson peak, $\delta z=-0.049$, $\omega=0.011$. (c) A typical fracton, $\delta z=-0.049$, $\omega=0.0007$.}}\label{app_mode}
\end{figure}

\begin{figure}[h!]
\centering
\includegraphics[width=.8\columnwidth]{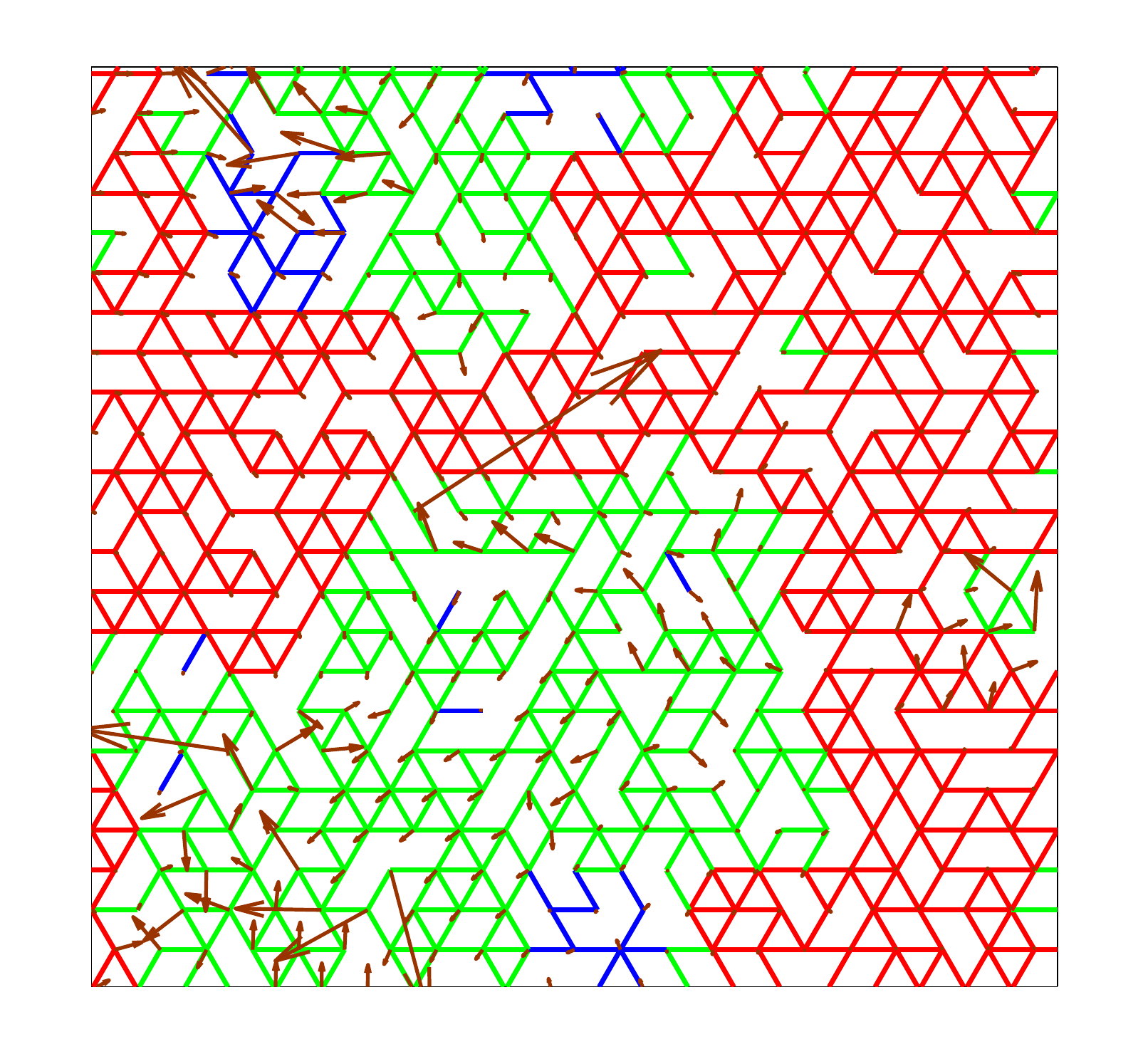}
\caption{\small{Correlation between a low frequency fractal mode and isostatic clusters. A network configuration ($\delta z=-0.042$) is shown with its springs in the over-constrained regions colored in red, in the isostatic regions colored in green, and in the floppy regions colored in blue. A typical fracton ($\omega=5\times10^{-4}$) specified in this configuration is plotted on top.
}}\label{app_coupl}
\end{figure}

% fractons
%We have claimed that the low frequency vibrational modes that are suppressed under cooling are ``fractons''~\cite{Alexander82,Feng84,Nakayama94}. They 
``Fractons'' are different from either the low frequency Debye modes or the anomalous modes on the boson peak, as shown in Fig.~\ref{app_mode}. They (Fig.~\ref{app_mode}(c)) are localized and random compared to the Debye modes (Fig.~\ref{app_mode}(a)), and concentrated on a fractal sets with sharp boundaries, unlike the extended anomalous modes (Fig.~\ref{app_mode}(b)). %All features are in line with the idea of ``fractons'', we thus conclude that the modes largely reduced under cooling when $\alpha>0$ are ``fractons''. 
% what fractons correspond to
The ``fractons'' are associated with the collective motion of large isostatic or nearly isostatic regions as shown in Fig.~\ref{app_coupl}. %Their small frequency corresponds to the energy induced by a small penetration of the displacement field into the stressed regions on the boundary (). %To get a better physical picture of the ``fractons'', here are some hints from density of states of networks 
%For the reason why $\alpha$ makes such a big difference, we strongly recommend the readers with the companion paper on this model~\cite{Yan14}. Here, we simply give an explanation on what fractal structures that ``fractons'' correspond to in the elastic network. 
%In effective medium theory~\cite{Wyart08,During13,DeGiuli14}, no isostatic cluster exists except $z=z_c$, no fracton but only anomalous density of states $D(\omega)\sim\omega^0$ is predicted.
%In summary, the low frequency localized fractons, reflecting large isostatic or nearly isostatic regions, are suppressed under cooling, and only contribute to the specific heat at relatively high temperature ($T\sim\min(1,10^3\alpha)$). 

\SkipTocEntry %\tocless
\section{Vibrational entropy contribution}
\label{app_C5}
The structure the elastic potential evolve with temperature in the liquid phase of the adaptive network model. Freezing into a glass phase eliminates this variability and leads to a contribution to the jump of specific heat~\cite{Wyart10}. Our model currently ignore the vibrational part of the specific heat, which incorporates that the shape of the inherent structure evolve with temperature - not only its bottom energy. We estimate this contribution from vibrations in this subsection, and argue that is is not significant for the models we consider.

The vibrational entropy includes both linear $\omega>0$ and floppy $\omega=0$ vibration modes~\cite{Wyart10}:
\be
s_{vib}(T)=[1-n_r(T)]\int\rd\omega D(\omega,T)\ln\frac{eT}{\hbar\omega}+f(T)\ln\Lambda
\label{app_svib}
\ee
$\Lambda$ sets a cutoff volume for floppy modes, which is approximately the atomic spacing measured in the Lindemann's length: $\Lambda\approx(1/0.15)^d$, of order $10^3$ in 3D~\cite{Lindemann10}. $f$ is the floppy mode density, dual to the number of redundant constraints density $f(T)=-\delta z/z+n_r(T)$ and thus $\partial f(T)/\partial T=\partial n_r(T)/\partial T$. The jump of specific heat follows:
%\begin{widetext}
\begin{multline}
\Delta c_{vib}=\left.T_g\frac{\partial n_r(T)}{\partial T}\right|_{T_g}\left[\ln\Lambda-\int\rd\omega D_{T_g}(\omega)\ln\frac{eT_g}{\hbar\omega}\right]\\+[1-n_r(T_g)]\int\rd\omega\left.T_g\frac{\partial D_T(\omega)}{\partial T}\right|_{T_g}\ln\frac{eT_g}{\hbar\omega}
\label{app_cvib}
\end{multline}
%\end{widetext}
%which gives the contribution from vibrational entropy to the jump of specific heat. %due to freezing of $n_r$ and $D(\omega)$ in glass transition. 
The derivatives on $\ln T$ in Eq.(\ref{app_svib}), continuous at the glass transition, have been subtracted. 

We estimate the upper limit of the vibrational contribution. 
\textcircled{1} The first term in Eq.(\ref{app_cvib}): Debye frequency $\omega_D$ sets the upper limit of the integral in the bracket, $-\ln({eT_g}/{\hbar\omega_D})$. As the glass transition temperature $T_g$ and Debye temperature $\theta_D=\hbar\omega_D/k_B$ are usually of the same order, the bracket in the first term is dominated by $\ln\Lambda$. 
From Eqs.(\ref{4_noweak}), we have ${\partial n_r}/{\partial\ln T}|_{T_g}\approx\frac{1}{2}n_{ex}(T_g)\lesssim\frac{1}{2}n_0\sqrt{T_g}\lesssim0.02\sqrt{\alpha}$, and $\ln\Lambda\approx5$ in 2D. Compared to the specific heat values, which are of order one shown in Fig.~\ref{4_cp}, and the scalings of the minima $-0.1/\ln\alpha$ given in \cite{Yan13}, the contribution, $0.1\sqrt{\alpha}$, is insignificant if $0<\alpha<0.1$. %When $\alpha=0$, we consider rigid networks, $\delta z>0$, with transitions at finite $T_g$. A temperature scale $T_f=(\delta z/zn_0(z)\ln\Lambda)^2$ emerges from balancing the configurational $\delta z/z$ and vibrational $n_0\sqrt{T}\ln\Lambda$ contributions. When $T_g\ll T_f$, the contribution from vibrations is negligible. Based on the shoving model $T_g\propto G\propto\delta z/z$, our simplification works for $\delta z/z\gtrsim(n_0(z)\ln\Lambda)^2\approx0.01$. %In real glasses, this contribution is not important, especially for fragile ones, where $\Delta c$ is several times larger than $1k_B/\text{bead}$.

\textcircled{2} The second term in Eq.(\ref{app_cvib}): The upper limit of the bracket is $1$. %, as $r_0$ sets the lower bound of $n_r(T)$. 
% $\int\rd\omega\frac{\partial D_T(\omega)}{\partial\ln T}\ln\frac{eT}{\hbar\omega}$: 
%When $\alpha=0$, the density of states changes little under cooling and the integral is then neglected. For $\alpha>0$, r
Replacing $\ln({eT}/{\hbar\omega})$ with its upper limit $\ln\Lambda$, we simplify the integral to $\int\rd\omega T{\partial D}/{\partial T}$. %, a same term from differentiating Eq.(13), which contributes little at $T_g$ as discussed in Section C. 
We can estimate the upper limit of the derivative in the integral approximately by ${\Delta n_T}/{\Delta\ln T}$, where $\Delta n_T$ is the number density of the modes reduced under cooling. $\Delta n_T\approx0.2\int_0^{0.01}\omega^{-0.25}\rd\omega\approx0.01$, roughly the number fraction of ``fractons'' suppressed under cooling. Together, the upper limit of the contribution of the second term is $\Delta n_T/\ln10\times\ln\Lambda\approx0.03$, which is moderate compared to the values of order one. % or $1k_B/\text{bead}$ in real glasses.% On one hand, we know the low frequency fractal modes changes a lot, but their number is small. On the other hand, we indeed see extensive number of modes changes, but only for the frequency near Debye frequency, in right panel of Fig.~\ref{app_doscp}. So it's better to estimate on two regions separately: Reduced number of ``fractons'', $n_T\approx0.2\int_0^{0.01}\omega^{-1/4}\rd\omega\approx0.01$, gives an upper limit $n_T\ln\Lambda/\ln10\approx0.03$; Enhanced modes, $n_T\approx0.1\times1=0.1$, end up with an upper limit $n_T\ln1/\ln10\approx0.04$. Again the contribution from the vibrational entropy is moderate compare to the jump value $\sim1k_B/\text{bead}$.

Therefore, the vibrational entropy contributes mildly to the jump of specific heat, and does not change the qualitative behavior of $\Delta c$ in our model of network glasses. %In other glasses, where the number of constraints depends on the temperature~\cite{Gupta09,Bauchy11}, the vibrational entropy may dominate the contribution to the jump of specific heat and the dynamics is very fragile~\cite{Wyart10}. 

%% file: appendices/appd_marginal.tex
\SkipTocEntry %\tocless
\chapter{Dynamics and correlations among soft excitations in marginally stable glasses}
\appcaption{\thechapter \space Dynamics and correlations among soft excitations in marginally stable glasses} %this needs to be here to show up in the ''List of appendices'' which is required if you have more than one appendix
\label{app_marginal}

\SkipTocEntry
\section{Stability criterion}
\label{app_D1}
\begin{figure}[h!]
\centering
\includegraphics[width=.8\columnwidth]{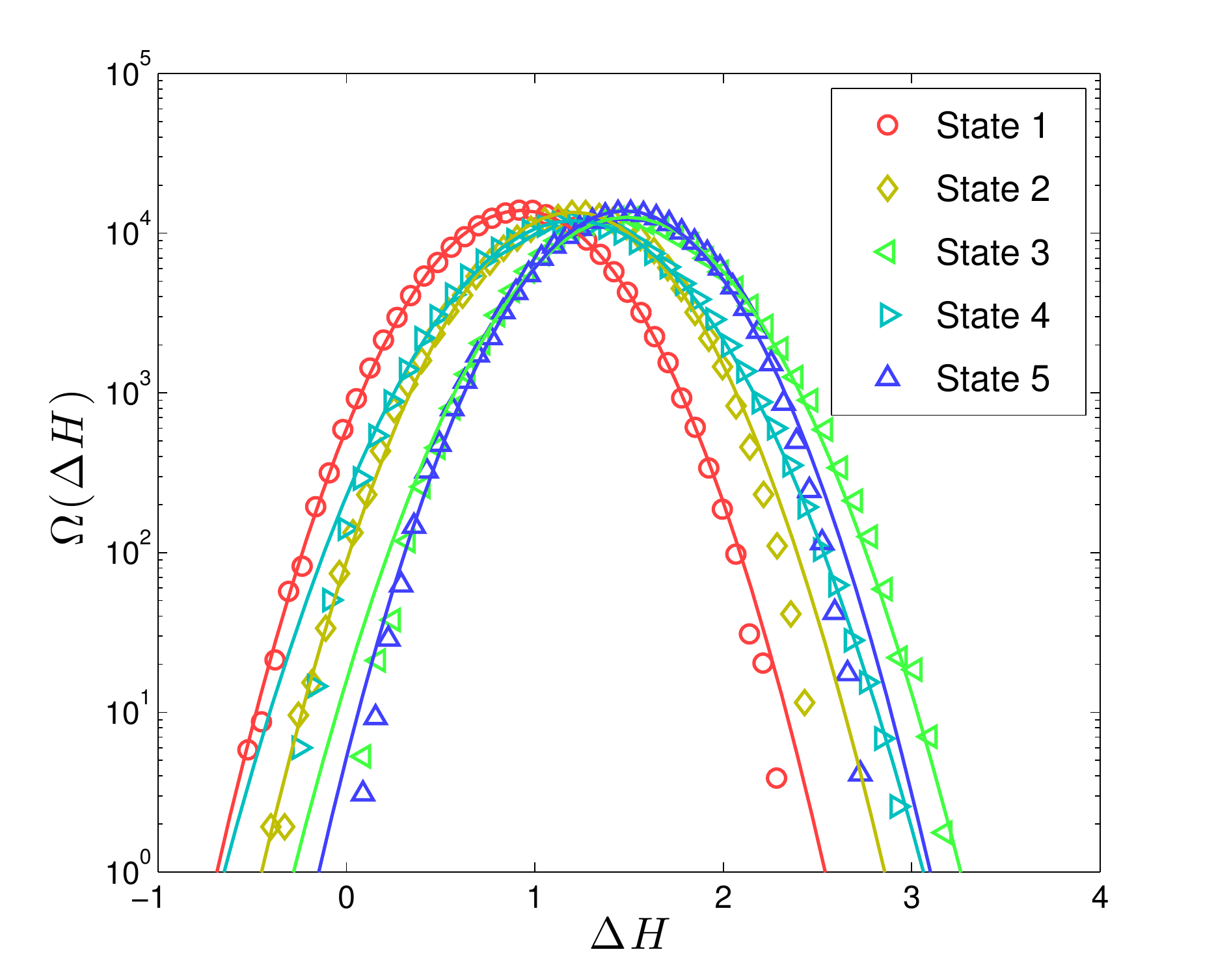}
\caption{\small{Histogram of excitations  with given energy change $\Delta H$ for different metastable states along the hysteresis curve, $m=8$, $m'=16$, and $N=3000$. }}\label{app_config}
\end{figure}

Consider flipping $m$ spins selected from the set of the $m'$ ($m'>m$) least stable spins. %This reservoir of spin combinations introduces the fluctuation of couplings that a certain set(s) of spins can have a net constructive coupling among each other though the average coupling of soft spins is frustrated. 
 %Instead of $m$ least stable spins, consider flipping $m$ spins chosen from the reservoir of $m'$ least stable spins, $m'\ll N$. 
We make the approximation that the exchange energy of such multi-flip excitations is a random Gaussian variable when $m$ is large, as supported by Fig.~\ref{app_config}. The mean and the variance of $\Delta\mh$ are: 
%In marginal stable states of SK model, the spins with smaller local energies are not necessarily less stable. The mean effect of the couplings among least stable spins is frustrating, which implies $\langle JSS\rangle<0$, so the mean energy change with flipping $m$ least stable spins are always positive, and therefore invulnerable to avalanches. We consider the marginality is achieved between the balance of the mean and fluctuation: though the mean effect makes the state stable, the fluctuation instead makes the state marginal. Specifically we consider flipping $m$ spins in a set of $m'$ spins, though the mean of combinations shows positive energy change, by chance a few combinations (of order one) is about to have negative energy change $\dH\sim-o(\langle\dH\rangle)$. 
%\be
%\dH=2\sum_{i\in m'}^m\lambda_i-2\sum_{i,j\in m'}^mJ_{ij}S_iS_j
%\ee
%We do a REM-like argument with assuming the distribution of energy change is Gaussian, so we need to find the mean and the variance of $\dH$.
\begin{subequations}
\begin{align}
\mu&\equiv\langle\dH\rangle=2m\left(\ml-m\mjs\right),
\label{app_mu}\\%gin{multline}\\
\sigma^2&\equiv\langle\dH^2\rangle-\langle\dH\rangle^2=8m^2/{N}, %\frac{\mll-\ml^2}{m}+
\label{app_sigma}
\end{align}
\end{subequations}%nd{multline}
where we have neglected the contribution from the non-diagonal terms of $\langle\sum Jss\sum Jss\rangle$. { We also omitted the fluctuations of the sum $\sum_i \lambda_i$, since that sum is anyhow always positive, and at large $m$ its fluctuations are smaller than its expectation value.} %which scale as $(m/N)^{2/(1+\theta)}/m$, and are thus subdominant for $\theta=1$ and $m\gg 1$.} %The numerical pre-factors are omitted in above expressions. %Terms with $i\neq j=k\neq l$ are assumed to vanish. %the scalings of $\ml$ in terms of $m$, $m'$ and $N$, 
The maximal stability in the set of $m'$ spins is
\[
\begin{aligned}
m'&=N\int_0^{\lm}\rho(\lambda)\rd\lambda,\\
\end{aligned}
\]
and the mean values of the local stability, $\ml$, and the correlation $\mjs$ are by definition,
\[
\begin{aligned}
\ml&\equiv\int_0^{\lm}\rho(\lambda)\lambda\rd\lambda/\int_0^{\lm}\rho(\lambda)\rd\lambda,\\
\mjs&\equiv-\frac{1}{2}\int_0^{\lm}\rd\lambda\int_0^{\lm}\rd\lambda'C(\lambda,\lambda')/\lm^2.
\end{aligned}
\]
% For the local stability with a pseudo-gap, Eq.(4), the average stability of the $m'$ spins is $\ml\sim\lm\sim(m'/N)^{1/(1+\theta)}$. %A rough scaling estimation gives $\ml\sim\lm$. %but the fluctuations of the local stabilities do not contribute to negative energy change, so we neglect $\mll-\ml^2$.
%The average correlation among the $m'$ spins from assumption, Eq.(7), is $\mjs\approx-{\rm const}/N^{\delta}\lm^{\gamma}\sim-m'^{-\gamma/(1+\theta)}N^{-\delta+\gamma/(1+\theta)}$. 
%we consider only $m\gtrsim N^{(1-\theta)/2}$ when $\theta<1$, where the pseudo-gap appears. Such a lower boundary would be $1$ for $m$ when $\theta\geq1$. %, is of the same order as $\lm$ when $\theta=1$. 
Here $C(\lambda,\lambda')$ is the correlation between the spins at the finite local stabilities $\lambda>0$ and $\lambda'>0$, defined as 
\begin{multline}
C(\lambda,\lambda')\equiv-2\langle J_{ij}s_is_j\rangle|_{\lambda_i=\lambda,\lambda_j=\lambda'}\\
\equiv \frac{1}{N^2\rho(\lambda)\rho(\lambda')}\sum_{i,j}\delta(\lambda_i-\lambda)\delta(\lambda_j-\lambda')(-2J_{ij}s_is_j).
\label{app_coupl}
\end{multline}
In the above $\rho(\lambda)=\sum_{i}\delta(\lambda_i-\lambda)/N$ is the  density of stabilities. 

$C(\lambda,\lambda')$ is a symmetric function, and continuous except for  the singular point $\lambda=\lambda'=0$. Defining the correlation $C(\lambda) \equiv C(\lambda,\lambda'=0)$ between a stable ($\lambda>0$) and a soft spin ($\lambda'=0$), we find that it 
 behaves as a power-law, described by Eq.(\ref{6_correlation}) in {\it Chapter 6}. As far as scaling is concerned we thus expect $C(\lambda,\lambda')\sim C(\max[\lambda,\lambda'])$. 
 %{\bf Put it as section B and adjust maintext}
 In Section B we find numerically that $C(\lambda,\lambda')\approx C(\sqrt{\lambda^2+\lambda'^2})$. 

Assuming the pseudo-gap distribution of Eq.(\ref{6_pseudogap}) together with Eq.(\ref{6_correlation}) of {\it Chapter 6}, one finds for $1\ll m<m'\ll N$:
\[
\begin{aligned}
\mu&=\frac{2m^{3/2}}{{N}^{1/(1+\theta)}}\lp a^{1/2}\frac{m'^{1/(1+\theta)}}{m^{1/2}}+\frac{b^{1/2}}{N^{\delta-(1+\gamma)/(1+\theta)}}\frac{m^{1/2}}{m'^{\gamma/(1+\theta)}}\rp,
%\sigma^2&={8m^2}/{N},%\lp c-\frac{d}{N^{2\delta-1-2\gamma/(1+\theta)}}\frac{m}{m'^{2\gamma/(1+\theta)}}\rp
\end{aligned}
\]
where $a$ and $b$ are numerical prefactors. { Note that the variation of $\mu$ from states to state in Fig.~\ref{app_config} is presumably a consequence of the small values of $m,m'$ used there.}
%{\bf The latter still depend on the metastable state considered, as the data of Fig.~\ref{app_config} shows.}
%We can fit the numeric factors from the simulation data later.
%As the couplings $Jss$ are independent, we expect and numerically find Gaussian distributions of energy changes, for $m'>m\gg1$, as shown in Fig.~\ref{app_config}. 
%Requesting the variance to be positive definite, we immediately get $\delta\geq1/2+\gamma/(1+\theta)$.
 
Among all excitations of $m$ out of $m'$ spins, the number of sets that lower the total energy, $\Delta\mh<0$, is
\[
\Omega(m,m')=\binom{m'}{m}\Phi\lp\frac{\mu}{\sqrt{2}\sigma}\rp\approx\frac{m'!}{m!(m'-m)!}\exp\lp-\frac{\mu^2}{2\sigma^2}\rp,
\]
where $\Phi$ is the complementary error function.

We define a free energy as the logarithm of $\Omega$, %and discuss the two cases separately,
\begin{eqnarray}
\label{app_free}
&&f(r)\equiv-\frac{1}{m}\ln\Omega\sim -r\ln r+(r-1)\ln(r-1) \\
&&+\frac{a}{4}r^{2/(1+\theta)}\lp\frac{m}{N}\rp^{\frac{1-\theta}{1+\theta}}+\frac{b}{4}r^{-2\gamma/(1+\theta)}\frac{m^{1-2\gamma/(1+\theta)}}{N^{2\delta-1-2\gamma/(1+\theta)}},\nonumber
\end{eqnarray}
where $r\equiv m'/m$. 
{The cross term ($\sim \langle\lambda\rangle_{m'} \langle Jss\rangle_{m'})$ in $\mu^2$ has been neglected, as it cannot diverge faster than the terms in the second line of Eq. (\ref{app_free}).} A positive $f(r)$ implies no unstable excitations of the initial state in the limit $m\gg 1$. Stability is thus achieved if either term in the second line of Eq.~(\ref{app_free}) diverges as $m\to \infty$. This requires that either $\theta\geq1$, or $\gamma\leq1$ and $\delta\leq1$, as explained in {\it Chapter 6}. 

\begin{figure}[h!]
\centering
\includegraphics[width=0.8\columnwidth]{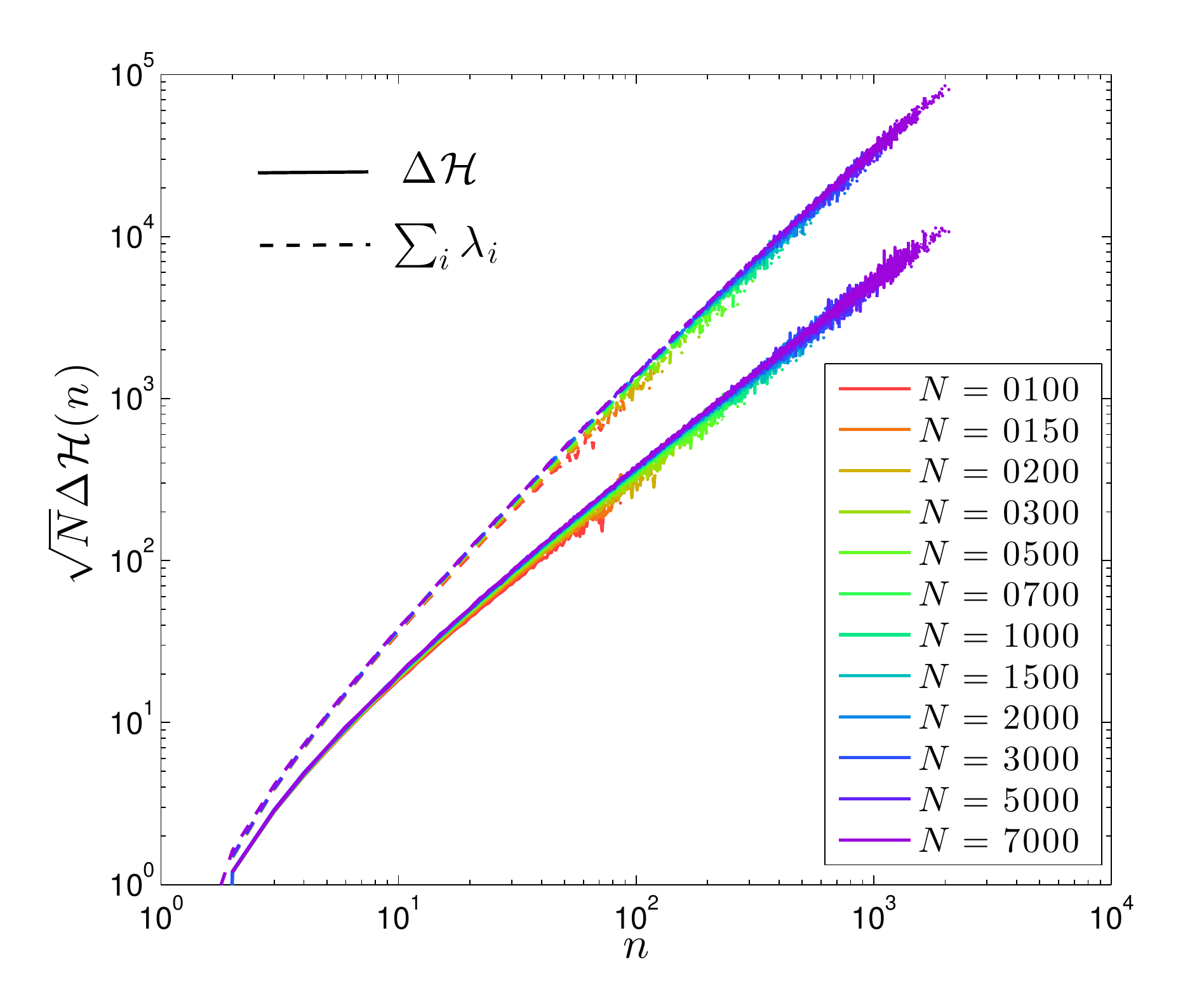}
\caption{ \small{ { The lower data set in solid lines is} the total energy  $\Delta \mh(n)$  dissipated in avalanches of size $n$. { The upper data set in dashed lines}  is the sum of local stabilities (before the avalanche) of spins that are going to flip in the avalanche, $\sum_{i\,  {\rm flip}}\lambda_i$. This shows that the dissipated energy is vanishingly small as compared to the na\"ive sum over local  stabilities, as $n\to\infty$, since the two curves scale as different power laws with $n$. %{\bf Is the reader to understand that the right panel shows that the upper data set scales as $n^{1.5}$? If so it should be stated. What is the power law of the lower set? Should we not mention it?} Right: The total stability $\sum_{i\,  {\rm flip}}\lambda_i$ versus the average rank $m'$ of flipped spins in general excitations. The solid line shows the expected slope $1.5$ in log-log scale. %{\bf is this the same data as on top on the left? If not, what is different? (left: avalanches vs. right: static excitations?}
}}\label{app_dee}
\end{figure}

In the marginal case, $\theta=\gamma=\delta=1$, %We consider the saddle point value $f(r^*)$, which gives the maximal number of unstable sets of spin flipping, as
%\be
%f'(r)=\ln\frac{r}{r-1}-\frac{a}{2c}+\frac{bc+2\sqrt{ab}d+ad^2/c}{2(cr-d)^2}
%\ee
%When $r\to1^{+}$, the first term dominates, $f'(r)\to+\infty$, when $r\to\infty$, $f'(r)\to-\frac{a}{2c}<0$. And take the second derivative,
the free energy becomes: 
\be
\label{app_freeb}
f(r)\approx -r\ln r+(r-1)\ln(r-1)+\frac{ar^2+b+2\sqrt{ab}r}{4r}.
%f''(r)=-\frac{1}{r(r-1)}-\frac{bc^2+2\sqrt{ab}dc+ad^2}{(cr-d)^3}<0
\ee
An interesting finding is that $f(r)$ is minimized by a finite ratio $r^*$, independently of the set size $m$. $r^*$  is an estimate of the optimal volume $m'^*= r^* m$ for finding energy lowering subsets of size $m$. 

The fact that stability is controlled by the ratio $r$  instead of the absolute values of $m$ or $m'$ is consistent with the observation that the dynamics proceeds via power-law avalanches with no scales (a fact implied by the argument of Ref.~\cite{Muller14}). Indeed in the marginal case multi-flip excitations can be slightly unstable (as illustrated in Fig.~\ref{app_config}), and can be triggered as the magnetic field is increased. Marginality is apparent when analyzing these avalanches: We find that the energy dissipated in avalanches is much smaller  than the na\"ive estimate which sums all local stabilities of spins that are going to flip in the avalanche, $\sum_{i\,  {\rm flip}}\lambda_i$. This reflects the fact that the total energy change in the avalanche, $\Delta\mh$, is a result of a near cancellation of several terms, as discussed in {\it Chapter 6}, verified in Fig.~\ref{app_dee}. 

%
%As the flipping set in an avalanche must lower the energy, this size independent optimal ratio is also accommodated with the scale-free avalanches excited from a marginal stable state. 
%In marginal, the number of excitations lowering the energy is about to vanish, thus the Gaussian tail does not penetrate deep into the negative region, see Fig.~\ref{app_config}. 

%{\bf I exchanged sections B and C}
\SkipTocEntry
\section{Two-point correlation}
\label{app_D2}
%We have seen the correlation emerges from the spin flipping and the diffusion of other spins. 
%The correlation among the spins at any finite local stabilities $\lambda>0$ and $\lambda'>0$ is defined as, %The idea is as follows: as the coupling $X_{ij}$ between spin $s_i$ and $s_j$ does not change as long as none of them flips, then we trace back in time till one of the spin at the boundary and extend the correlations $C(0,\lambda)$ to $C(\lambda,\lambda')$.
%%We define this correlation mathematically as,
%\begin{multline}
%C(\lambda,\lambda')\equiv-2\langle J_{ij}s_is_j\rangle|_{i=\lambda,j=\lambda'}\\
%\equiv \frac{1}{N^2\rho(\lambda)\rho(\lambda')}\sum_{i,j}\delta(\lambda_i-\lambda)\delta(\lambda_j-\lambda')(-2J_{ij}s_is_j)
%%\equiv-2\lim_{\rd\lambda\to0}\frac{1}{\nu(\lambda)\nu(\lambda')}\sum_{\lambda\leq i<\lambda+\rd\lambda}\sum_{\lambda'\leq j<\lambda'+\rd\lambda}J_{ij}s_is_j
%\label{app_coupl}
%\end{multline}
%where $\rho(\lambda)=\sum_{i}\delta(\lambda_i-\lambda)/N$ is the number density of spins at $\lambda$, and $\delta(x)$ is Dirac delta function. 

\begin{figure}[h!]
\centering
\includegraphics[width=.48\columnwidth]{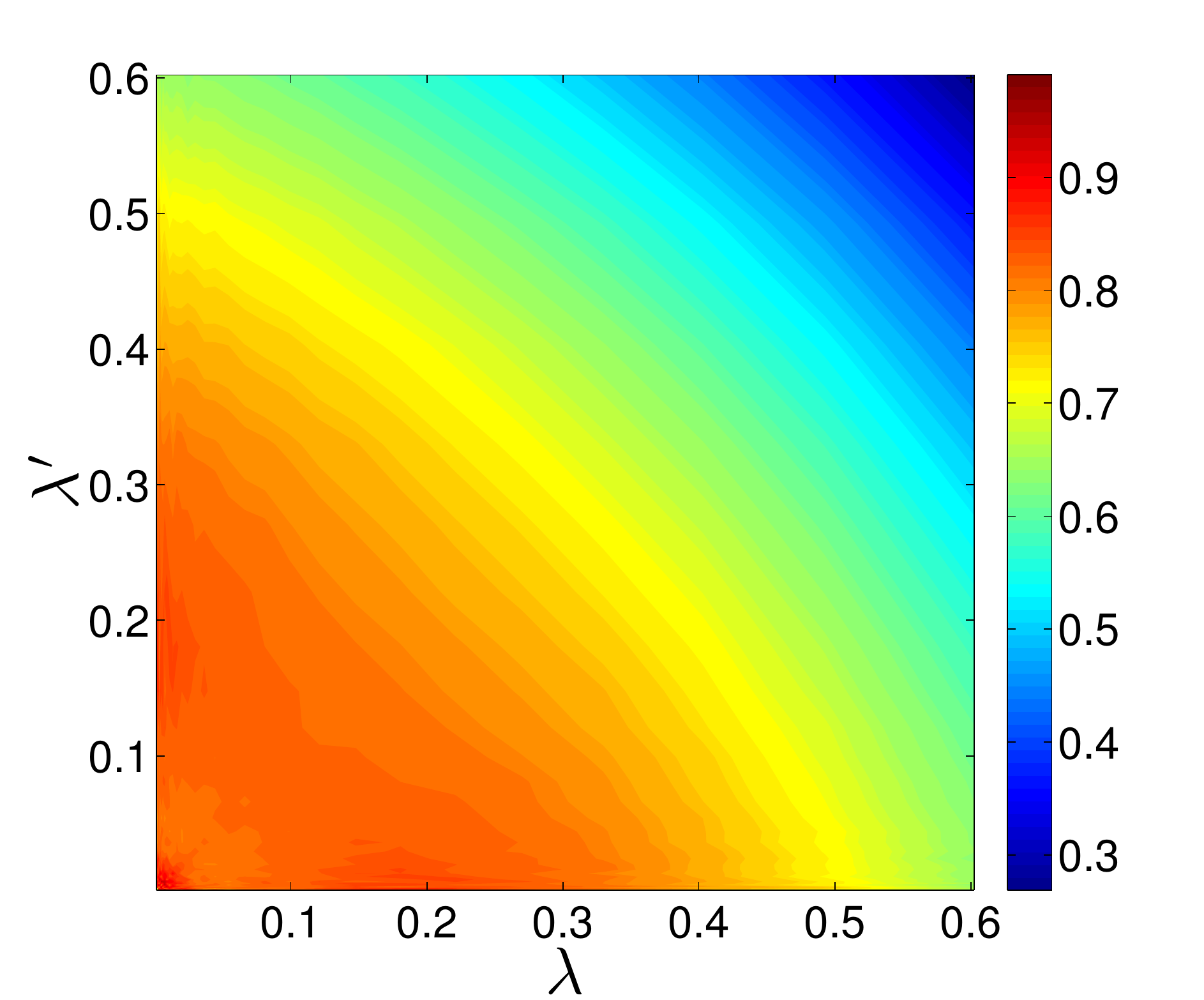}
\includegraphics[width=.48\columnwidth]{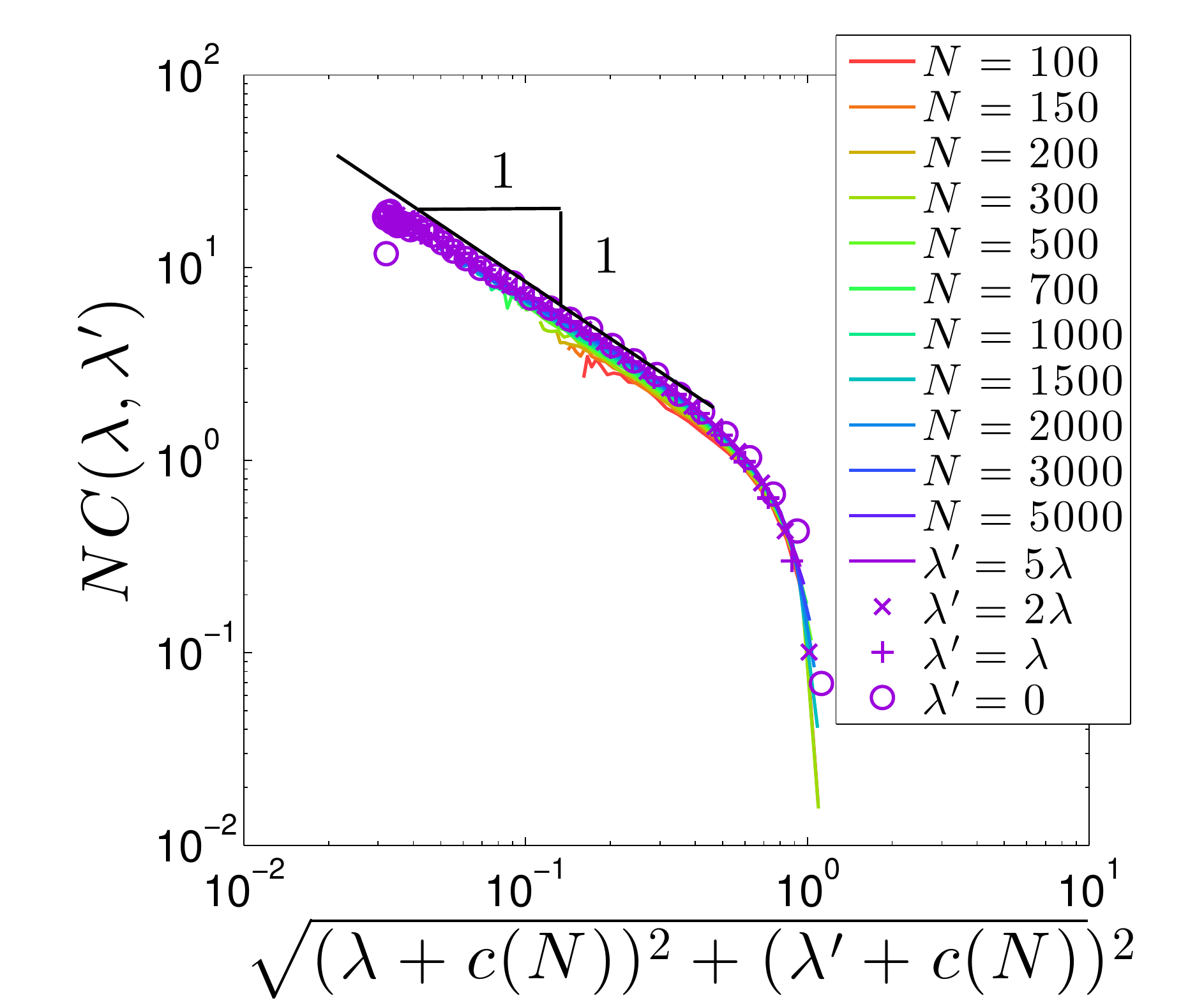}
\caption{\small{Left: the quantity $NC(\lambda,\lambda')\sqrt{(\lambda+c(N))^2+(\lambda'+c(N))^2}$ is numerically computed for various $\lambda$ and $\lambda'$, and behaves nearly as a constant (as the color code indicates, this quantity only changes by a factor 3 in the entire range considered. Right: Correlation $C(\lambda,\lambda')$, for $\lambda'=5\lambda$ with different system sizes $N$ and for different directions $\lambda'=a\lambda$, $a=0,1,2,5$ with $N=5000$.}}\label{app_correl}
\end{figure}

%This correlation can be solved in principle with considering all higher order correlations~\cite{Horner07,YanNP}.
 We numerically measured the two-point correlation, and found $NC(\lambda,\lambda')=D/\sqrt{[\lambda+c(N)]^2+[\lambda'+c(N)]^2}$ as long as $\lambda,\lambda'\ll1$. Here, $c(N)=1.1\sqrt{\ln N/N}$ is a finite size correction which vanishes in the thermodynamic limit. This result is illustrated in Fig.~\ref{app_correl}. When finite size effects are negligible this implies that $C(\lambda,\lambda')\approx  C(\sqrt{\lambda^2+\lambda'^2)}$. 
 %by changing $\lambda$ to $\sqrt{\lambda^2+\lambda'^2}$, shown in the right panel of Fig.~\ref{app_correl}.

%In principle, one can write similar diffusion equations for this bond-bond correlation, using a diagram language and then solve for the truncated equation for $K$ and the spin-spin correlation $C$ at finite $\lambda$. 

\SkipTocEntry
\section{Dynamical Constraints on $\theta$}
\label{app_D3}

In {\it Chapter 6}, we  argue that the density of local stabilities $\rho(\lambda)$ satisfies a Fokker-Planck (FP) equation of the type:
\be
\label{app_001}
\partial_t\rho(\lambda,t)=-\partial_{\lambda}\,\left[v(\lambda,t)-\partial_{\lambda}D(\lambda,t)\right]\rho(\lambda,t),
\ee
with a reflecting boundary at $\lambda=0$. In addition, we show that  correlations emerge dynamically, which in the steady state take the form:
\be
\label{app_002}
C(\lambda)N=v_{\rm ss}(\lambda)=D\partial_{\lambda}\rho_{\rm ss}(\lambda)/\rho_{\rm ss}(\lambda).
\ee
{These results seem to imply that the fact that correlations are necessary to obtain a steady state, is not constraining the latter in any way. Indeed, any function $\rho_{\rm ss}(\lambda)$ could in principle appear as a steady state, as long as the correlations satisfy Eq.~(\ref{app_002}).} This is true in particular for any scaling function $\rho_{\rm ss}(\lambda)\sim \lambda^\theta$, whatever the value of $\theta$. However, we now argue that only the case $\theta=1$ is a viable solution in SK model. 

{\it Excluding $\theta<1$:}  Our FP description only applies beyond the discretization scale of the kicks due to flipping spins, which are of order $J\sim 1/\sqrt{N}$. 
In particular, from its definition, $C(\lambda)$ must be bounded by $1/\sqrt{N}$. Taking this into account, Eq.~(\ref{app_002}) should be modified to:
\be
\label{app_003}
v_{\rm ss}(\lambda)\approx \min\{D\partial_{\lambda}\rho_{\rm ss}(\lambda)/\rho_{\rm ss}(\lambda)\sim 1/\lambda, \sqrt{N}\}.
\ee
This modification has no effect when $\theta\geq 1$, since in that case $\lambda_{\min}\sim N^{-1/(1+\theta)}\geq 1/\sqrt{N}$. 
In contrast, pseudo-gaps with $\theta<1$ have $\lambda_{\min}\ll 1/\sqrt{N}$. To maintain such a pseudo-gap in a stationary state, one would require correlations much larger than what the discreteness of the model allows. Pseudo-gaps with $\theta<1$ are thus not admissible solutions of Eqs.~(\ref{app_001},~\ref{app_003}).

%Dynamically, the pseudo-gap exponent can only be the marginal one, $\theta=1$, in the SK model. Based on the Fokker Planck equation and the stationarity, we have shown that the correlation, $C(\lambda)=2\theta/N\lambda$. First, the correlation can not be greater than the typical coupling among spins, $J_{\rm typ}\sim1/\sqrt{N}$. However, assuming a pseudo-gap profile with $\theta<1$ is achieved in stationary, there would be $N^{(1-\theta)/2}$, diverging number of spins in the range $\lambda\lesssim1/\sqrt{N}$ where the drift needs to be much stronger than $J_{\rm typ}$ to reinforce the pseudo-gap profile. The contradiction implies that if a stable state with $\theta<1$ is achieved, a small perturbation immediately drives a diverging number of spins to be unstable, and $\theta\geq1$ when approaching the steady state. 
%

{\it Excluding $\theta>1$:}
In this case, $\lambda_{\min}\gg 1/\sqrt{N}\sim J$. Thus when one spin flips, the second least stable spin will not flip in general, and avalanches are typically of size unity \cite{Muller14}. It can easily be shown that in that case, our assumption (ii) in {\it Chapter 6} is violated:  
the number of flips per spin along the loop would be small (in fact it would even vanish in the thermodynamic limit, which is clearly impossible). In terms of our FP description, the motion of the spin stabilities due to other flips would be small in comparison with the motion of the stabilities inbetween avalanches, due to changes of the magnetic field.   Making the crude assumption that the magnetization is random for any $\lambda$, the change of external magnetic field leads to an additional diffusion term in the Fokker-Planck equation:
\be
\label{app_fpwithd}
\partial_t\rho(\lambda,t)=-\partial_{\lambda}(v-D\partial_{\lambda})\rho(\lambda,t)+D_h\partial_{\lambda}^2\rho(\lambda,t),
\ee
where the term $D_h$ is related to the typical field increment $h_{\rm min}\sim \lambda_{\rm min}$ required  to trigger an avalanche. Indeed $D_{h}\sim Nh_{\rm min}^2\sim N^{(\theta-1)/(\theta+1)}\gg D\sim 1$. Under these circumstances, Eq.~(\ref{app_002}) does not hold. The dynamics would be a simple diffusion with reflecting boundary, whose only stationary solution corresponds to $\theta=0$, violating our hypothesis $\theta>1$.  
%Thus the last term of Eq.(\ref{app_fpwithd}) provides a restoring force toward 
%dominated dynamics flattens the distribution. As soon as the pseudo-gap is filled up to $\theta=1$, this diffusion contribution becomes sub-dominant and the dynamics is dominated by the transient dynamics concentrated in the main text. In stationary, a typical pseudo-gap profile must thus converge to $\theta=1$.

\SkipTocEntry
\section{Analogy to a $d$ dimensional random walk}
\label{app_D4}
Consider a non-biased random walk in $d$ dimension,
\be
\label{app_drw}
\vec{x}(t+\rd t) = \vec{x}(t) + \sqrt{{2D}\rd t}\vec{\eta}(t),
\ee
where $D$ is the diffusion constant, and $\vec{\eta}$ is a random Gaussian vector. Then the probability density $P(\vec{x},t)$ satisfies the Fokker-Planck equation~\cite{Risken96},
\[
\partial_{t}P(\vec{x},t) = D\nabla^2P(\vec{x},t).
\]

The process is angle-independent, so $P(\vec{x},t)$ satisfies, 
\be
\label{app_bfp}
\partial_tP(r,t) = D\partial_r^2P(r,t)+D\frac{d-1}{r}\partial_rP(r,t),
\ee
where $r=|\vec{x}|$.
However, $P(r,t)$ is not a probability density with respect to $r$. Including the Jacobian of the change of variables, the probability density, $\rho(r,t)\equiv\Omega_dr^{d-1}P(r,t)$, satisfies the corresponding Fokker-Planck equation, 
\be
\label{app_ffp}
\partial_t\rho(r,t) = -\partial_r\lp D\frac{d-1}{r}-D\partial_r\rp\rho(r,t),
\ee
with $\Omega_d$ the solid angle in $d$ dimensions. This Fokker-Planck equation %, Eq.~(\ref{app_ffp}), 
for the radial component of a $d$-dimensional unbiased random walk is exactly the same as the diffusion equation, Eq.(\ref{6_fp}), with reflecting boundary at $\lambda=0$ and diverging drift $v(\lambda)=\theta/\lambda$ at small $\lambda$. The case of pseudo-gap exponent $\theta=1$ in the spin model thus maps to a two-dimensional random walk.

%{\bf Please clean the references here as well.}

%% file: bibliography/bibliography.tex
\printbibliography